\documentclass[12pt]{article}
\usepackage{amsmath,amssymb,amsfonts}
\usepackage{color,graphicx,cite,soul}
\usepackage{array,booktabs,longtable}
\usepackage{caption,microtype,url}

\newcommand{\gsim}
{\;\raisebox{-.3em}{$\stackrel{\displaystyle >}{\sim}$}\;}

\newcommand\Code[1]{\ensuremath{\texttt{#1}}}
\newcommand\Var[1]{\ensuremath{\mathit{#1}}}

\newcommand\Vg{\Var{g}}

\newcommand\al{\alpha}

\newcommand\tb{\tan\beta}
\newcommand\TB{t_\beta}

\newcommand\LP{\left(}
\newcommand\RP{\right)}
\newcommand\LB{\left[}
\newcommand\RB{\right]}
\newcommand\LV{\left\{}
\newcommand\RV{\right\}}

\renewcommand\Re{\mathop{\mathrm{Re}}}

\newcommand\ReDiag{\mathop{%
  \raise .5pt\hbox{[}%
  \widetilde{\mathrm{Re}}%
  \raise .5pt\hbox{]}}}
\newcommand\ReOffDiag{\mathop{%
  \raise .5pt\hbox{$\llbracket$}%
  \widetilde{\mathrm{Re}}%
  \raise .5pt\hbox{$\rrbracket$}}}
\newcommand\SE[1]{\Sigma_{#1}}

\newcommand\DRbar{\ensuremath{\smash{\overline{\mathrm{DR}}}}}
\newcommand\MSbar{\ensuremath{\overline{\mathrm{MS}}}}
\newcommand\matr[1]{\mathbf{#1}}

\newcommand\cL{{\cal L}}

\newcommand\SW{s_\mathrm{w}}
\newcommand\CW{c_\mathrm{w}}
\newcommand\MW{M_W}
\newcommand\MZ{M_Z}

\newcommand\MHp{M_{H^\pm}}

\newcommand\Ab{A_b}
\newcommand\At{A_t}

\newcommand\Sf{{\tilde f}}

\newcommand\Se{\mathrm{\tilde e}}
\newcommand\Fe{\mathrm{e}}

\newcommand\mse[1]{m_{\Se_{#1}}}

\newcommand\Stop[1]{{\tilde t_{#1}}}

\newcommand\Sbot[1]{{\tilde b_{#1}}}

\newcommand\cpri{c^\prime}

\newcommand\npri{n^\prime}

\newcommand\dZm[1]{\delta\matr{Z}_{#1}}
\newcommand\dbZm[1]{\delta\matr{\breve Z}_{#1}}
\newcommand\dBZm[1]{\delta\matr{\breve{\bar Z}}{}_{#1}}

\newcommand\dTB{\delta\TB}
\newcommand\ino[1]{\tilde\chi_{#1}}

\newcommand\chapm[1]{\ino{#1}^\pm}
\newcommand\champ[1]{\ino{#1}^\mp}
\newcommand\chap[1]{\ino{#1}^+}
\newcommand\cham[1]{\ino{#1}^-}
\newcommand\cha{\chapm}
\newcommand\mcha[1]{m_{\chapm{#1}}}

\newcommand\neu[1]{\ino{#1}^0}
\newcommand\mneu[1]{m_{\neu{#1}}}

\newcommand\refeq[1]{Eq.~(\ref{#1})}

\newcommand\refta[1]{Tab.~\ref{#1}}
\newcommand\refse[1]{Sect.~\ref{#1}}

\newcommand\citere[1]{Ref.~\cite{#1}}
\newcommand\citeres[1]{Refs.~\cite{#1}}

\newcommand\eg{e.g.\ }
\newcommand\ie{i.e.\ }
\newcommand\wrt{w.r.t.\ }

\newcommand{\CP}{{\cal CP}}
\newcommand{\cp}{{\CP}}
\newcommand{\os}{\mathrm{os}}

\newcommand{\onel}{one-loop}
\newcommand{\tev}{\,\, \mathrm{TeV}}
\newcommand{\gev}{\,\, \mathrm{GeV}}
\newcommand{\mev}{\,\, \mathrm{MeV}}

\newcommand{\He}{h_1}
\newcommand{\Hz}{h_2}
\newcommand{\Hd}{h_3}

\newcommand{\eecc}{\ensuremath{e^+e^- \to \chapm{c} \champ{\cpri}}}
\newcommand{\eecece}{e^+e^- \to \chap1 \cham1}
\newcommand{\eececz}{e^+e^- \to \chapm1 \champ2}
\newcommand{\eeczcz}{e^+e^- \to \chap2 \cham2}
\newcommand{\eecmcp}{e^+e^- \to \cham{c} \chap{\cpri}}
\newcommand{\eecpcm}{e^+e^- \to \chap{c} \cham{\cpri}}
\newcommand{\eecepczm}{e^+e^- \to \chap1 \cham2}
\newcommand{\eecemczp}{e^+e^- \to \cham1 \chap2}

\newcommand{\eenn}{\ensuremath{e^+e^- \to \neu{n} \neu{\npri}}}
\newcommand{\eenene}{e^+e^- \to \neu1 \neu1}
\newcommand{\eenznz}{e^+e^- \to \neu2 \neu2}
\newcommand{\eendnd}{e^+e^- \to \neu3 \neu3}
\newcommand{\eenvnv}{e^+e^- \to \neu4 \neu4}
\newcommand{\eenenz}{e^+e^- \to \neu1 \neu2}
\newcommand{\eenend}{e^+e^- \to \neu1 \neu3}
\newcommand{\eenenv}{e^+e^- \to \neu1 \neu4}
\newcommand{\eenznd}{e^+e^- \to \neu2 \neu3}
\newcommand{\eenznv}{e^+e^- \to \neu2 \neu4}
\newcommand{\eendnv}{e^+e^- \to \neu3 \neu4}

\newcommand\FA{\texttt{FeynArts}}
\newcommand\FC{\texttt{FormCalc}}
\newcommand\LT{\texttt{LoopTools}}

\newcommand\FT{\texttt{FeynTools}}

\newcommand\fb{\ensuremath{\mbox{fb}}}
\newcommand\ab{\ensuremath{\mbox{ab}}}

\newcommand\iab{\ensuremath{\ab^{-1}}}

\newcommand{\Scs}{$\mathcal S$}

\newcommand{\sig}{\sigma}
\newcommand{\sigfull}{\sigma_{\text{full}}}
\newcommand{\sigtree}{\sigma_{\text{tree}}}
\newcommand{\sigloop}{\sigma_{\text{loop}}}

\newcommand{\sigvirt}{\sigma_{\text{virt}}}
\newcommand{\sigsoft}{\sigma_{\text{soft}}}
\newcommand{\sighard}{\sigma_{\text{hard}}}
\newcommand{\sigcoll}{\sigma_{\text{coll}}}

\def\order#1{\ensuremath{{\cal O}(#1)}}
\def\reffi#1{\mbox{Fig.~\ref{#1}}}
\def\reffis#1{\mbox{Figs.~\ref{#1}}}

\def\ga{\gamma}
\def\de{\delta}
\def\la{\lambda}

\def\phiAt{\varphi_{\At}}

\def\phimu{\varphi_{\mu}}
\def\phiMe{\varphi_{M_1}}
\def\phiMz{\varphi_{M_2}}

\def\MSUSY{M_{\text{SUSY}}}
\def\MSL{M_{\tilde L}}
\def\MSE{M_{\tilde E}}

\definecolor{Orange}{named}{Orange}
\definecolor{Purple}{named}{Purple}
\definecolor{Lightblue}{cmyk}{0.9,0.1,0.1,0.3}
\definecolor{dgelborange}{cmyk}{0.,0.3,0.5, 0.}
\definecolor{Lila}{rgb}{0.5,0.,1}

\graphicspath{{figs/}}

\evensidemargin \oddsidemargin
\allowdisplaybreaks
\begin{document}
\thispagestyle{empty}

\def\thefootnote{\fnsymbol{footnote}}

\begin{flushright}
\mbox{}
IFT--UAM/CSIC--17-013 
\end{flushright}

\vspace{0.5cm}

\begin{center}

{\large\sc 
{\bf Chargino and Neutralino Production at \boldmath{$e^+e^-$} Colliders}} 

\vspace{0.4cm}

{\large\sc {\bf in the Complex MSSM: A Full One-Loop Analysis}}

\vspace{1cm}

{\sc
S.~Heinemeyer$^{1,2,3}$%
\footnote{email: Sven.Heinemeyer@cern.ch}%
~and C.~Schappacher$^{4}$%
\footnote{email: schappacher@kabelbw.de}%
}

\vspace*{.7cm}

{\sl
$^1$Campus of International Excellence UAM+CSIC, 
Cantoblanco, 28049, Madrid, Spain 

\vspace*{0.1cm}

$^2$Instituto de F\'isica Te\'orica (UAM/CSIC), 
Universidad Aut\'onoma de Madrid, \\ 
Cantoblanco, 28049, Madrid, Spain

\vspace*{0.1cm}

$^3$Instituto de F\'isica de Cantabria (CSIC-UC), 
39005, Santander, Spain

\vspace*{0.1cm}

$^4$Institut f\"ur Theoretische Physik,
Karlsruhe Institute of Technology, \\
76128, Karlsruhe, Germany (former address)
}

\end{center}

\vspace*{0.1cm}

\begin{abstract}
\noindent
For the search for charginos and neutralinos in the Minimal Supersymmetric 
Standard Model (MSSM) as well as for future precision analyses of these 
particles an accurate knowledge of their production and decay properties
is mandatory. 
We evaluate the cross sections for the chargino and neutralino production 
at $e^+e^-$ colliders in the MSSM with complex parameters (cMSSM). 
The evaluation is based on a full one-loop calculation of the production 
mechanisms \eecc\ and \eenn\, including soft and hard photon radiation.  
We mostly restricted ourselves to a version of our renormalization scheme 
which is valid for $|M_1| < |M_2|, |\mu|$ and $M_2 \neq \mu$ to simplify
the analysis, even though we are able to switch to other parameter
regions and correspondingly different renormalization schemes.
The dependence of the chargino/neutralino cross sections on the relevant 
cMSSM parameters is analyzed numerically.  We find sizable contributions 
to many production cross sections.  They amount roughly 10-20\% of the 
tree-level results, but can go up to 40\% or higher in extreme cases. 
Also the complex phase dependence of the one-loop corrections was found
non-negligible.  The full one-loop contributions are thus crucial for 
physics analyses at a future linear $e^+e^-$ collider such as the ILC or 
CLIC.
\end{abstract}


\def\thefootnote{\arabic{footnote}}
\setcounter{page}{0}
\setcounter{footnote}{0}

\newpage


\section{Introduction}
\label{sec:intro}

One of the important tasks at the LHC is to search for physics beyond the 
Standard Model (SM), where the Minimal Supersymmetric Standard Model 
(MSSM)~\cite{mssm,HaK85,GuH86} is one of the leading candidates.
Two related important tasks are the investigation of the mechanism of
electroweak symmetry breaking, including the identification of the underlying 
physics of the Higgs boson discovered at 
$\sim 125\gev$~\cite{ATLASdiscovery,CMSdiscovery}, as well as the production 
and measurement of the properties of Cold Dark Matter (CDM).  Here the MSSM 
offers a natural candidate for CDM, the Lightest Supersymmetric Particle 
(LSP), the lightest neutralino,~$\neu{1}$~\cite{EHNOS} (see below).
These three (related) tasks will be the top priority in the future program of 
particle physics.  

Supersymmetry (SUSY) predicts two scalar partners for all SM fermions as well
as fermionic partners to all SM bosons. 
Contrary to the case of the SM, in the MSSM two Higgs doublets are required.
This results in five physical Higgs bosons instead of the single Higgs
boson in the SM. These are the light and heavy $\cp$-even Higgs bosons, $h$
and $H$, the $\cp$-odd Higgs boson, $A$, and the charged Higgs bosons,
$H^\pm$.
In the MSSM with complex parameters (cMSSM) the three neutral Higgs
bosons mix~\cite{mhiggsCPXgen,Demir,mhiggsCPXRG1,mhiggsCPXFD1}, 
giving rise to the $\cp$-mixed states $\He, \Hz, \Hd$.
The neutral SUSY partners of the (neutral) Higgs and electroweak gauge
bosons are the four neutralinos, $\neu{1,2,3,4}$. The corresponding
charged SUSY partners are the charginos, $\cha{1,2}$.

If SUSY is realized in nature and the scalar quarks and/or the gluino
are in the kinematic reach of the LHC, it is expected that these
strongly interacting particles are copiously produced. 
On the other hand, SUSY particles that interact only via the electroweak
force, \eg the charginos and neutralinos, have a much smaller
production cross section at the LHC. Correspondingly, the LHC
discovery potential as well as the current experimental bounds are
substantially weaker. 

At a (future) $e^+e^-$ collider charginos and neutralinos, depending on
their masses and the available center-of-mass energy, could be produced and
analyzed in detail. Corresponding studies can be found for the ILC in
\citeres{ILC-TDR,teslatdr,ilc,LCreport} and for CLIC in
\citeres{CLIC,LCreport}. 
(Results on the combination of LHC and ILC results can be found in 
\citere{lhcilc}.) Such precision studies will be crucial to determine
their nature and the underlying (SUSY) parameters.

In order to yield a sufficient accuracy, one-loop corrections to the 
various chargino/neutralino production and decay modes have to be considered.
Full one-loop calculations in the cMSSM for various chargino/neutralino 
decays in the cMSSM have been presented over the last 
years~\cite{LHCxC,LHCxN,LHCxNprod}.
One-loop corrections for their production from the decay of Higgs bosons
(at the LHC or ILC/CLIC) can be found in \citere{HiggsDecayIno}. 
In this paper we take the next step and concentrate on the chargino/neutralino
production at $e^+e^-$ colliders, \ie we calculate,
\begin{align}
\label{eq:eecc}
&\sig(\eecc) \qquad (c,\cpri = 1,2)\,, \\
\label{eq:eenn}
&\sig(\eenn) \qquad (n,\npri = 1,2,3,4)\,.
\end{align}
Our evaluation of the two channels (\ref{eq:eecc}) and (\ref{eq:eenn}) 
is based on a full one-loop calculation, \ie including electroweak (EW) 
corrections, as well as soft and hard QED radiation. 
The renormalization scheme employed is the same one as for the decay 
of charginos/neutralinos~\cite{LHCxC,LHCxN,LHCxNprod}.
Consequently, the predictions for the production and decay can be 
used together in a consistent manner.

Results for the cross sections (\ref{eq:eecc}) and (\ref{eq:eenn}) at 
various levels of sophistication have been obtained over the last three
decades. 
Tree-level results were published for \eecc\ and \eenn\ in the MSSM
with real parameters (rMSSM) in \citeres{Bartl:1985fk,Bartl:1986hp}.
Tree-level results for the cMSSM for \eenn\ (using a ``projector
formalism'') were presented in \citere{Gounaris:2002pj}.  Results for
$\CP$-odd observables with \eecc\ ($c \neq c^{\prime}$) were shown in 
\citere{Osland:2007xw} (including ``selected box contributions'') and 
extended to the full contributions in \citeres{RoKa2007,OsKaRoVe2007}.
Vertex corrections to \eecc\ in the rMSSM including the contributions of
$t/\Stop{}/b/\Sbot{}$ were evaluated in \citere{Diaz:1997kv}, using an
\MSbar\ renormalization scheme. The results including all quark/squark
contributions were shown in \citere{Kiyoura:1998yt} (claiming
differences to \citere{Diaz:1997kv}).
Full one-loop corrections in the rMSSM for \eecc\ were first presented
in \citere{Blank:2000uc} and later in \citere{Diaz:2002rr}. The
inclusion of multi-photon emission and the implementation into an event
generator was presented in \citeres{Kilian:2006cj,Robens:2006np}.
\eenn\ and \eecc\ were calculated at the full one-loop level in the
rMSSM in \citere{Blank:2000fqa}, and later also in 
\citere{Diaz:2001vm} (but without including a numerical analysis).
Full one-loop results for \eenn\ in the rMSSM were shown in
\citere{Oller:2004br}, where the soft SUSY-breaking parameter $M_2$ and
the Higgs mixing parameter $\mu$ were renormalized on-shell (and only
results for $\eenenz$ and $\eenznz$ were analyzed numerically).
The latter results were extended to \eenn\ and \eecc\ in the cMSSM
in \citere{Oller:2005xg}, but only real parameters have been considered.
Subsequently, full one-loop results in the cMSSM for \eenn\ and \eecc\ 
were obtained in \citeres{FrHo2004,Fritzsche:2004ek,dissTF,Diaz:2009um}, 
but only real parameters were included in the phenomenological analysis.
Finally, in \citere{Bharucha:2012nx} the effects of imaginary and 
absorptive parts have been analyzed for \eecc, and for a precise cMSSM
parameter extraction from experiment, full one-loop corrections to \eenn\ 
and \eecc\ were presented (for three benchmark points) in
\citere{Bharucha:2012ya}. The differences in our renormalization in the
chargino/neutralino sector from the previous two papers are discussed 
in our \citere{LHCxN}.

\medskip

In this paper we present for the first time a full and consistent 
one-loop calculation in the cMSSM for chargino and neutralino production 
at $e^+e^-$ colliders.  We take into account soft and hard QED radiation 
and the treatment of collinear divergences.
Again, here it is crucial to stress that the same renormalization scheme 
as for the decay of charginos/neutralinos~\cite{LHCxC,LHCxN,LHCxNprod} 
(and for the production of charginos/neutralinos from Higgs-boson 
decays~\cite{HiggsDecayIno}) has been used. 
Consequently, the predictions for the production and decay can be used 
together in a consistent manner (\eg in a global phenomenological 
analysis of the chargino/neutralino sector at the one-loop level. 
We analyze all processes \wrt the most relevant parameters, including the 
relevant complex phases.  In this way we go substantially beyond the 
existing analyses (see above).
In \refse{sec:calc} we very briefly review the renormalization of the
relevant sectors of the cMSSM and give details as regards the calculation.
In \refse{sec:comparisons} various comparisons with results from other
groups are given.  The numerical results for the production channels 
(\ref{eq:eecc}) and (\ref{eq:eenn}) are presented in \refse{sec:numeval}.
The conclusions can be found in \refse{sec:conclusions}.  
\vfill


\subsection*{Prolegomena}

We use the following short-hands in this paper:
\begin{itemize}

\item \FT\ $\equiv$ \FA\ + \FC\ + \LT.

\item full = tree + loop.

\item $\SW \equiv \sin\theta_W$, $\CW \equiv \cos\theta_W$.

\item $\TB \equiv \tb$.

\end{itemize}
They will be further explained in the text below.


\section{Calculation of diagrams}
\label{sec:calc}

In this section we give some details regarding the renormalization
procedure and the calculation of the tree-level and higher-order 
corrections to the production of charginos and neutralinos in $e^+e^-$ 
collisions.  The diagrams and corresponding amplitudes have been obtained 
with \FA\ (version 3.9) \cite{feynarts}, using the MSSM model file 
(including the MSSM counterterms) of \citere{MSSMCT}. 
The further evaluation has been performed with \FC\ (version 9.5) 
and \LT\ (version 2.13) \cite{formcalc}.


\subsection{The complex MSSM}
\label{sec:renorm}

The cross sections (\ref{eq:eecc}) and (\ref{eq:eenn}) are calculated 
at the one-loop level, including soft and hard QED radiation; see the 
next section.  This requires the simultaneous renormalization of the 
gauge-boson sector, the fermion/sfermion sector as well as the 
chargino/neutralino sector of the cMSSM.  
We give a few relevant details as regards these sectors and their 
renormalization.  More details and the application to Higgs boson
and SUSY particle decays can be found in 
\citeres{HiggsDecaySferm,HiggsDecayIno,MSSMCT,SbotRen,Stop2decay,%
Gluinodecay,Stau2decay,LHCxC,LHCxN,LHCxNprod}.
Similarly, the application to Higgs-boson production cross sections at
$e^+e^-$  colliders are given in \citeres{HiggsProd,HpProd}.

The renormalization of the fermion/sfermion and gauge-boson sectors 
follows strictly \citere{MSSMCT} and references therein 
(see especially \citere{mhcMSSMlong}). 
This defines in particular the counterterm $\dTB$, as well as the 
counterterms for the $Z$~boson mass, $\de\MZ^2$, and for the sine 
of the weak mixing angle, $\de\SW$ 
(with $\SW = \sqrt{1 - \CW^2} = \sqrt{1 - \MW^2/\MZ^2}$, where $\MW$ 
and $\MZ$ denote the $W$~and $Z$~boson masses, respectively).

For the fermion sector we use the default values as given in 
\citere{MSSMCT}.  In the slepton sector we use the on-shell (OS)
scheme \Code{OS[1]}, \ie in the notation of \cite{MSSMCT}%
\footnote{
  Accidentally, for our parameter set (see \refta{tab:para}) 
  the renormalization scheme \Code{OS[2]} leads to unacceptable 
  large loop corrections.
}%
:
\begin{align}
\Code{\$SfScheme[2,\,\Vg]}~ &\Code{= OS[1]} & \qquad &
\text{on-shell scheme with $\mse{1g}$ OS, $Y_{\Fe_g}$ OS}\,. \notag
\end{align}

The chargino/neutralino sector is also described in detail in 
\citere{MSSMCT} and references therein; see in particular
\citeres{Stop2decay,LHCxC,LHCxN,LHCxNprod}.
In this paper we use mostly the \Code{CCN[1]} scheme (\ie OS 
conditions for the two charginos and the lightest neutralino), as 
implemented in the \FA\ model file \Code{MSSMCT.mod} \cite{MSSMCT}.
Also some \Code{CNN[$c,n,n'$]} schemes (OS conditions for one chargino
and two neutralinos, as implemented in \Code{MSSMCT.mod}) have been 
used for a few comparative calculations, as will be detailed below.
Either scheme fixes three out of six chargino/neutralino masses to be 
on-shell.  The other three masses then acquire a finite shift. 
The one-loop masses of the remaining charginos/neutralinos are 
obtained from the tree-level ones via the shifts \cite{dissAF}:
\begin{alignat}{2}
\label{eq:Deltamcha}
\Delta \mcha{c} 
&= -\Re \bigg\{\mcha{c} \LP \SE{\cha{c}}^L(\mcha{c}^2) 
         + \frac{1}{2} \LB 
           \dZm{\cha{}}^L + \dBZm{\cha{}}^L 
                       \RB_{cc} \RP \notag \\
&\hspace{1.6cm} + \SE{\cha{c}}^{SL}(\mcha{c}^2) 
    - \frac{1}{2} \mcha{c} \LB 
      \dZm{\cha{}}^R + \dBZm{\cha{}}^R \RB_{cc}
    - \LB \delta \matr{M}_{\cha{}} \RB_{cc} \bigg\}\,, \\
\Delta \mneu{n} 
&= -\Re \bigg\{\mneu{n} \LP \SE{\neu{n}}^L(\mneu{n}^2) 
         + \frac{1}{2} \LB 
           \dZm{\neu{}}^R + \dbZm{\neu{}}^R 
                       \RB_{nn} \RP \notag \\
&\hspace{1.6cm} + \SE{\neu{n}}^{SL}(\mneu{n}^2) 
    - \frac{1}{2} \mneu{n} \LB 
      \dZm{\neu{}}^L + \dbZm{\neu{}}^L \RB_{nn}
    - \LB \delta \matr{M}_{\neu{}} \RB_{nn} \bigg\}
\label{eq:Deltamneu}
\end{alignat}
with $c = 1,2;\, n = 1,2,3,4$, where the renormalization constants 
$\dZm{}$, $\dbZm{}$, $\dBZm{}$ and $\delta \matr{M}$ can be found 
in section 3.4 of \citere{MSSMCT}.
For all externally appearing chargino/neutralino masses the (shifted) 
``on-shell'' masses are used:
\begin{align}
\mcha{c}^\os = \mcha{c} + \Delta \mcha{c}\,, \qquad
\mneu{n}^\os = \mneu{n} + \Delta \mneu{n}\,.
\label{eq:minoOS}
\end{align}
In order to yield UV-finite results the tree-level values $\mcha{c}$
and/or $\mneu{n}$ for all internally appearing chargino/neutralino masses 
in loop calculations are used.  Renormalizing the two charged states OS
(as done in CCN schemes), \ie ensuring that they have the same mass at the 
tree- and at the loop level is (in general) crucial for the cancellation 
of the IR divergencies. On the other hand, CNN schemes are IR divergent
if an externally appearing chargino is not chosen OS.

The \Code{CCN[1]} scheme defines in particular the counterterm $\de\mu$, where 
$\mu$ denotes the Higgs mixing parameter.  This scheme yields numerically 
stable results for $|M_1| < |M_2|, |\mu|$ and $M_2 \neq \mu$, \ie the 
lightest neutralino is bino-like and defines the counterterm for
$M_1$~\cite{LHCxC,LHCxN,LHCxNprod,onshellCNmasses}.  In the numerical 
analysis this mass pattern holds. Switching to a different mass pattern,
\eg with $|M_2| < |M_1|$ and/or $M_2 \sim \mu$ requires one to switch 
to a different renormalization scheme~\cite{MSSMCT,onshellCNmasses}. 
While these schemes are implemented into the \FA/\FC\ framework~\cite{MSSMCT},
so far no automated choice of the renormalization scheme has been devised. 
For simplicity we stick (mostly) to the \Code{CCN[1]} scheme with a 
matching choice of SUSY parameters; see \refse{sec:paraset}.


\subsection{Contributing diagrams}
\label{sec:diagrams}

\begin{figure}
\begin{center}
\framebox[14cm]{\includegraphics[width=0.21\textwidth]{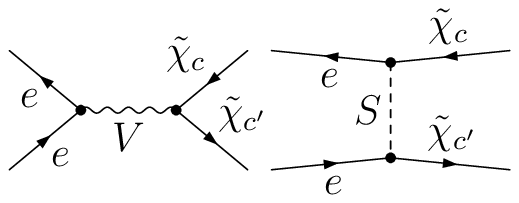}}
\framebox[14cm]{\includegraphics[width=0.67\textwidth]{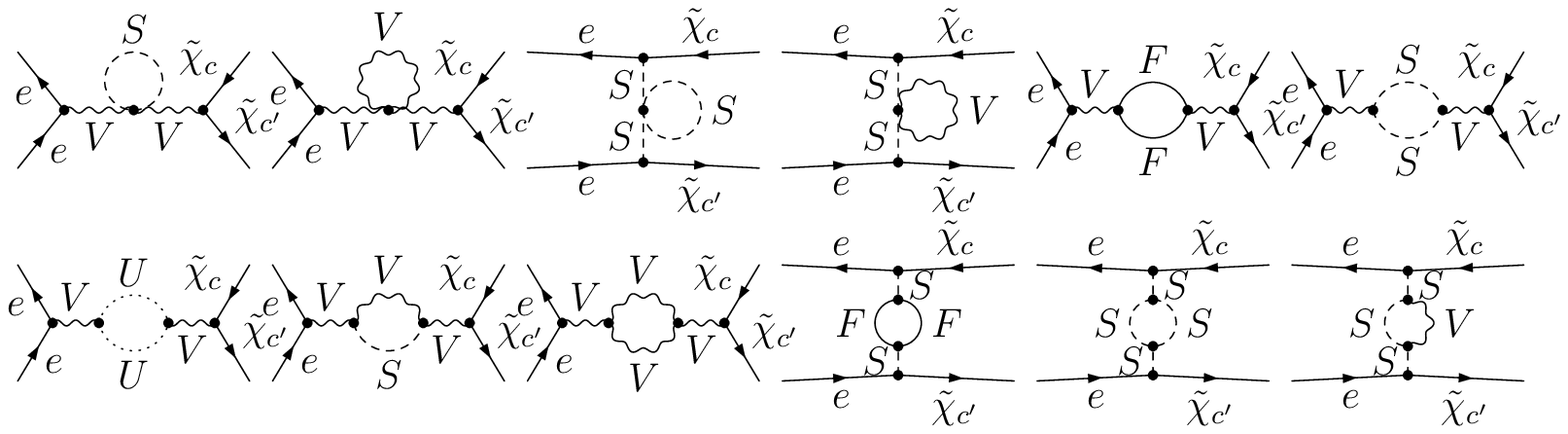}}
\framebox[14cm]{\includegraphics[width=0.67\textwidth]{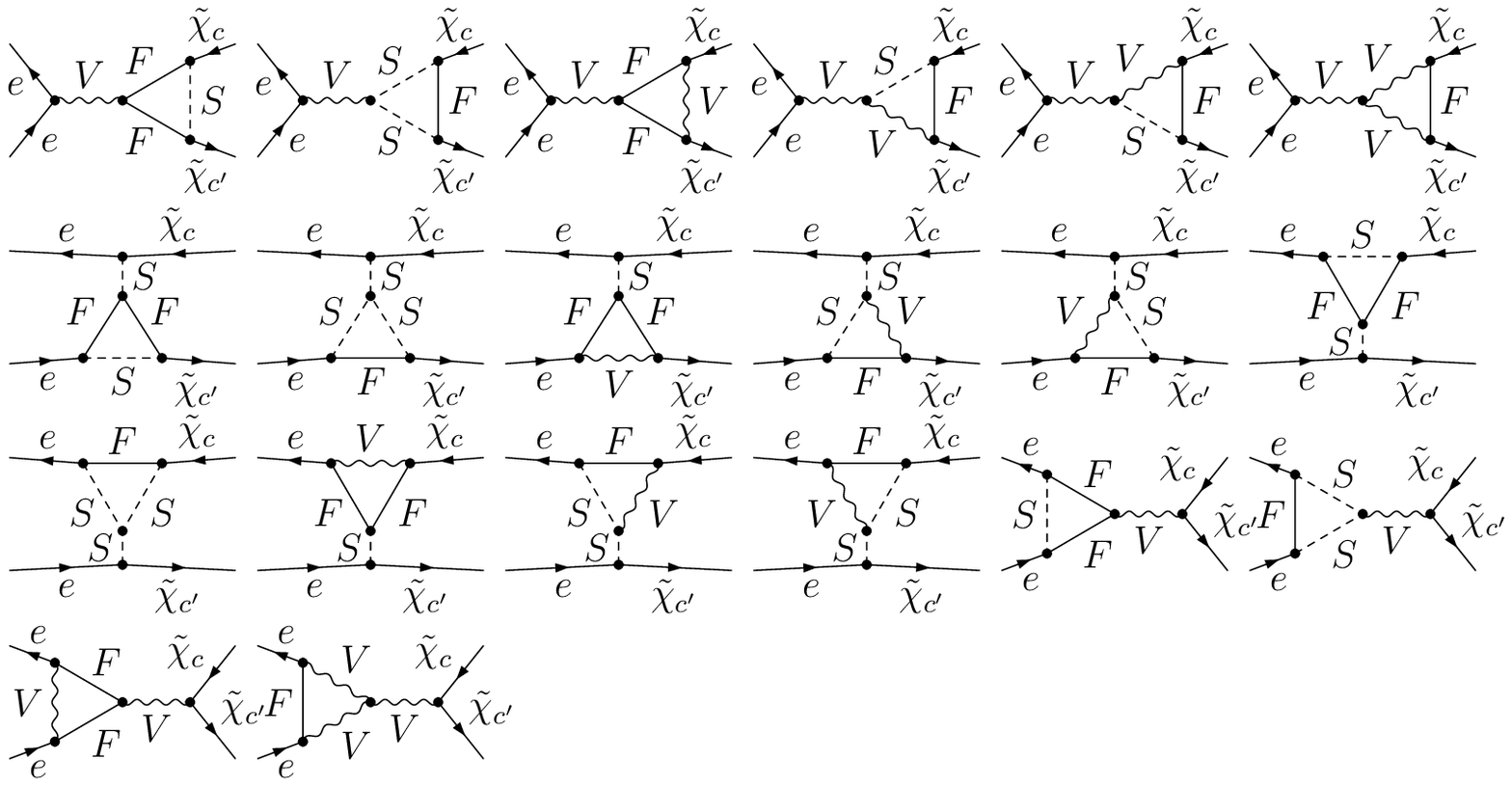}}
\framebox[14cm]{\includegraphics[width=0.67\textwidth]{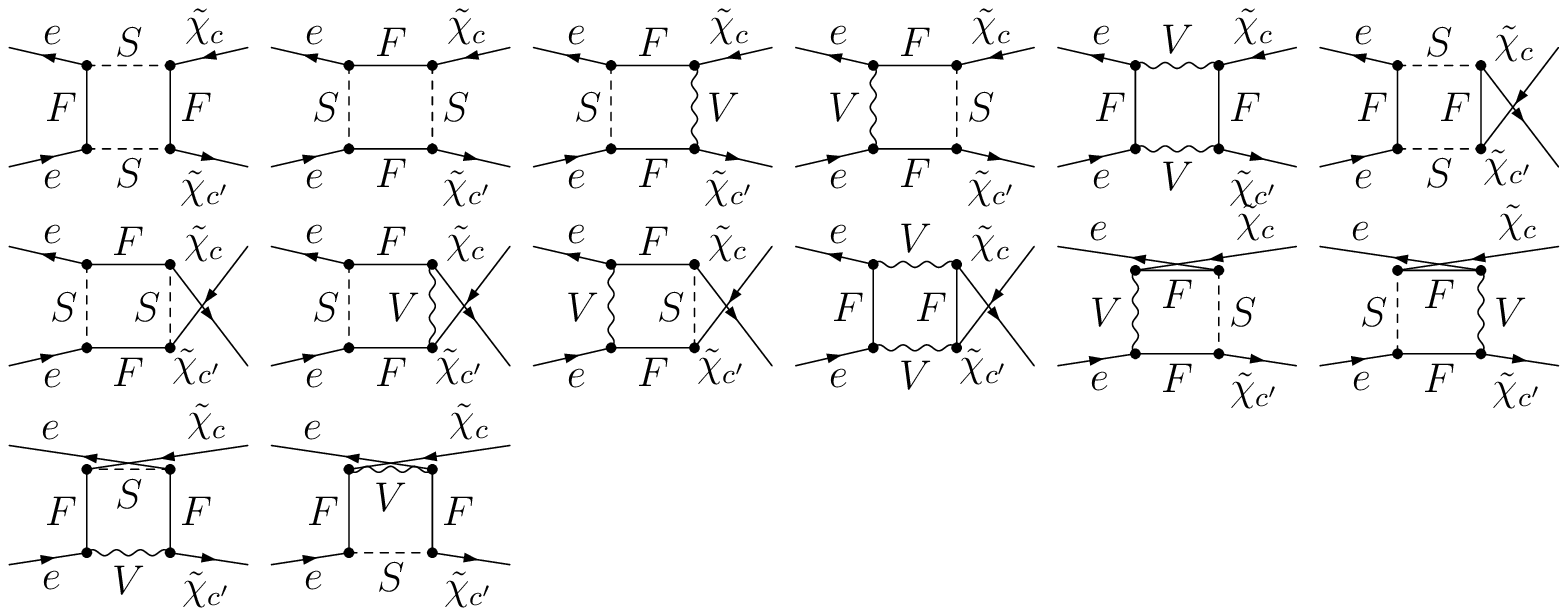}}
\framebox[14cm]{\includegraphics[width=0.67\textwidth]{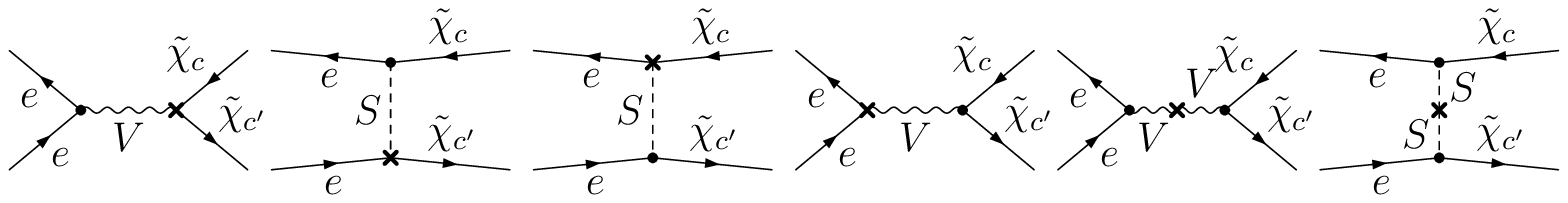}}
\caption{
  Generic tree, self-energy, vertex, box, and counterterm diagrams 
  for the process \eecc\ ($c,\cpri = 1,2$). 
  $F$ can be a SM fermion, chargino or neutralino; 
  $S$ can be a sfermion or a Higgs/Goldstone boson; 
  $V$ can be a $\ga$, $Z$ or $W^\pm$. 
  It should be noted that electron--Higgs couplings are neglected.  
}
\label{fig:CCiagrams}
\end{center}
\end{figure}

\begin{figure}
\begin{center}
\framebox[15cm]{\includegraphics[width=0.32\textwidth]{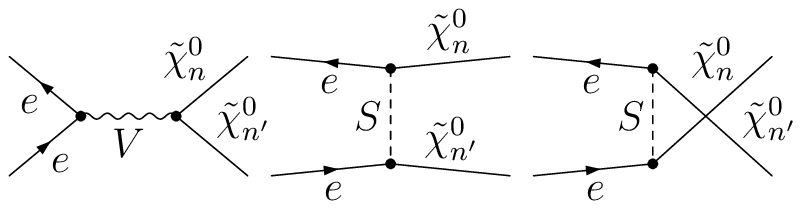}}
\framebox[15cm]{\includegraphics[width=0.67\textwidth]{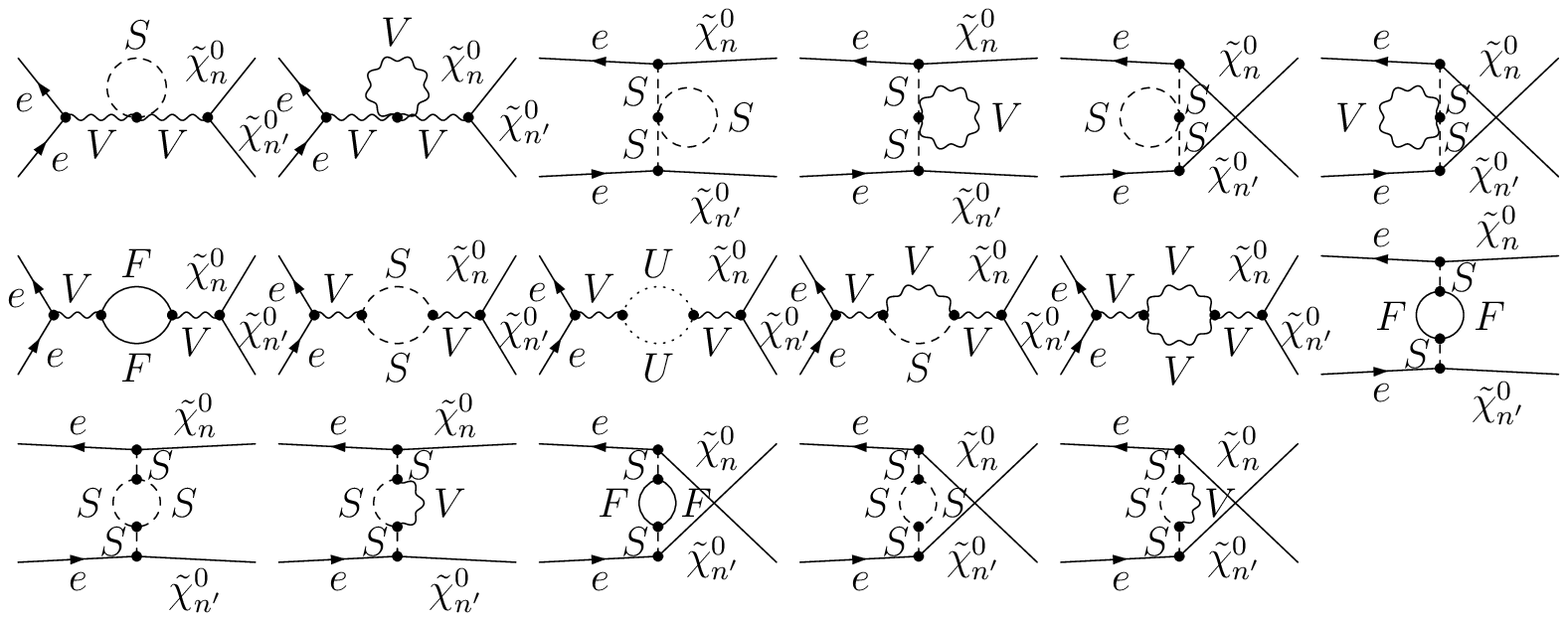}}
\framebox[15cm]{\includegraphics[width=0.67\textwidth]{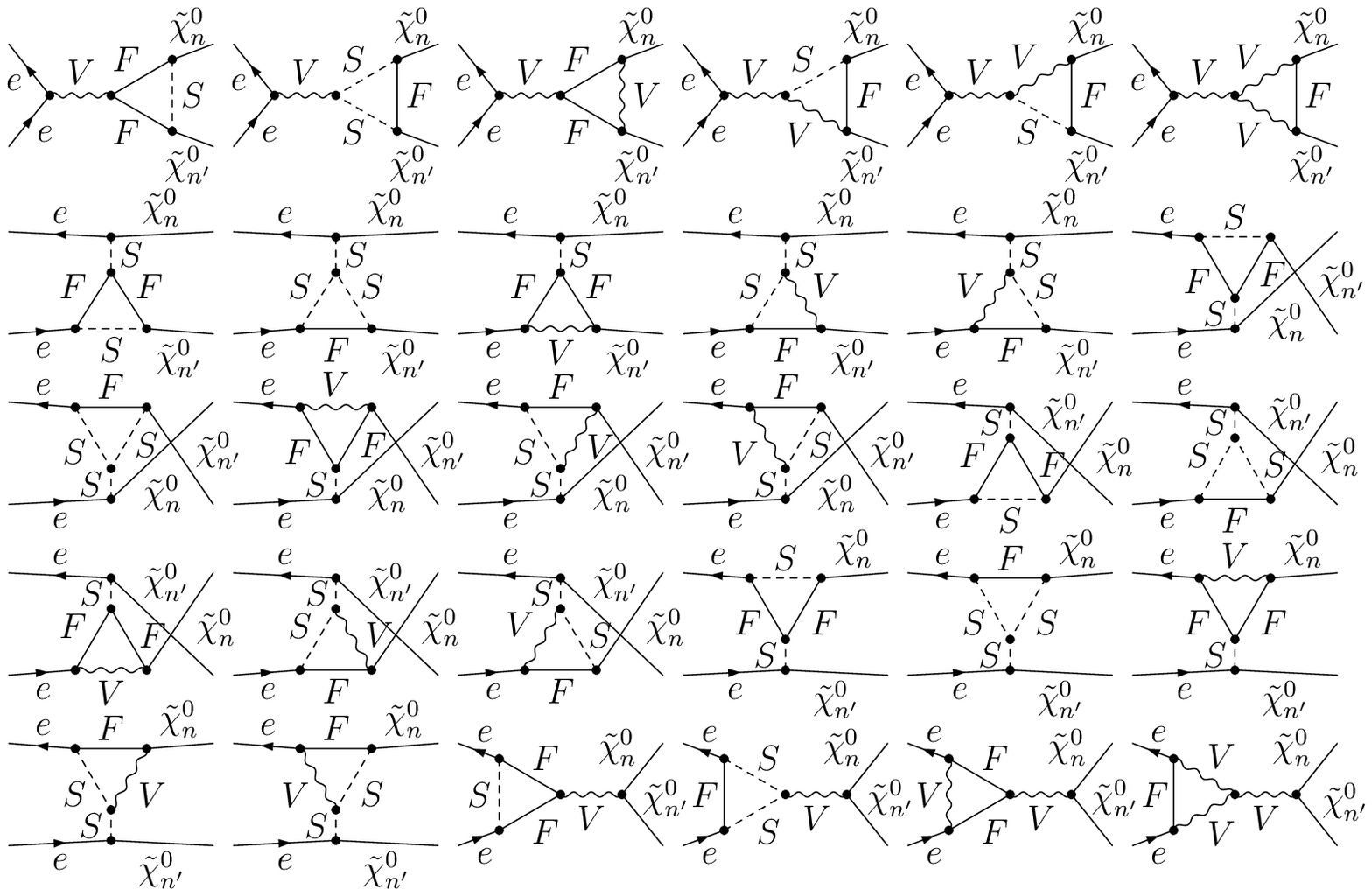}}
\framebox[15cm]{\includegraphics[width=0.67\textwidth]{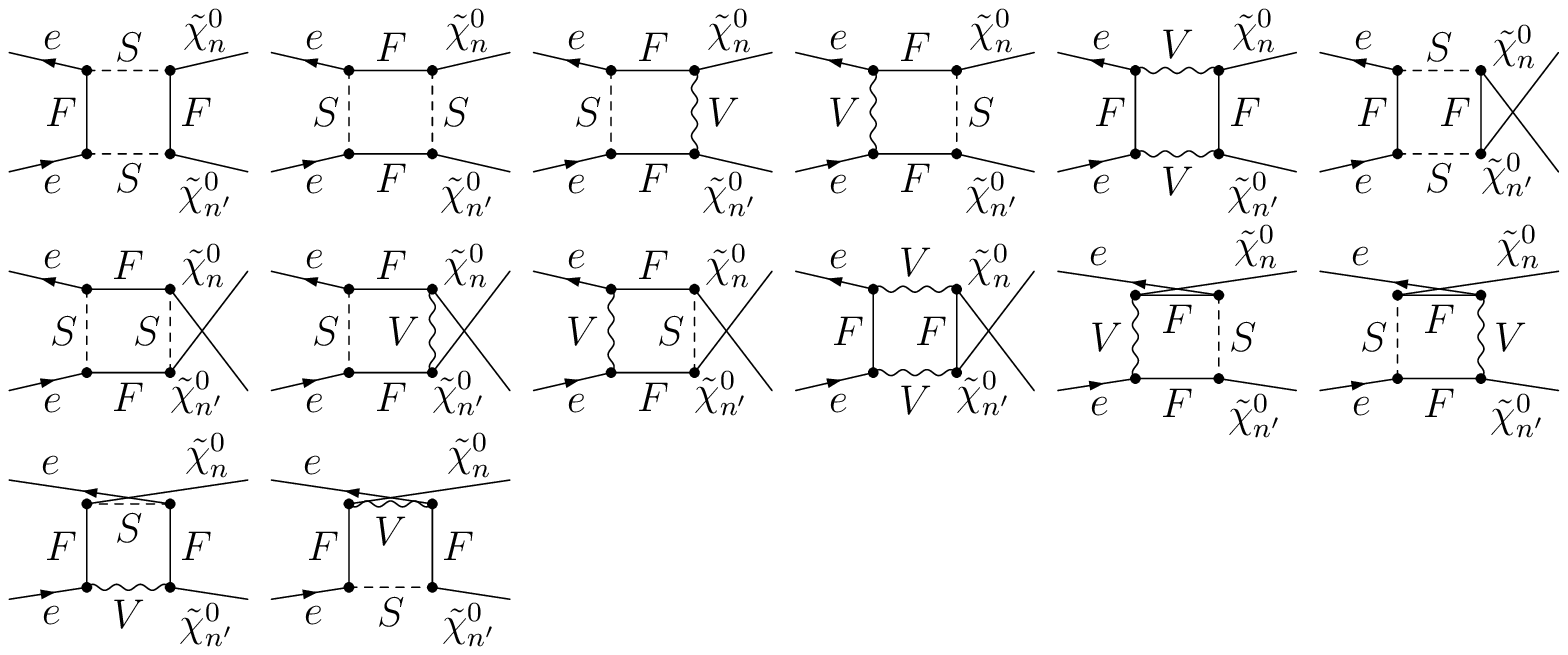}}
\framebox[15cm]{\includegraphics[width=0.67\textwidth]{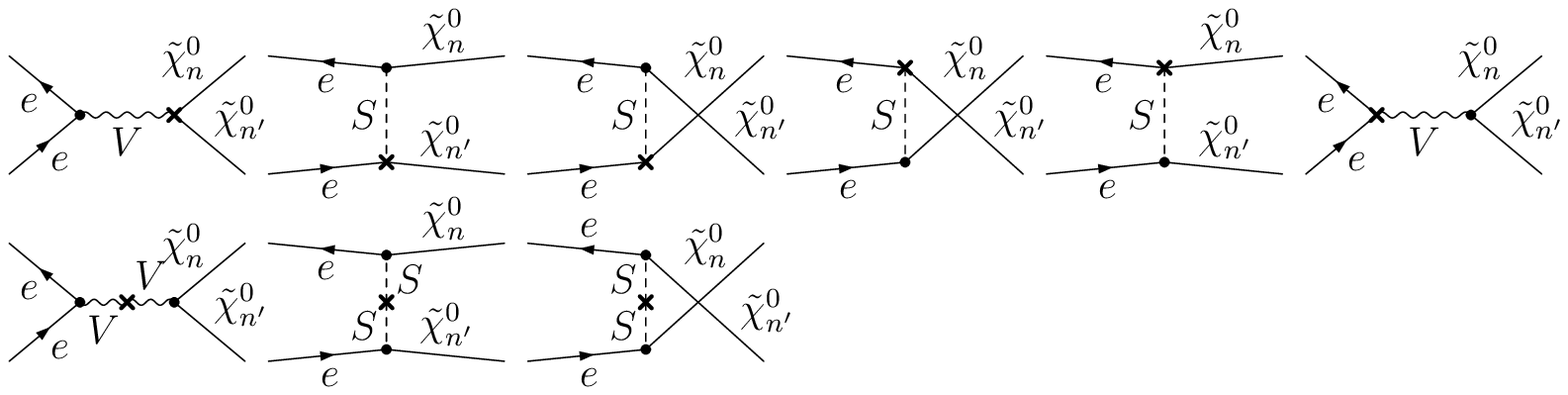}}
\caption{
  Same as \reffi{fig:CCiagrams}, but for the 
  process \eenn\ ($n,\npri = 1,2,3,4$). 
}
\label{fig:NNdiagrams}
\end{center}
\end{figure}

Sample diagrams for the process \eecc\ are shown in \reffi{fig:CCiagrams} 
and for the process \eenn\ in \reffi{fig:NNdiagrams}.
Not shown are the diagrams for real (hard and soft) photon radiation. 
They are obtained from the corresponding tree-level diagrams by attaching a 
photon to the (incoming/outgoing) electron or chargino.
The internal particles in the generically depicted diagrams in 
\reffis{fig:CCiagrams} and \ref{fig:NNdiagrams} are labeled as follows: 
$F$ can be a SM fermion $f$, chargino $\cha{c}$ or neutralino 
$\neu{n}$; $S$ can be a sfermion $\Sf_s$ or a Higgs (Goldstone) boson 
$h^0, H^0, A^0, H^\pm$ ($G, G^\pm$); $U$ denotes the ghosts $u_V$;
$V$ can be a photon $\ga$ or a massive SM gauge boson, $Z$ or $W^\pm$. 
We have neglected all electron--Higgs couplings and terms proportional 
to the electron mass whenever this is safe, \ie except when the electron 
mass appears in negative powers or in loop integrals.
We have verified numerically that these contributions are indeed totally 
negligible.  For internally appearing Higgs bosons no higher-order
corrections to their masses or couplings are taken into account; 
these corrections would correspond to effects beyond one-loop order.%
\footnote{
  We found that using loop corrected Higgs boson masses 
  in the loops leads to a UV divergent result.
}

Moreover, in general, in \reffis{fig:CCiagrams} and \ref{fig:NNdiagrams}
we have omitted diagrams with self-energy type corrections of external 
(on-shell) particles.  While the contributions from the real parts of the 
loop functions are taken into account via the renormalization constants 
defined by OS renormalization conditions, the contributions coming from 
the imaginary part of the loop functions can result in an additional (real) 
correction if multiplied by complex parameters.  In the analytical and 
numerical evaluation, these diagrams have been taken into account via the 
prescription described in \citere{MSSMCT}. 

Within our one-loop calculation we neglect finite width effects that 
can help to cure threshold singularities.  Consequently, in the close 
vicinity of those thresholds our calculation does not give a reliable
result.  Switching to a complex mass scheme \cite{complexmassscheme} 
would be another possibility to cure this problem, but its application 
is beyond the scope of our paper.

The tree-level formulas $\sigtree(\eecc)$ and $\sigtree(\eenn)$ are rather 
lengthy and can be found elsewhere~\cite{Bartl:1985fk,Bartl:1986hp}.
Concerning our evaluation of $\sig(\eecc)$ we define:
\begin{align}
\label{eeCCsum}
\sig(\eecc) \equiv \sig(\eecpcm) + \sig(\eecmcp) \qquad 
\forall\; c \neq c^{\prime}\,, 
\end{align}
if not indicated otherwise.  Differences between the two charge conjugated 
processes can appear at the loop level when complex parameters are taken 
into account, as will be discussed in \refse{sec:eecc}. 
We furthermore define the $\CP$ asymmetry $A_{12}$ for the non-diagonal
chargino production (see \citere{RoKa2007} for details),
\begin{align}
\label{A12}
A_{12} := \frac{\sigfull(\eecepczm) - \sigfull(\eecemczp)}
              {\sigtree(\eecepczm) + \sigtree(\eecemczp)}\,.
\end{align}
This asymmetry will be used for a comparison with previous calculations
and for an evaluation of the effects of the complex phases.


\subsection{Ultraviolet, infrared and collinear divergences}

As regularization scheme for the UV divergences we have used constrained 
differential renormalization~\cite{cdr}, which has been shown to be 
equivalent to dimensional reduction~\cite{dred} at the \onel\ 
level~\cite{formcalc}. 
Thus the employed regularization scheme preserves SUSY~\cite{dredDS,dredDS2}
and guarantees that the SUSY relations are kept intact, \eg that the gauge 
couplings of the SM vertices and the Yukawa couplings of the corresponding 
SUSY vertices also coincide to \onel\ order in the SUSY limit. 
Therefore no additional shifts, which might occur when using a different 
regularization scheme, arise. All UV divergences cancel in the final result.

\begin{figure}
\centering
\includegraphics[width=0.49\textwidth,height=7.5cm]{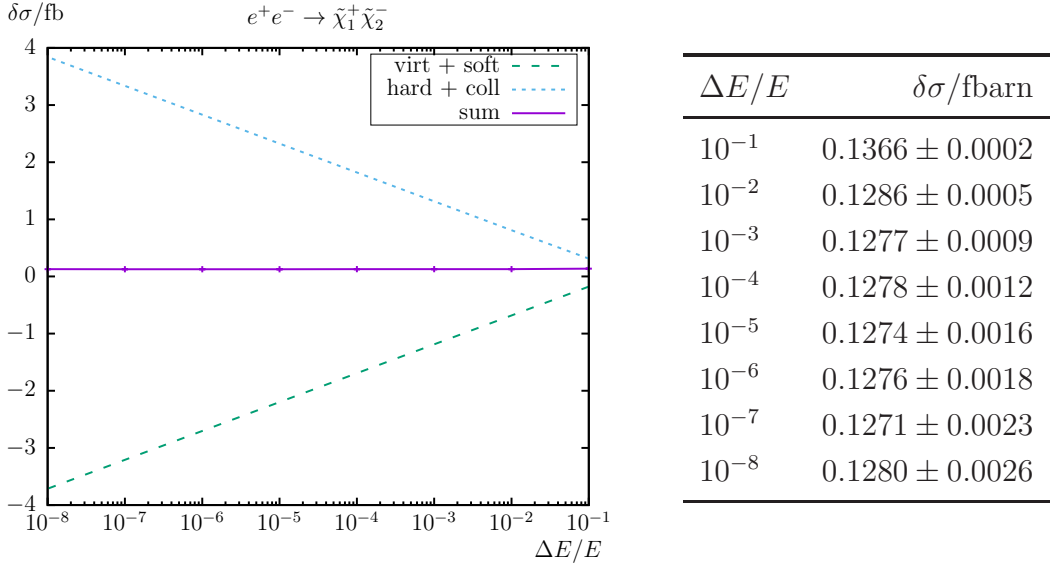}
\vspace{1em}
\begin{minipage}[b]{0.4\textwidth}
\centering
\begin{tabular}[b]{lr}
\toprule
$\Delta E/E$ & $\delta\sig$/fbarn \\
\midrule
$10^{-1}$ & $ 0.1366 \pm 0.0002$ \\
$10^{-2}$ & $ 0.1286 \pm 0.0005$ \\
$10^{-3}$ & $ 0.1277 \pm 0.0009$ \\
$10^{-4}$ & $ 0.1278 \pm 0.0012$ \\
$10^{-5}$ & $ 0.1274 \pm 0.0016$ \\
$10^{-6}$ & $ 0.1276 \pm 0.0018$ \\
$10^{-7}$ & $ 0.1271 \pm 0.0023$ \\
$10^{-8}$ & $ 0.1280 \pm 0.0026$ \\
\bottomrule
\end{tabular}
\vspace{2em}
\end{minipage}
\caption{\label{fig:collE}
  Phase space slicing method.  The different contributions to 
  the one-loop corrections $\delta\sig(\eececz)$ for our input 
  parameter scenario \Scs\ (see \refta{tab:para} below) 
  as a function of $\Delta E/E$ with 
  fixed $\Delta \theta/\text{rad} = 10^{-2}$.
}
\end{figure}

Soft photon emission implies numerical problems in the phase space 
integration of radiative processes.  The phase space integral diverges 
in the soft energy region where the photon momentum becomes very small,
leading to infrared (IR) singularities.  Therefore the IR divergences from 
diagrams with an internal photon have to cancel with the ones from the 
corresponding real soft radiation.  We have included the soft photon contribution 
via the code already implemented in \FC\ following the description given 
in \citere{denner}.  The IR divergences arising from the diagrams involving 
a photon are regularized by introducing a photon mass parameter, $\la$. 
All IR divergences, \ie all divergences in the limit $\la \to 0$, cancel 
once virtual and real diagrams for one process are added. 
We have numerically checked that our results do not depend on $\la$ or 
on $\Delta E = \delta_s E = \delta_s \sqrt{s}/2$ defining the energy 
cut that separates the soft from the hard radiation. As one can see
from the example in \reffi{fig:collE} this holds for several orders of 
magnitude.  Our numerical results below have been obtained for fixed 
$\delta_s = 10^{-3}$.

Numerical problems in the phase space integration of the radiative 
process arise also through collinear photon emission. Mass singularities 
emerge as a consequence of the collinear photon emission off massless
particles.  But already very light particles (such as electrons) can 
produce numerical instabilities.
For the treatment of collinear singularities in the photon radiation off 
initial state electrons and positrons we used the 
\textit{phase space slicing method}~\cite{slicing}, which is not (yet) 
implemented in \FC\ and therefore we have developed and implemented the 
code necessary for the evaluation of collinear contributions; see also
\citeres{HiggsProd,HpProd}.

\begin{figure}
\centering
\includegraphics[width=0.49\textwidth,height=7.5cm]{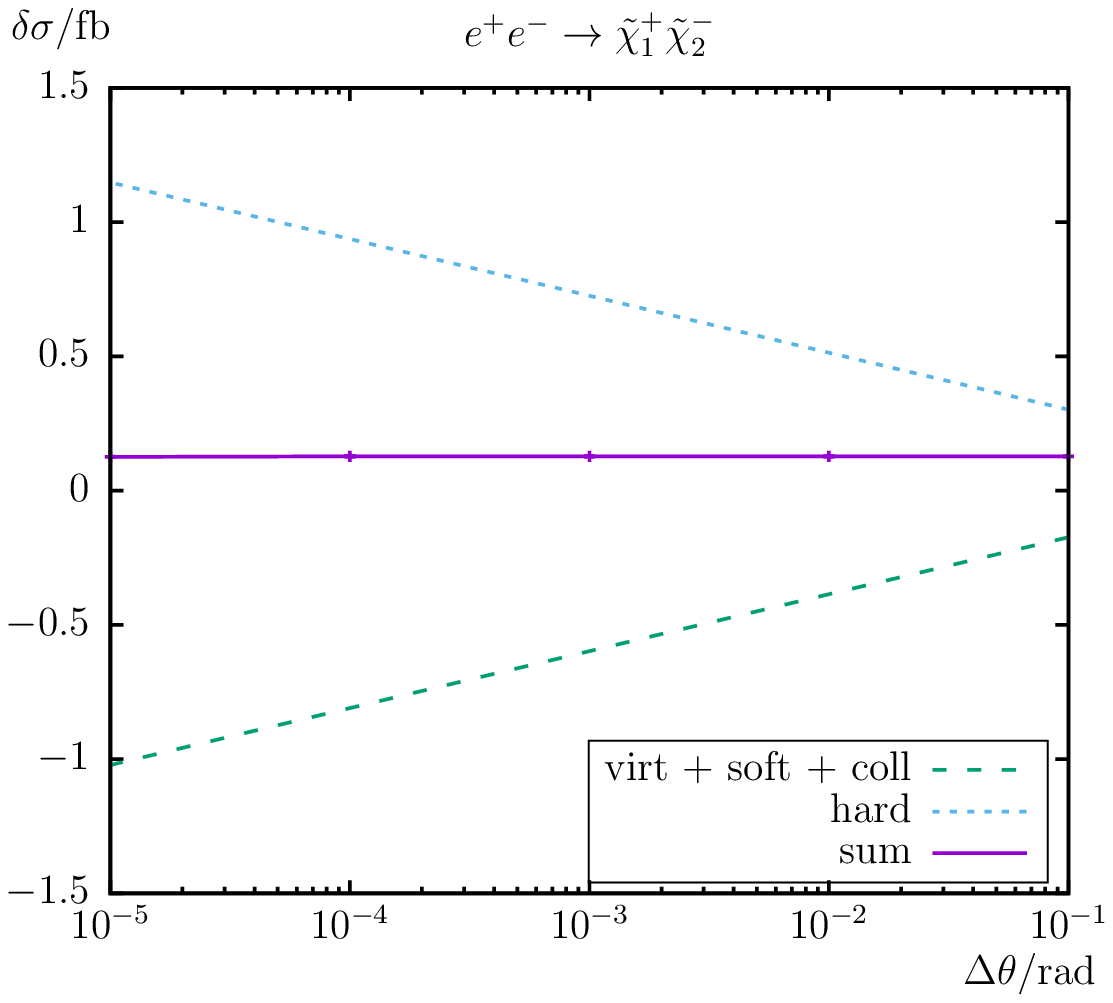}
\begin{minipage}[b]{0.4\textwidth}
\centering
\begin{tabular}[b]{lr}
\toprule
$\Delta \theta$/rad & $\delta\sig$/fbarn \\
\midrule
$10^{ 0}$ & $ 0.1080 \pm 0.0012$ \\
$10^{-1}$ & $ 0.1276 \pm 0.0010$ \\
$10^{-2}$ & $ 0.1277 \pm 0.0009$ \\
$10^{-3}$ & $ 0.1277 \pm 0.0009$ \\
$10^{-4}$ & $ 0.1275 \pm 0.0010$ \\
$10^{-5}$ & $ 0.1259 \pm 0.0011$ \\
$10^{-6}$ & $ 0.0714 \pm 0.0013$ \\
\bottomrule
\end{tabular}
\vspace{2em}
\end{minipage}
\caption{\label{fig:collT}
  Phase space slicing method.  The different contributions to 
  the one-loop corrections $\delta\sig(\eececz)$ for our input 
  parameter scenario \Scs\ (see \refta{tab:para} below) 
  as a function of $\Delta\theta/\text{rad}$ with 
  fixed $\Delta E/E = 10^{-3}$.
}
\end{figure}

In the phase space slicing method, the phase space is divided into
regions where the integrand is finite (numerically stable) and 
regions where it is divergent (or numerically unstable).
In the stable regions the integration is performed numerically, whereas
in the unstable regions it is carried out (semi-) analytically using 
approximations for the collinear photon emission.

The collinear part is constrained by the angular cut-off parameter 
$\Delta\theta$, imposed on the angle between the photon and the
(in our case initial state) electron/positron.

The differential cross section for the collinear photon radiation 
off the initial state $e^+e^-$ pair corresponds to a convolution
\begin{align}
\text{d}\sigcoll(s) = \frac{\alpha}{\pi} \int_0^{1-\delta_s} \text{d}z\,
  \text{d}\sigtree(\sqrt{z s}) \LV \LB 2\, \ln \LP 
  \frac{\Delta \theta \sqrt{s}}{2\, m_e} \RP - 1 \RB P_{ee}(z) + 1 - z \RV\,,
\end{align}
with $P_{ee}(z) = (1 + z^2)/(1 - z)$
denoting the splitting function of a photon from the initial $e^+e^-$ pair.
The electron momentum is reduced (because of the radiated photon) by 
the fraction $z$ such that the center-of-mass frame of the hard process 
receives a boost.  The integration over all possible factors $z$ is 
constrained by the soft cut-off $\delta_s = \Delta E/E$, to prevent 
over-counting in the soft energy region.

We have numerically checked that our results do not depend on the angular 
cut-off parameter $\Delta\theta$ over several orders of magnitude; 
see the example in \reffi{fig:collT}. Our numerical results below have 
been obtained for fixed $\Delta \theta/\text{rad} = 10^{-2}$.

The one-loop corrections of the differential cross section are decomposed 
into the virtual, soft, hard, and collinear parts as follows:
\begin{align}
\text{d}\sigloop = \text{d}\sigvirt(\la) + 
                   \text{d}\sigsoft(\la, \Delta E) + 
                   \text{d}\sighard(\Delta E, \Delta\theta) + 
                   \text{d}\sigcoll(\Delta E, \Delta\theta)\,.
\end{align}
The hard and collinear parts have been calculated via Monte Carlo 
integration algorithms of the \texttt{CUBA} library \cite{cuba} as 
implemented in \FC~\cite{formcalc}.


\section{Comparisons}
\label{sec:comparisons}

In this section we present the comparisons with results from other groups 
in the literature for chargino/neutralino production in $e^+e^-$ collisions.
These comparisons were mostly restricted to the MSSM with real parameters.
The level of agreement of such comparisons (at one-loop order) depends on 
the correct transformation of the input parameters from our renormalization 
scheme into the schemes used in the respective literature, as well as on the 
differences in the employed renormalization schemes as such.
In view of the non-trivial conversions and the large number of comparisons 
such transformations and/or change of our renormalization prescription is 
beyond the scope of our paper.

\begin{itemize}

\item
In \citeres{Bartl:1985fk,Bartl:1986hp} the processes \eecc\ and \eenn\ 
have been calculated in the rMSSM at tree level.  Because our tree-level 
results are in good agreement with other groups (see below), we omitted a 
comparison with \citeres{Bartl:1985fk,Bartl:1986hp}.

\begin{figure}
\begin{center}
\begin{tabular}{c}
\includegraphics[width=0.48\textwidth,height=6cm]{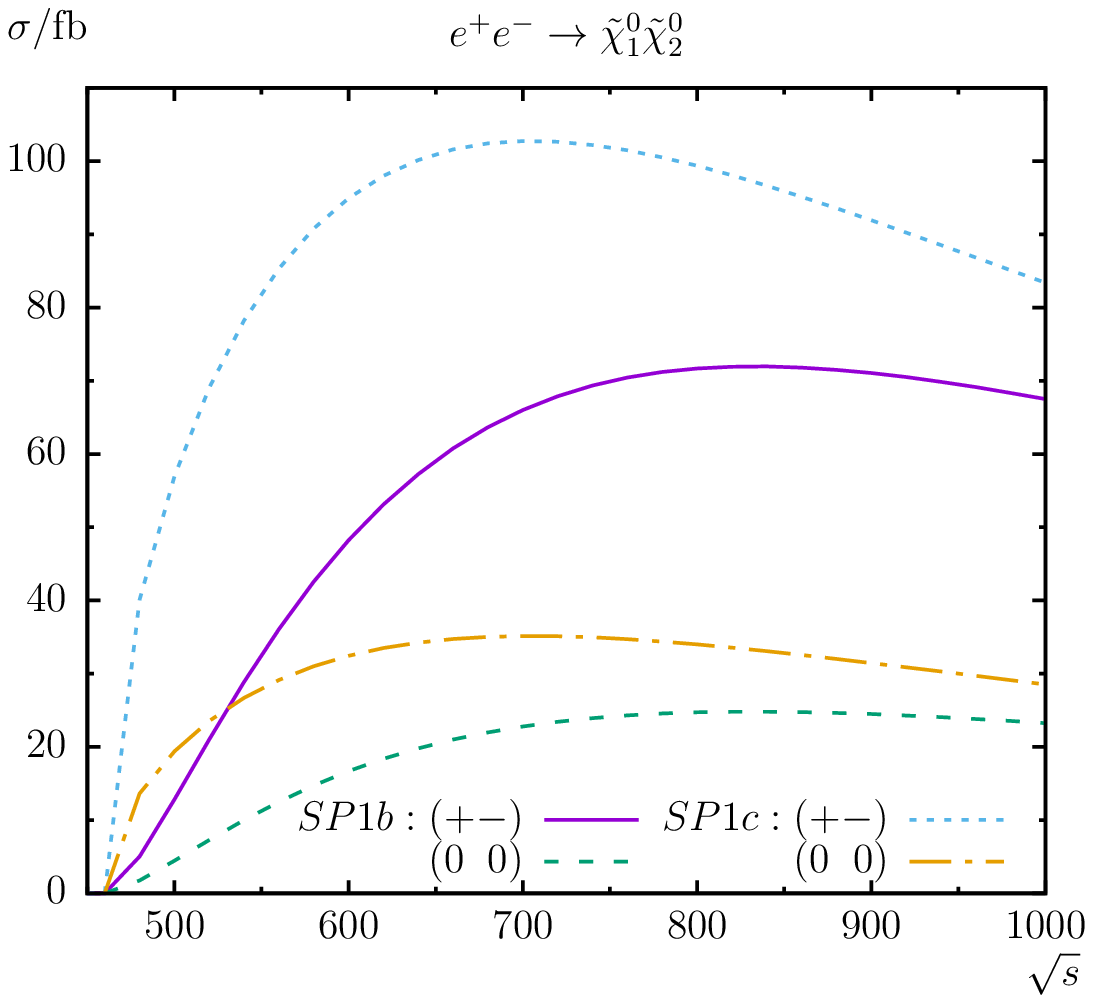}
\includegraphics[width=0.48\textwidth,height=6cm]{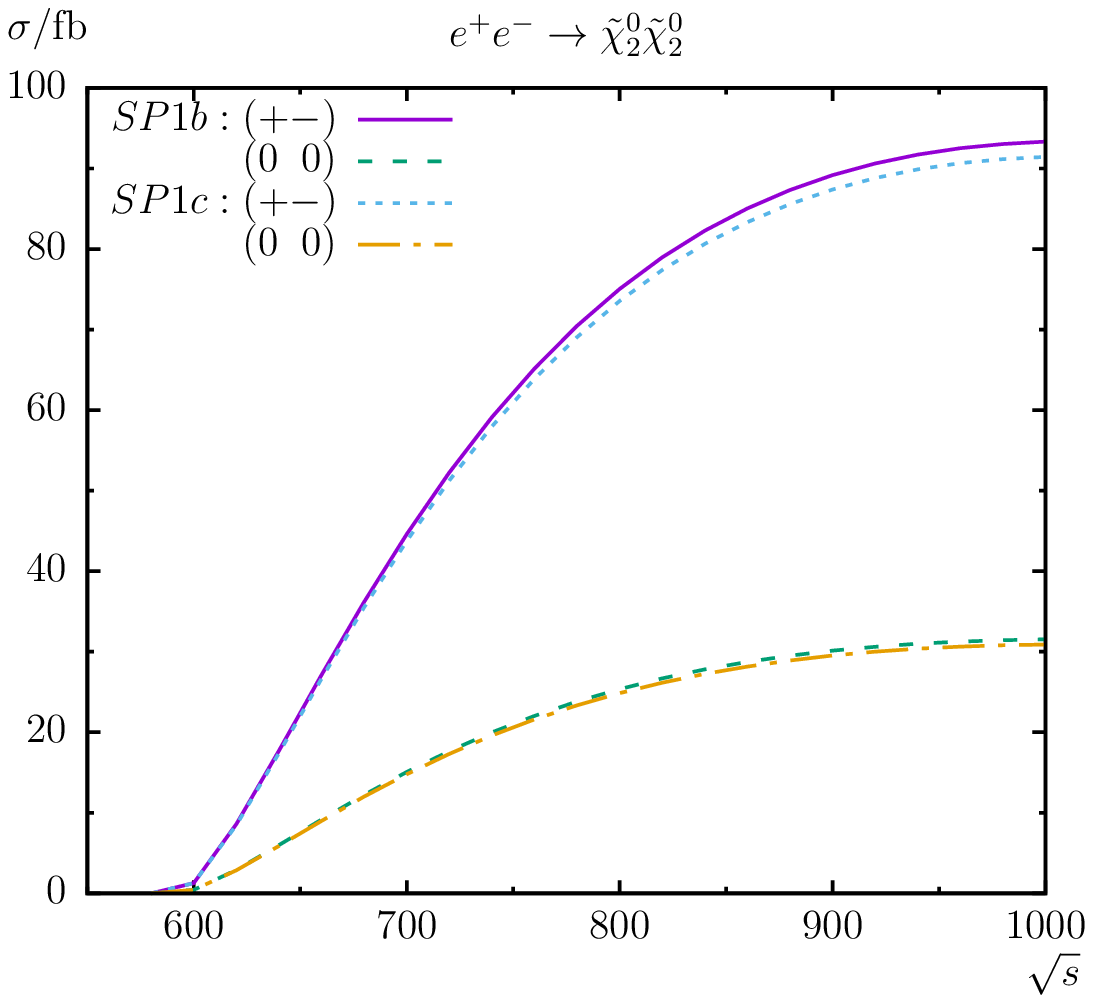}
\end{tabular}
\caption{\label{fig:GoMo2002}
  Comparison with \citere{Gounaris:2002pj} for $\sig(\eenn)$.
  Tree cross sections are shown with parameters chosen according 
  to \citere{Gounaris:2002pj} as a function of $\sqrt{s}$.
  The left (right) plot shows cross sections for $\eenenz$ 
  ($\eenznz$).
  $(+ -)$ denotes $(+0.60, -0.85)$ polarization of the 
  positrons/electrons, whereas $(0\;\; 0)$ denotes unpolarized 
  positrons/electrons.
}
\end{center}
\end{figure}

\item
Tree-level results in the cMSSM for (polarized) \eenn\ (using a 
``projector formalism'') were presented in \citere{Gounaris:2002pj}. 
As input we used their parameter sets ``$SP1b$'' and ``$SP1c$'', but it 
should be noted that they gave no SM input parameters. 
In \reffi{fig:GoMo2002} we show our calculation in comparison to their 
Figs.~6a,b where we find good agreement with their results.
The small differences can be explained with the different SM input 
parameters.

\item
%
In \citere{Osland:2007xw} the process \eecc\ ($c \neq c^{\prime}$) has 
been computed in the cMSSM (including ``selected box contributions'') and 
extended to the full contributions in \citeres{RoKa2007,OsKaRoVe2007}.  
We performed a comparison with \citere{RoKa2007} using their input 
parameters (as far as possible).  They also used (older versions of) \FT\ 
for their calculations.  We find good agreement with their Fig.~3; 
as can be seen in our \reffi{fig:RoKa2007}, where we show the $\CP$-odd 
observable $A_{12}$; see \refeq{A12}.
While the box contributions to $A_{12}$ and the full results are in good 
agreement, the self-energy and vertex contributions differ significantly. 
However, here it should be noted that we have included the absorptive 
parts from self-energy type contributions via additional renormalization 
constants (see \citeres{MSSMCT, Stop2decay}) and not via the self-energy 
diagrams by themselves, which explains the large differences in the pure 
``self'' and ``vert'' parts.  But in combination the results are in 
agreement as expected.
It should also be noted that $A_{12}$ is very sensitive to the input 
parameters, explaining the small differences in the box and full results.

\begin{figure}
\begin{center}
\begin{tabular}{c}
\includegraphics[width=0.48\textwidth,height=6cm]{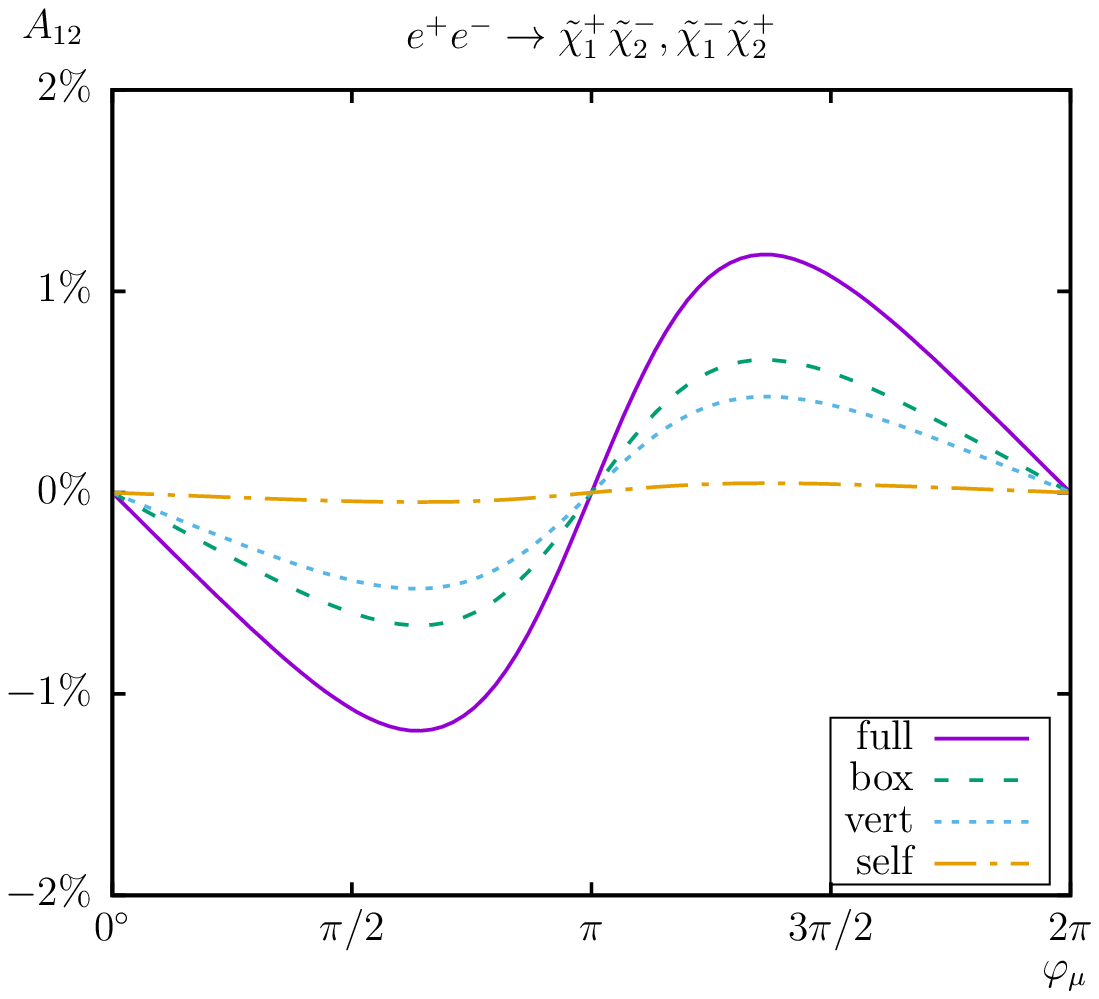}
\includegraphics[width=0.48\textwidth,height=6cm]{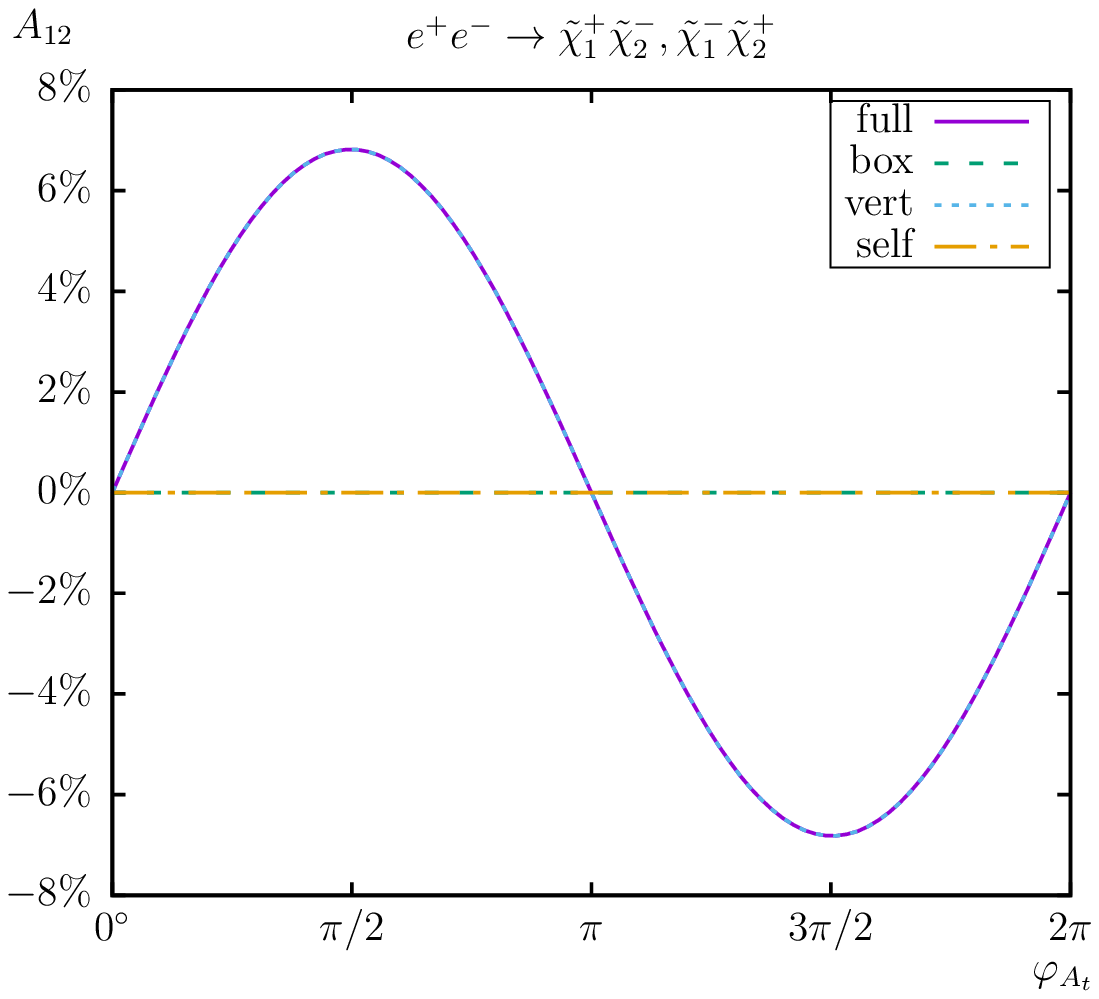}
\end{tabular}
\caption{\label{fig:RoKa2007}
  Comparison with \citere{RoKa2007} for $\sig(\eecc)$.
  $\CP$-odd observables $A_{12}$ are shown within scenario (A) chosen 
  according to \citere{RoKa2007}.
  The left (right) plot shows $A_{12}$ (see \refeq{A12}) with 
  $\phimu$ ($\phiAt$) varied and the different contributions from the 
  box, vertex, and self-energy corrections including absorptive parts 
  via renormalization constants.  
}
\end{center}
\end{figure}

\begin{figure}
\begin{center}
\begin{tabular}{c}
\includegraphics[width=0.48\textwidth,height=6cm]{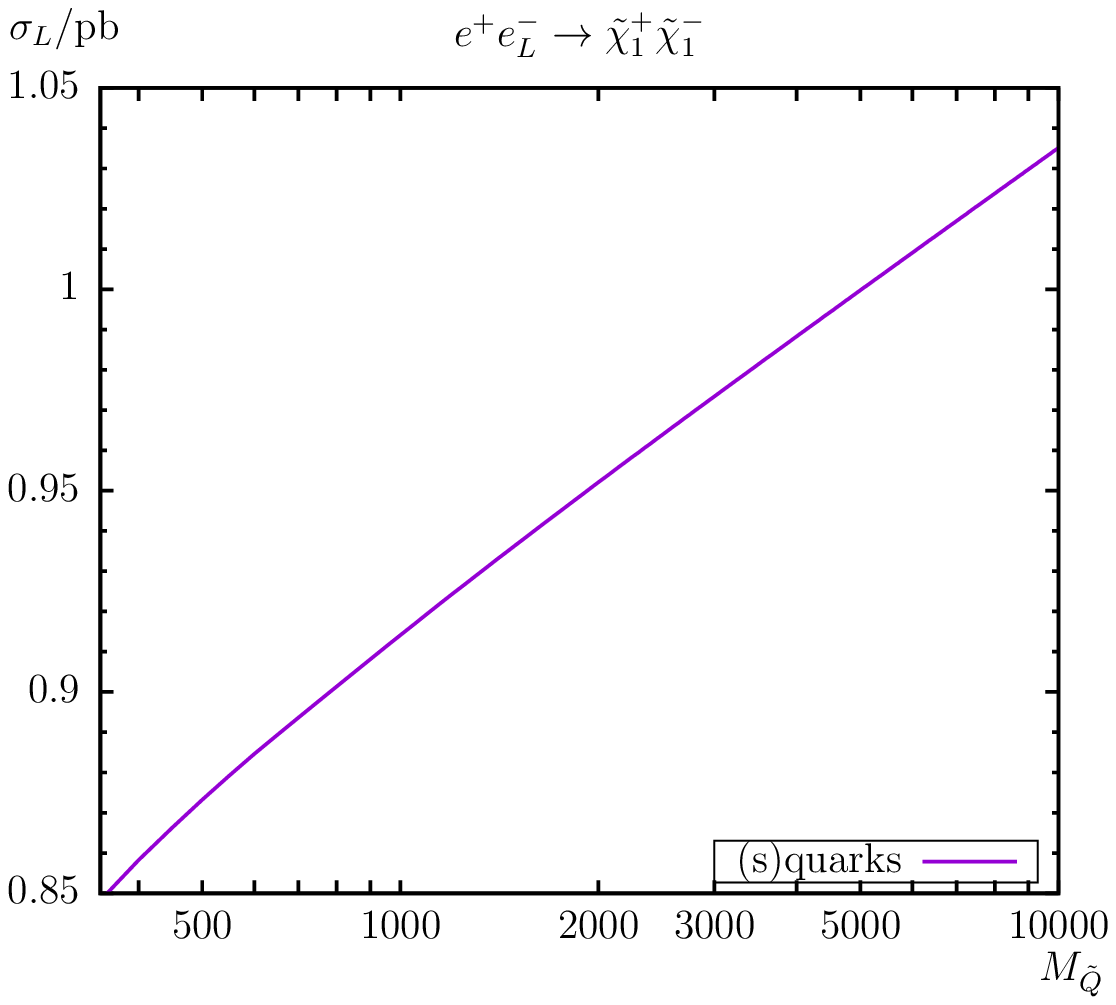}
\includegraphics[width=0.48\textwidth,height=6cm]{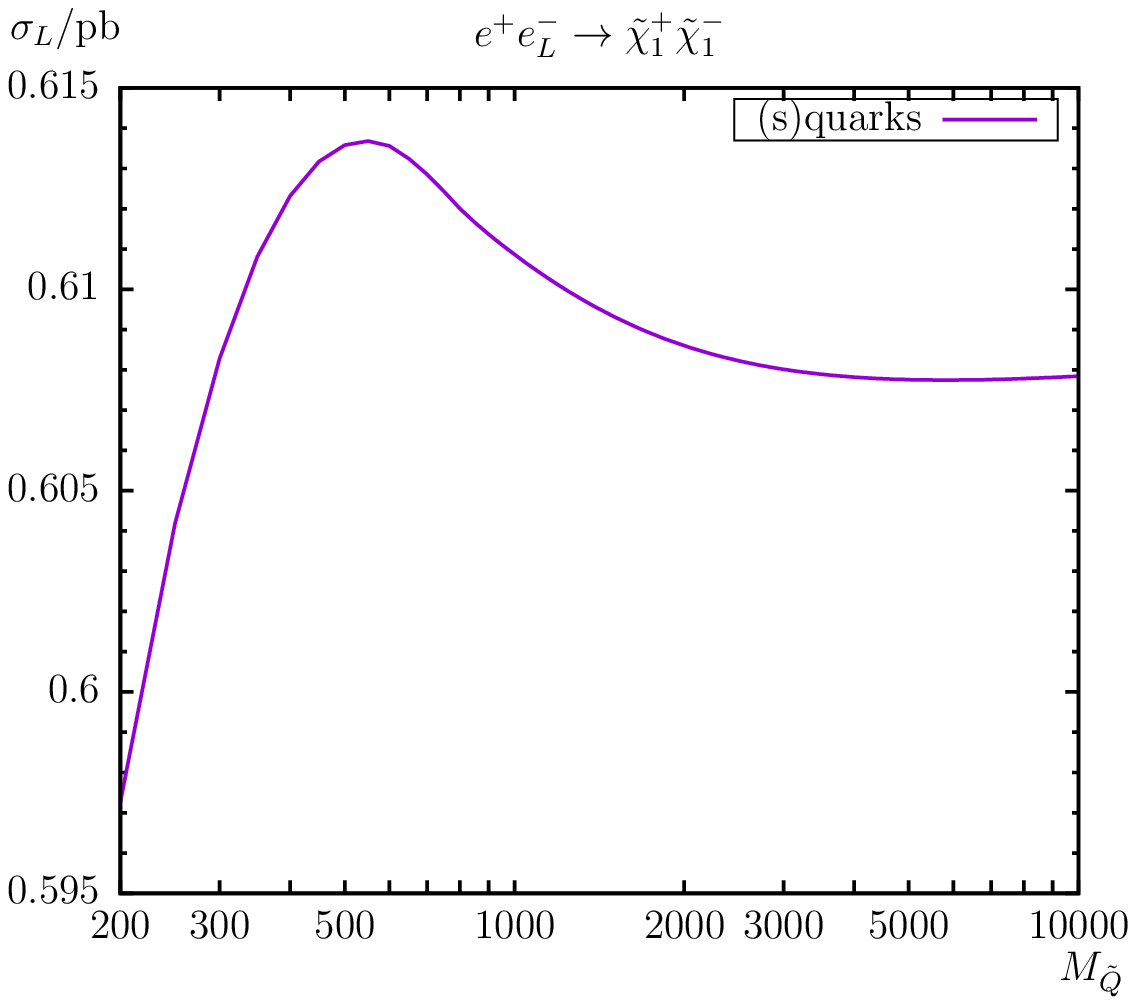}
\end{tabular}
\caption{\label{fig:KiNoPiYa1998}
  Comparison with \citere{Kiyoura:1998yt} for 
  $\sig(e^+e^-_L \to \chap1\cham1)$.
  The left (right) plot shows cross sections with $M_{\tilde Q}$
  varied for left-handed electrons and the parameter set G1 (H1) 
  according to \citere{Kiyoura:1998yt}.
}
\end{center}
\end{figure}

\item
Radiative corrections to chargino production in electron--positron collisions 
in the rMSSM were analyzed in \citere{Diaz:1997kv}.  The vertex corrections to 
\eecc\ in the approximation of $t/\Stop{}/b/\Sbot{}$ contributions were 
evaluated, using an \MSbar\ renormalization scheme.  It should be noted that 
\citere{Kiyoura:1998yt} (see the next item) claimed differences to 
\citere{Diaz:1997kv}.  In addition this paper is (more or less) a prelude to 
\citeres{Diaz:2002rr,Diaz:2009um}, therefore we omitted a comparison with 
\citere{Diaz:1997kv}.

\item
In \citere{Kiyoura:1998yt} the process \eecc\ including all quark/squark
contributions in the self-energy and vertex corrections has been calculated 
in the rMSSM.  It should be noted that the authors claimed that the 
calculation of the cross section including only $t/\Stop{}/b/\Sbot{}$ loops 
as presented in \citere{Diaz:1997kv} is not a reasonable approximation in 
general.
We used their input parameters (\ie scenarios G1 and H1) as far as possible 
(no SM parameters have been given)%
\footnote{
  As SM parameters we chose the PDG values from 1998.
}
and reproduced Fig.~1(a) and Fig.~3 of \citere{Kiyoura:1998yt}, where
$e^+e^-_L \to \chap1\cham1$ had been evaluated. Our results are shown in 
\reffi{fig:KiNoPiYa1998}.  As in \citere{Kiyoura:1998yt} we also
include only the (s)quark contributions for this comparison.  
We are in very good qualitative agreement and the loop corrections 
differ numerically less than $2\%$.  The reasons for these small
differences can again be found in the different renormalization schemes 
and SM input parameters. 

\item
Full one-loop corrections in the rMSSM for \eecc\ were presented in 
\citere{Blank:2000uc}.  Because \citere{Blank:2000uc} is only an extract
from \citere{Blank:2000fqa} (see the corresponding item below), we omitted 
a comparison with \citere{Blank:2000uc}.

\item
In \citere{Diaz:2002rr} the (weak) one-loop contributions of the rMSSM to the
process \eecc\ have been calculated, \ie neglecting the pure QED corrections
involving photon loops and radiation.
The calculation has been performed within the \DRbar\ scheme for polarized 
electrons and charginos.  We used their input parameters (benchmark point C
model) as far as possible and 
reproduced Fig.~3 and Fig.~4 of \citere{Diaz:2002rr} in our 
\reffi{fig:DiRo2002}.  While we are in good qualitative agreement the
loop corrections differ numerically.  
Besides the different renormalization schemes the main reason is that we 
must keep the QED corrections for UV finiteness in our on-shell scheme.
Although we subtracted the leading QED logarithms $\propto \text{ln}(s/m_e^2)$ 
(by hand) for the comparison the differences are quite large, rendering this 
comparison not significant.

\begin{figure}
\begin{center}
\begin{tabular}{c}
\includegraphics[width=0.48\textwidth,height=6cm]{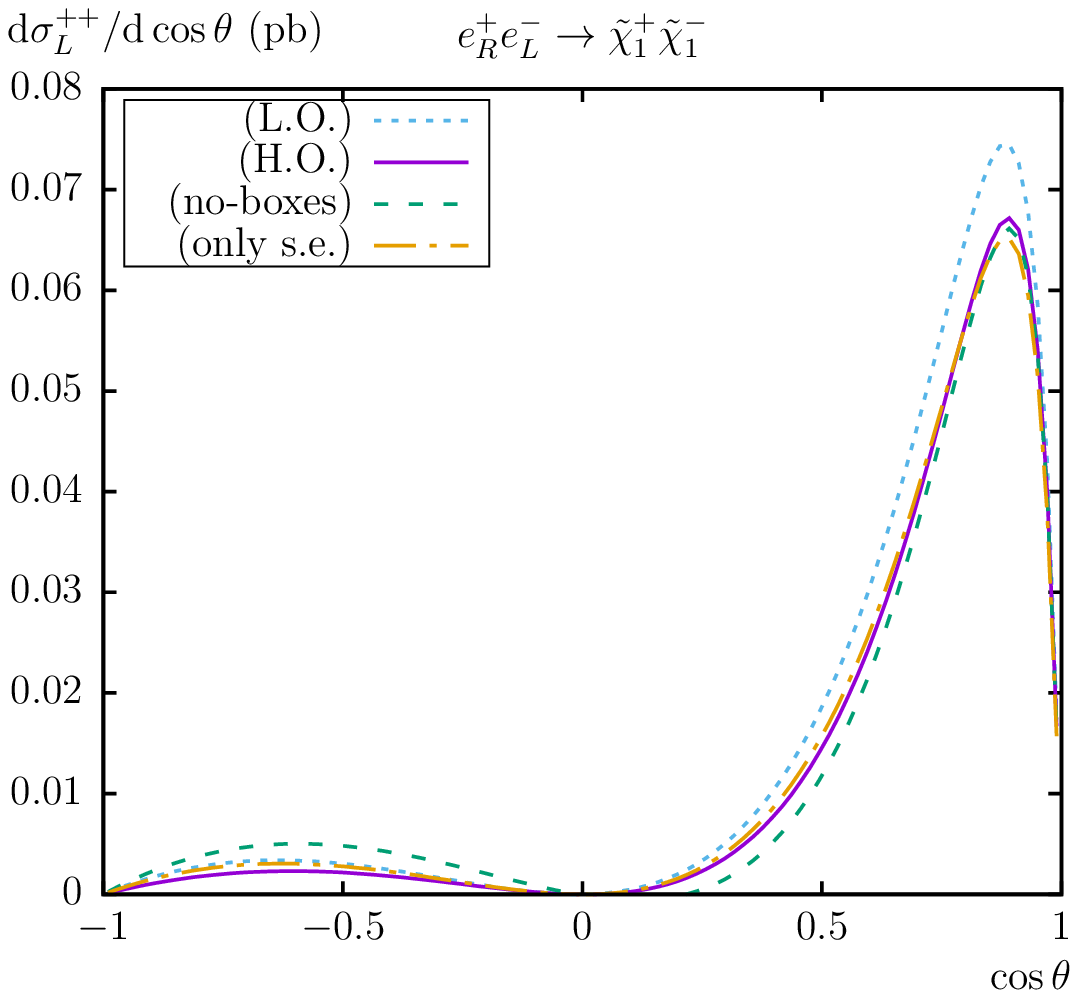}
\includegraphics[width=0.48\textwidth,height=6cm]{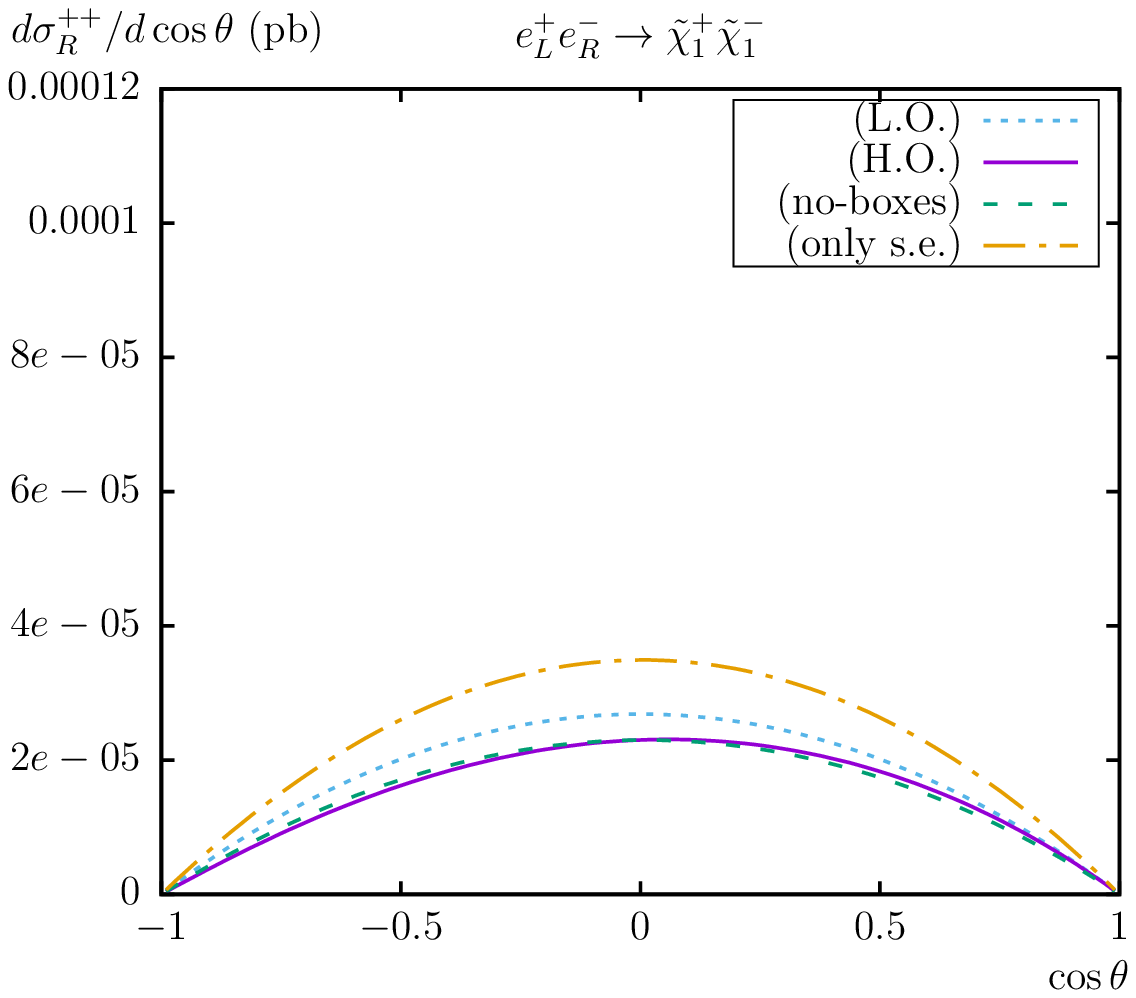}
\end{tabular}
\caption{\label{fig:DiRo2002}
  Comparison with \citere{Diaz:2002rr} for $\sig(\eecece)$.
  The left (right) plot shows differential cross sections with 
  $\cos\theta$ varied for left-handed electrons, right-handed
  positrons (right-handed electrons, left-handed positrons) and 
  charginos with positive helicity within the benchmark point C model
  according to \citere{Diaz:2002rr}.
}
\end{center}
\end{figure}

\begin{figure}
\begin{center}
\begin{tabular}{c}
\includegraphics[width=0.48\textwidth,height=6cm]{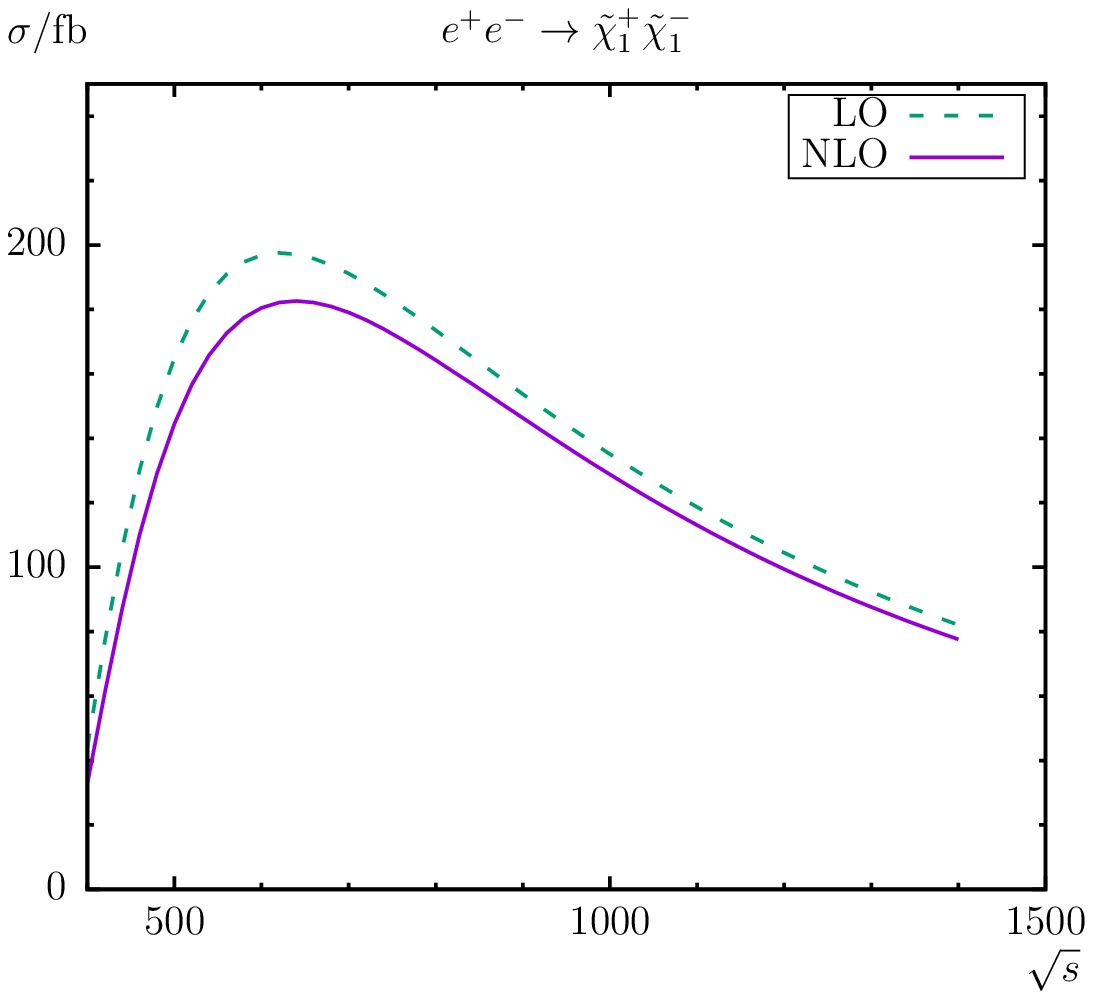}
\includegraphics[width=0.48\textwidth,height=6cm]{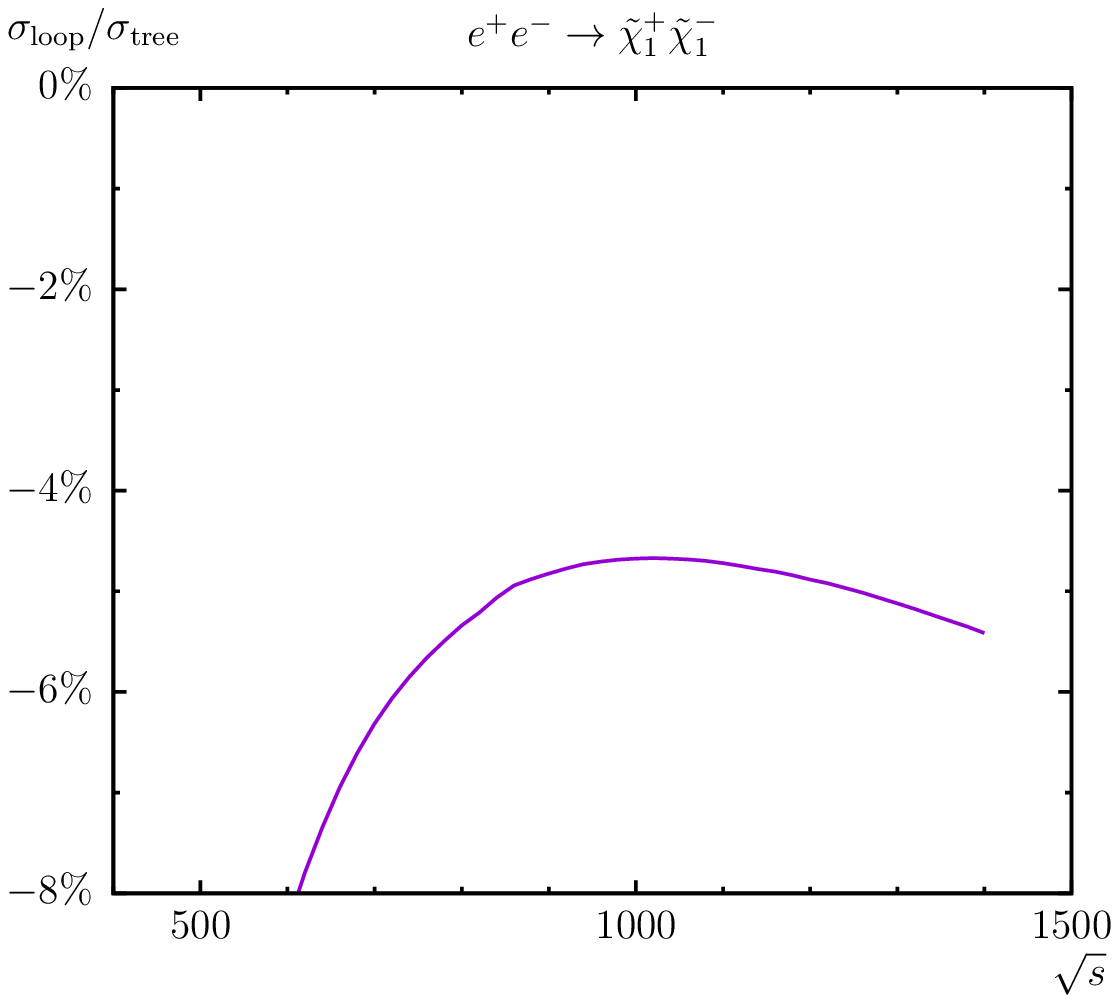}
\end{tabular}
\caption{\label{fig:KiReRo2006}
  Comparison with \citere{Kilian:2006cj} for $\sig(\eecece)$.
  The left (right) plot shows cross sections (relative corrections) 
  with $\sqrt{s}$ (in GeV) varied within the on-shell parameter set 
  SPS$1a^{\prime}$.
}
\end{center}
\end{figure}

\item
The inclusion of multi-photon emission in \eecc\ and the implementation into 
an NLO event generator was presented in \citere{Kilian:2006cj,Robens:2006np}.
As input parameters they used the SUSY parameter point SPS$1a^{\prime}$; see
\citere{SPS1a}.  We also used the parameter point SPS$1a^{\prime}$ but
translated from the \DRbar\ to on-shell values and reproduced successfully 
Fig.~7 of \citere{Kilian:2006cj} in our \reffi{fig:KiReRo2006}.  Our one-loop 
results are in reasonable agreement with the ones in \citere{Kilian:2006cj} 
within $\pm 3\%$.  The small difference can be easily explained with  
the different renormalization schemes, slightly different input parameters, 
and the different treatment of the photon bremsstrahlung, where they have 
included multi-photon emission while we kept our calculation at $\order\al$.

\item
\citere{Blank:2000fqa} is the source of \citere{Blank:2000uc}
(see the corresponding previous item), dealing with chargino and neutralino
production in the rMSSM.  A comparison with Fig.~6.13 of 
\citere{Blank:2000fqa} is given in our \reffi{fig:Bl2000}, where we show 
$\tilde{\Delta} = (\sigloop-\sigma_{\text{ISR}})/\sigtree$ as a function of
$\TB$ for two numerical scenarios (used in the original Fig.~6.13).%
\footnote{
  It should be noted that $\sigma_{\text{ISR}}$ denotes the 
  (large) initial state radiation $\propto \text{ln}(s/m_e^2)$, 
  whereas the hard and collinear photon radiation had been 
  neglected in \citere{Blank:2000fqa}.
}
Using their input parameters and our \Code{CNN[2,1,3]} scheme 
(which appears closest to their renormalization scheme) we are in rather 
good agreement for $\MSUSY \ge 500\gev$, while only in very rough 
agreement for $\MSUSY = 200\gev$.  This can be explained with the 
different renormalization schemes, especially with the different 
renormalization of $\TB$, which in \citere{Blank:2000fqa} is defined 
via the imaginary part of the $AZ$~self-energy.  Furthermore, 
$\tilde{\Delta}$ is very sensitive to the loop corrections.

\begin{figure}
\begin{center}
\begin{tabular}{c}
\includegraphics[width=0.48\textwidth,height=6cm]{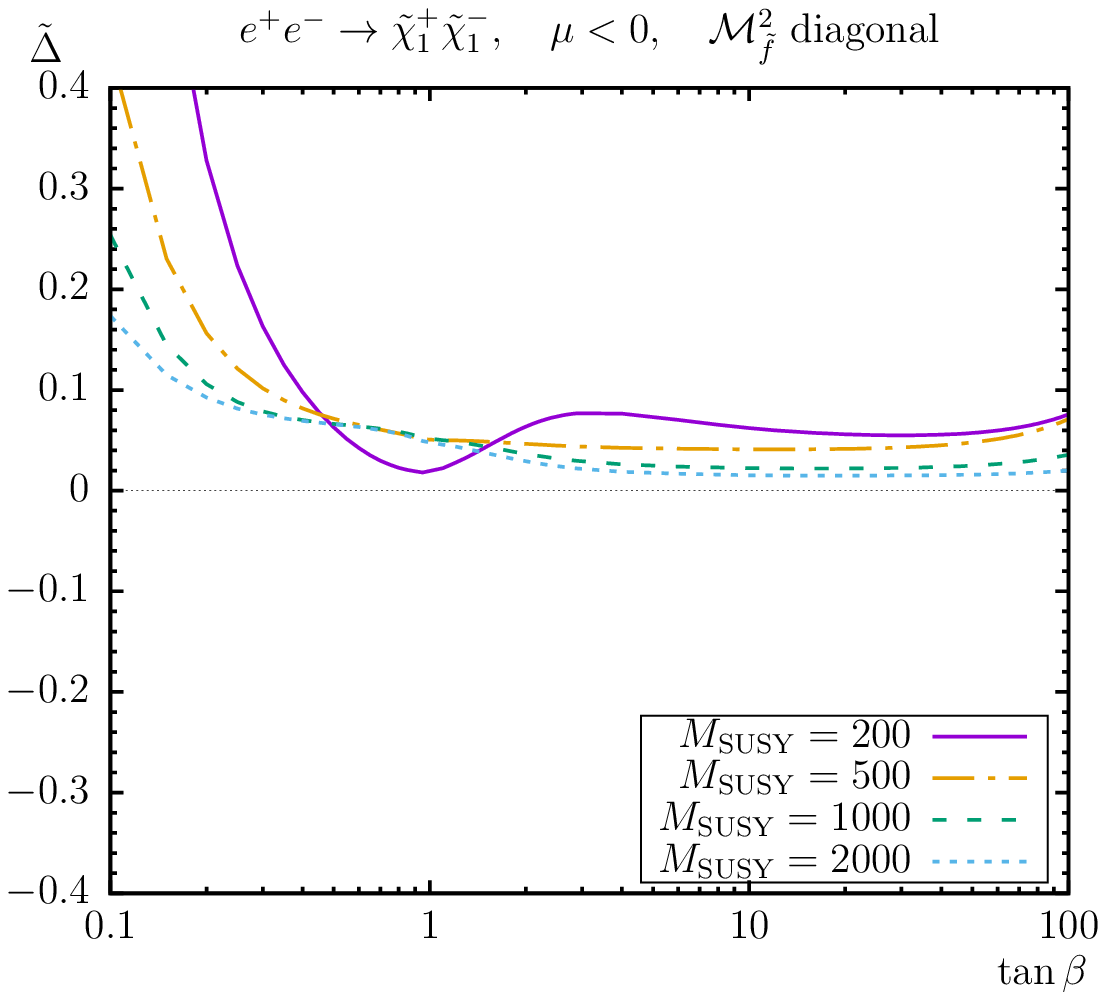}
\includegraphics[width=0.48\textwidth,height=6cm]{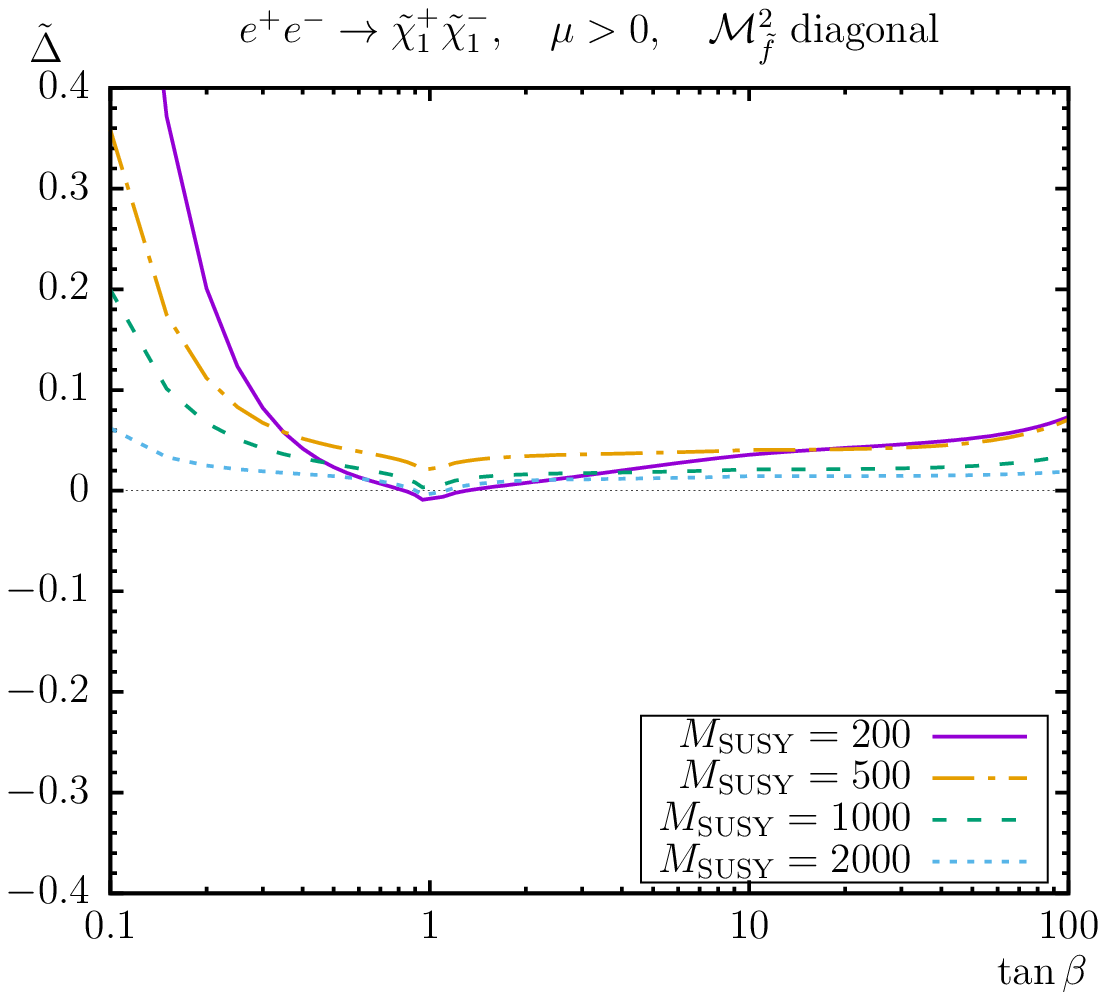}
\end{tabular}
\caption{\label{fig:Bl2000}
  Comparison with \citere{Blank:2000fqa} for $\sig(\eecece)$.
  The left (right) plot shows the relative corrections $\tilde{\Delta}$ 
  with $\TB$ varied for $\mu < 0$ ($\mu > 0$) within the parameter set 
  of \citere{Blank:2000fqa}.
}
\end{center}
\end{figure}

\item
\citere{Diaz:2001vm} deals with ``prototype graphs'' for radiative 
corrections to polarized chargino or neutralino production in 
electron--positron annihilation.  This paper contains no numerical analysis, 
rendering a comparison impossible.

\item
In \citere{Oller:2004br} the ``full'' one-loop corrections to neutralino pair 
production in the rMSSM were analyzed numerically including QED corrections.
They used (older versions of) \FT\ for their calculations but implemented 
their own on-shell renormalization procedure. 
In their analysis they show ``full'' $\order\al$ corrections but without 
initial state radiation.  The authors extended their analysis to \eenn\ and 
\eecc\ in \citere{Oller:2005xg} in the cMSSM, making some improvements also 
concerning the photon radiation.
Therefore, we skip the comparison with \citere{Oller:2004br}, but focus on 
\citere{Oller:2005xg}.  We used their input parameters (\ie the \textit{real} 
on-shell parameter set of the SUSY parameter point SPS$1a^{\prime}$, 
see \citere{SPS1a}) as far as possible and reproduced their Figs.~7--11 in 
our \reffi{fig:OlEbMa2005}.  Qualitatively we are in good agreement with 
\citere{Oller:2005xg}, but our (relative) one-loop results are numerically 
only roughly in agreement with their results within $\pm 25\%$. 
The differences can be explained (besides the different renormalization 
schemes) with the fact that they used an ``$\al(Q)$'' scheme, that yields 
particularly large corrections, \wrt our $\al(0)$ scheme (these 
effects were known already for a long time, see \citeres{eennHWiener,eennH}, 
where the different renormalizations even yielded a different sign of the 
one-loop corrections).  They also included higher order contributions into 
their initial state radiation while we kept our calculation at $\order\al$.

\begin{figure}
\begin{center}
\begin{tabular}{c}
\includegraphics[width=0.48\textwidth,height=6cm]{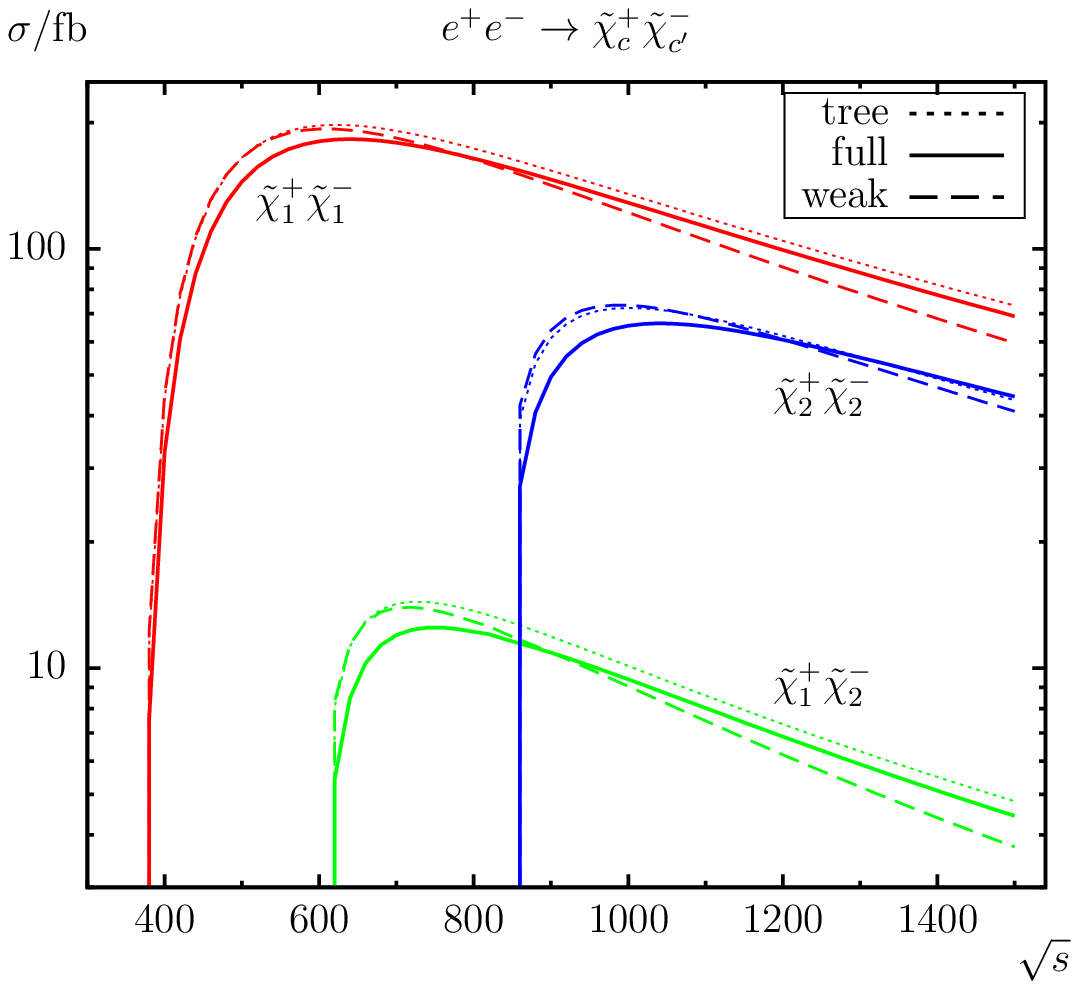}
\includegraphics[width=0.48\textwidth,height=6cm]{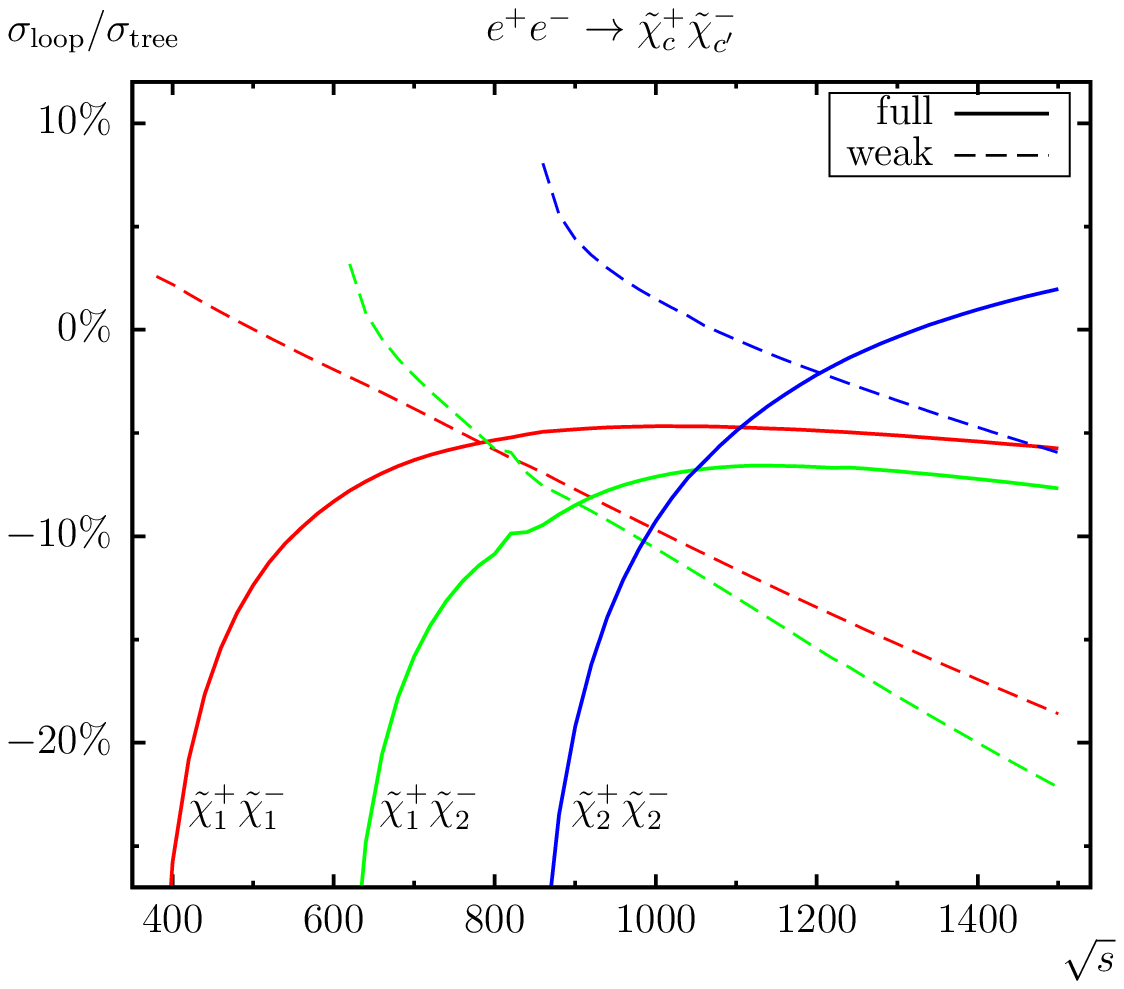}
\\[1em]
\includegraphics[width=0.48\textwidth,height=6cm]{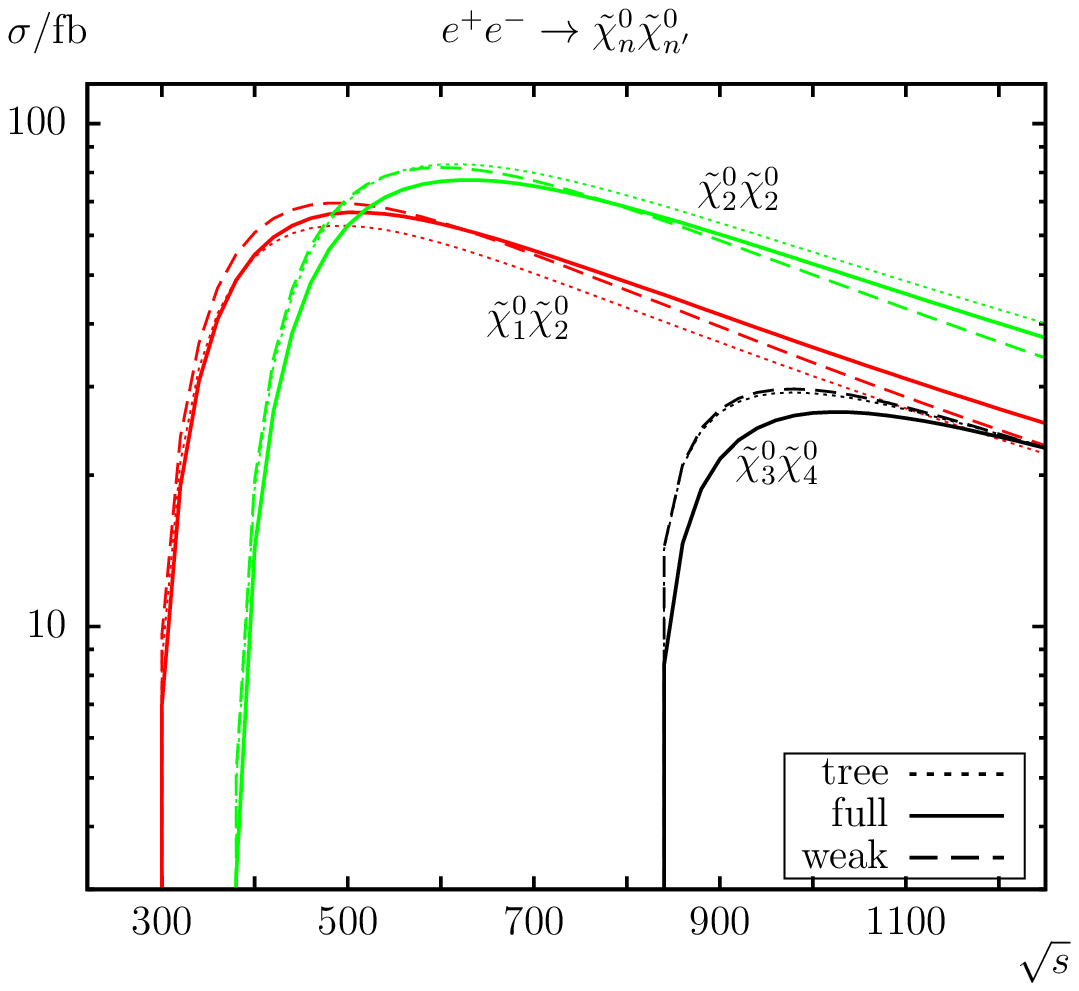}
\includegraphics[width=0.48\textwidth,height=6cm]{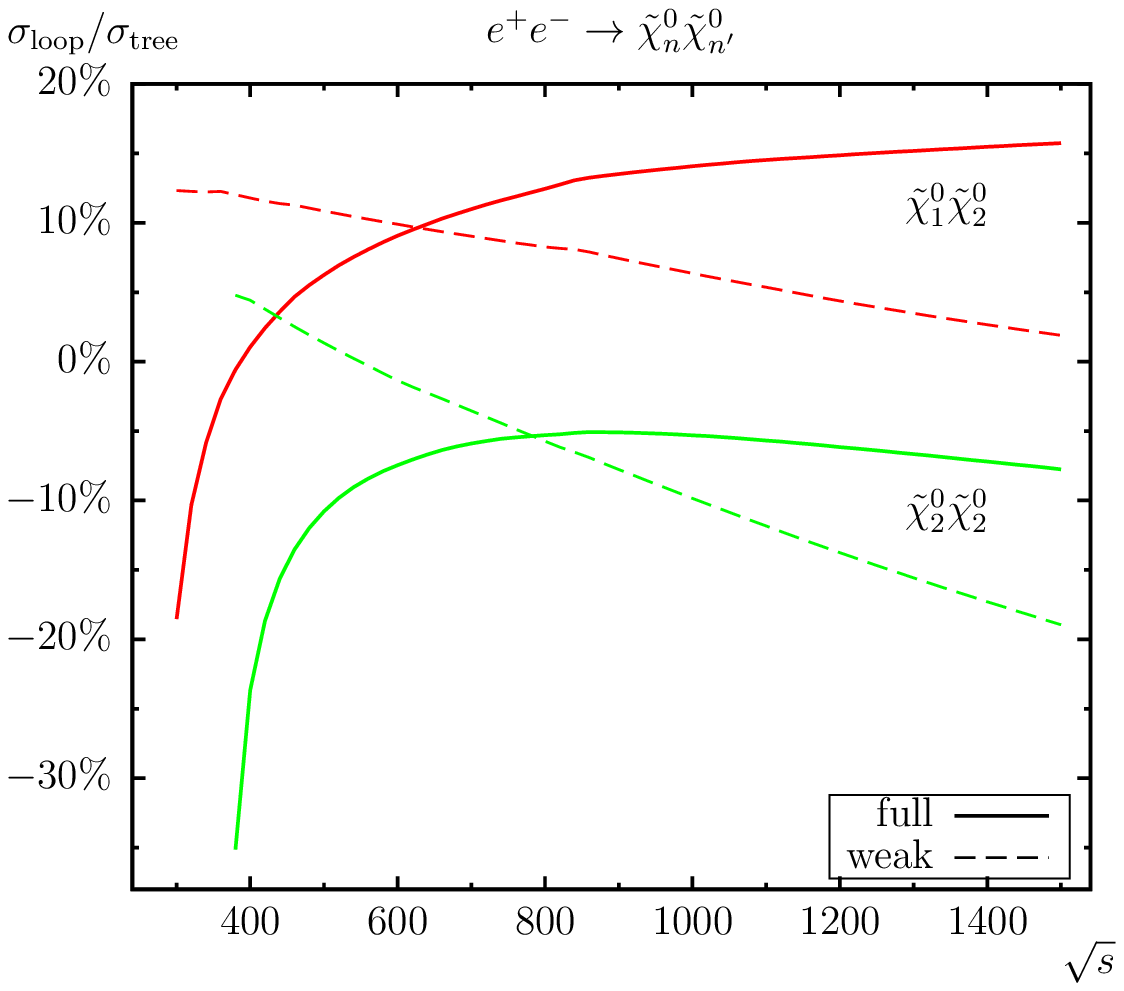}
\end{tabular}
\caption{\label{fig:OlEbMa2005}
  Comparison with \citere{Oller:2005xg} for $\sig(\eecc)$ (upper row) 
  and $\sig(\eenn)$ (lower row).
  The left (right) plot shows cross sections (relative corrections) with 
  $\sqrt{s}$ (in GeV) varied within the on-shell parameter set of
  SPS$1a^{\prime}$.
}
\end{center}
\end{figure}

\item
In \citere{FrHo2004,Fritzsche:2004ek,dissTF} the processes \eecc\
$(c,\cpri = 1,2)$ and \eenn\ $(n,\npri = 1,2,3,4)$ have been calculated in 
the cMSSM, but only real parameters were included in the phenomenological 
analysis.  Unfortunately, in \citeres{FrHo2004,Fritzsche:2004ek} not 
sufficient information about their input parameters where given, rendering 
a comparison impossible.  On the other hand, both papers are contained in 
\citere{dissTF}.  For the comparison with \citere{dissTF} we successfully
reproduced their Tab.~7.1 (see our \refta{tab:dissTF}) and their
Figs.~6.7, 7.2 and 7.3 (see our \reffi{fig:dissTF}, where we show some 
examples).  The (expected) \textit{small} differences at \order{1\%} are 
likely caused by the slightly different renormalization scheme.
An exception is $\eendnd$, where the tree-level cross section is 
accidentally very small, resulting in a larger deviation of the one-loop
corrections.

\begin{table}
\caption{\label{tab:dissTF}
  Comparison of the tree and one-loop corrected neutralino and 
  chargino masses with \citere{dissTF} in the \Code{CCN[1]} scheme. 
  All masses are in GeV.
}
\centering
\begin{tabular}{llrrrrrrr}
\toprule  
 & & $\mneu{1}$ & $\mneu{2}$ & $\mneu{3}$ & $\mneu{4}$ 
& $\mcha{1}$ & $\mcha{2}$ \\
\midrule
                               & \citere{dissTF} 
& 97.75 & 184.55 & 405.10 & 420.46 & 184.20 & 421.24 \\
\raisebox{1.5ex}[-1.5ex]{tree} & \FT             
& 97.75 & 184.55 & 405.09 & 420.46 & 184.20 & 421.24 \\
\midrule
                               & \citere{dissTF} 
& 97.75 & 184.44 & 407.46 & 419.45 & 184.20 & 421.24 \\
\raisebox{1.5ex}[-1.5ex]{loop} & \FT             
& 97.75 & 184.43 & 407.44 & 419.45 & 184.20 & 421.24 \\
\bottomrule
\end{tabular}
\end{table}

\begin{figure}
\begin{center}
\begin{tabular}{c}
\includegraphics[width=0.48\textwidth,height=6cm]{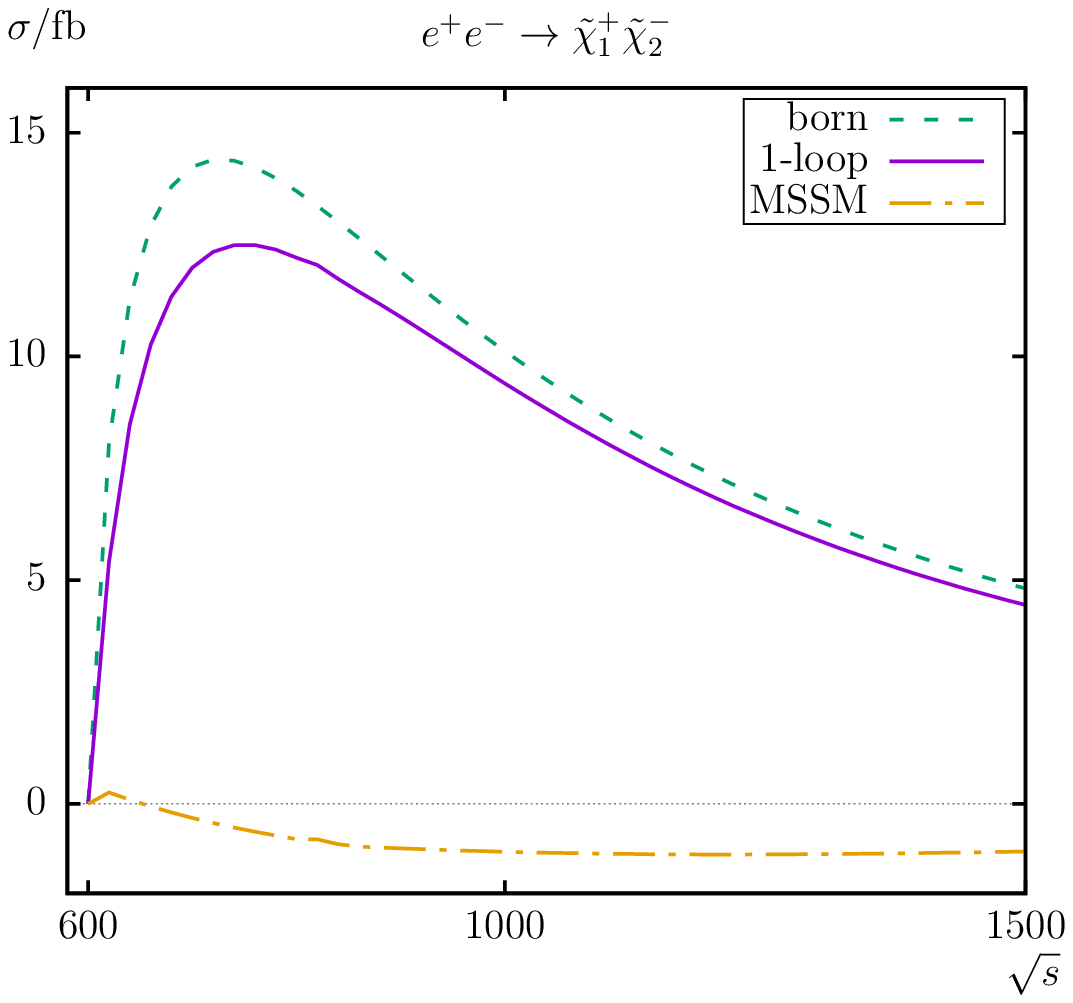}
\includegraphics[width=0.48\textwidth,height=6cm]{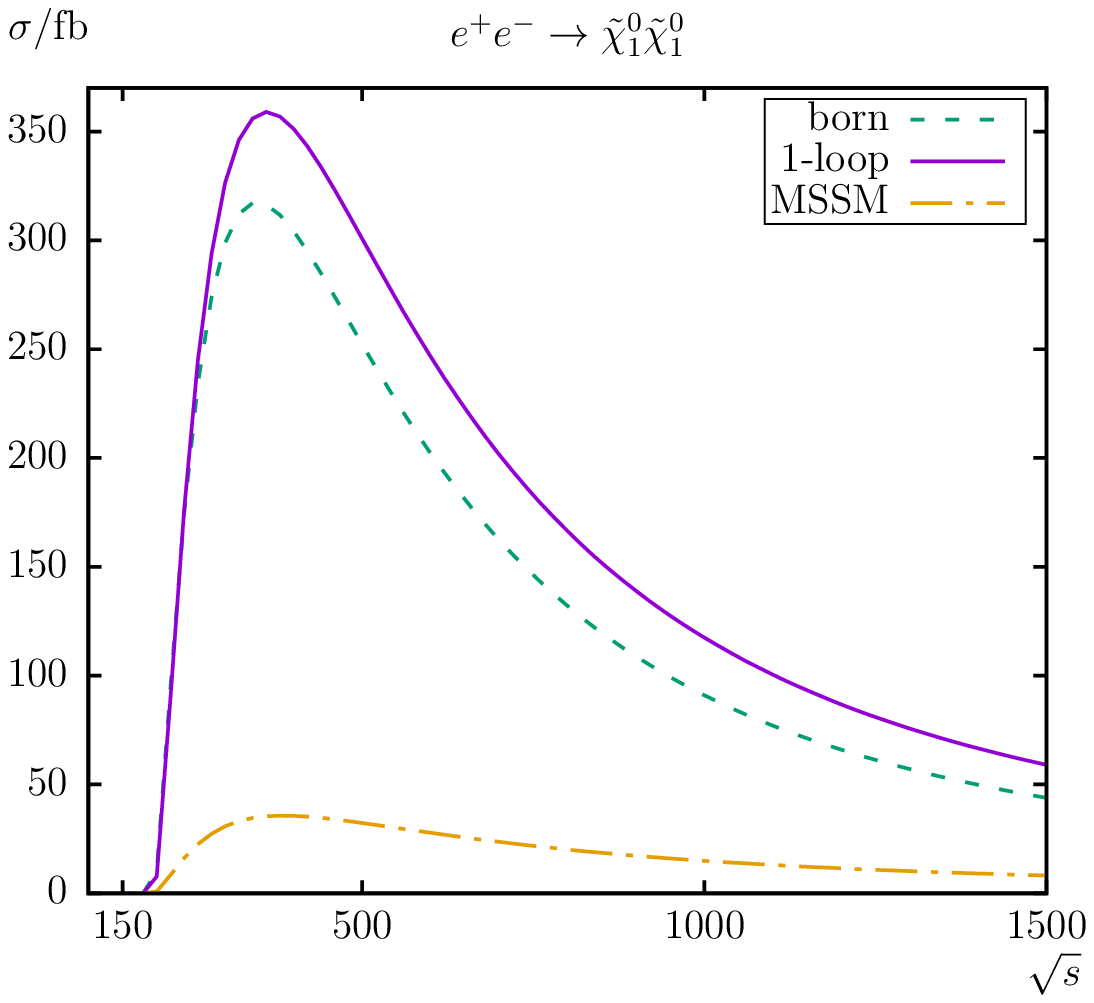}
\\[1em]
\includegraphics[width=0.48\textwidth,height=6cm]{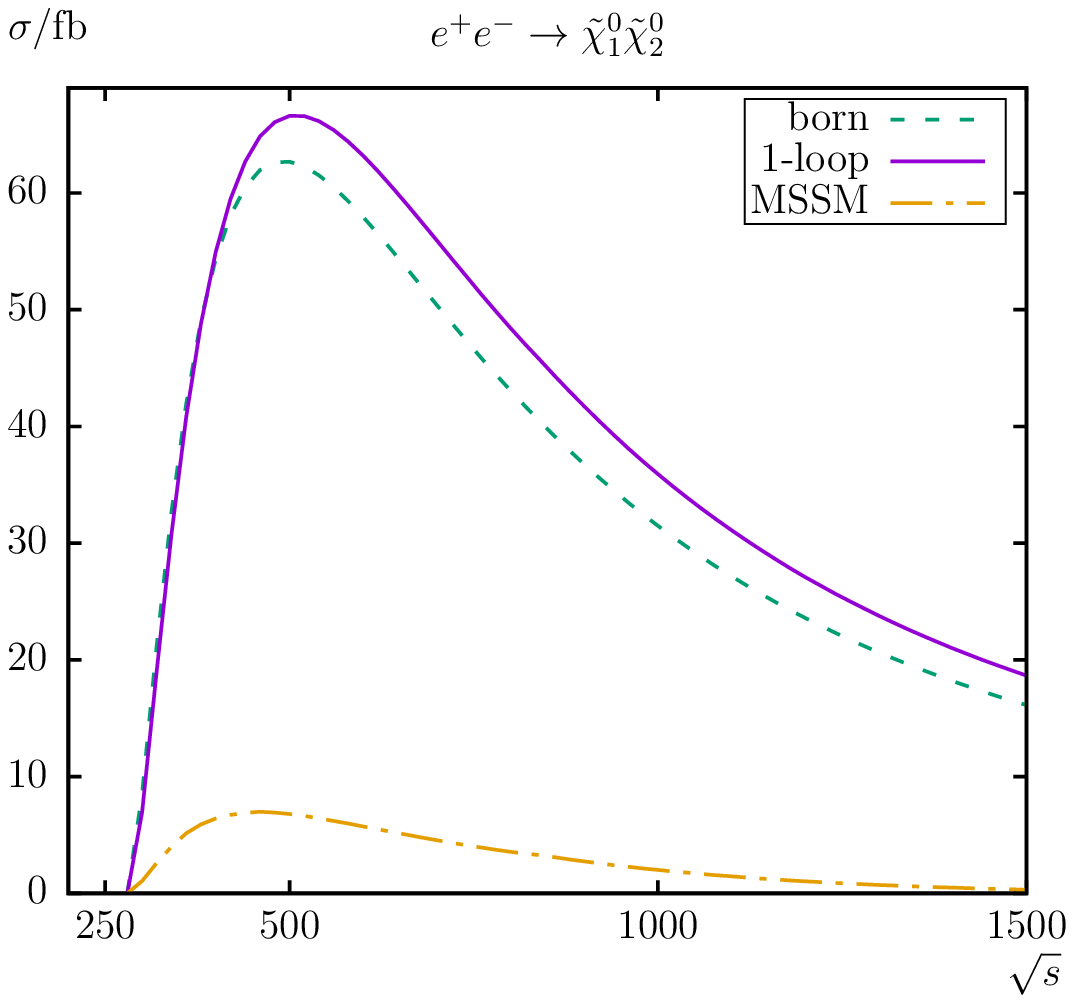}
\includegraphics[width=0.48\textwidth,height=6cm]{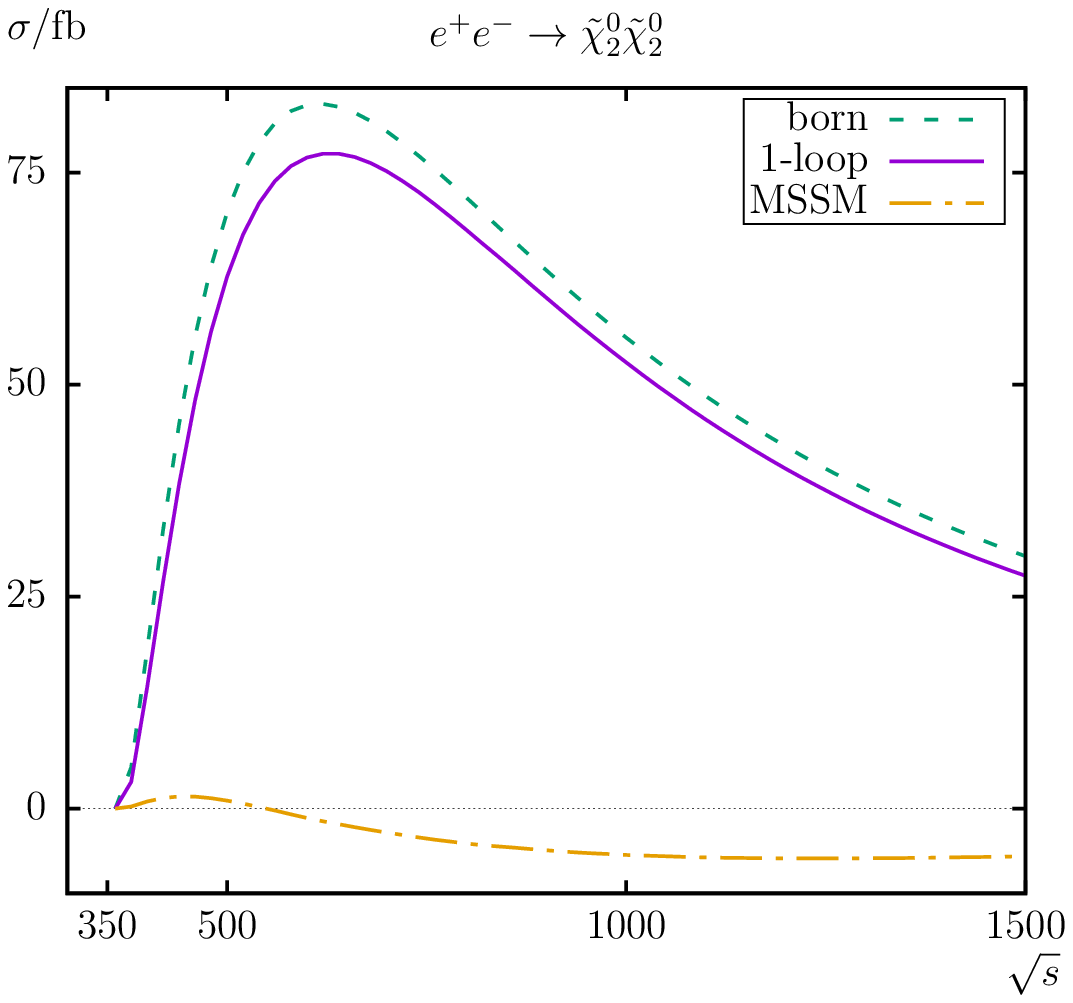}
\\[1em]
\includegraphics[width=0.48\textwidth,height=6cm]{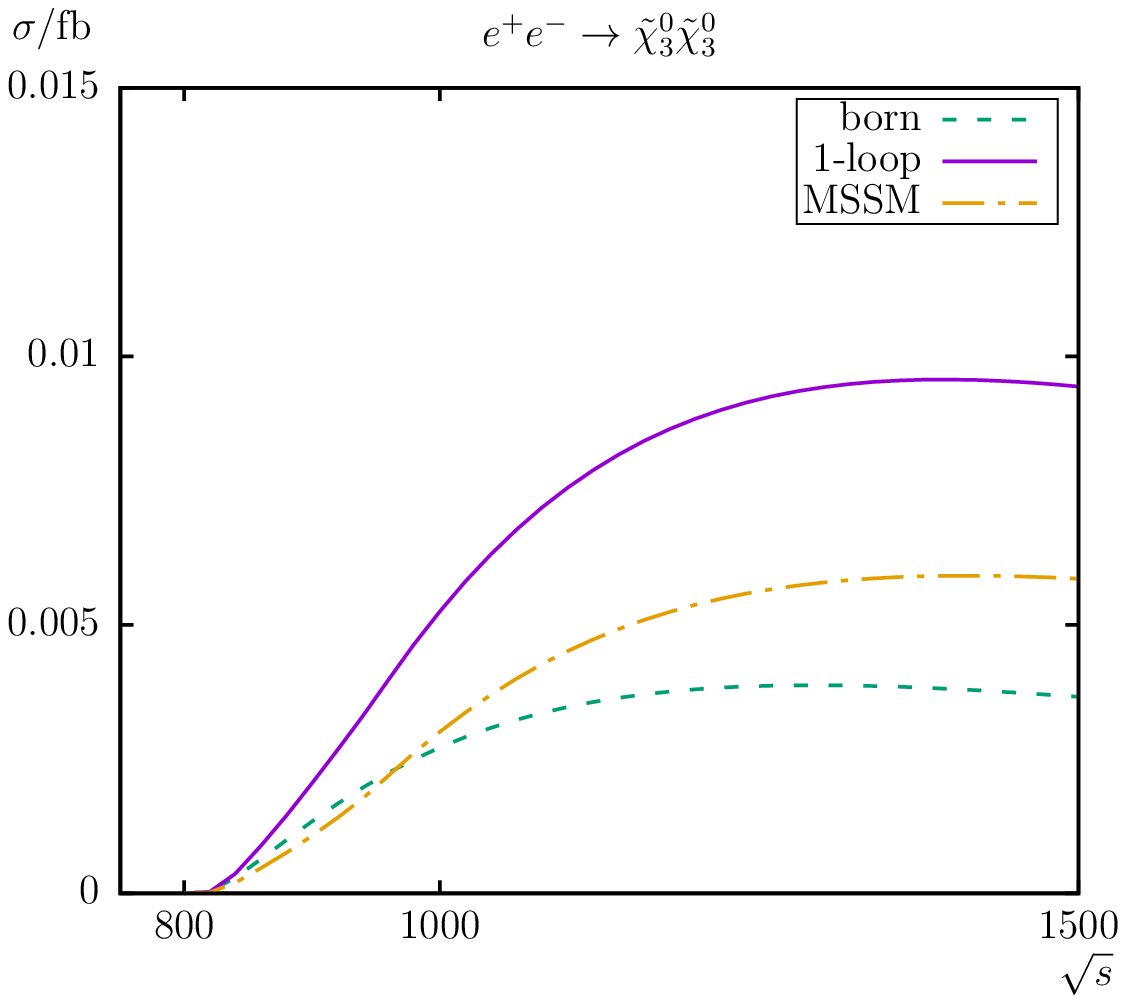}
\includegraphics[width=0.48\textwidth,height=6cm]{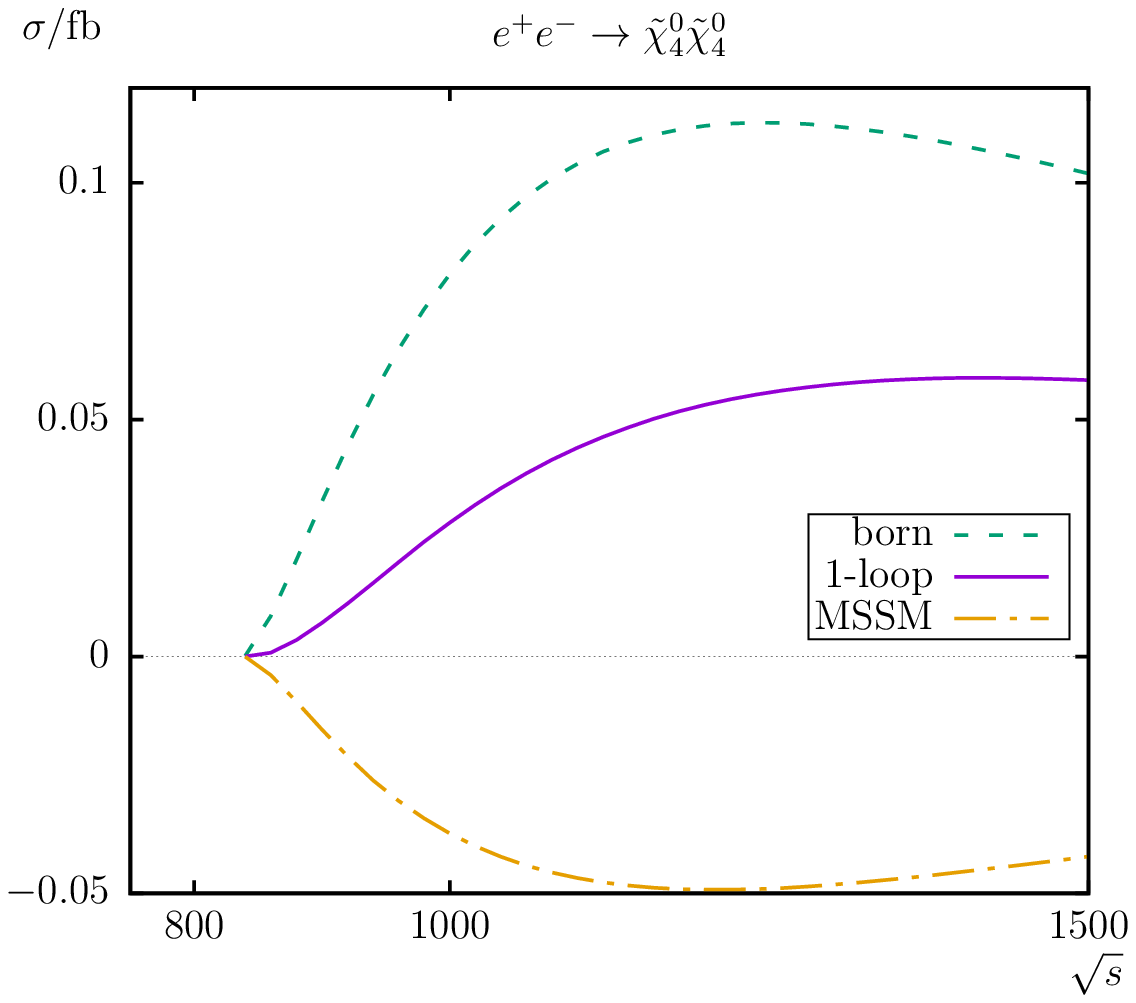}
\end{tabular}
\caption{\label{fig:dissTF}
  Comparison with \citere{dissTF} for $\sig(\eecc)$ and $\sig(\eenn)$.
  As example some of the cross sections (in fb) are shown with the 
  on-shell parameter set SPS$1a^{\prime}$ chosen according to 
  \citere{dissTF}, varied with $\sqrt{s}$ (in GeV). 
}
\end{center}
\end{figure}

\item
Finally, \citere{Diaz:2009um} is (more or less) an extension to 
\citere{Diaz:2002rr} (see the corresponding previous item), dealing with
polarized electrons and charginos and with multi-photon bremsstrahlung
in the rMSSM. The authors claimed that they are in reasonable agreement with 
\citere{Oller:2005xg} within $\pm 2\%$.  We used their input parameters 
as far as possible and reproduced their Fig.~6 in our \reffi{fig:DiRiRo2009}.
The relative corrections agree (away from the production threshold) better 
than $\pm 10\%$. The differences arise for the same reasons as already
described in the comparison with \citere{Oller:2005xg}, see the corresponding 
item above.

\item
The effects of imaginary and absorptive parts were analyzed for \eecc\ in
\citere{Bharucha:2012nx}.  The differences in the renormalization of the 
chargino/neutralino sector between \citere{Bharucha:2012nx} and our work 
are discussed in \citere{LHCxN}.
The chargino/neutralino production in the cMSSM at the full one-loop level 
has been numerically compared with \citere{Bharucha:2012nx} using their 
latest \FA\ model file implementation.  We found overall agreement better 
than 4\% (in the most cases better than 1\%) in the loop corrections for 
real and complex parameters.%
\footnote{
  It should be noted that the original code used for 
  \citere{Bharucha:2012nx} is no longer available~\cite{aoife}, 
  where we found significant numerical differences with the 
  results shown in \citere{Bharucha:2012nx}.
}

\item
For the precise extraction of the underlying SUSY parameters, in
\citere{Bharucha:2012ya} \eecc\ has been calculated at the full one-loop 
level for three cMSSM benchmark points.  As in the previous point, the 
differences in the renormalization of the chargino/neutralino sector 
between \citere{Bharucha:2012ya} and our work are discussed in 
\citere{LHCxN}.  Again, using their latest \FA\ model file implementation,  
we found overall agreement better than 2\% in the loop corrections.
But we found significant numerical differences with the results shown in 
\citere{Bharucha:2012ya}, as already noted in the previous item.

\end{itemize}

To conclude, we found good agreement with the literature where
expected, and the encountered differences can be traced back to
different renormalization schemes, corresponding mismatches in the input
parameters and small differences in the SM parameters. After comparing
to the existing literature we would like to stress again that here we
present for the first time a full one-loop calculation of $\sig(\eenn)$
and $\sig(\eecc)$ in the cMSSM, using the scheme that was employed
successfully already for the full one-loop decays of the (produced)
charginos and neutralinos. The two calculations can readily be used
together for the full production and decay chain.

\begin{figure}
\begin{center}
\begin{tabular}{c}
\includegraphics[width=0.48\textwidth,height=6cm]{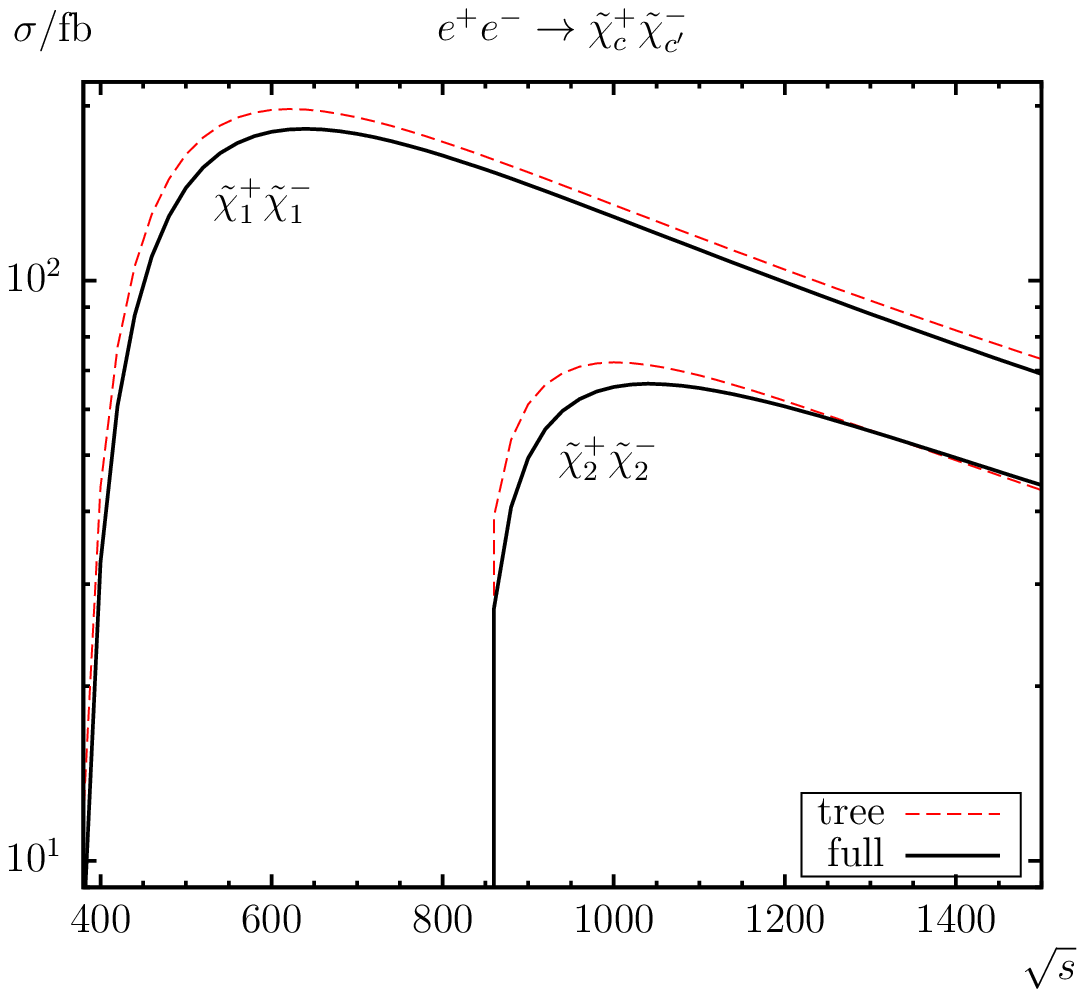}
\\[1em]
\includegraphics[width=0.48\textwidth,height=6cm]{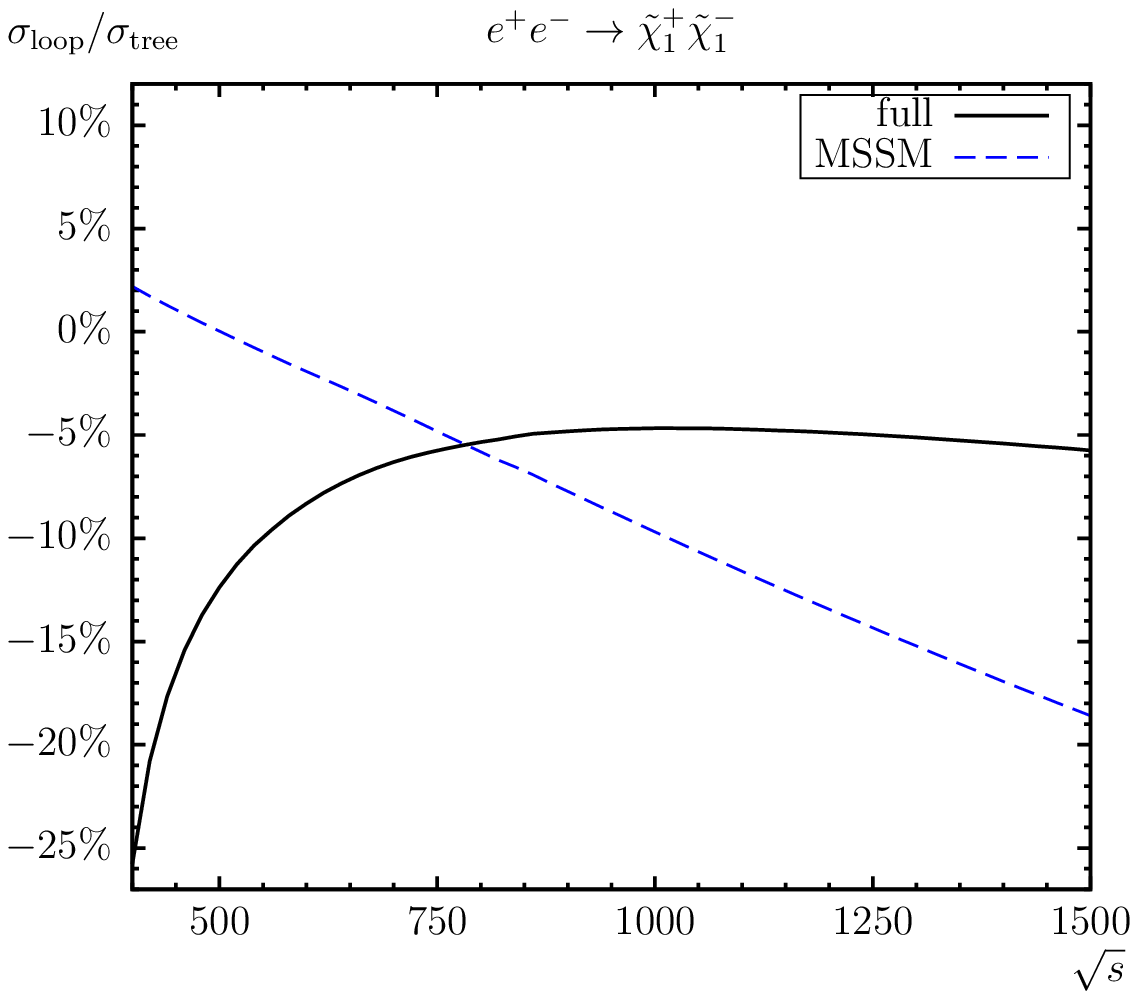}
\includegraphics[width=0.48\textwidth,height=6cm]{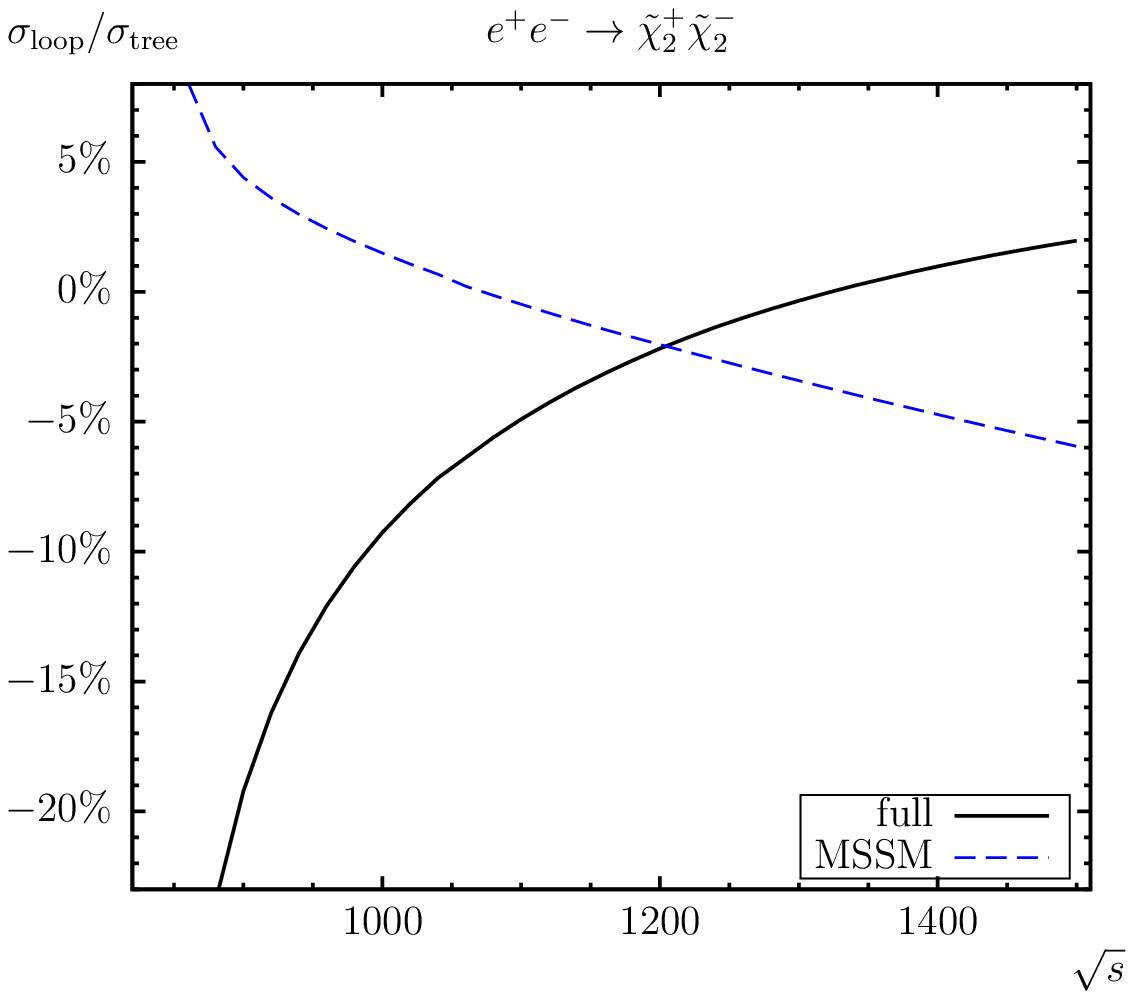}
\end{tabular}
\caption{\label{fig:DiRiRo2009}
  Comparison with \citere{Diaz:2009um} for $\sig(\eecc)$.
  The upper (lower) plot(s) shows cross sections (relative corrections) 
  with $\sqrt{s}$ (in GeV) varied within the SUSY parameter point 
  SPS$1a^{\prime}$.
}
\end{center}
\end{figure}


\section{Numerical analysis}
\label{sec:numeval}

In this section we present our numerical analysis of chargino/neutralino
production at $e^+e^-$ colliders in the cMSSM. 
In the figures below we show the cross sections at the tree level 
(``tree'') and at the full one-loop level (``full''), which is the cross 
section including \textit{all} one-loop corrections as described in 
\refse{sec:calc}.  
The \Code{CCN[1]} scheme (\ie OS conditions for the two charginos and 
the lightest neutralino) has been used for most evaluations. 
For comparative calculations also some \Code{CNN[$c,n,n'$]} schemes 
(OS conditions for one chargino and two neutralinos) have been used, 
as indicated below.

We first define the numerical scenario for the cross section evaluation. 
Then we start the numerical analysis with the cross sections of \eecc\
($c,\cpri = 1,2$) in \refse{sec:eecc}, evaluated as a function of $\sqrt{s}$ 
(up to $3\tev$, shown in the upper left plot of the respective figures), 
$\mu$ (starting at $\mu = 240\gev$ up to $\mu = 2000\gev$, shown in the 
upper right plots), $M_{\tilde L, \tilde E}$ (from 200 to 2000 GeV, lower 
left or middle plots) and $\phiMe$ 
(between $0^{\circ}$ and $360^{\circ}$, lower right or middle plots).
In some cases also the $\TB$ dependence is shown.
Then we turn to the processes \eenn\ ($n,\npri = 1,2,3,4$) in 
\refse{sec:eenn}.  All these processes are of particular interest for 
ILC and CLIC analyses~\cite{ILC-TDR,teslatdr,ilc,CLIC} 
(as emphasized in \refse{sec:intro}).


\subsection{Parameter settings}
\label{sec:paraset}

The renormalization scale $\mu_R$ has been set to the center-of-mass energy, 
$\sqrt{s}$.  The SM parameters are chosen as follows; see also \cite{pdg}:
\begin{itemize}

\item Fermion masses (on-shell masses, if not indicated differently):
\begin{align}
m_e    &= 0.5109989461\mev\,, & m_{\nu_e}    &= 0\,, \notag \\
m_\mu  &= 105.6583745\mev\,,  & m_{\nu_{\mu}} &= 0\,, \notag \\
m_\tau &= 1776.86\mev\,,      & m_{\nu_{\tau}} &= 0\,, \notag \\
m_u &= 71.03\mev\,,          & m_d         &= 71.03\mev\,, \notag \\ 
m_c &= 1.27\gev\,,           & m_s         &= 96.0\mev\,, \notag \\
m_t &= 173.21\gev\,,         & m_b         &= 4.66\gev\,.
\end{align}
According to \citere{pdg}, $m_s$ is an estimate of a so-called 
"current quark mass" in the \MSbar\ scheme at the scale 
$\mu \approx 2\gev$.  $m_c \equiv m_c(m_c)$ is the "running" mass 
in the \MSbar\ scheme and $m_b$ is the $\Upsilon(1S)$ bottom quark mass. 
$m_u$ and $m_d$ are effective parameters, calculated through the 
hadronic contributions to
\begin{align}
\Delta\alpha_{\text{had}}^{(5)}(M_Z) &= 
      \frac{\alpha}{\pi}\sum_{f = u,c,d,s,b}
      Q_f^2 \Bigl(\ln\frac{M_Z^2}{m_f^2} - \frac 53\Bigr) \approx 0.02764\,.
\end{align}

\item Gauge-boson masses\index{gaugebosonmasses}:
\begin{align}
M_Z = 91.1876\gev\,, \qquad M_W = 80.385\gev\,.
\end{align}

\item Coupling constant\index{couplingconstants}:
\begin{align}
\alpha(0) = 1/137.035999139\,.
\end{align}
\end{itemize}

The SUSY parameters are chosen according to the scenario \Scs, shown in 
\refta{tab:para}.  This scenario is viable for the various cMSSM 
chargino/neutralino production modes, \ie not picking specific parameters 
for each cross section.  They are in particular in agreement with the 
chargino and neutralino searches of ATLAS~\cite{ATLAS-CN} and 
CMS~\cite{CMS-CN}.

\begin{table}[t]
\caption{\label{tab:para}
  MSSM default parameters for the numerical investigation; all parameters 
  (except of $\TB$) are in GeV.  The values for the trilinear sfermion 
  Higgs couplings, $A_{t,b,\tau}$ are chosen to be real and such that charge- 
  and/or color-breaking minima are avoided \cite{ccb}.  (It should be noted 
  that we chose common values for the three sfermion generations.)
  For the charginos and neutralinos we show the tree-level values as well as 
  their OS masses in the \Code{CCN[1]}, \Code{CNN[1,1,3]}, \Code{CNN[1,1,4]}, 
  \Code{CNN[2,1,2]}, and \Code{CNN[2,1,3]} scheme as obtained from 
  \refeq{eq:minoOS}. Also shown are some values for the complex phase 
  $\phiMe$ in the \Code{CCN[1]} scheme. 
}
\centering
\begin{tabular}{lrrrrrrrrrrrr}
\toprule
Scen. & $\sqrt{s}$ & $\TB$ & $\mu$ & $\MHp$ & $M_{\tilde Q, \tilde U, \tilde D}$ & 
$M_{\tilde L, \tilde E}$ & $|\At|$ & $\Ab$ & $A_{\tau}$ & 
$|M_1|$ & $M_2$ & $M_3$ \\ 
\midrule
\Scs & 1000 & 10 & 450 & 500 & 1500 & 1500 & 2000 & $|\At|$ &
$M_{\tilde L}$ & $\mu$/4 & $\mu$/2 & 2000 \\
\bottomrule
\end{tabular}

\vspace{0.5em}

\begin{tabular}{lrrrrrr}
\toprule
& $\mcha1$ & $\mcha2$ & $\mneu1$ & $\mneu2$ & $\mneu3$ & $\mneu4$ \\
\midrule
tree                  & 212.760 & 469.874 & 110.434 & 213.002 & 455.162 & 469.226 \\
\midrule
\Code{CCN[1]}         & 212.760 & 469.874 & 110.434 & 212.850 & 455.195 & 469.560 \\
\midrule
$\phiMe = 90^{\circ}$  & 212.760 & 469.874 & 111.356 & 212.722 & 455.628 & 468.972 \\
$\phiMe = 180^{\circ}$ & 212.760 & 469.874 & 112.274 & 212.593 & 456.105 & 468.340 \\
\midrule
\Code{CNN[1,1,3]}     & 212.760 & 469.844 & 110.434 & 212.850 & 455.162 & 469.530 \\
\Code{CNN[1,1,4]}     & 212.760 & 469.539 & 110.434 & 212.850 & 454.837 & 469.226 \\
\Code{CNN[2,1,2]}     & 212.912 & 469.874 & 110.434 & 213.002 & 455.184 & 469.561 \\
\Code{CNN[2,1,3]}     & 213.214 & 469.874 & 110.434 & 213.303 & 455.162 & 469.563 \\
\bottomrule
\end{tabular}
\end{table}

It should be noted that higher-order corrected Higgs boson masses do not 
enter our calculation.%
\footnote{
  Since we work in the MSSM with complex parameters, 
  $\MHp$ is chosen as input parameter, and higher-order 
  corrections affect only the neutral Higgs boson spectrum; 
  see \citere{chargedmhiggs2L} for the most recent evaluation.
}
However, we ensured that over larger parts of the parameter space the 
lightest Higgs boson mass is around $\sim 125 \pm 3\gev$ to indicate the
phenomenological validity of our scenarios.  
In our numerical evaluation we will show the variation with $\sqrt{s}$,
$\mu$, $\MSL = \MSE$, and $\phiMe$, the phase of $M_1$. 
The dependence of $\TB$ turned out to be rather small, therefore 
we show it only in a few cases, where it is of special interest.

Concerning the complex parameters, some more comments are in order.
Potentially complex parameters that enter the chargino/neutralino
production cross sections at tree-level are the soft
SUSY-breaking parameters $M_1$ and $M_2$ as well as the Higgs mixing
parameter~$\mu$.  However, when performing an analysis involving complex 
parameters it should be noted that the results for physical observables 
are affected only by certain combinations of the complex phases of the 
parameters $\mu$, the trilinear couplings $A_f$ and the gaugino mass 
parameters $M_{1,2,3}$~\cite{MSSMcomplphasen,SUSYphases}.
It is possible, for instance, to rotate the phase $\phiMz$ away.
Experimental constraints on the (combinations of) complex phases 
arise, in particular, from their contributions to electric dipole 
moments of the electron and the neutron (see \citeres{EDMrev2,EDMPilaftsis} 
and references therein), of the deuteron~\cite{EDMRitz} and of heavy 
quarks~\cite{EDMDoink}.
While SM contributions enter only at the three-loop level, due to its
complex phases the MSSM can contribute already at one-loop order.
Large phases in the first two generations of sfermions can only be 
accommodated if these generations are assumed to be very heavy 
\cite{EDMheavy} or large cancellations occur~\cite{EDMmiracle};
see, however, the discussion in \citere{EDMrev1}. 
A review can be found in \citere{EDMrev3}.
Recently additional constraints at the two-loop level on some $\CP$ 
phases of SUSY models have been investigated in \citere{Ya2013}.
Accordingly (using the convention that $\phiMz = 0$, as done in this paper), 
in particular, the phase $\phimu$ is tightly constrained~\cite{plehnix}, 
and we set it to zero. On the other hand, the bounds on the phases of the
third-generation trilinear couplings are much weaker. Consequently, 
the largest effects on the neutralino production cross sections at the 
tree-level are expected from the complex gaugino mass parameter $M_1$, 
\ie from $\phiMe$.  At the loop level the largest effects are expected 
from contributions involving large Yukawa couplings, and thus $\phiAt$ 
potentially has the strongest impact.  This motivates our choice of 
$\phiMe$ and $\phiAt$ as parameters to be varied.

Since now complex parameters can appear in the couplings, contributions 
from absorptive parts of self-energy type corrections on external legs 
can arise.  The corresponding formulas for an inclusion of these 
absorptive contributions via finite wave function correction factors can 
be found in \citeres{MSSMCT,Stop2decay}.

The numerical results shown in the next subsections are of course 
dependent on the choice of the SUSY parameters.  Nevertheless, they 
give an idea of the relevance of the full one-loop corrections.


\subsection{The process \boldmath{\eecc}}
\label{sec:eecc}

\begin{figure}[t]
\begin{center}
\begin{tabular}{c}
\includegraphics[width=0.48\textwidth,height=6cm]{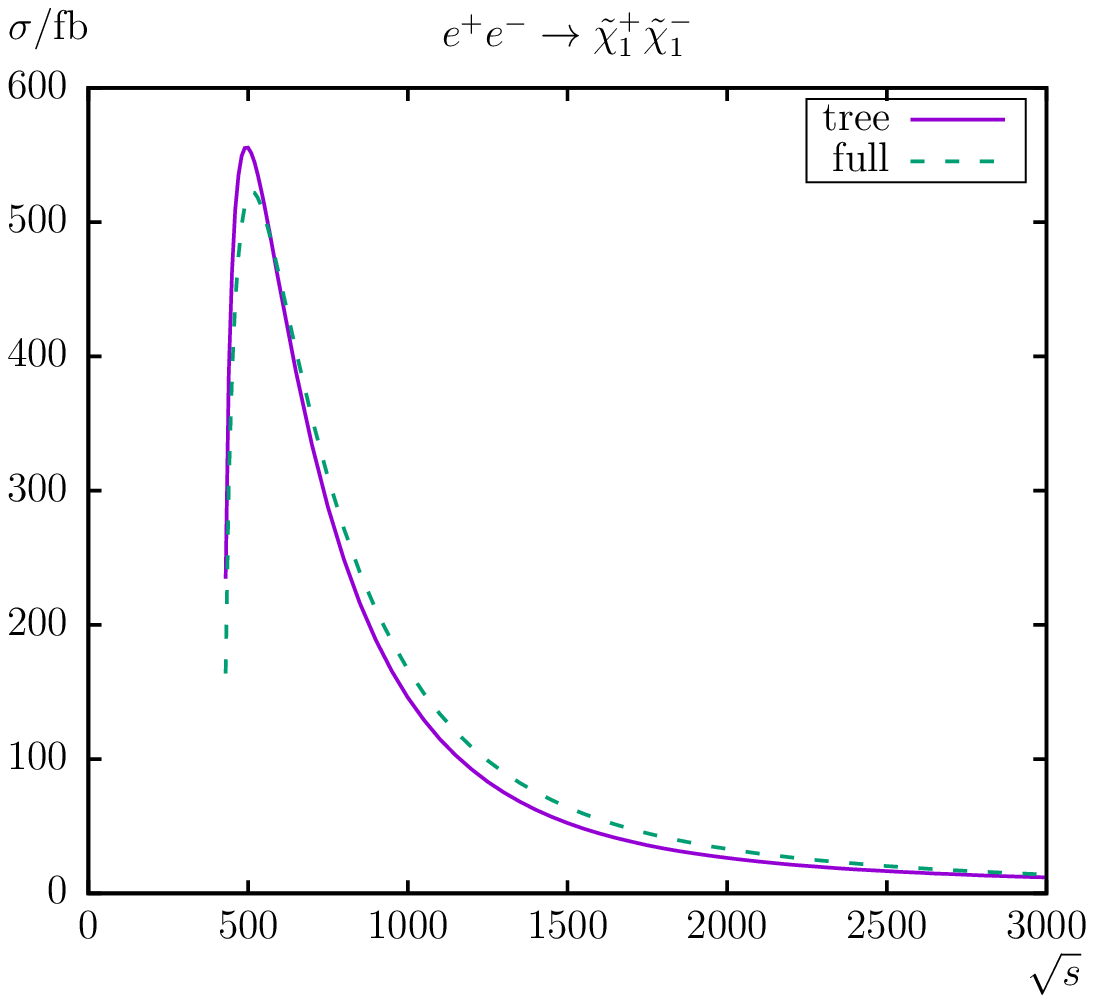}
\includegraphics[width=0.48\textwidth,height=6cm]{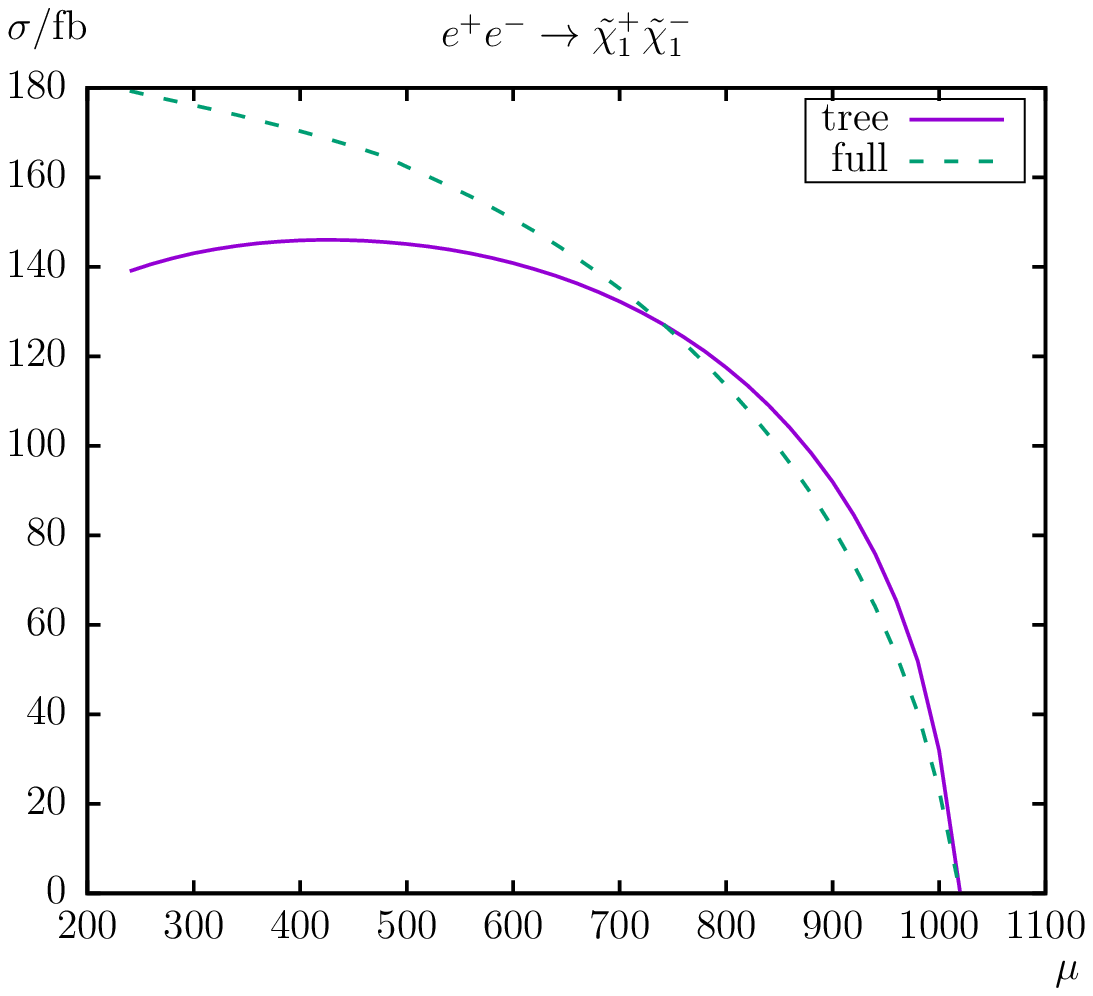}
\\[1em]
\includegraphics[width=0.48\textwidth,height=6cm]{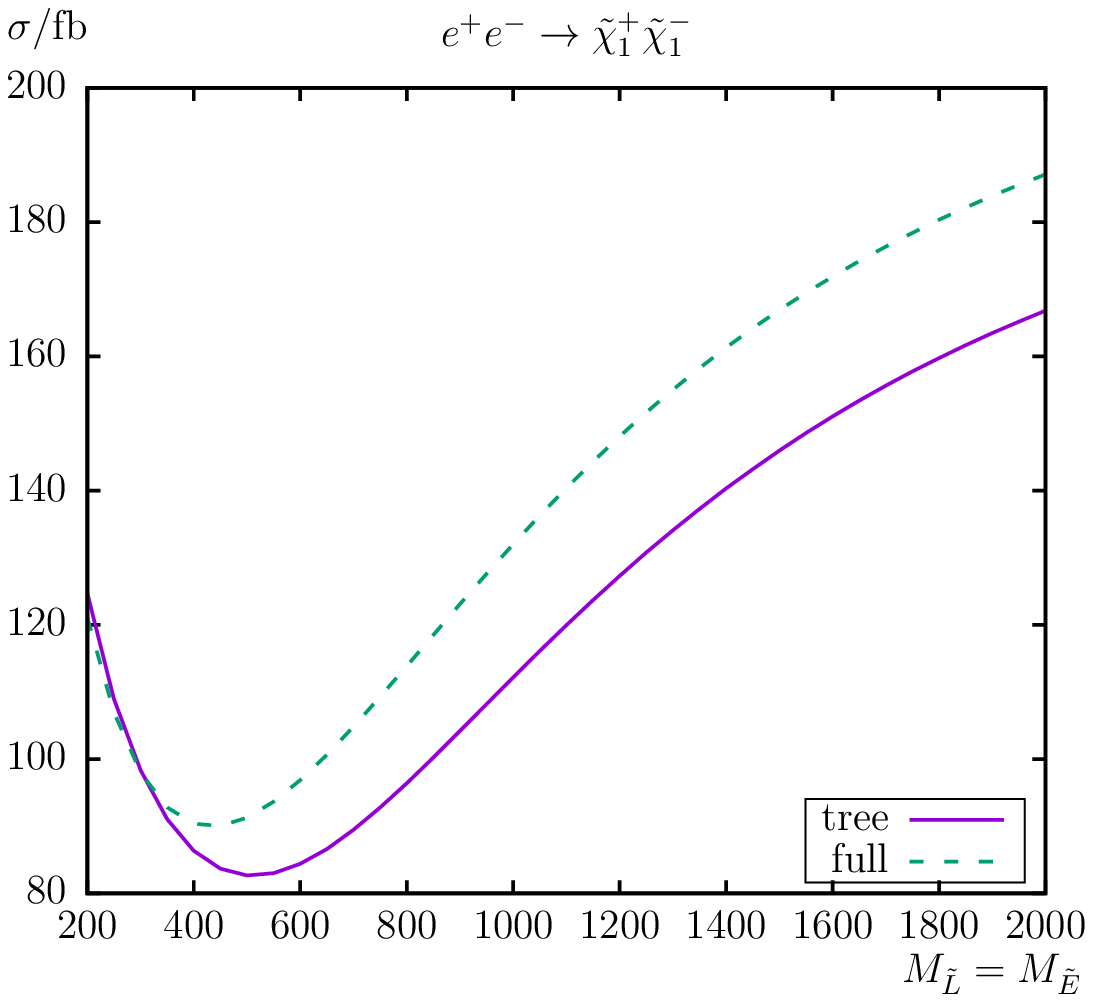}
\includegraphics[width=0.48\textwidth,height=6cm]{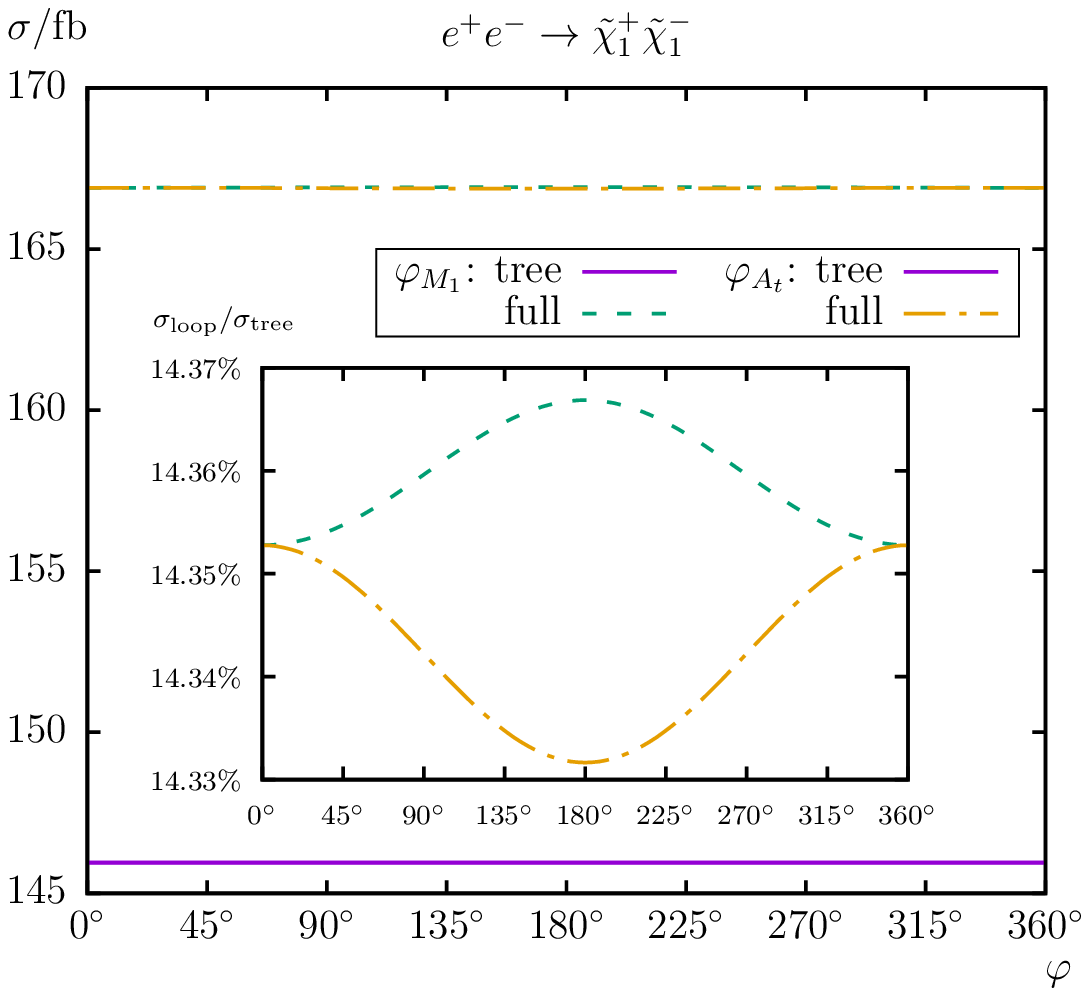}
\end{tabular}
\caption{\label{fig:eec1c1}
  $\sig(\eecece)$.
  Tree-level and full one-loop corrected cross sections are shown with 
  parameters chosen according to \Scs; see \refta{tab:para}.
  The upper plots show the cross sections with $\sqrt{s}$ (left) and 
  $\mu$ (right) varied;  the lower plots show $\MSL = \MSE$ (left) and 
  $\phiMe$, $\phiAt$ (right) varied.
}
\end{center}
\end{figure}

The process $\eecece$ is shown in \reffi{fig:eec1c1}. 
It should be noted that for $s \to \infty$ decreasing cross sections 
$\propto 1/s$ are expected; see \citere{Bartl:1985fk}.
If not indicated otherwise, unpolarized electrons and positrons are 
assumed.  We also remind the reader that $\sig(\eecc)$ denotes the 
sum of the two charge conjugated processes $\forall\; c \neq c^{\prime}$; 
see \refeq{eeCCsum}.

In the analysis of the production cross section as a function of $\sqrt{s}$ 
(upper left plot) we find the expected behavior: a strong rise close to the 
production threshold, followed by a decrease with increasing $\sqrt{s}$. 
We find a very small shift \wrt $\sqrt{s}$ around the production threshold. 
Away from the production threshold, loop corrections of $\sim -8\%$ at 
$\sqrt{s} = 500\gev$ and $\sim +14\%$ at $\sqrt{s} = 1000\gev$ are found 
in scenario \Scs\ (see \refta{tab:para}), with a ``tree crossing'' 
(\ie where the loop corrections become approximately zero and therefore 
cross the tree-level result) at $\sqrt{s} \approx 575\gev$.
The relative size of loop corrections increase with increasing $\sqrt{s}$ 
(and decreasing $\sig$) and reach $\sim +19\%$ at $\sqrt{s} = 3000\gev$.

With increasing $\mu$ in \Scs\ (upper right plot) we find a strong decrease 
of the production cross section, as can be expected from kinematics, 
discussed above.  The relative loop corrections in \Scs\ reach $\sim +30\%$ 
at $\mu = 240\gev$ (at the border of the experimental limit), $\sim +14\%$ 
at $\mu = 450\gev$ (\ie \Scs) and $\sim -30\%$ at $\mu = 1000\gev$. 
In the latter case these large loop corrections are due to the (relative) 
smallness of the tree-level results, which goes to zero for $\mu = 1020\gev$
(\ie the chargino production threshold).

The cross section as a function of $\MSL$ ($= \MSE$) is shown in the lower 
left plot of \reffi{fig:eec1c1}.  This mass parameter controls the
$t$-channel exchange of first generation sleptons at tree-level.
First a small decrease down to $\sim 90$~fb can be observed for 
$\MSL \approx 400\gev$.  For larger $\MSL$ the cross section rises up 
to $\sim 190$~fb for $\MSL = 2\tev$.
In scenario \Scs\ we find a substantial increase of the cross sections from 
the loop corrections.  They reach the maximum of $\sim +18\%$ at 
$\MSL \approx 850\gev$ with a nearly constant offset of about $20$~fb
for higher values of $\MSL$.

Due to the absence of $\phiMe$ in the tree-level production cross section 
the effect of this complex phase is expected to be small.  Correspondingly
we find that the phase dependence $\phiMe$ of the cross section in our
scenario is tiny.  The loop corrections are found to be nearly independent 
of $\phiMe$ at the level below $\sim +0.1\%$ in \Scs.
We also show the variation with $\phiAt$, which enter via final state 
vertex corrections.  While the variation with $\phiAt$ is somewhat larger 
than with $\phiMe$, it remains tiny and unobservable.

\begin{figure}[p]
\begin{center}
\begin{tabular}{c}
\includegraphics[width=0.48\textwidth,height=6cm]{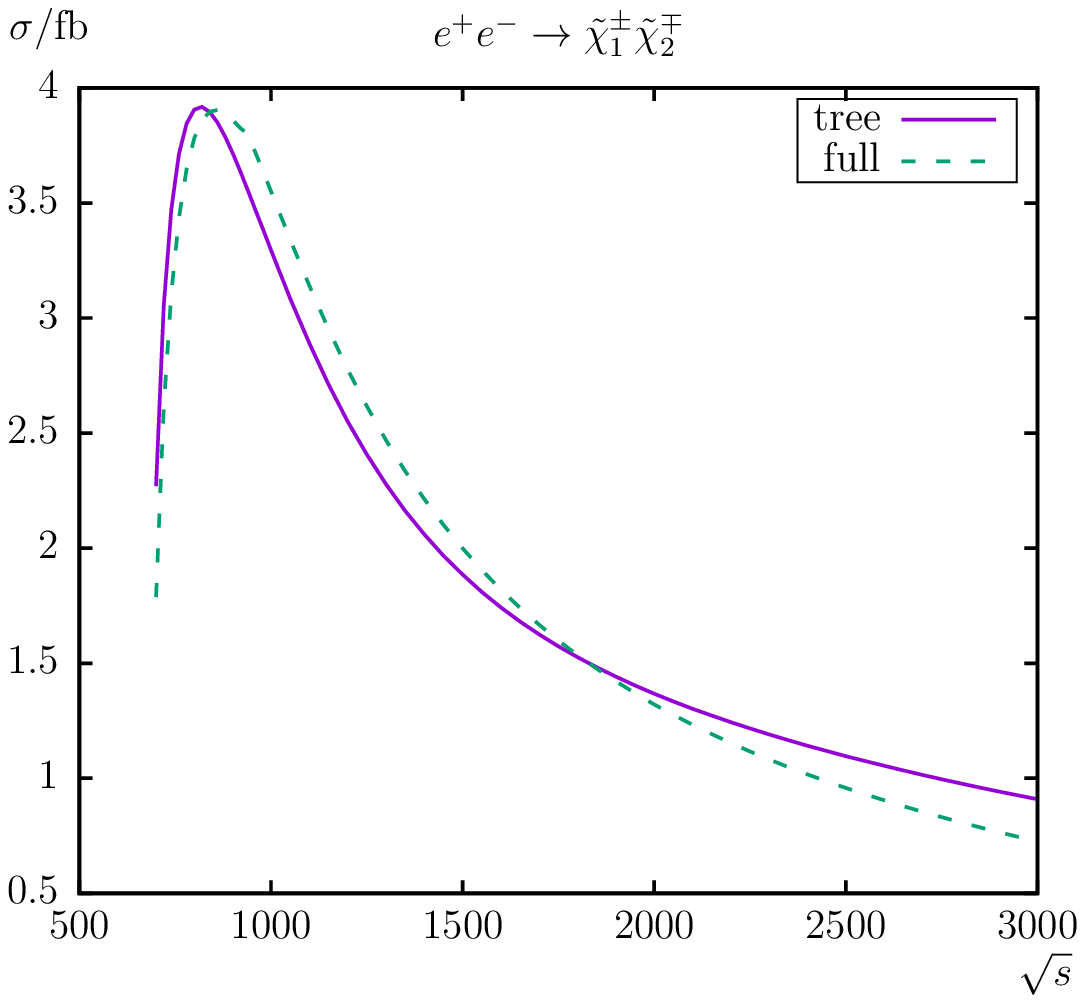}
\includegraphics[width=0.48\textwidth,height=6cm]{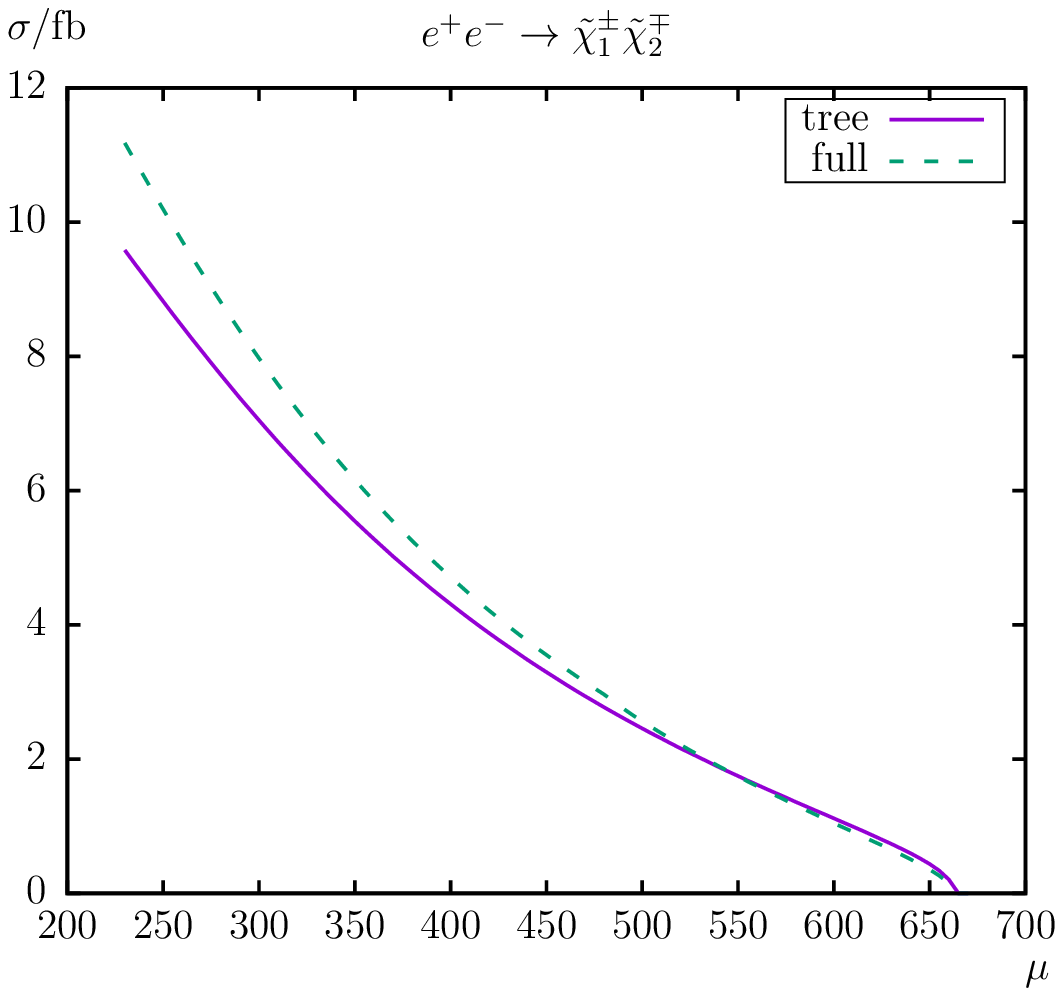}
\\[1em]
\includegraphics[width=0.48\textwidth,height=6cm]{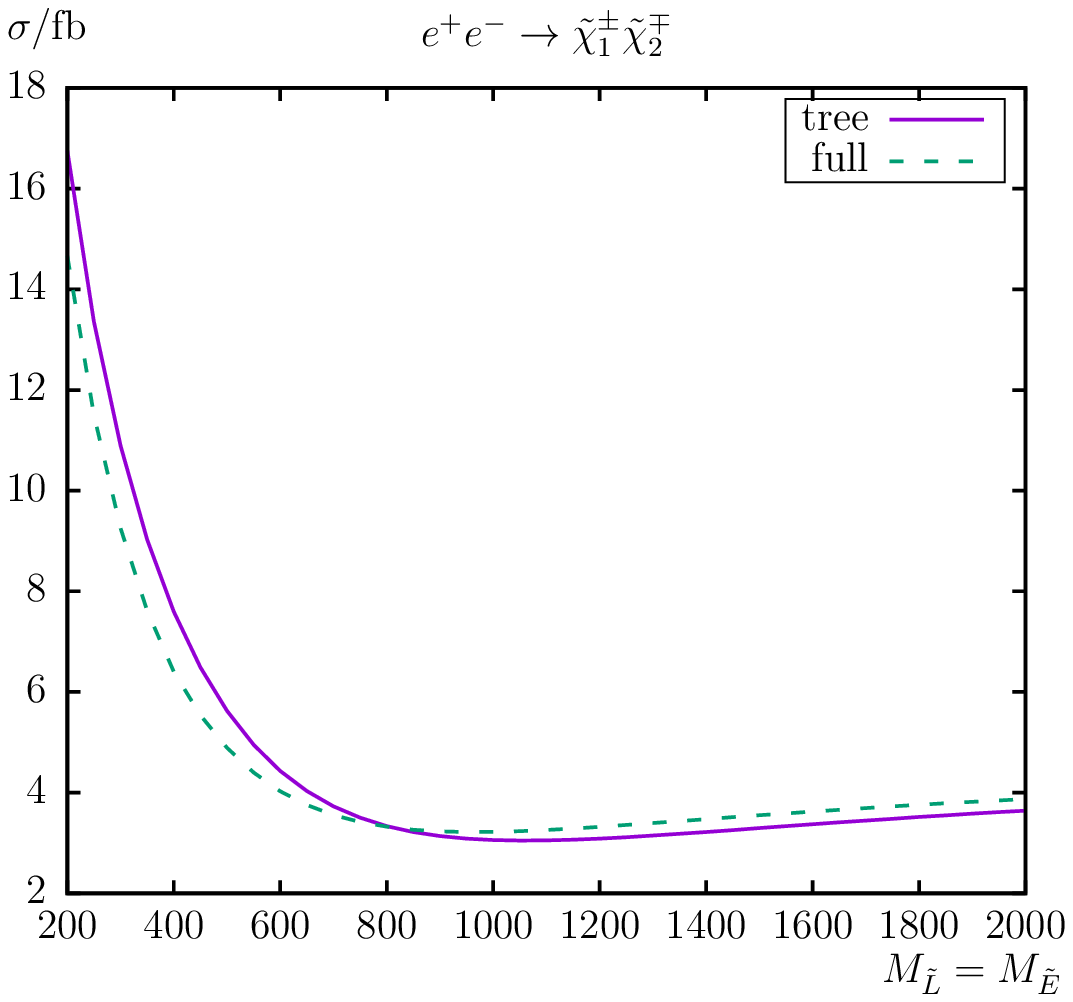}
\includegraphics[width=0.48\textwidth,height=6cm]{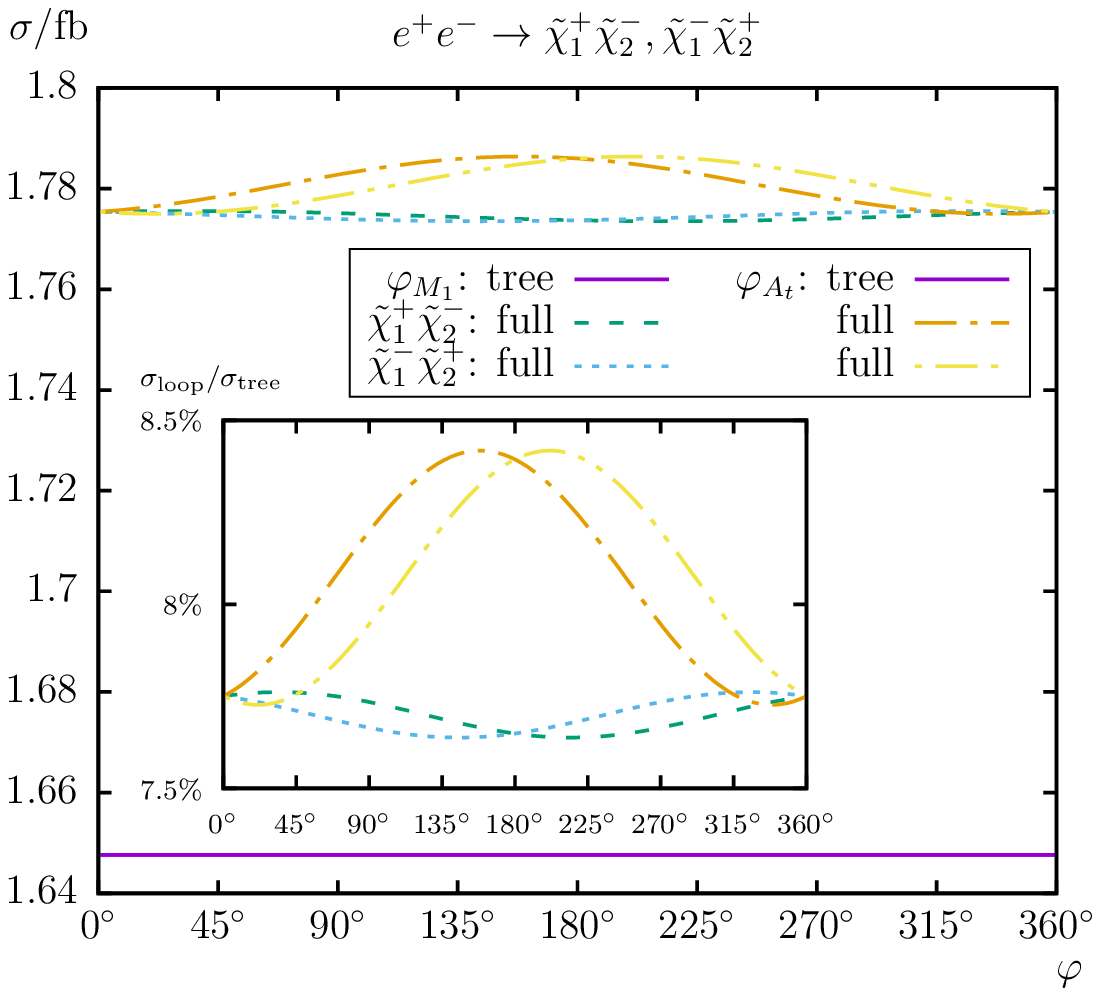}
\\[1em]
\includegraphics[width=0.48\textwidth,height=6cm]{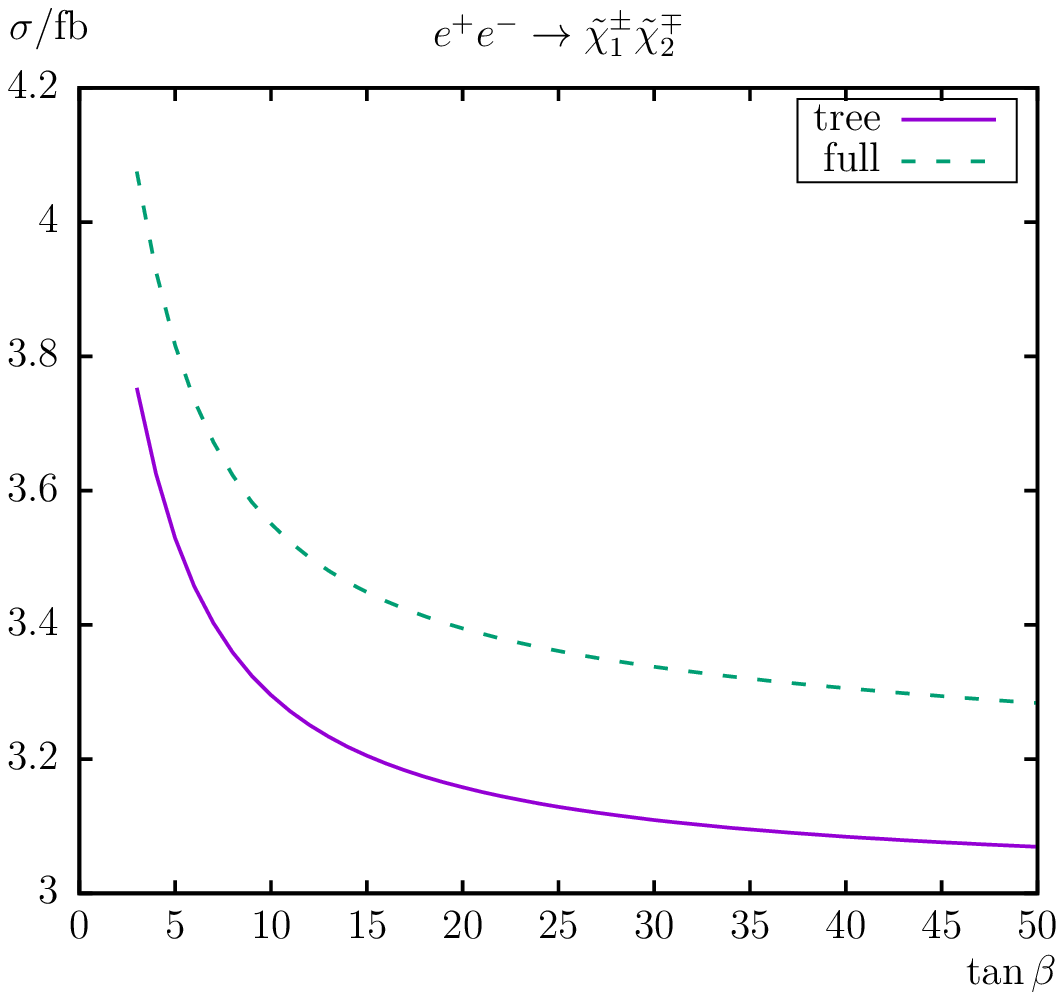}
\includegraphics[width=0.48\textwidth,height=6cm]{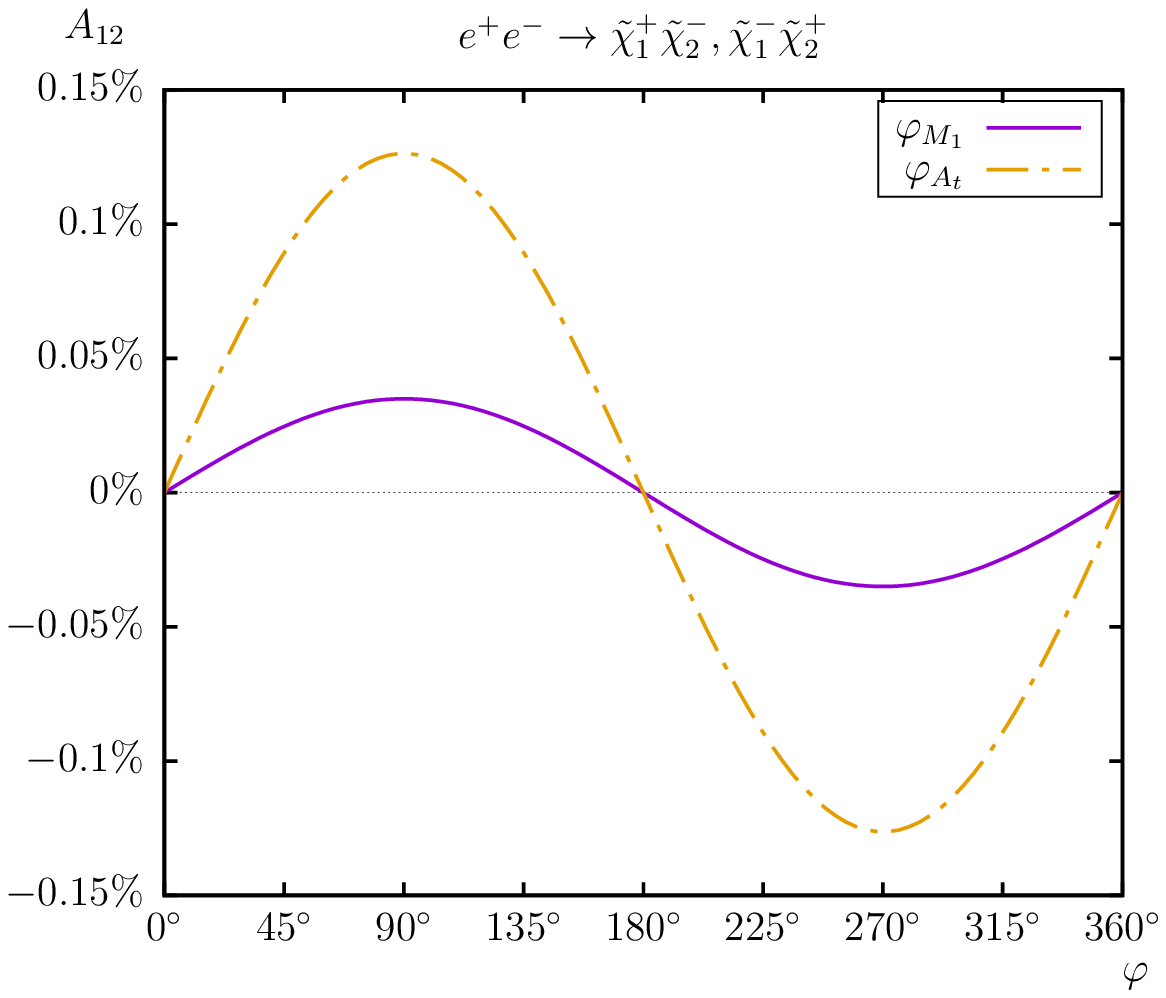}
\end{tabular}
\caption{\label{fig:eec1c2}
  $\sig(\eececz)$.
  Tree-level and full one-loop corrected cross sections are shown 
  with parameters chosen according to \Scs; see \refta{tab:para}.
  The upper plots show the cross sections with $\sqrt{s}$ (left) and 
  $\mu$ (right) varied; the middle plots show $\MSL = \MSE$ (left) and 
  $\phiMe, \phiAt$ (right) varied; the lower plot the variation with $\TB$
  (left) and the $\CP$-odd observable $A_{12}$ (right) varied with the 
  complex phases $\phiMe$ and $\phiAt$. 
  All masses and energies are in GeV.
}
\end{center}
\end{figure}

\medskip

In \reffi{fig:eec1c2} we present the cross sections $\eececz$.
In the analysis as a function of $\sqrt{s}$ (upper left plot) we find 
as before a tiny shift \wrt $\sqrt{s}$, where the position of
the maximum cross section shifts by about $+50\gev$.  The relative
corrections are found to be of $\sim +8\%$ at $\sqrt{s} = 1000\gev$ 
(\ie \Scs), and $\sim -20\%$ at $\sqrt{s} = 3000\gev$.  The peak 
(hardly visible in the dotted line) at $\sqrt{s} \approx 940\gev$ is 
the production threshold $\mcha2 + \mcha2 = \sqrt{s}$.

The dependence on $\mu$ (upper row, right plot) is nearly linear, and 
mostly due to kinematics.  The loop corrections are $\sim +17\%$ at 
$\mu = 240\gev$, $\sim +8\%$ at $\mu = 450\gev$ (\ie \Scs), and 
$\sim -27\%$ at $\mu = 660\gev$ where the tree level goes to zero
(\ie the chargino production threshold).
These large loop corrections are again due to the (relative) smallness of 
the tree-level results at $\mu \approx 660\gev$.

As a function of $\MSL$ (middle left plot) the cross section is rather 
flat for $\MSL \gsim 800\gev$.  The relative corrections increase from 
$\sim -16\%$ at $\mu = 400\gev$ to $\sim +8\%$ at $\mu = 1300\gev$, with 
a tree crossing at $\mu \approx 800\gev$.

The dependence on $\phiMe$ (middle right plot) is again very small, the 
loop corrections are found to be nearly independent of $\phiMe$ below 
the level of $\sim +0.1\%$.  We show separately the cross sections for 
$\eecepczm$ and $\eecemczp$. As the inlay shows they differ from each other, 
but only at an (experimentally indistinguishable) level of \order{0.1\%}.
In addition we also show here the dependence on $\phiAt$, which turns 
out to be substantially larger than the effects of $\phiMe$. 
They are found at the level of $\sim +0.7\%$, most likely below the 
level of observation.

For this production channel we also show the
variation with $\TB$ in the lower left plot of \reffi{fig:eec1c2}. 
From $\TB \approx 3$ and $\sig(\eececz) \sim 4.1$~fb the cross section
decreases to $\sig(\eececz) \sim 3.3$~fb at $\TB = 50$. The size of the
loop corrections varies from $\sim +8.6\%$ to $\sim +7\%$ from low to
high $\TB$.

In addition, here we show in the lower right plot of \reffi{fig:eec1c2} 
the $\CP$-odd observable $A_{12}$, see \refeq{A12}, varied with the
complex phases $\phiMe$ and $\phiAt$.
However, for our parameter set \Scs\ the $\CP$ asymmetries turn out to 
be very small, well below $\pm 1\%$, hardly measurable in future 
$e^+e^-$ collider experiments.

\medskip

\begin{figure}[t]
\begin{center}
\begin{tabular}{c}
\includegraphics[width=0.48\textwidth,height=6cm]{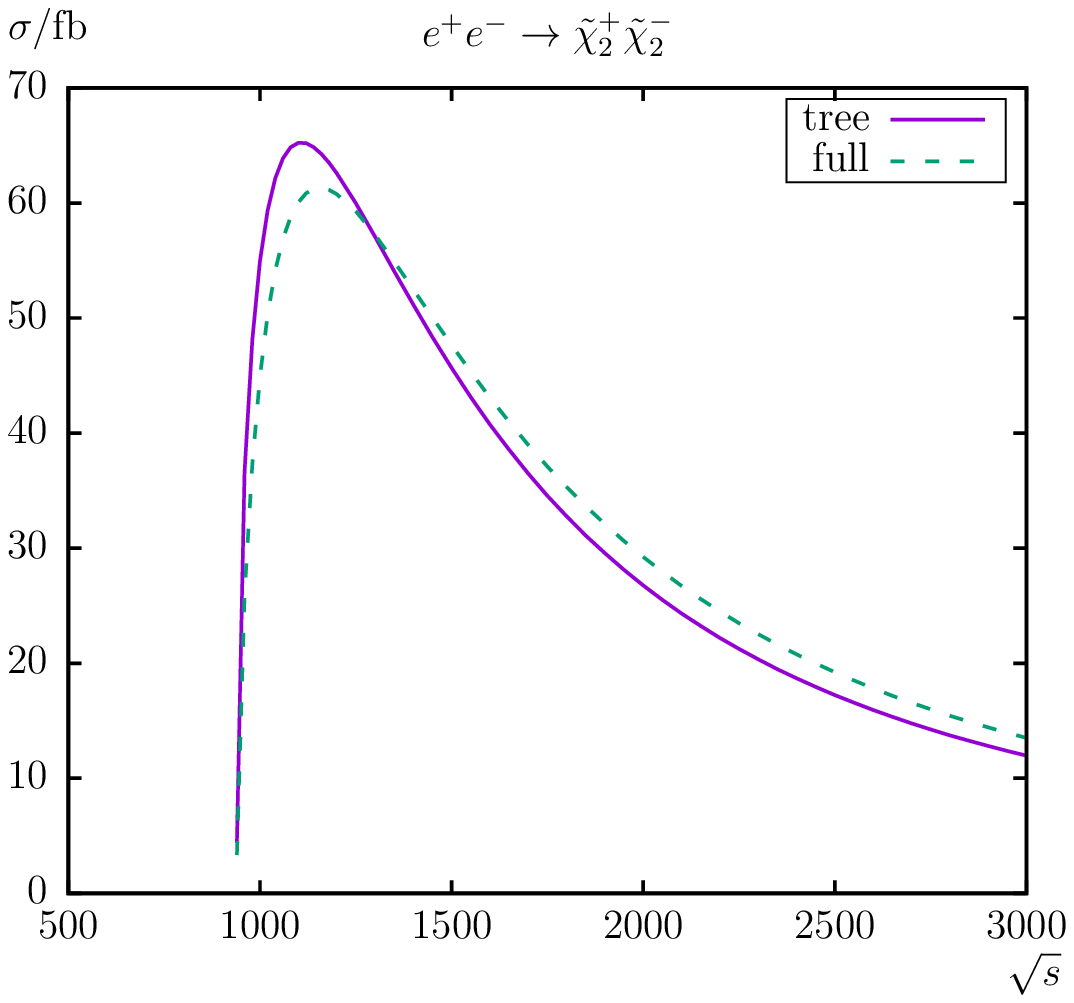}
\includegraphics[width=0.48\textwidth,height=6cm]{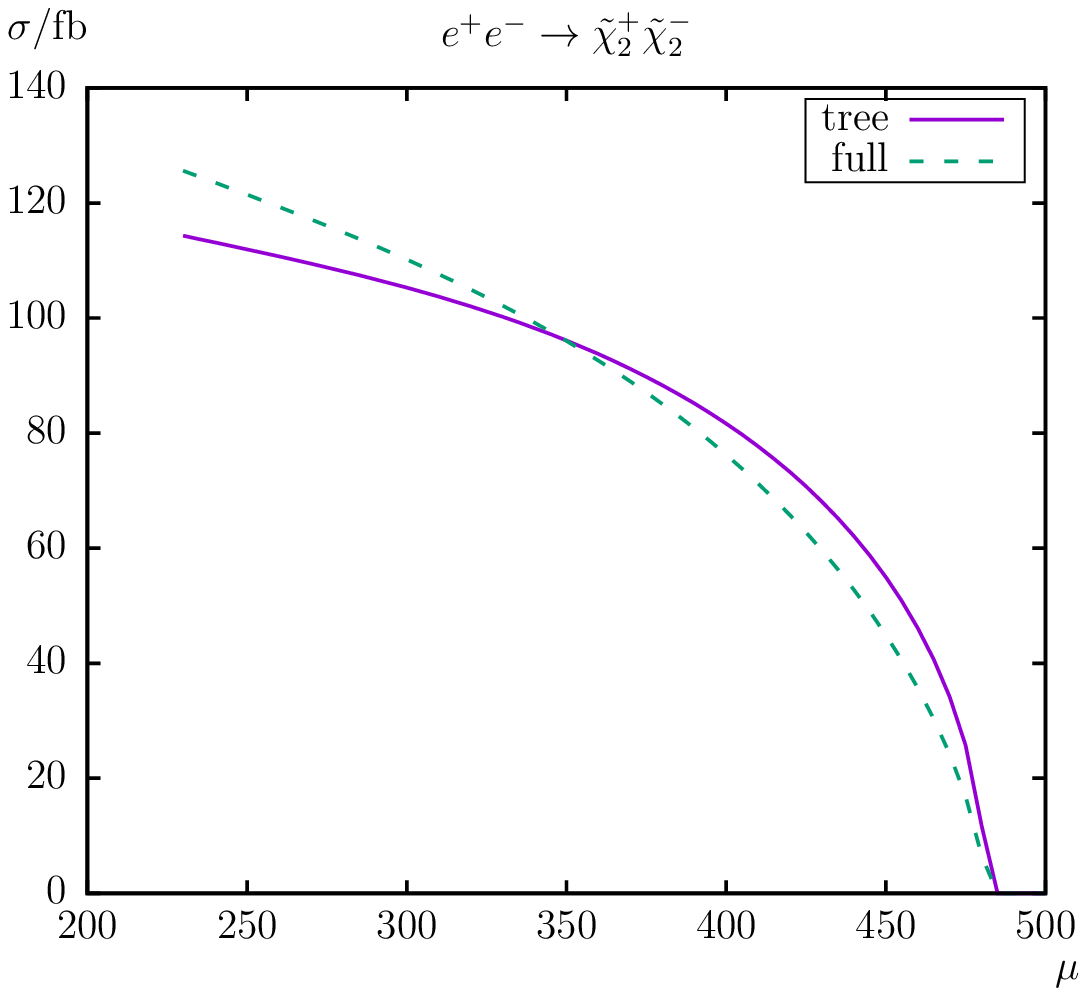}
\\[1em]
\includegraphics[width=0.48\textwidth,height=6cm]{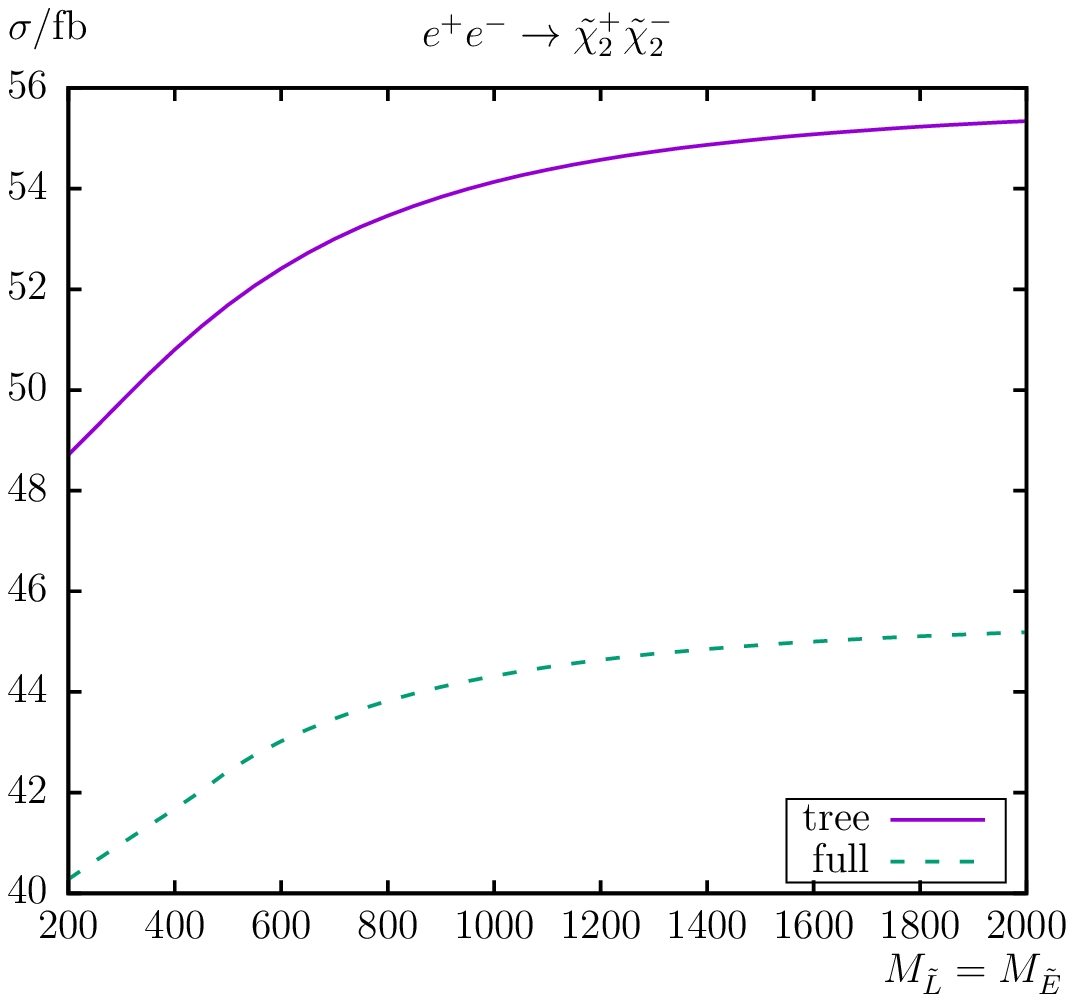}
\includegraphics[width=0.48\textwidth,height=6cm]{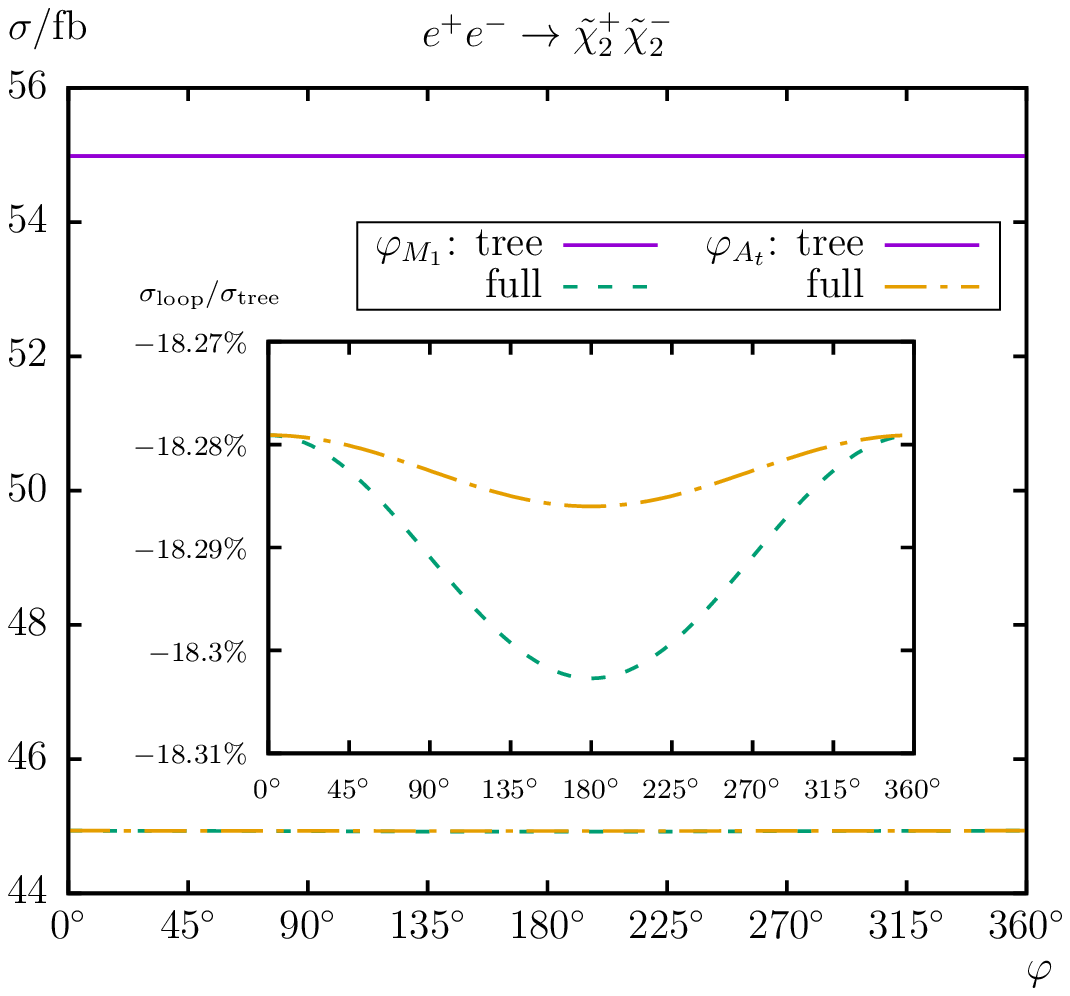}
\end{tabular}
\caption{\label{fig:eec2c2}
  $\sig(\eeczcz)$.
  Tree-level and full one-loop corrected cross sections are shown with 
  parameters chosen according to \Scs; see \refta{tab:para}.
  The upper plots show the cross sections with $\sqrt{s}$ (left) and 
  $\mu$ (right) varied;  the lower plots show $\MSL = \MSE$ (left) and 
  $\phiMe$, $\phiAt$ (right) varied.
}
\end{center}
\end{figure}

We finish the \eecc\ analysis in \reffi{fig:eec2c2} in which the results 
for $c = c^{\prime} = 2$ are displayed.  As a function of $\sqrt{s}$ (upper
left plot) we find a maximum of $\sim 60$~fb at $\sqrt{s} = 1200\gev$.  
The loop corrections are $\sim -19\%$ at $\sqrt{s} = 1000\gev$ 
(\ie \Scs), and $\sim +13\%$ at $\sqrt{s} = 3000\gev$.

In \Scs, but with $\mu$ varied (upper right plot) we find the highest 
values of $\sim 125$~fb at the lowest mass scales, going to zero for 
$\mu \approx 480\gev$ due to kinematics.  The relative corrections are 
$\sim +10\%$ at $\mu = 240\gev$ and $\sim -19\%$ at $\mu = 450\gev$ 
(\ie \Scs), with a tree crossing at $\mu \approx 350\gev$.

The cross section increases slowly with increasing $\MSL$ and the full 
corrections reach their maximum of $\sim 45$~fb at the highest values 
shown, $\MSL = 2000\gev$.  The relative corrections are nearly constant, 
increasing only from $\sim -17.3\%$ at $\MSL = 200\gev$ to $\sim -18.4\%$ 
at $\MSL = 2000\gev$.

The dependence on $\phiMe$ and $\phiAt$ (lower right plot) is again
tiny, the loop corrections are found to be nearly independent of $\phiMe$ 
and $\phiAt$ below the level of $\sim +0.1\%$, as shown explicitely in 
the inlay.

\medskip

Overall, for the chargino pair production we observed an decreasing 
cross section $\propto 1/s$ for $s \to \infty$; see \citere{Bartl:1985fk}. 
The full one-loop corrections are very roughly 10-20\% of the
tree-level results, but depend strongly on the size of $\mu$, where
larger values result even in negative loop corrections.
The cross sections are largest for $\eecece$ and $\eeczcz$ and roughly
smaller by one order of magnitude for $\eececz$. 
This is because there is no $\gamma\, \chapm{1} \champ{2}$ coupling 
at tree level in the MSSM.
The variation of the cross sections and of the $\CP$~asymmetry $A_{12}$
with $\phiMe$ or $\phiAt$ is found extremely small and the dependence on 
other phases were found to be roughly at the same level and have not been 
shown explicitely.


\subsection{The process \boldmath{\eenn}}
\label{sec:eenn}

\begin{figure}[t]
\begin{center}
\begin{tabular}{c}
\includegraphics[width=0.48\textwidth,height=6cm]{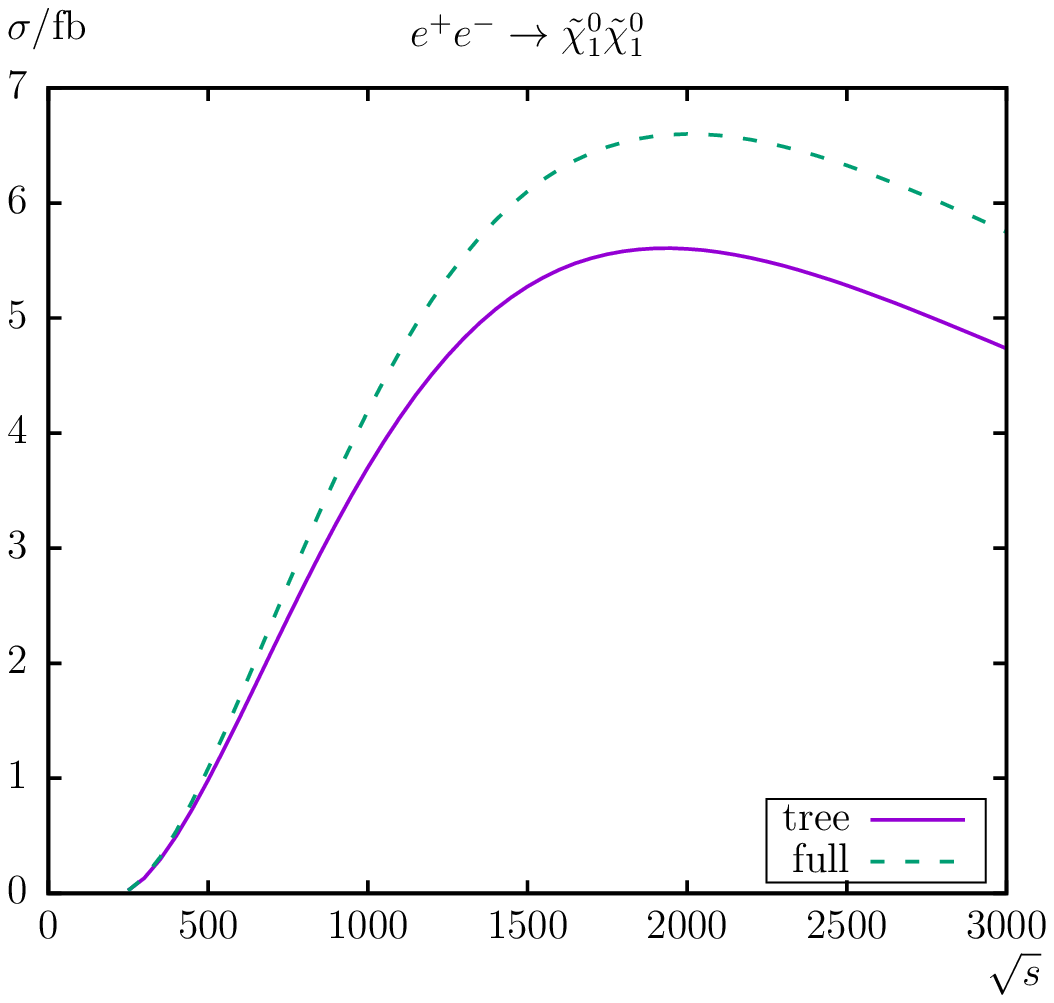}
\includegraphics[width=0.48\textwidth,height=6cm]{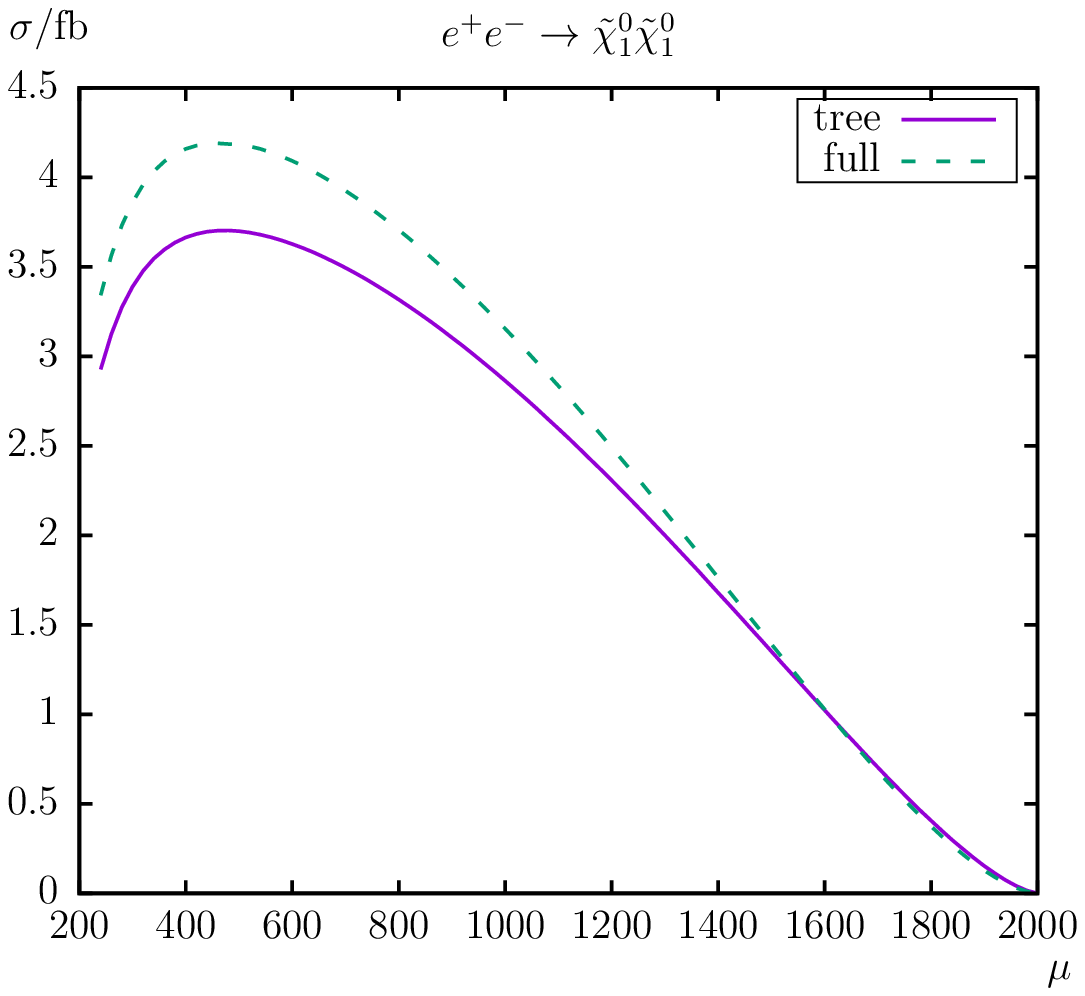}
\\[1em]
\includegraphics[width=0.48\textwidth,height=6cm]{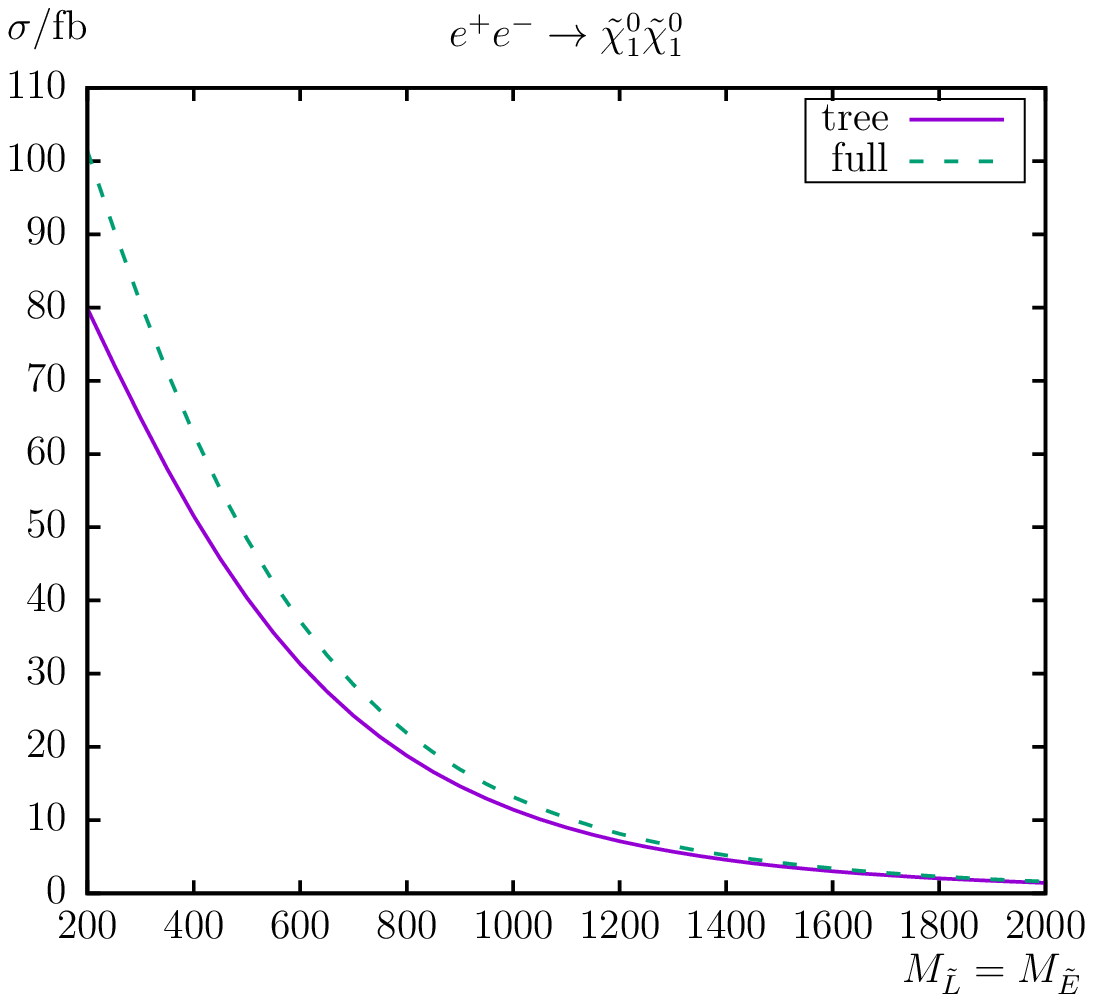}
\includegraphics[width=0.48\textwidth,height=6cm]{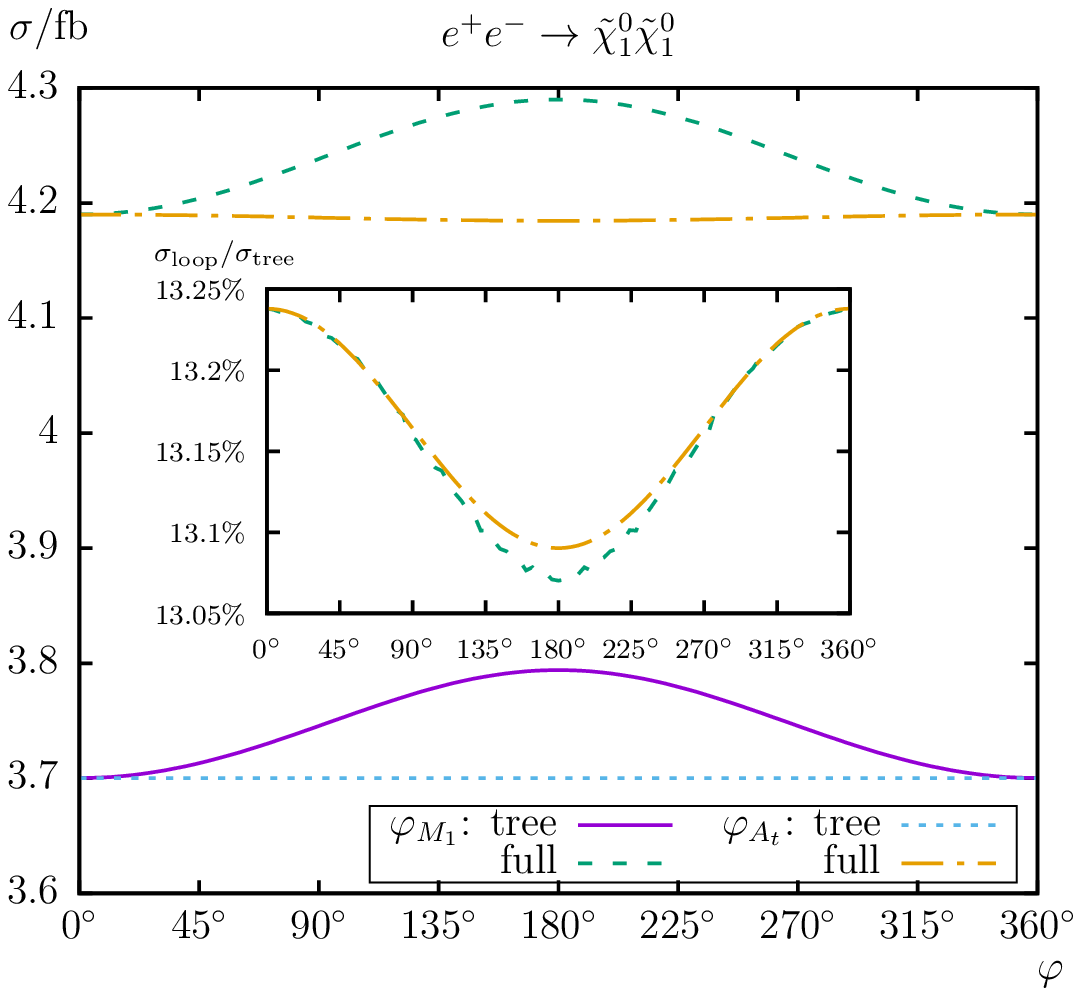}
\end{tabular}
\caption{\label{fig:een1n1}
  $\sig(\eenene)$.
  Tree-level and full one-loop corrected cross sections are shown with 
  parameters chosen according to \Scs; see \refta{tab:para}.
  The upper plots show the cross sections with $\sqrt{s}$ (left) and 
  $\mu$ (right) varied;  the lower plots show $\MSL = \MSE$ (left) and 
  $\phiMe$, $\phiAt$ (right) varied.
}
\end{center}
\end{figure}

\begin{figure}[t]
\begin{center}
\begin{tabular}{c}
\includegraphics[width=0.48\textwidth,height=6cm]{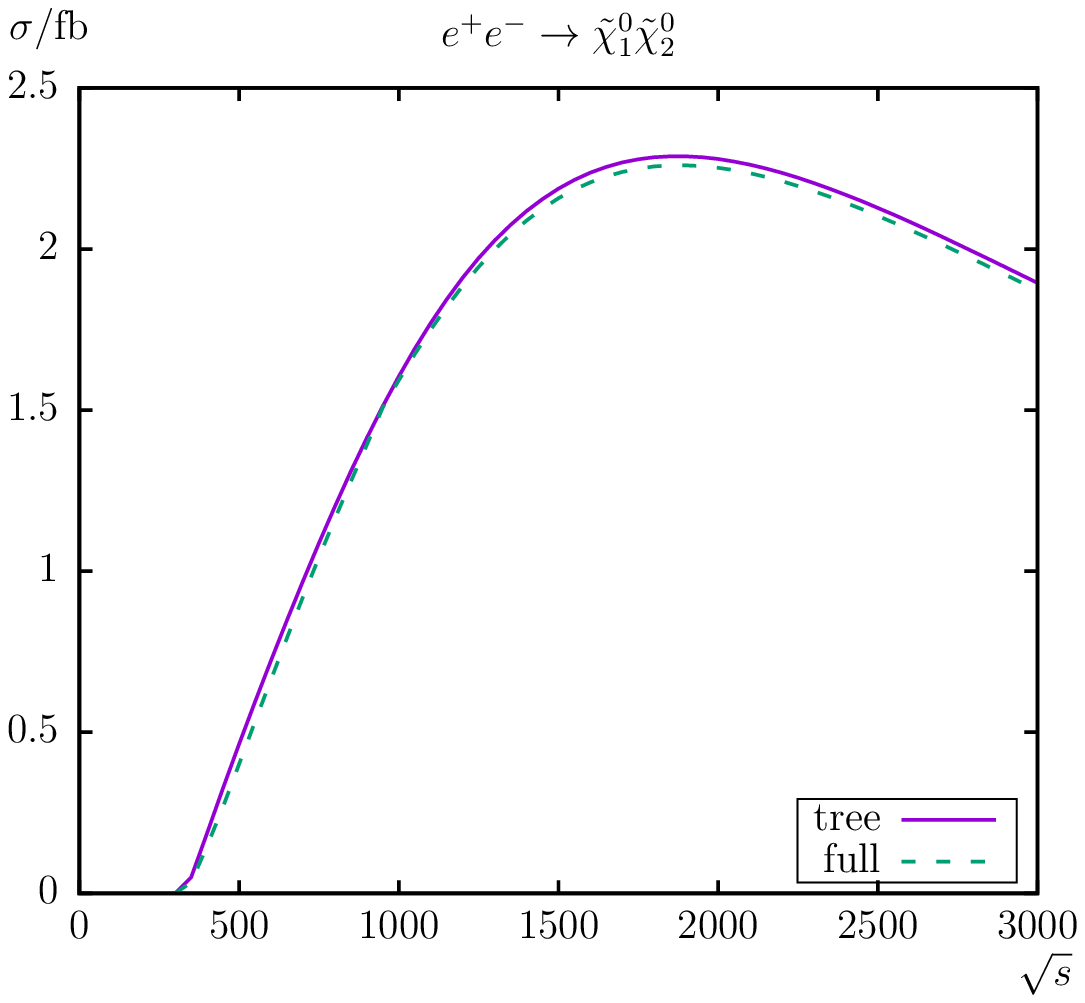}
\includegraphics[width=0.48\textwidth,height=6cm]{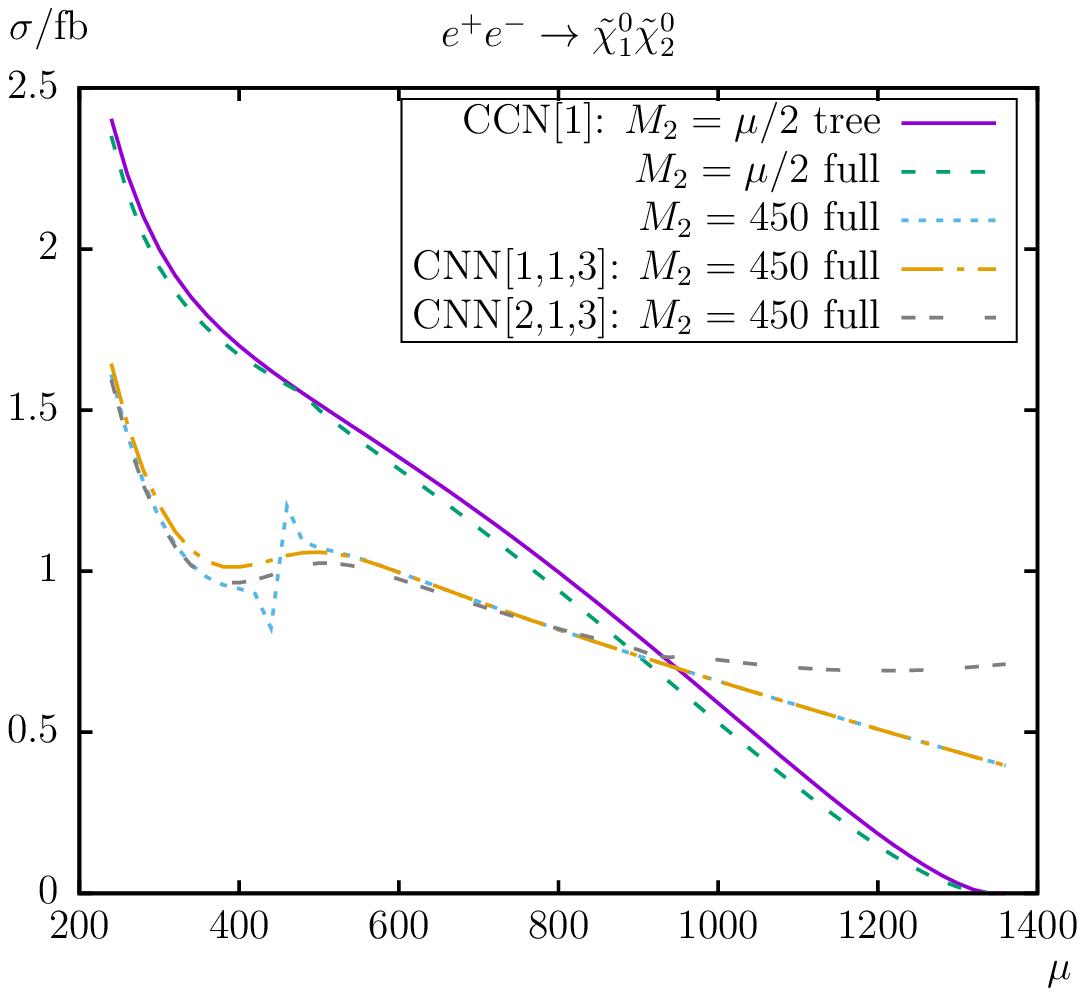}
\\[1em]
\includegraphics[width=0.48\textwidth,height=6cm]{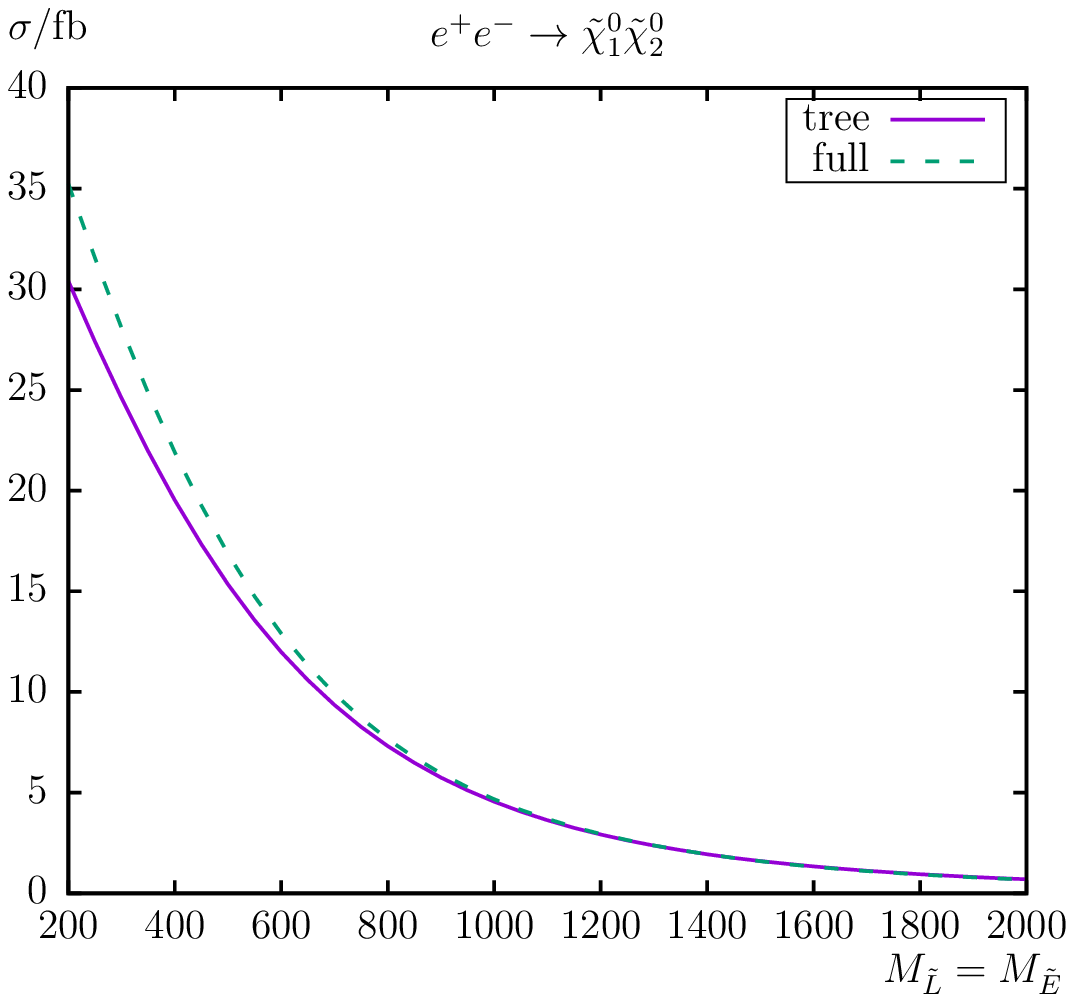}
\includegraphics[width=0.48\textwidth,height=6cm]{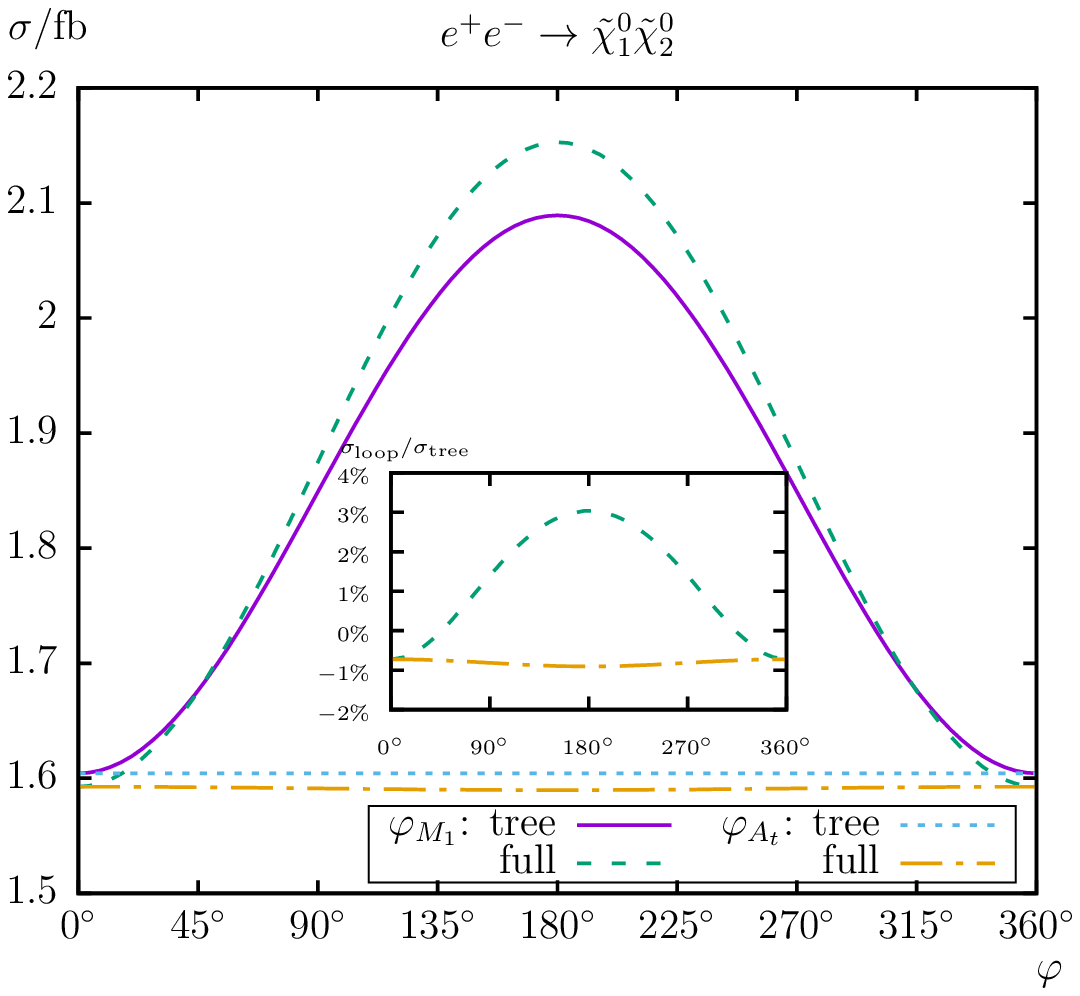}
\end{tabular}
\caption{\label{fig:een1n2}
  $\sig(\eenenz)$.
  Tree-level and full one-loop corrected cross sections are shown with 
  parameters chosen according to \Scs; see \refta{tab:para}.
  The upper plots show the cross sections with $\sqrt{s}$ (left) and 
  $\mu$ (right) varied;  the lower plots show $\MSL = \MSE$ (left) and 
  $\phiMe$, $\phiAt$ (right) varied.
}
\end{center}
\end{figure}

\begin{figure}
\begin{center}
\begin{tabular}{c}
\includegraphics[width=0.48\textwidth,height=6cm]{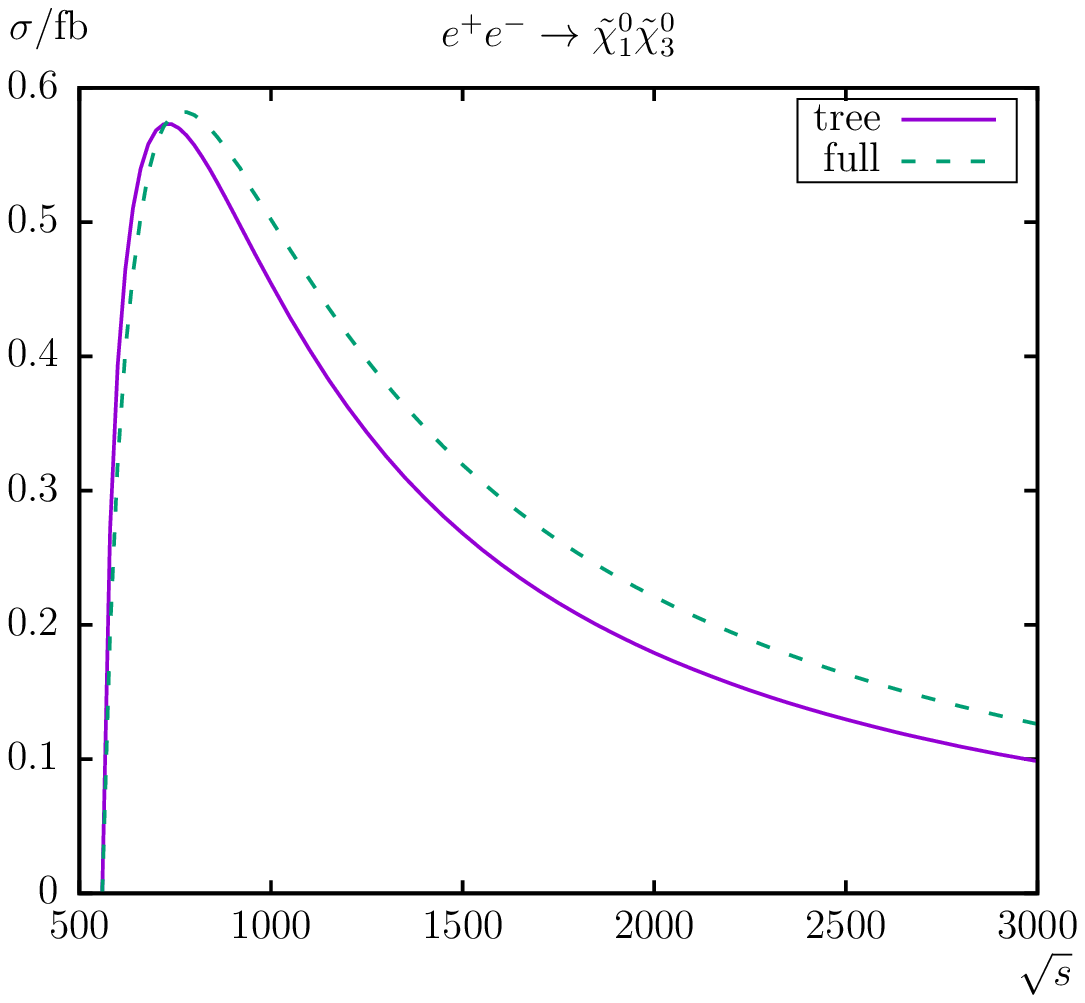}
\includegraphics[width=0.48\textwidth,height=6cm]{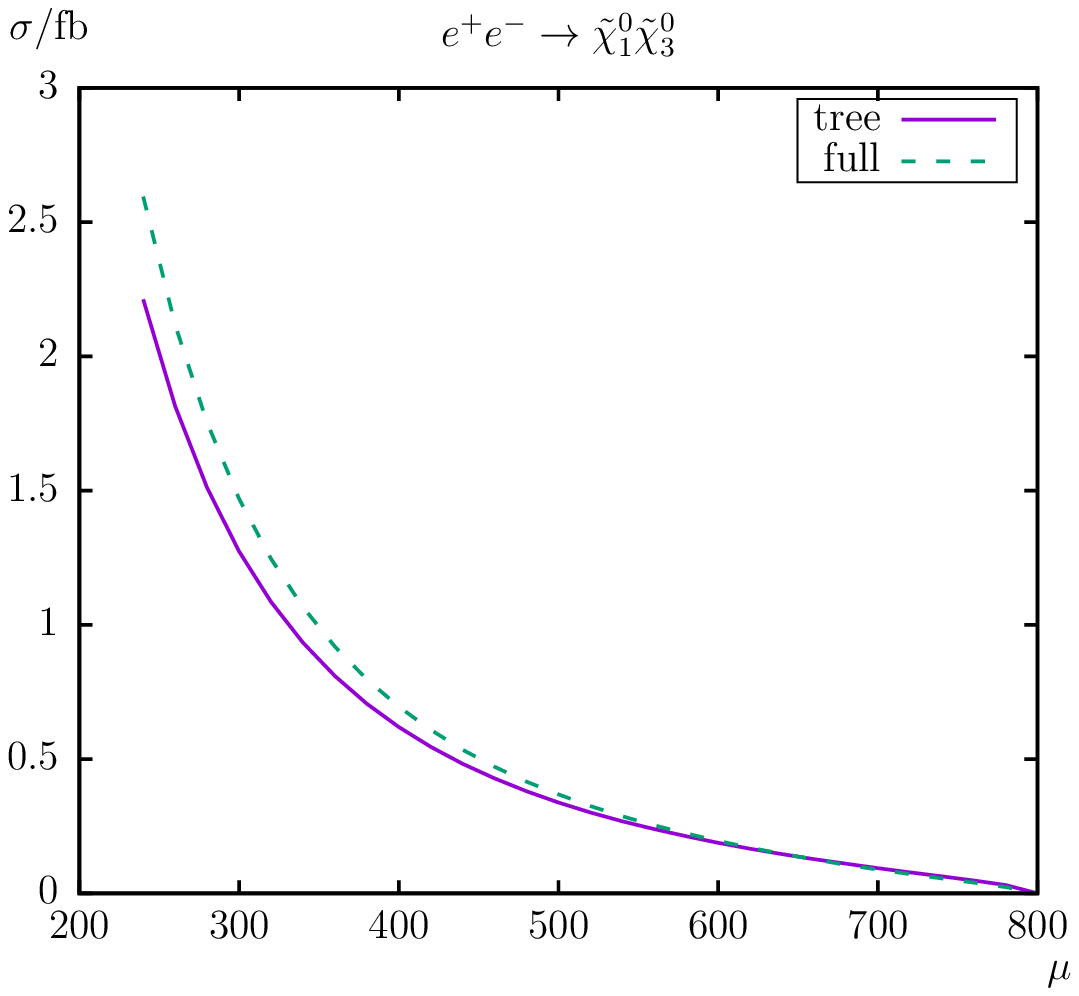}
\\[1em]
\includegraphics[width=0.48\textwidth,height=6cm]{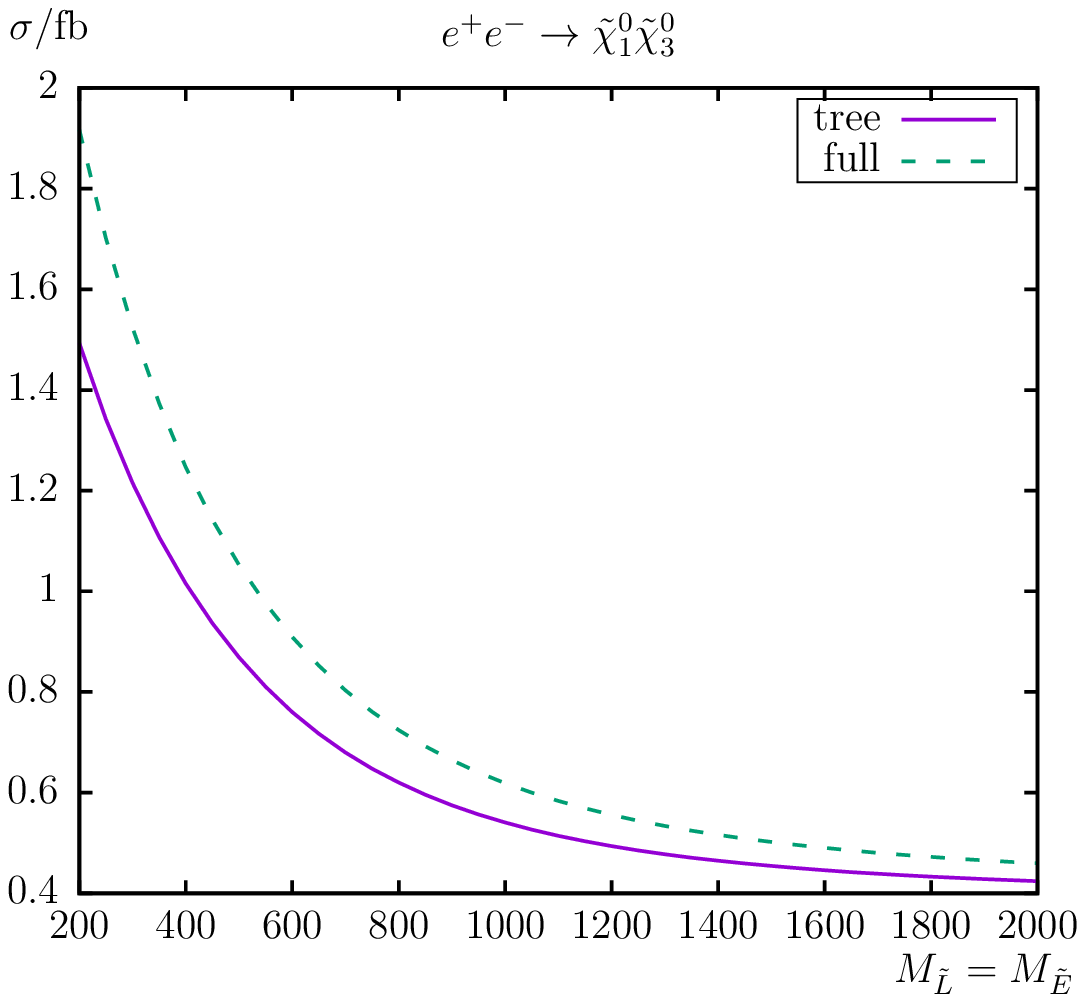}
\includegraphics[width=0.48\textwidth,height=6cm]{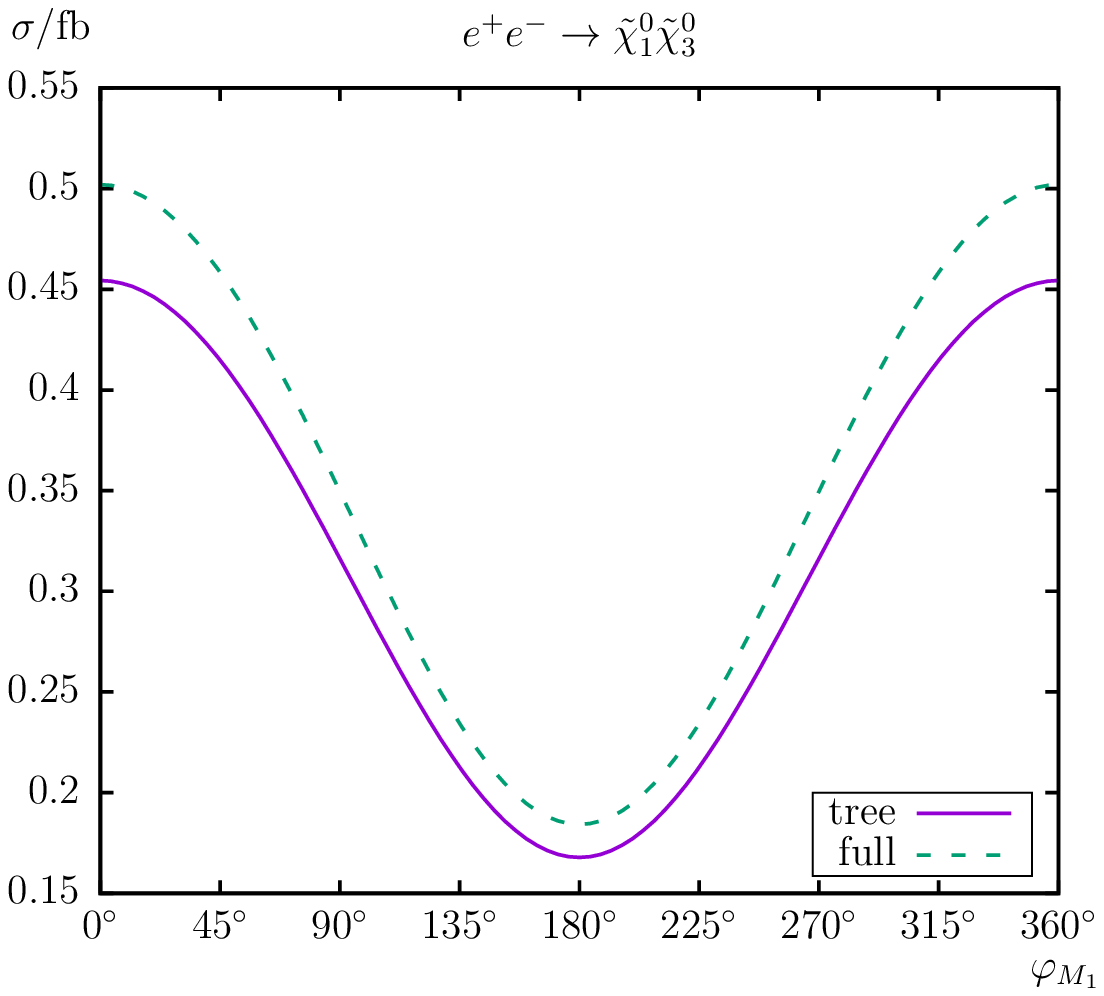}
\\[1em]
\includegraphics[width=0.48\textwidth,height=6cm]{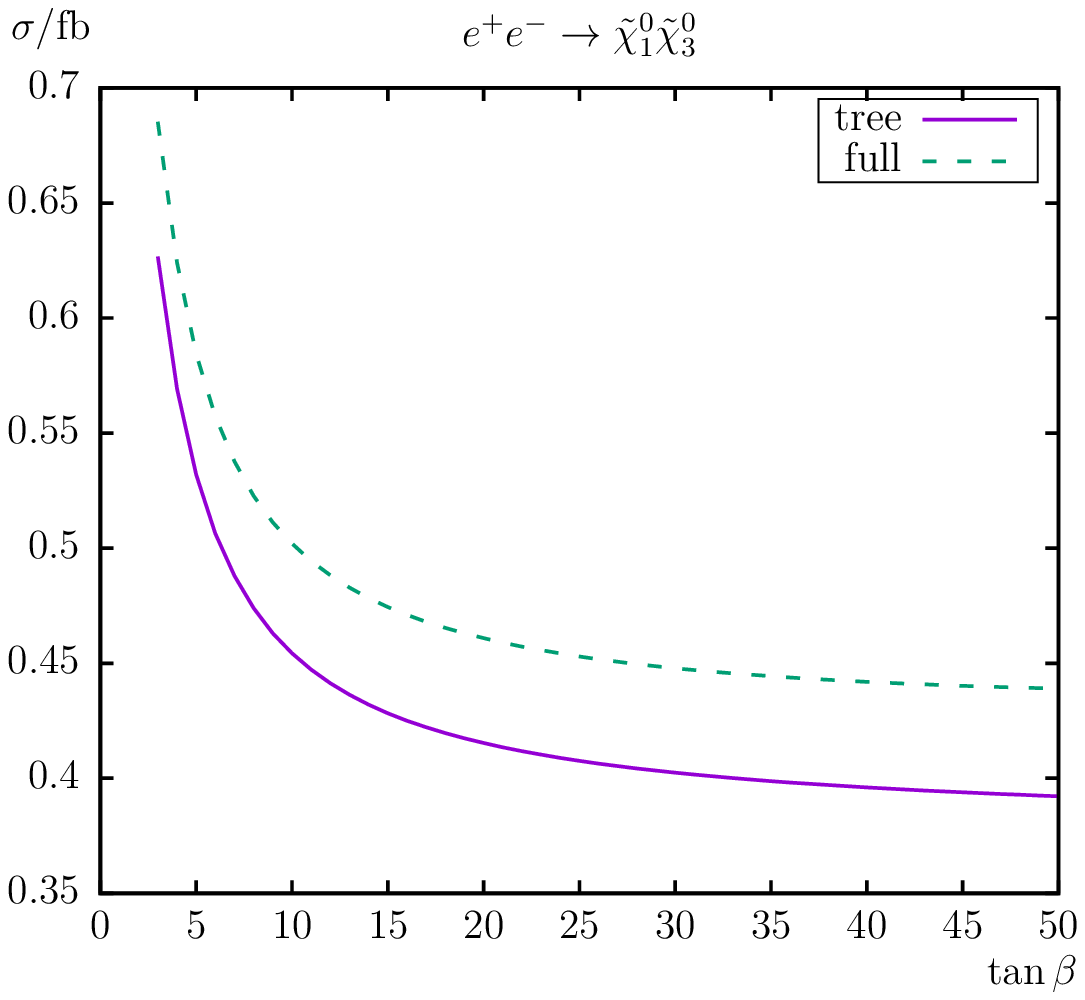}
\end{tabular}
\caption{\label{fig:een1n3}
  $\sig(\eenend)$.
  Tree-level and full one-loop corrected cross sections are shown with 
  parameters chosen according to \Scs; see \refta{tab:para}.
  The upper plots show the cross sections with $\sqrt{s}$ (left) and 
  $\mu$ (right) varied;  the middle plots show $\MSL = \MSE$ (left) and 
  $\phiMe$ (right) varied; the lower plot shows the variation with $\TB$.
}
\end{center}
\end{figure}

\begin{figure}[t]
\begin{center}
\begin{tabular}{c}
\includegraphics[width=0.48\textwidth,height=6cm]{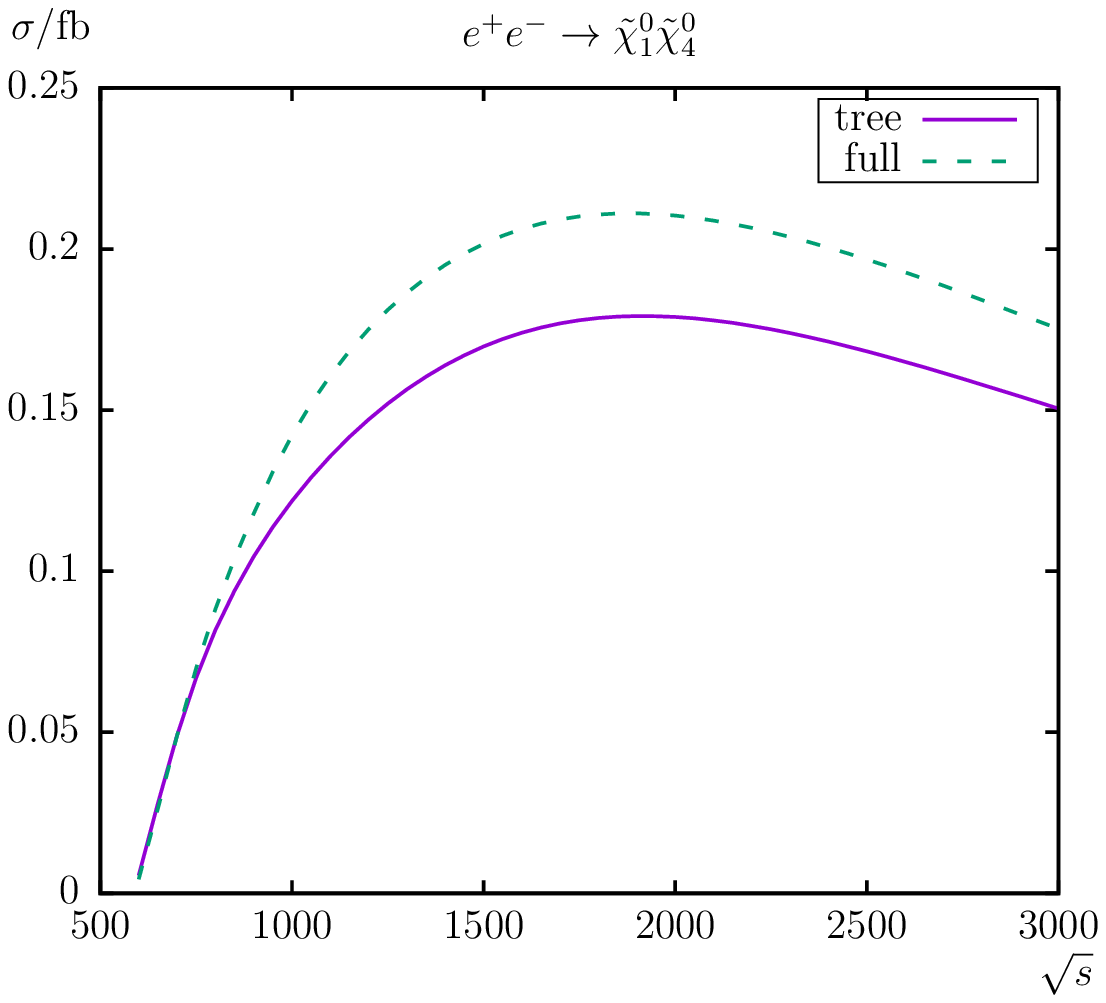}
\includegraphics[width=0.48\textwidth,height=6cm]{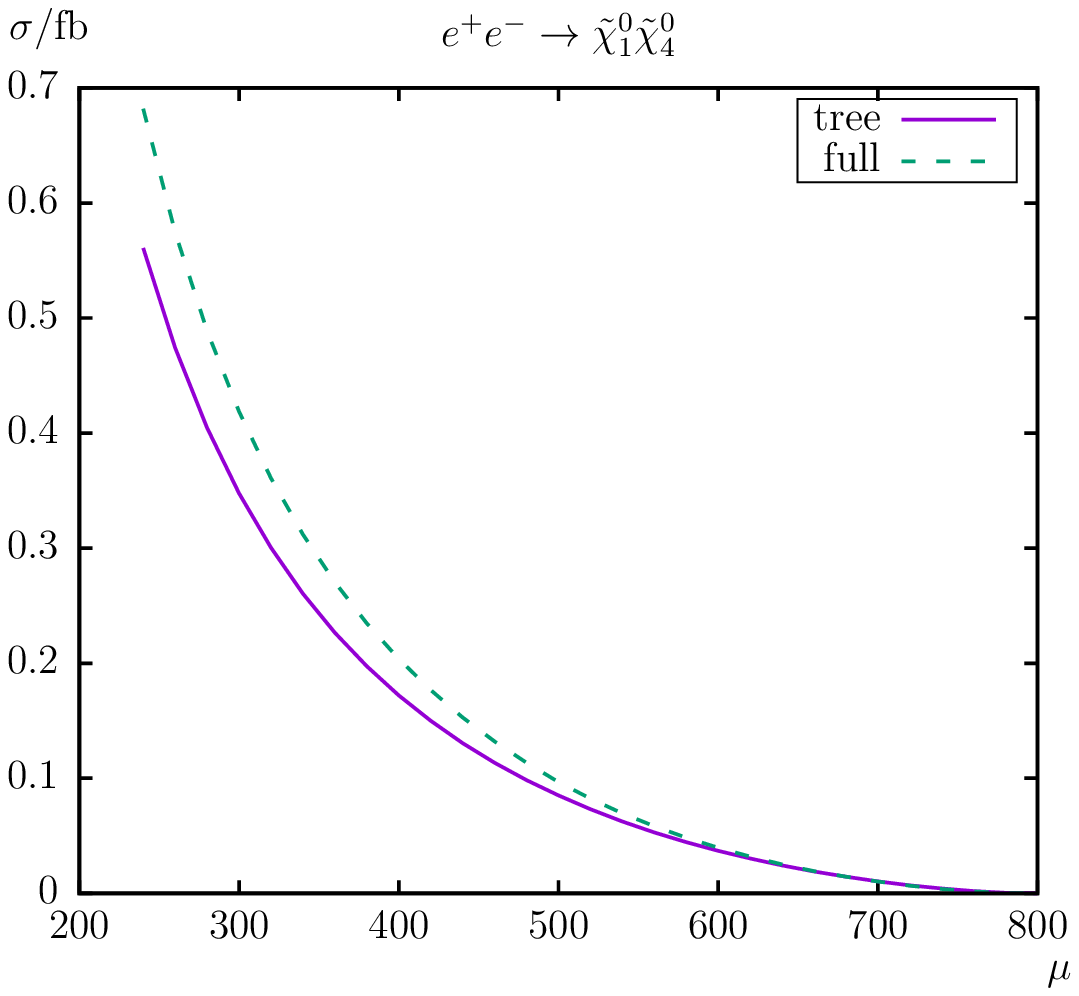}
\\[1em]
\includegraphics[width=0.48\textwidth,height=6cm]{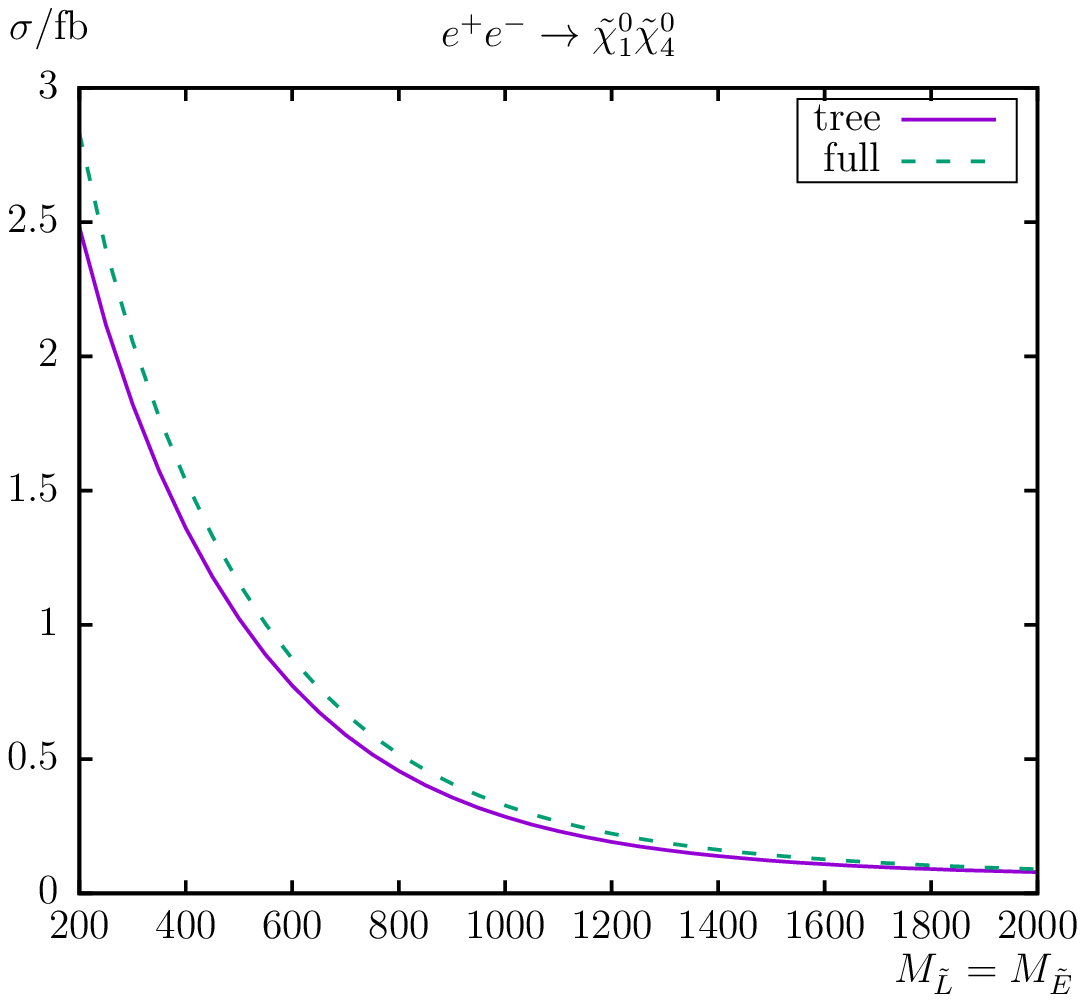}
\includegraphics[width=0.48\textwidth,height=6cm]{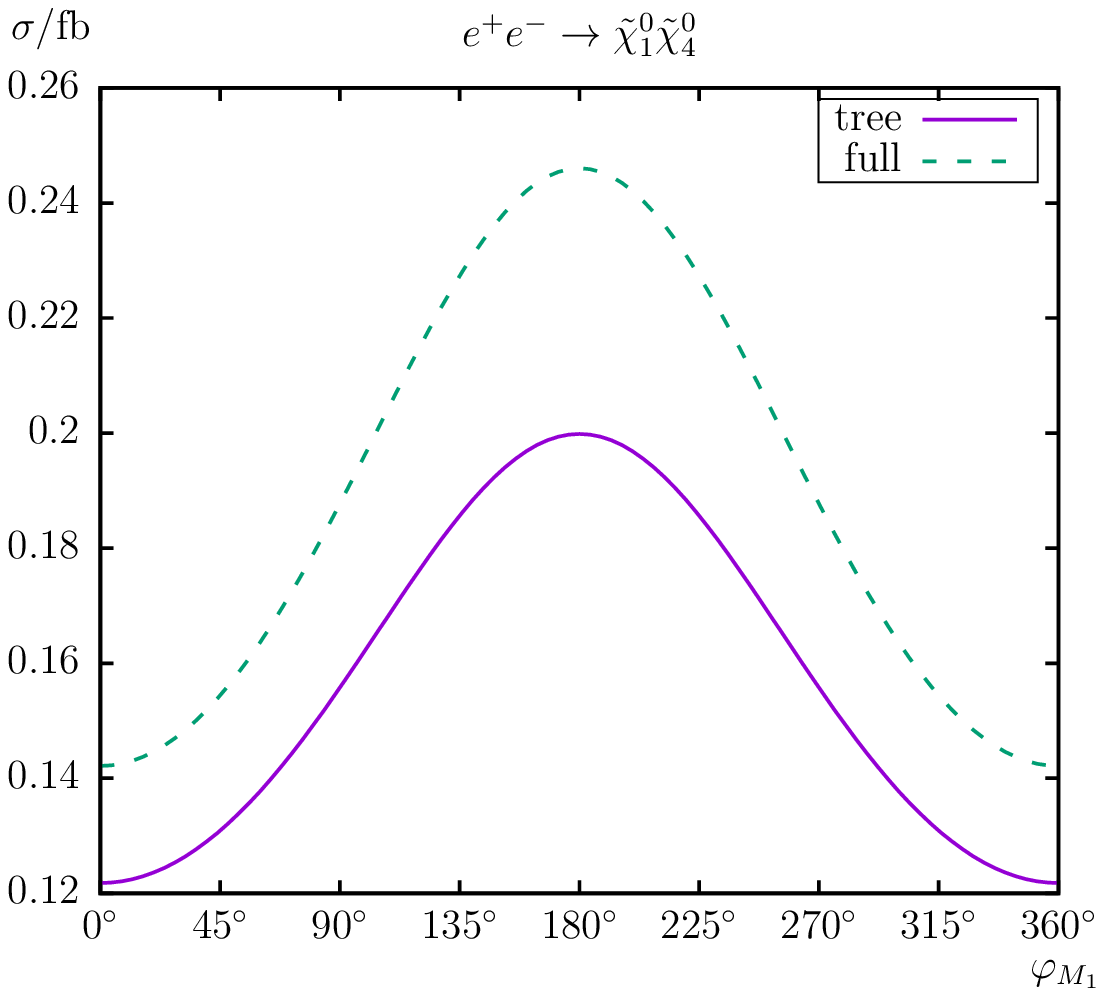}
\end{tabular}
\caption{\label{fig:een1n4}
  $\sig(\eenenv)$.
  Tree-level and full one-loop corrected cross sections are shown with 
  parameters chosen according to \Scs; see \refta{tab:para}.
  The upper plots show the cross sections with $\sqrt{s}$ (left) and 
  $\mu$ (right) varied;  the lower plots show $\MSL = \MSE$ (left) and 
  $\phiMe$ (right) varied.
}
\end{center}
\end{figure}

\begin{figure}[t]
\begin{center}
\begin{tabular}{c}
\includegraphics[width=0.48\textwidth,height=6cm]{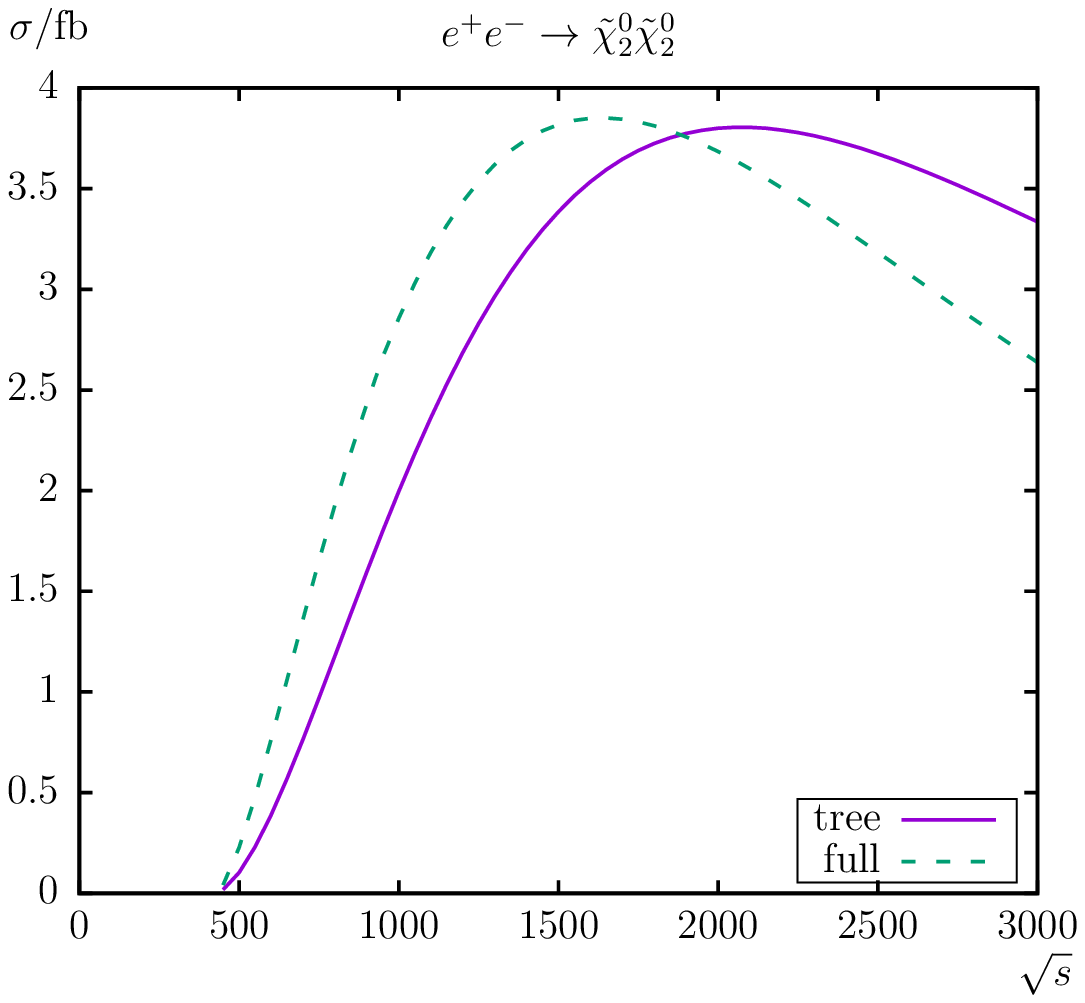}
\includegraphics[width=0.48\textwidth,height=6cm]{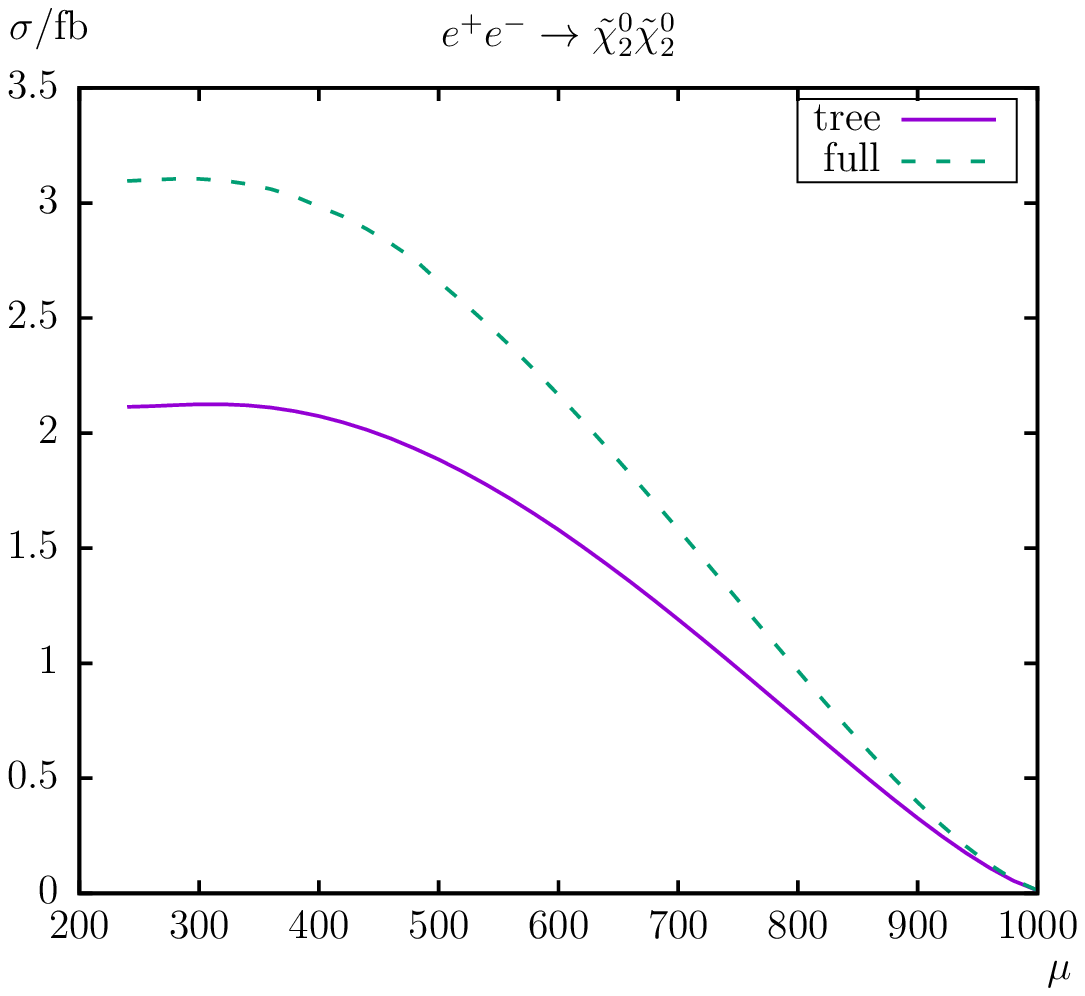}
\\[1em]
\includegraphics[width=0.48\textwidth,height=6cm]{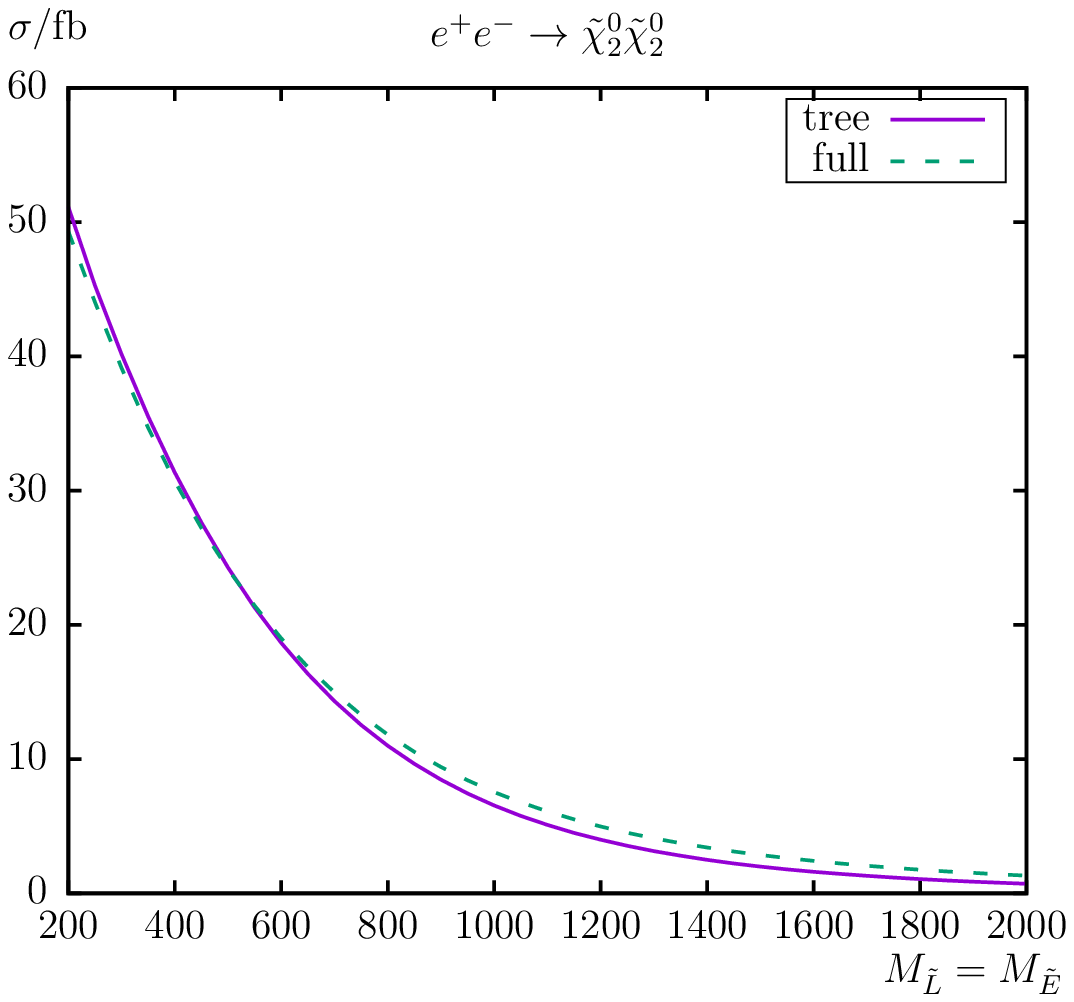}
\includegraphics[width=0.48\textwidth,height=6cm]{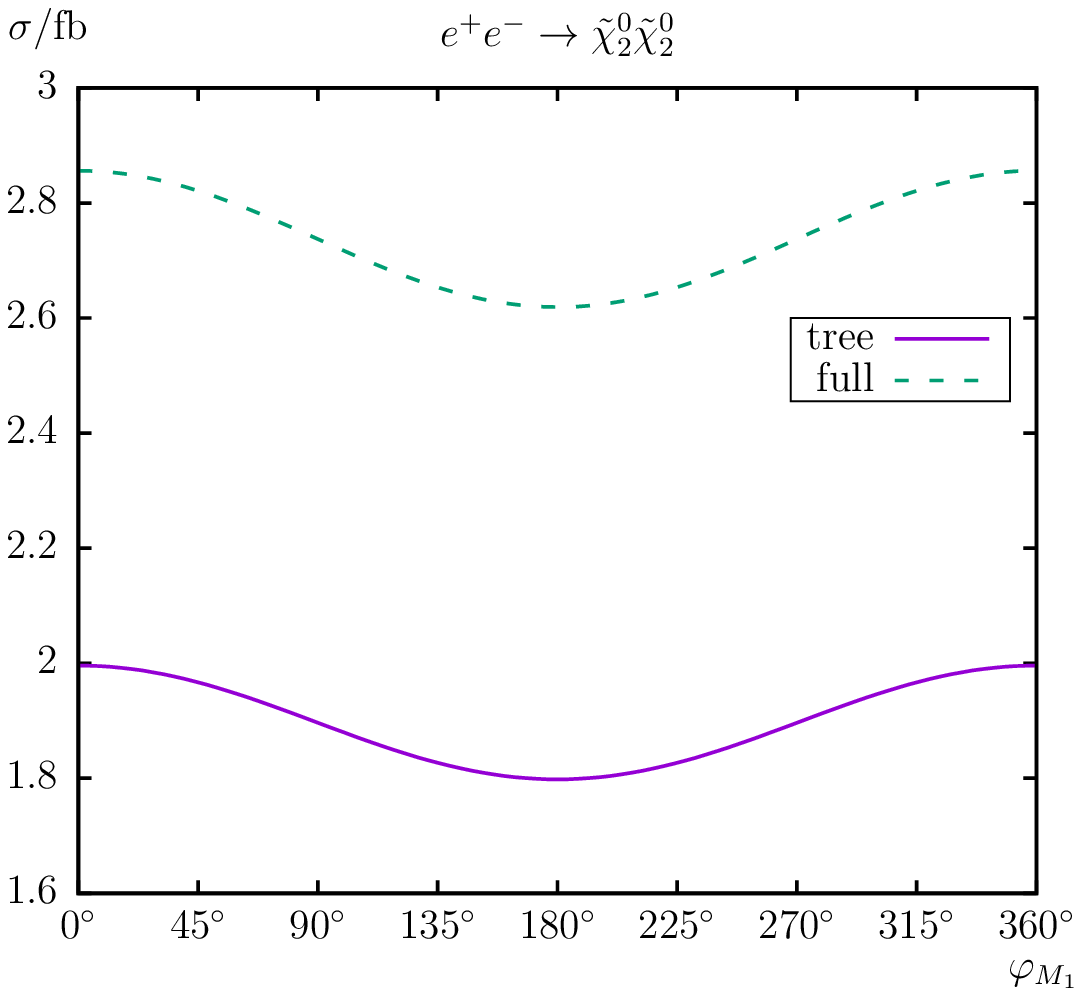}
\end{tabular}
\caption{\label{fig:een2n2}
  $\sig(\eenznz)$.
  Tree-level and full one-loop corrected cross sections are shown with 
  parameters chosen according to \Scs; see \refta{tab:para}.
  The upper plots show the cross sections with $\sqrt{s}$ (left) and 
  $\mu$ (right) varied;  the lower plots show $\MSL = \MSE$ (left) and 
  $\phiMe$ (right) varied.
}
\end{center}
\end{figure}

\begin{figure}[t]
\begin{center}
\begin{tabular}{c}
\includegraphics[width=0.48\textwidth,height=6cm]{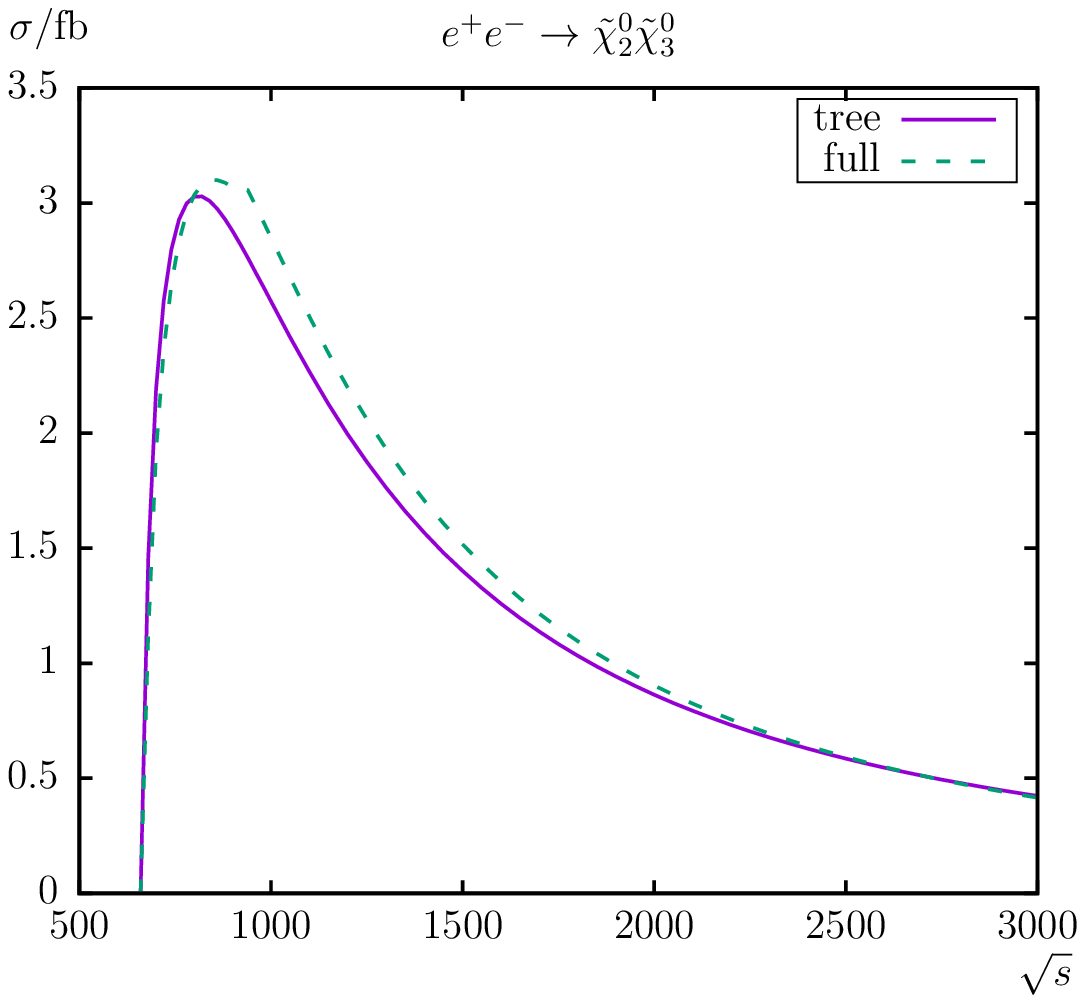}
\includegraphics[width=0.48\textwidth,height=6cm]{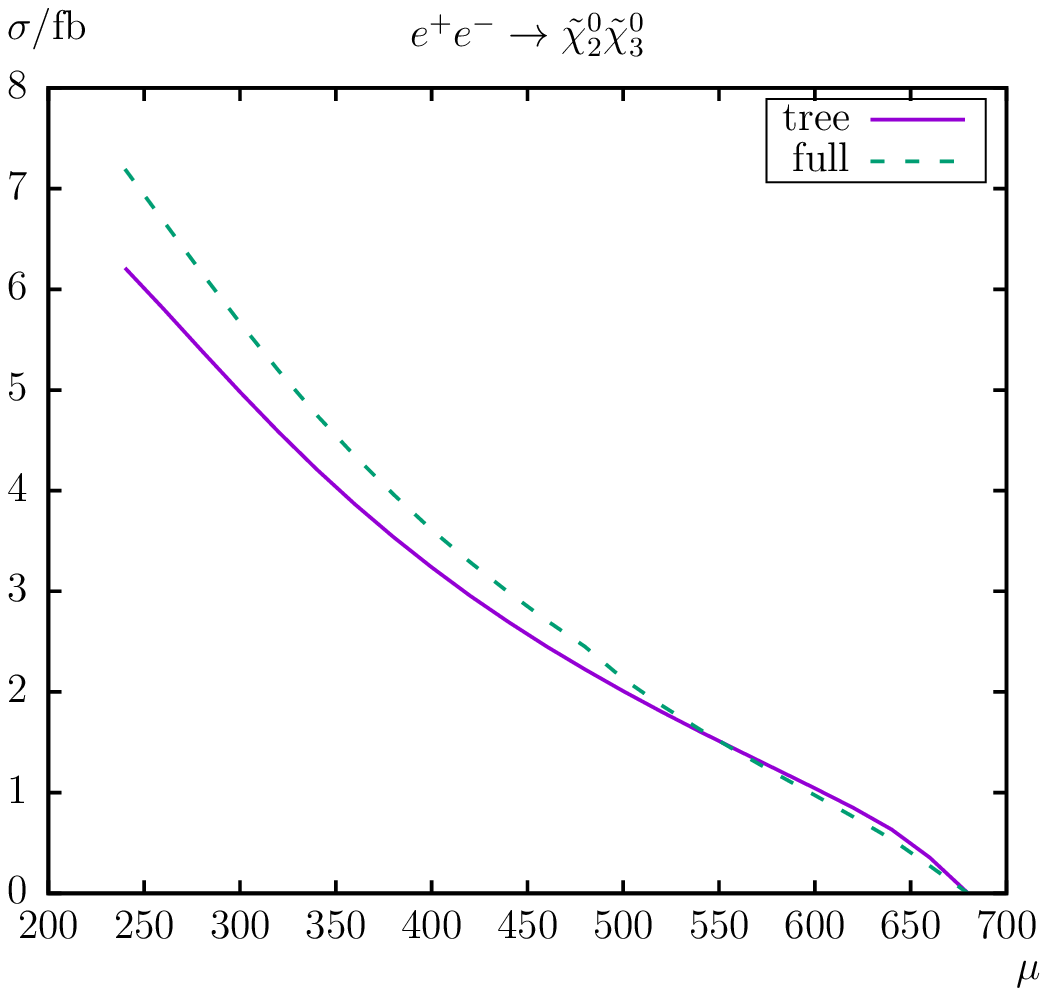}
\\[1em]
\includegraphics[width=0.48\textwidth,height=6cm]{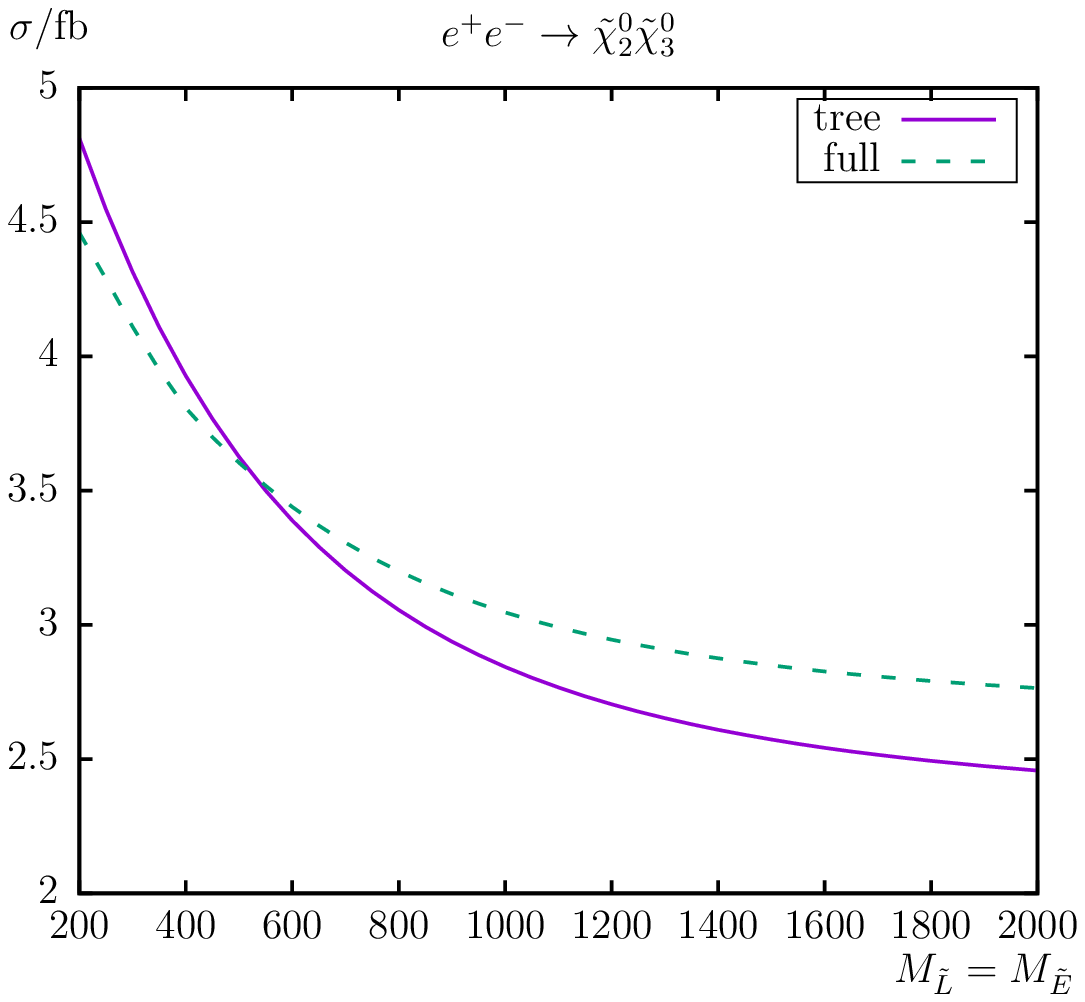}
\includegraphics[width=0.48\textwidth,height=6cm]{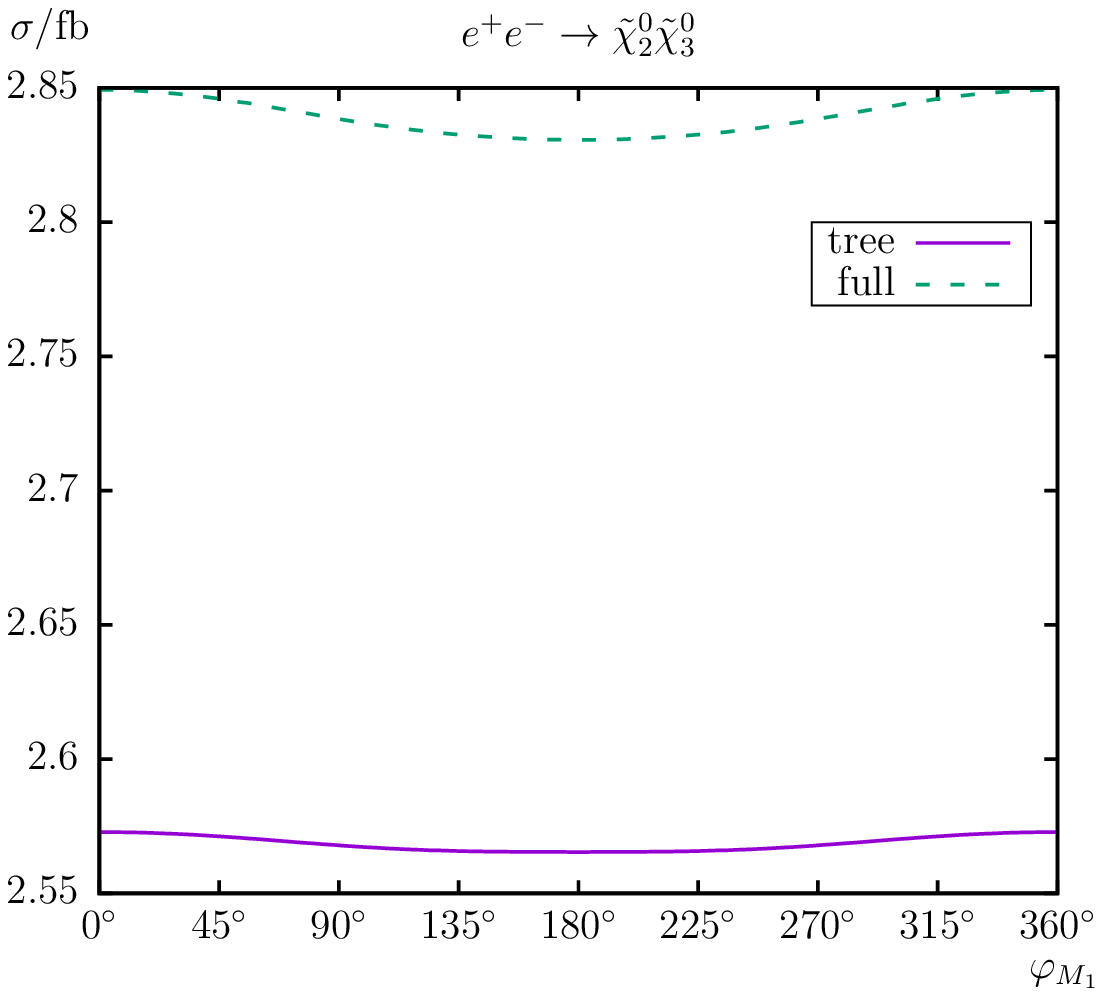}
\end{tabular}
\caption{\label{fig:een2n3}
  $\sig(\eenznd)$.
  Tree-level and full one-loop corrected cross sections are shown with 
  parameters chosen according to \Scs; see \refta{tab:para}.
  The upper plots show the cross sections with $\sqrt{s}$ (left) and 
  $\mu$ (right) varied;  the lower plots show $\MSL = \MSE$ (left) and 
  $\phiMe$ (right) varied.
}
\end{center}
\end{figure}

\begin{figure}
\begin{center}
\begin{tabular}{c}
\includegraphics[width=0.48\textwidth,height=6cm]{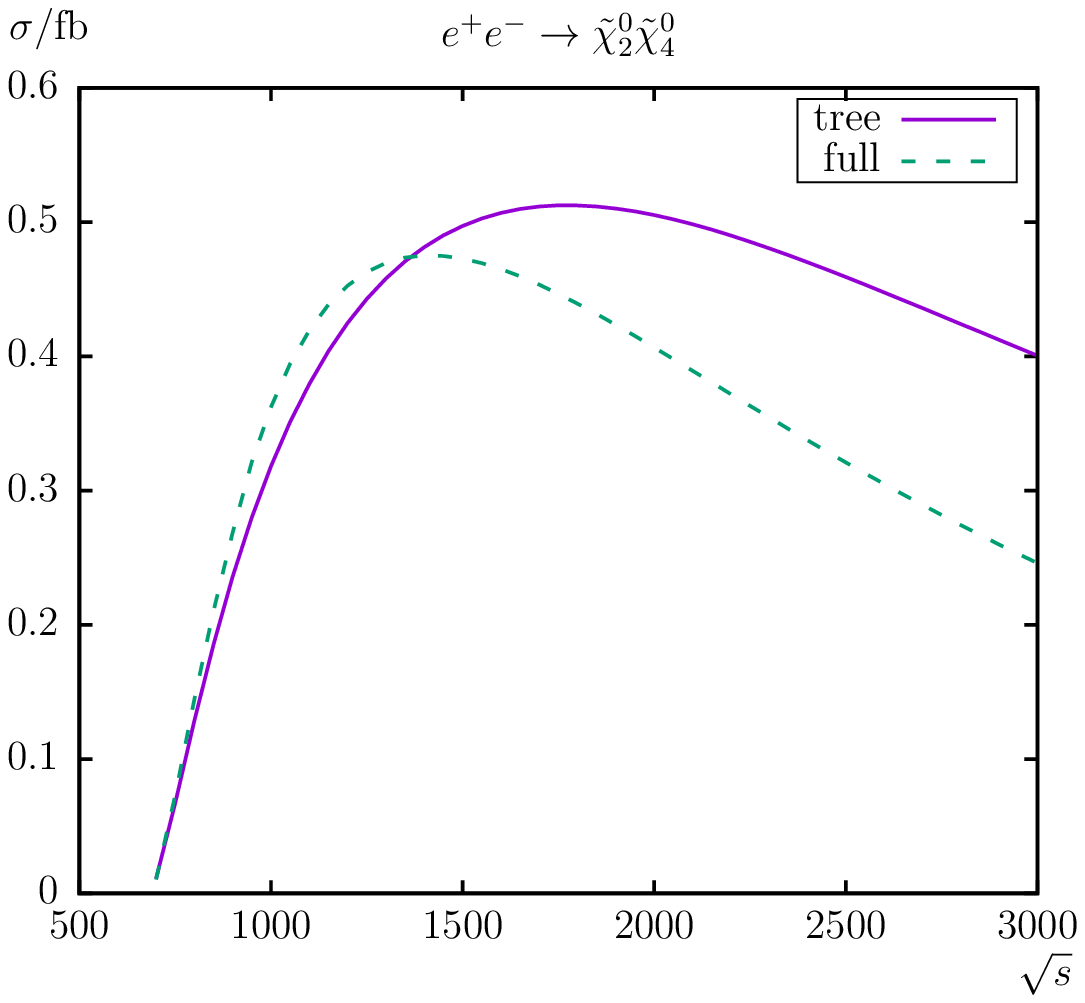}
\includegraphics[width=0.48\textwidth,height=6cm]{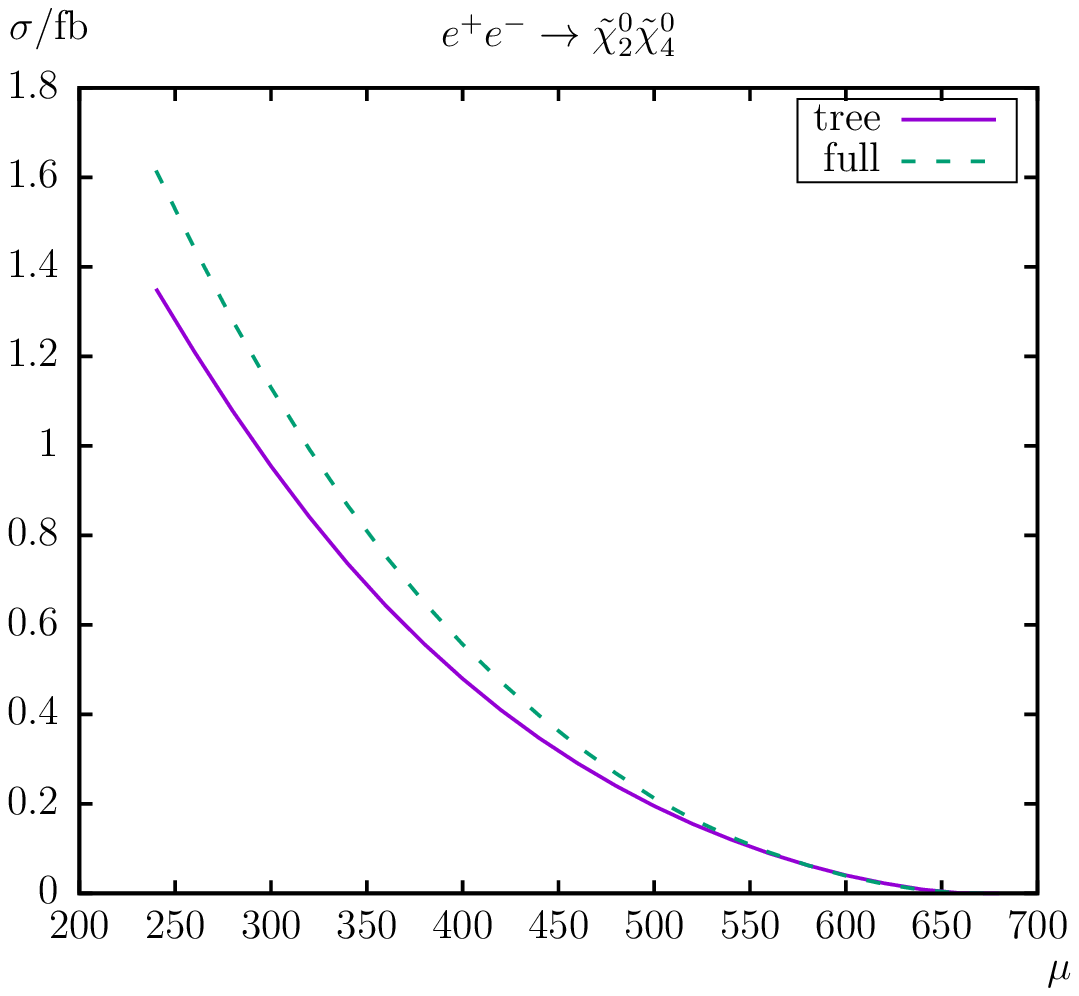}
\\[1em]
\includegraphics[width=0.48\textwidth,height=6cm]{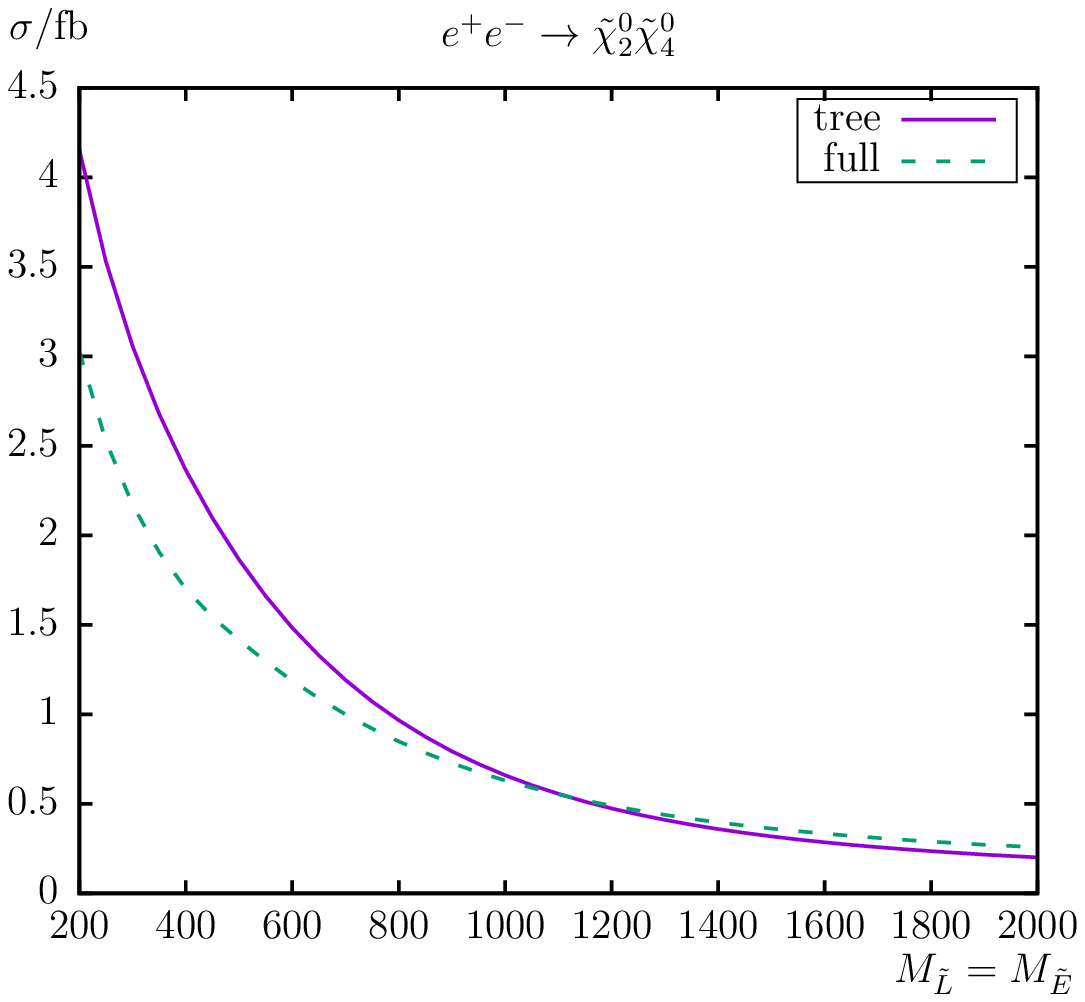}
\includegraphics[width=0.48\textwidth,height=6cm]{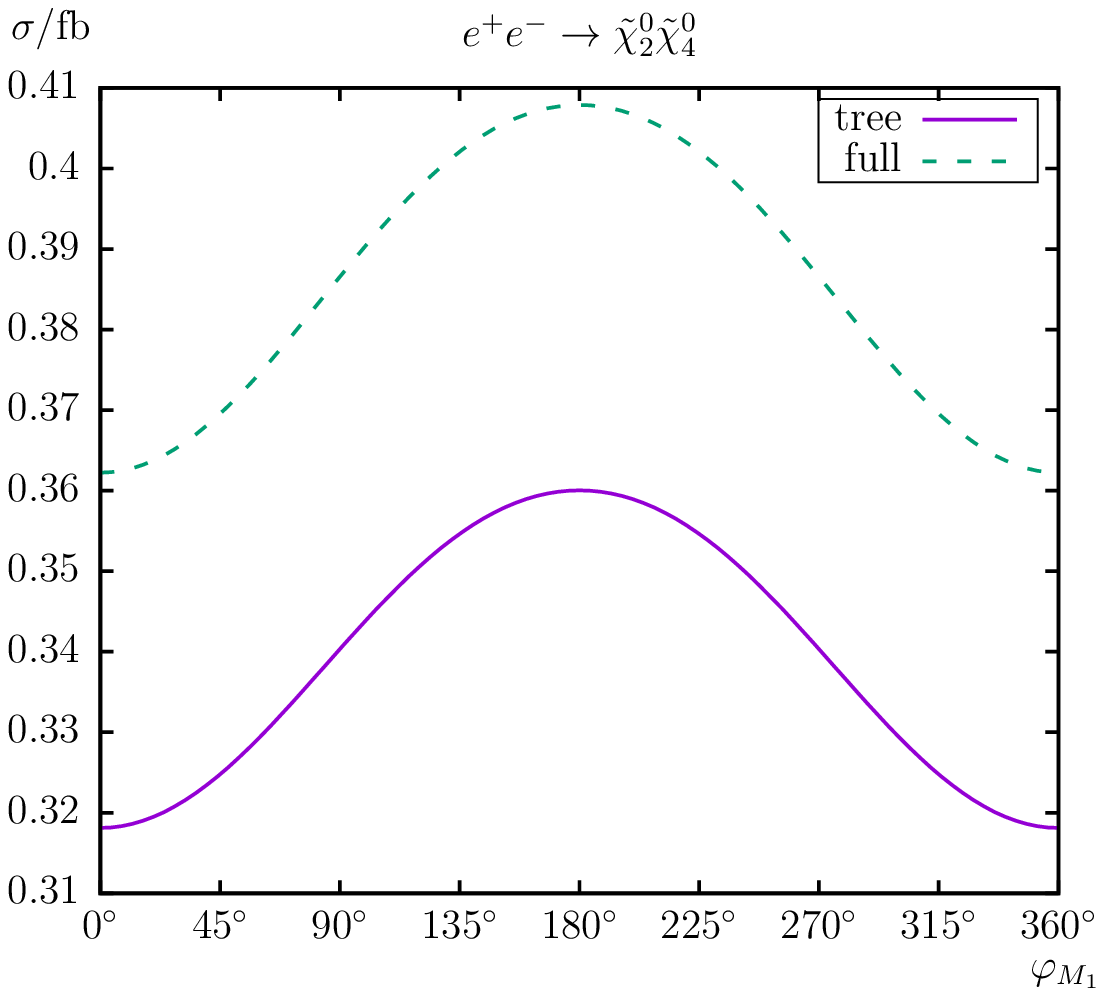}
\\[1em]
\includegraphics[width=0.48\textwidth,height=6cm]{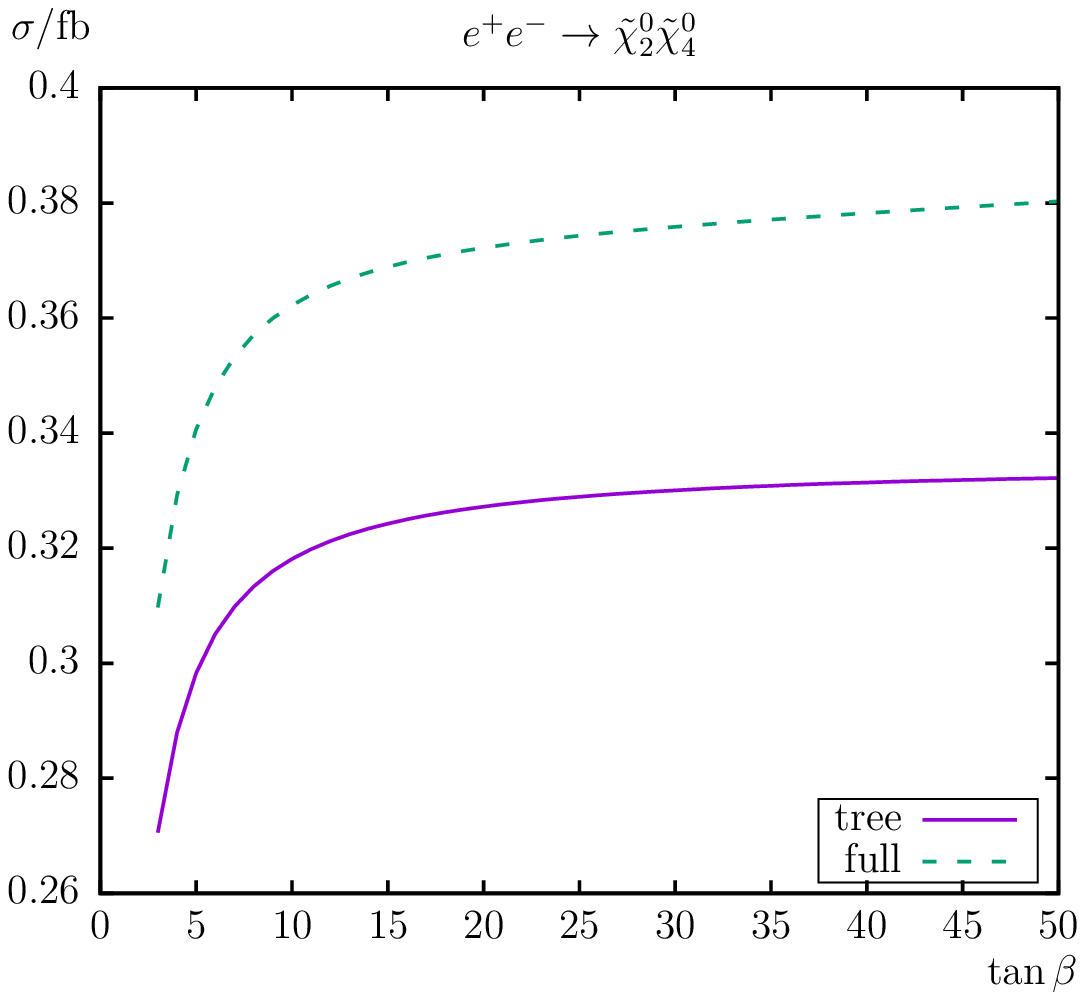}
\end{tabular}
\caption{\label{fig:een2n4}
  $\sig(\eenznv)$.
  Tree-level and full one-loop corrected cross sections are shown with 
  parameters chosen according to \Scs; see \refta{tab:para}.
  The upper plots show the cross sections with $\sqrt{s}$ (left) and 
  $\mu$ (right) varied;  the middle plots show $\MSL = \MSE$ (left) and 
  $\phiMe$ (right) varied, the lower plot shows the variation with $\TB$.
}
\end{center}
\end{figure}

\begin{figure}[t]
\begin{center}
\begin{tabular}{c}
\includegraphics[width=0.48\textwidth,height=6cm]{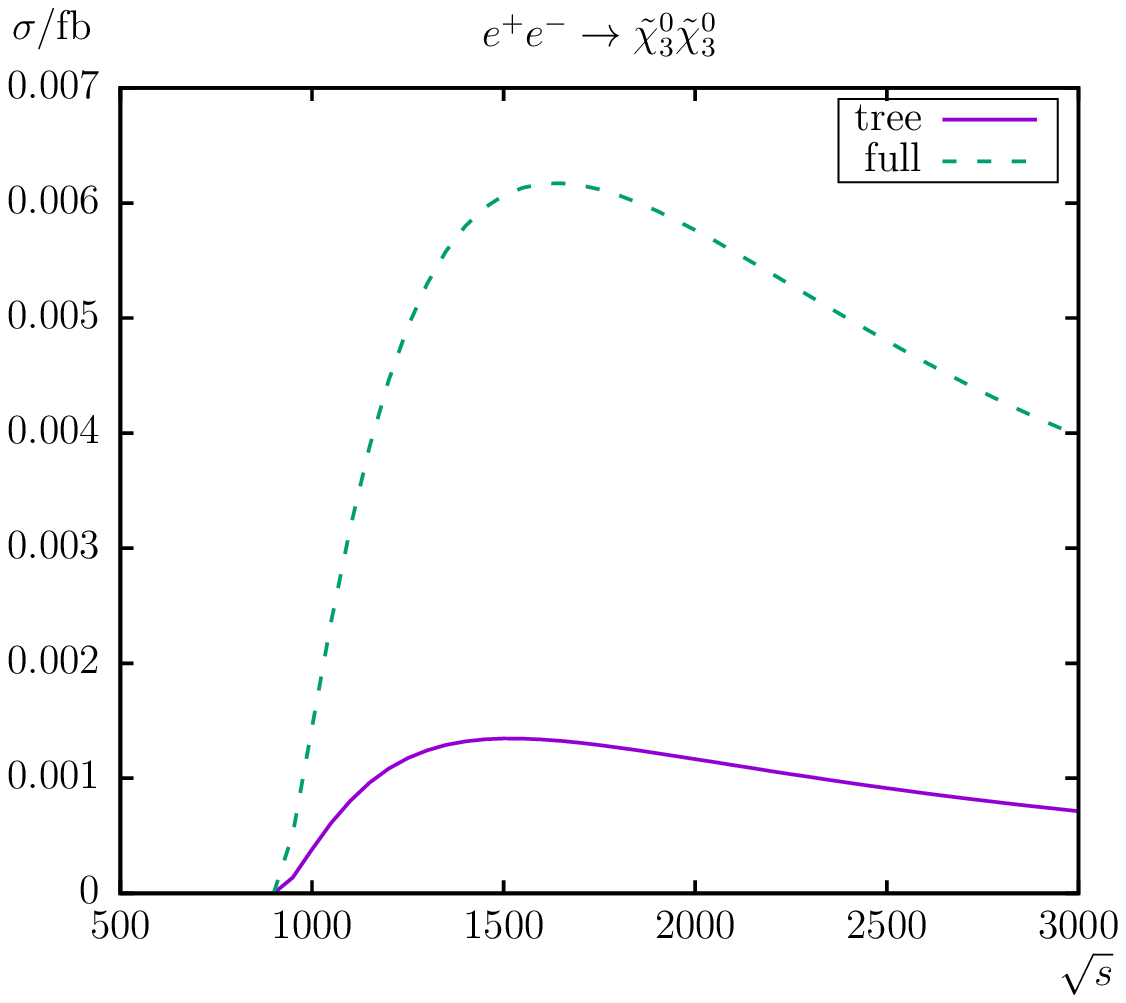}
\includegraphics[width=0.48\textwidth,height=6cm]{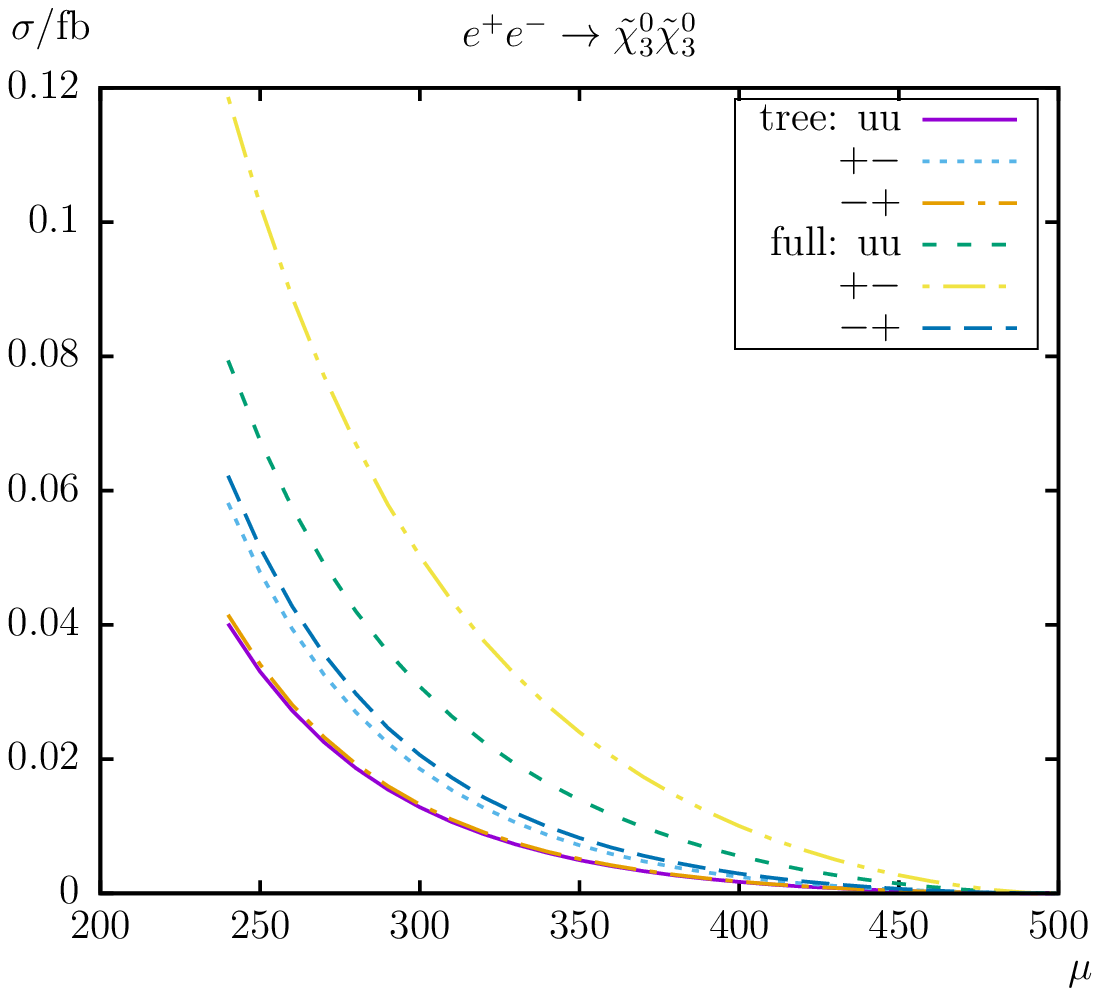}
\\[1em]
\includegraphics[width=0.48\textwidth,height=6cm]{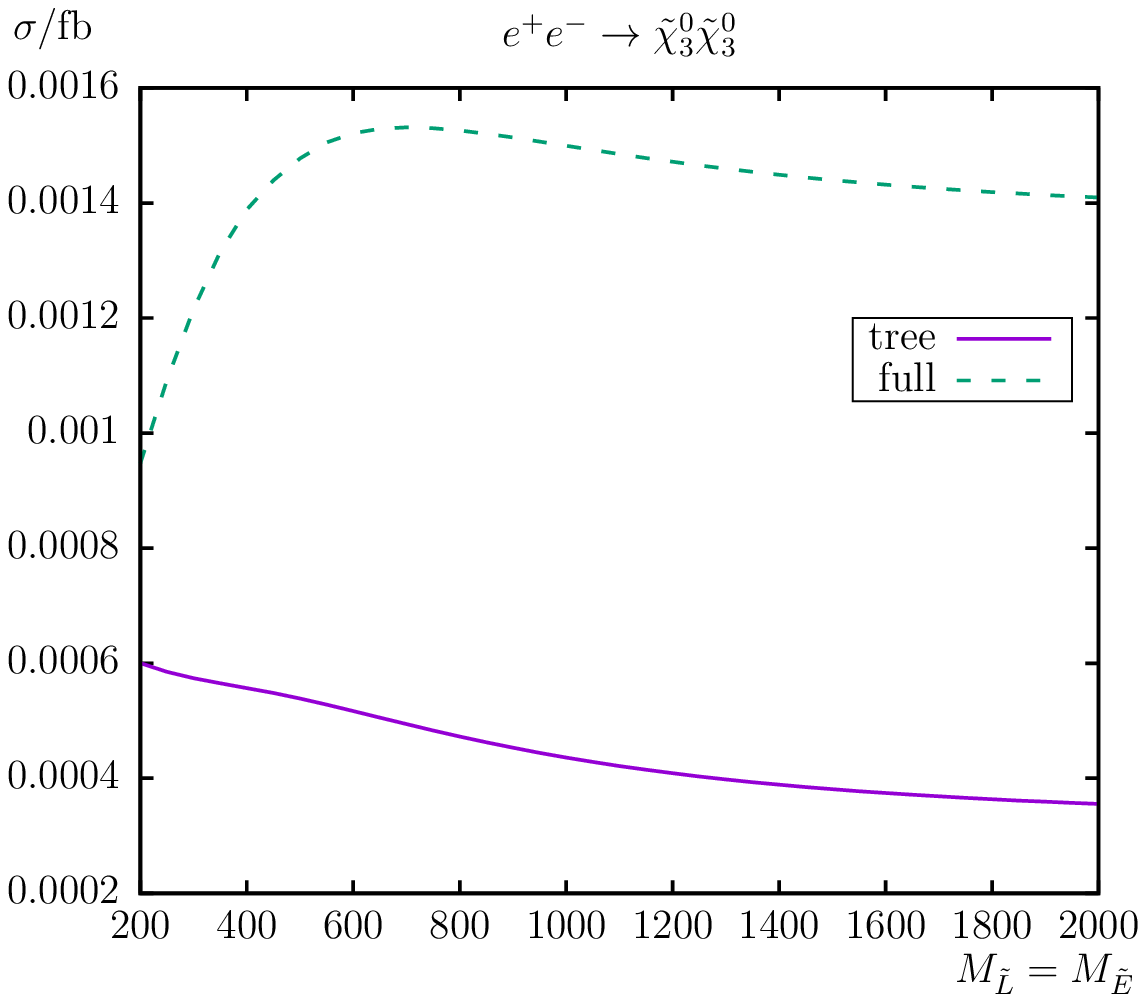}
\includegraphics[width=0.48\textwidth,height=6cm]{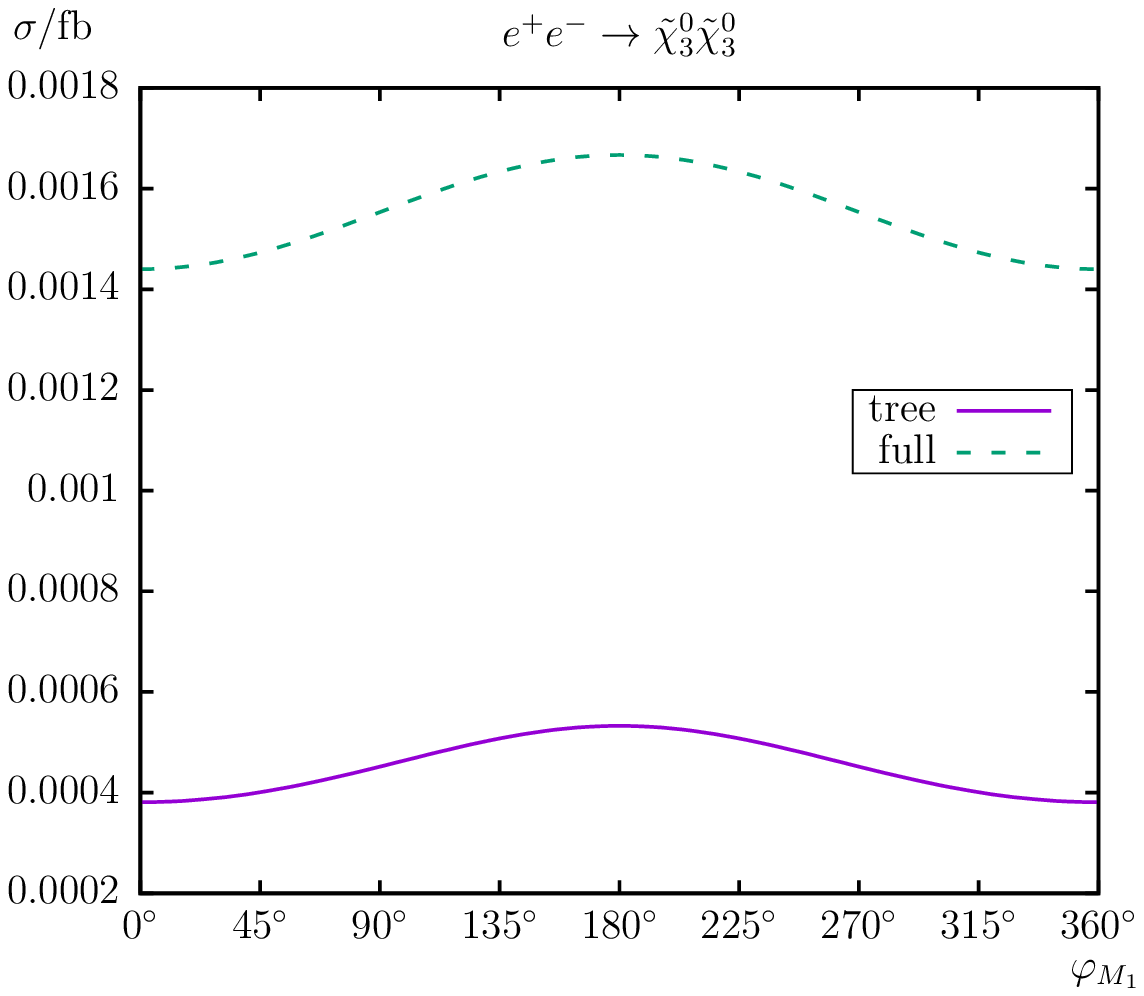}
\end{tabular}
\caption{\label{fig:een3n3}
  $\sig(\eendnd)$.
  Tree-level and full one-loop corrected cross sections are shown with 
  parameters chosen according to \Scs; see \refta{tab:para}.
  The upper plots show the cross sections with $\sqrt{s}$ (left) and 
  $\mu$ (right) varied;  the lower plots show $\MSL = \MSE$ (left) and 
  $\phiMe$ (right) varied. 
  u denotes unpolarized, $+$ right-, and $-$ left-circular 
  polarized electrons and/or positrons (see text).
}
\end{center}
\end{figure}

\begin{figure}[t]
\begin{center}
\begin{tabular}{c}
\includegraphics[width=0.48\textwidth,height=6cm]{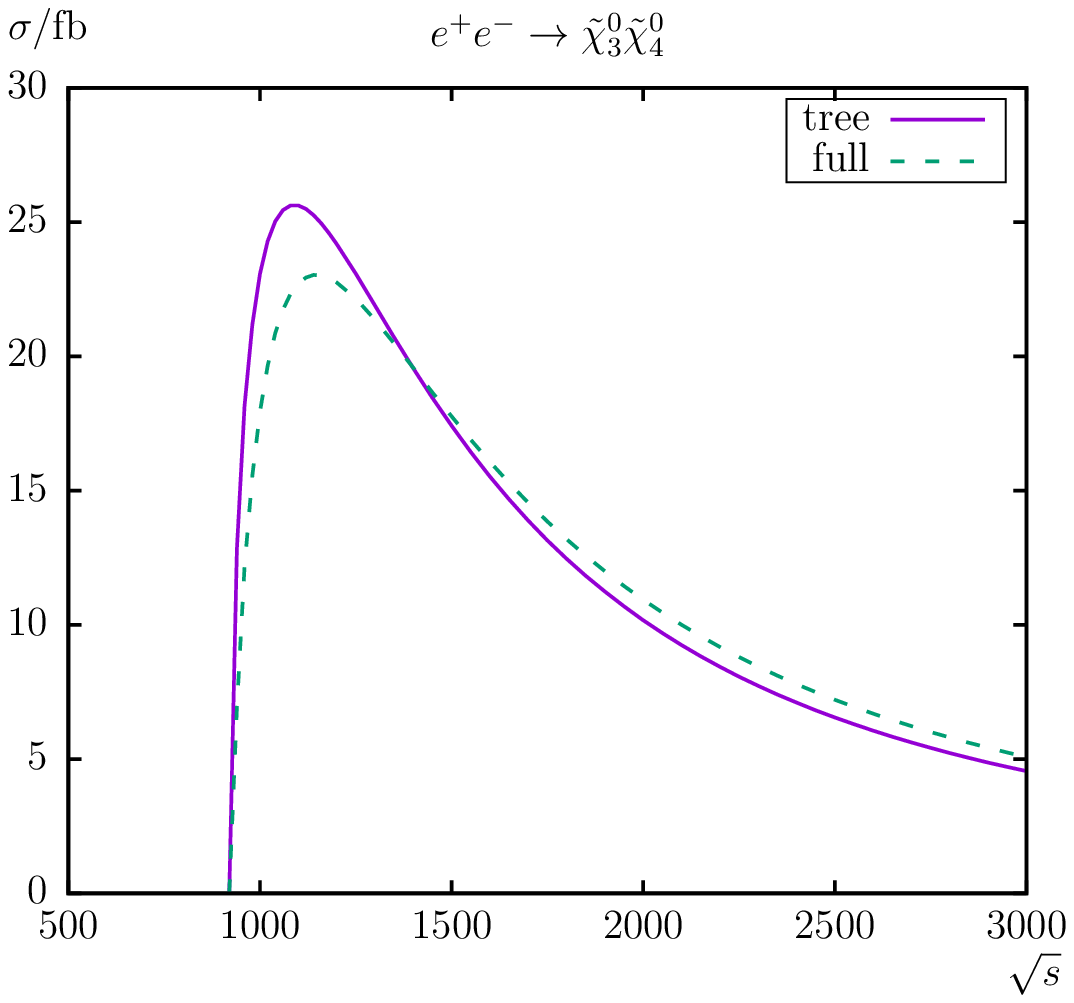}
\includegraphics[width=0.48\textwidth,height=6cm]{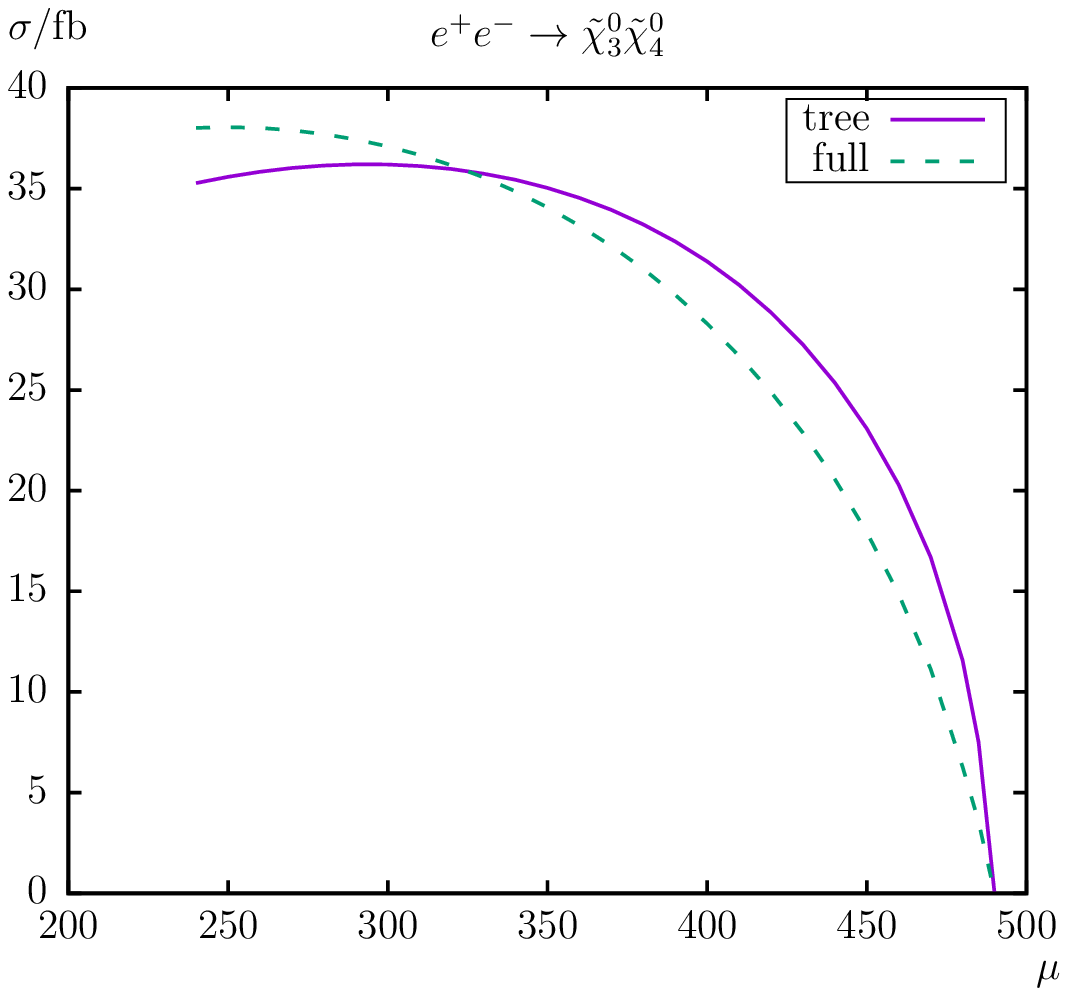}
\\[1em]
\includegraphics[width=0.48\textwidth,height=6cm]{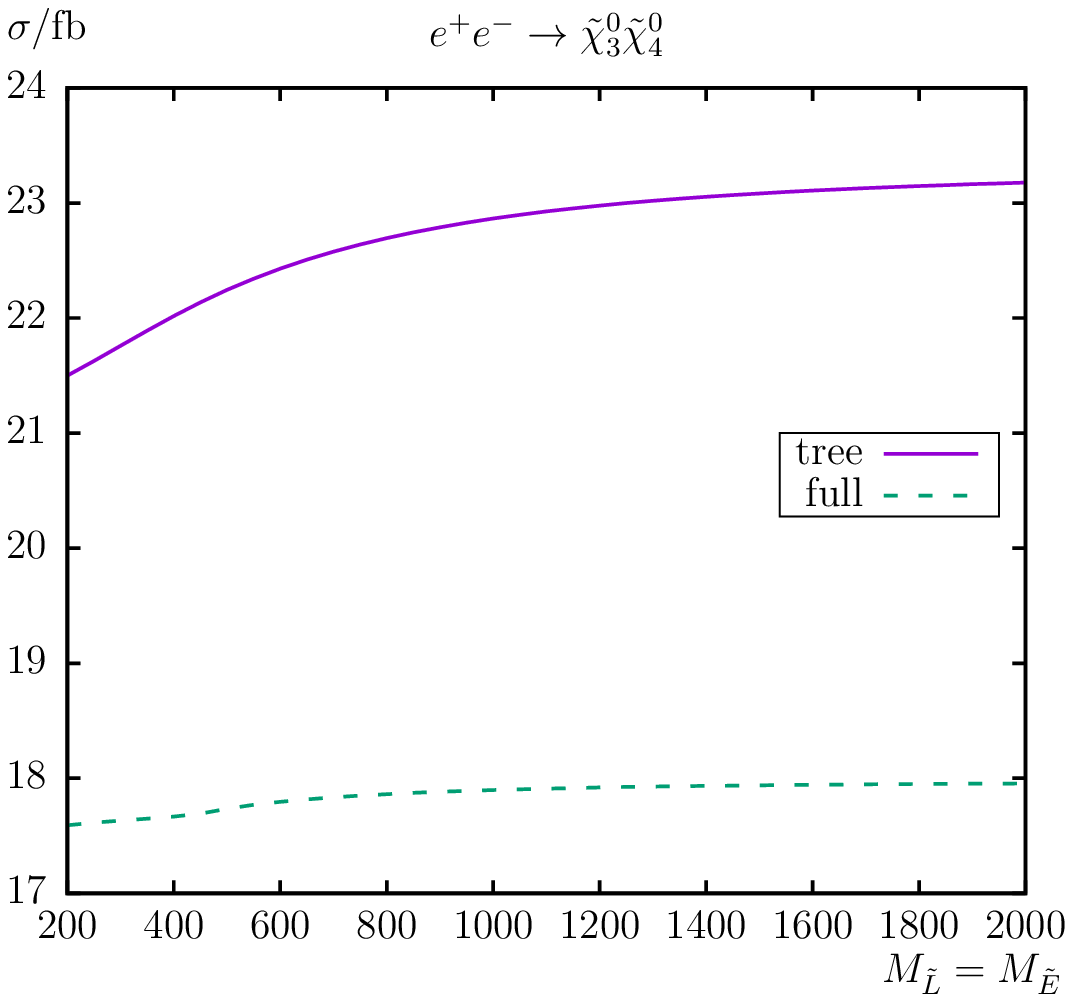}
\includegraphics[width=0.48\textwidth,height=6cm]{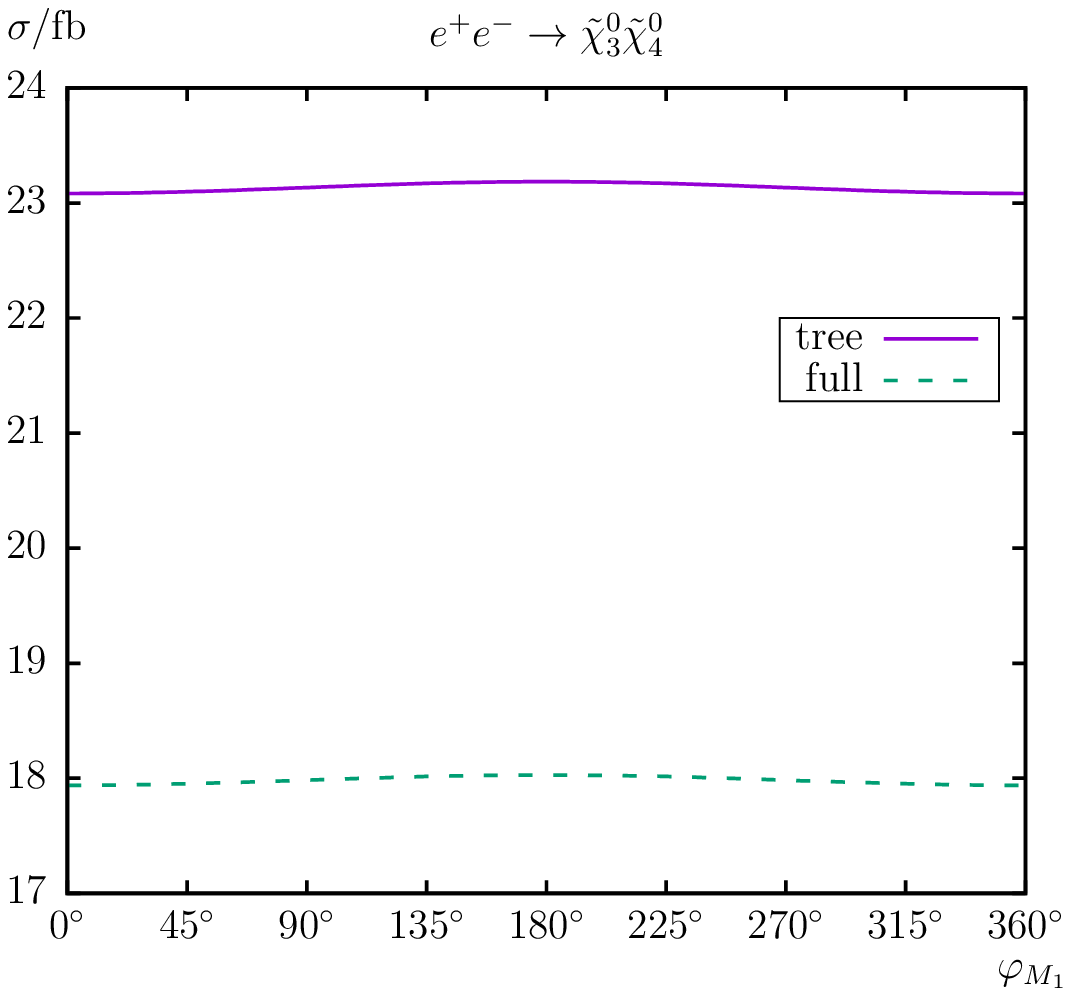}
\end{tabular}
\caption{\label{fig:een3n4}
  $\sig(\eendnv)$.
  Tree-level and full one-loop corrected cross sections are shown with 
  parameters chosen according to \Scs; see \refta{tab:para}.
  The upper plots show the cross sections with $\sqrt{s}$ (left) and 
  $\mu$ (right) varied;  the lower plots show $\MSL = \MSE$ (left) and 
  $\phiMe$ (right) varied.
}
\end{center}
\end{figure}

\begin{figure}[t]
\begin{center}
\begin{tabular}{c}
\includegraphics[width=0.48\textwidth,height=6cm]{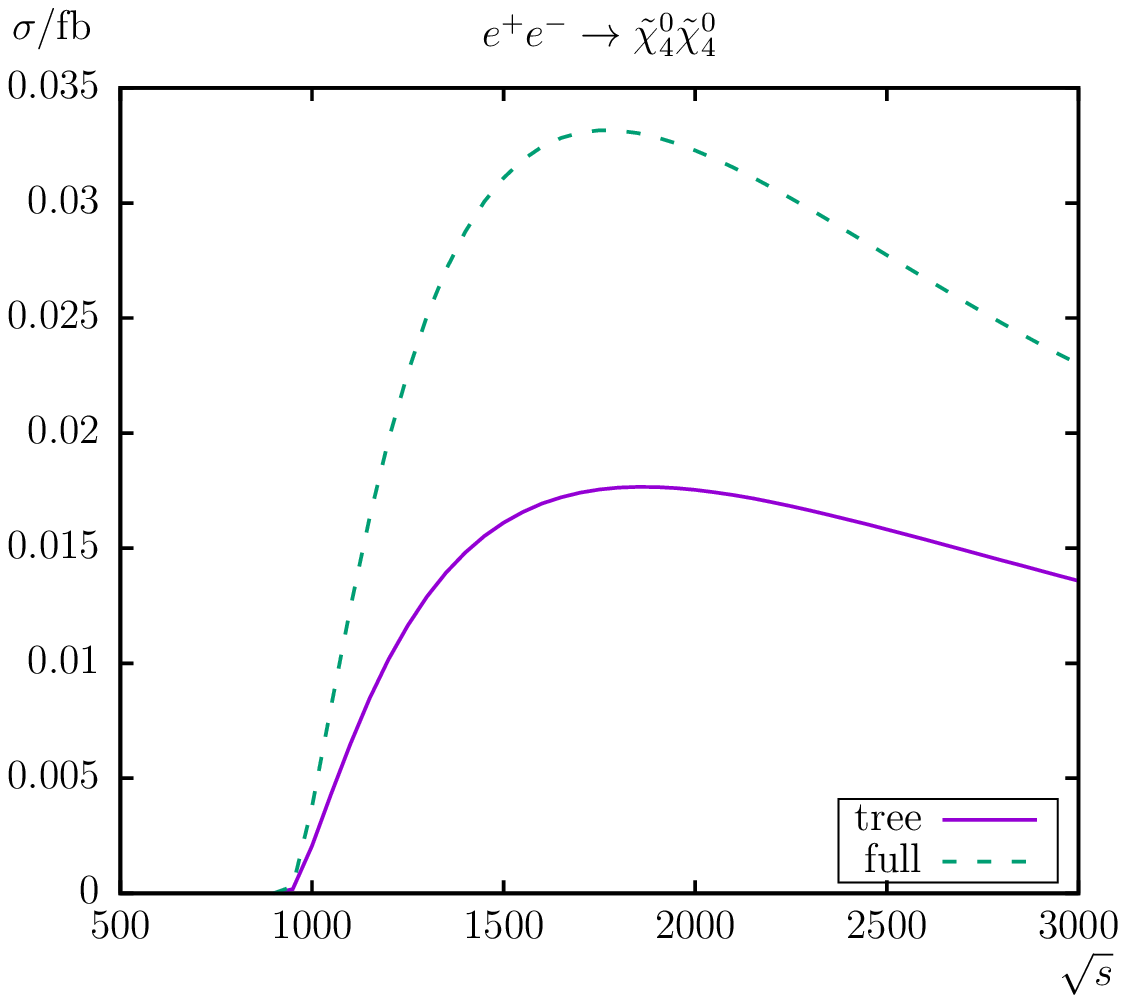}
\includegraphics[width=0.48\textwidth,height=6cm]{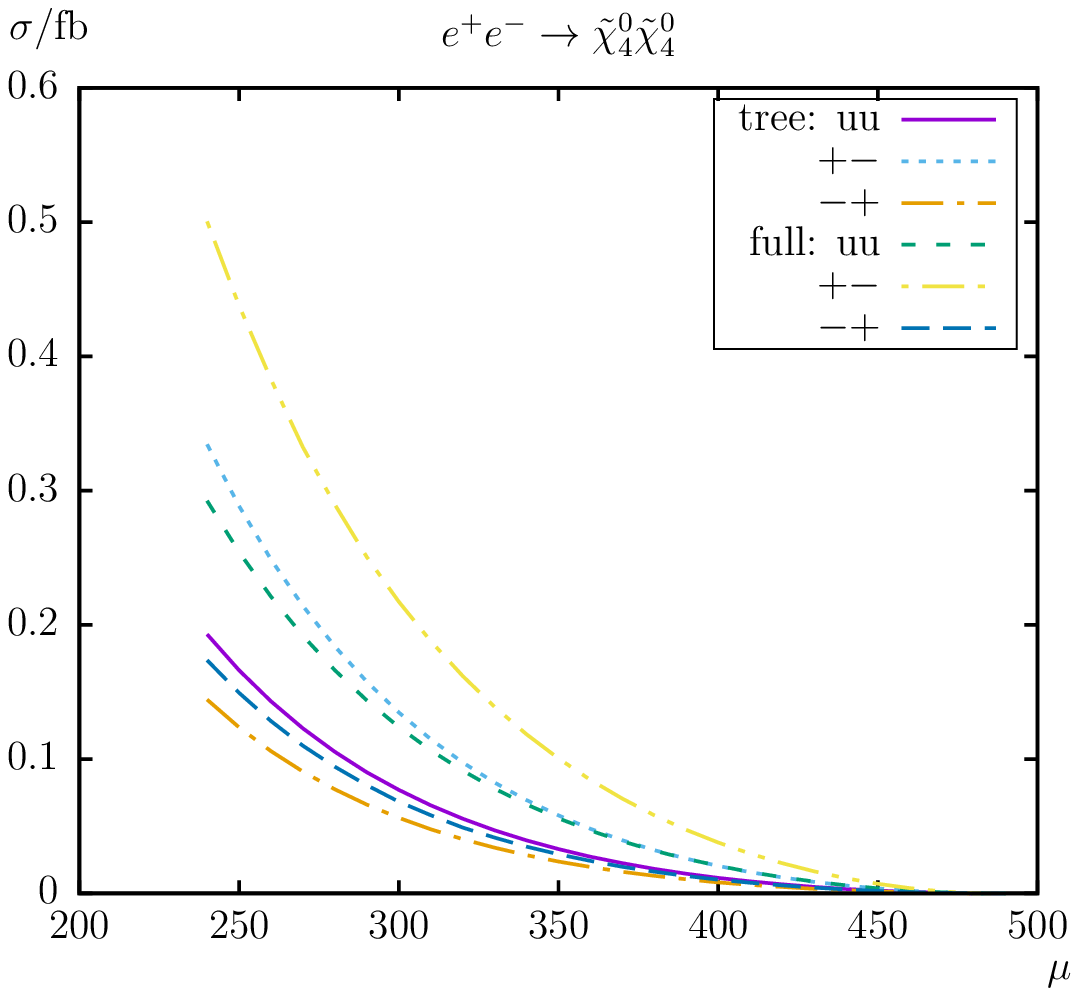}
\\[1em]
\includegraphics[width=0.48\textwidth,height=6cm]{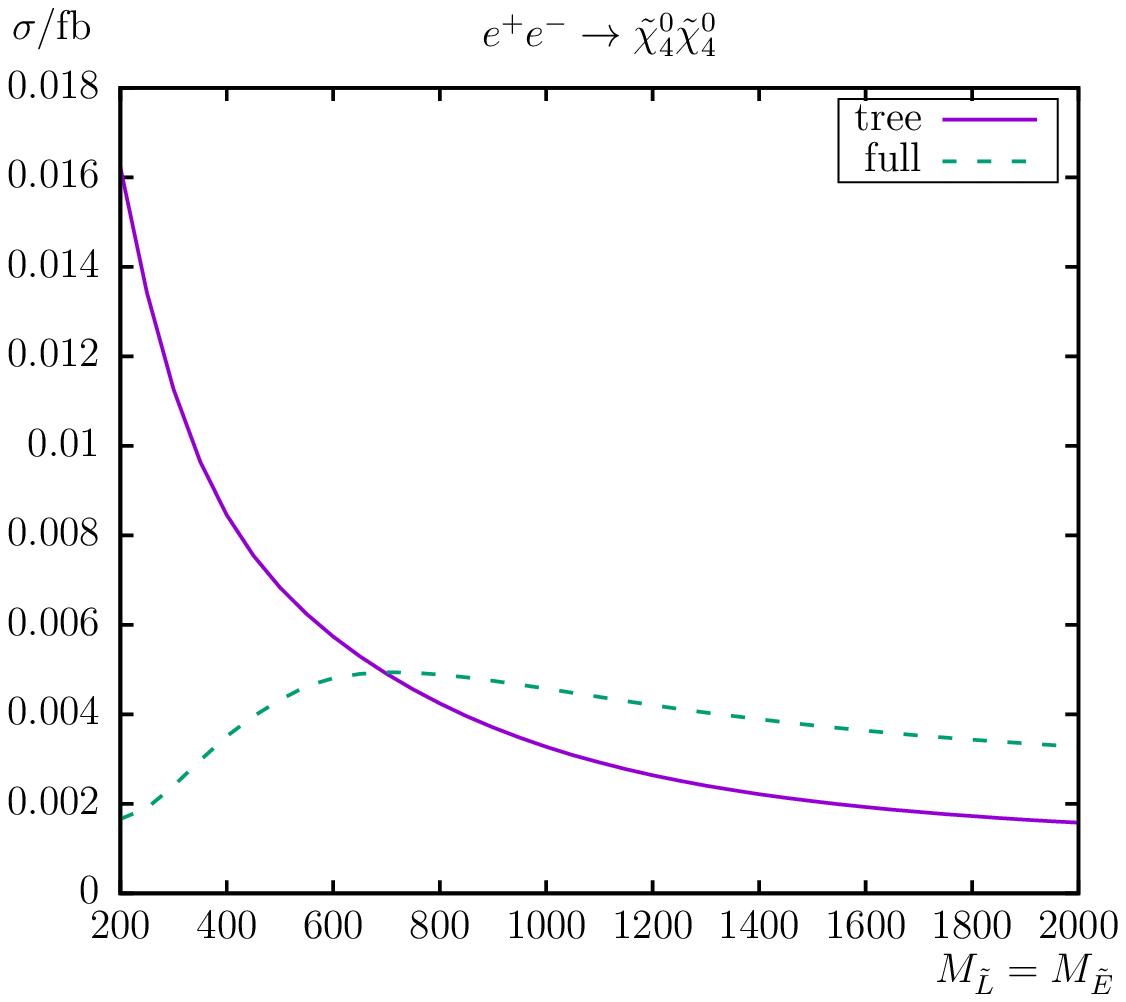}
\includegraphics[width=0.48\textwidth,height=6cm]{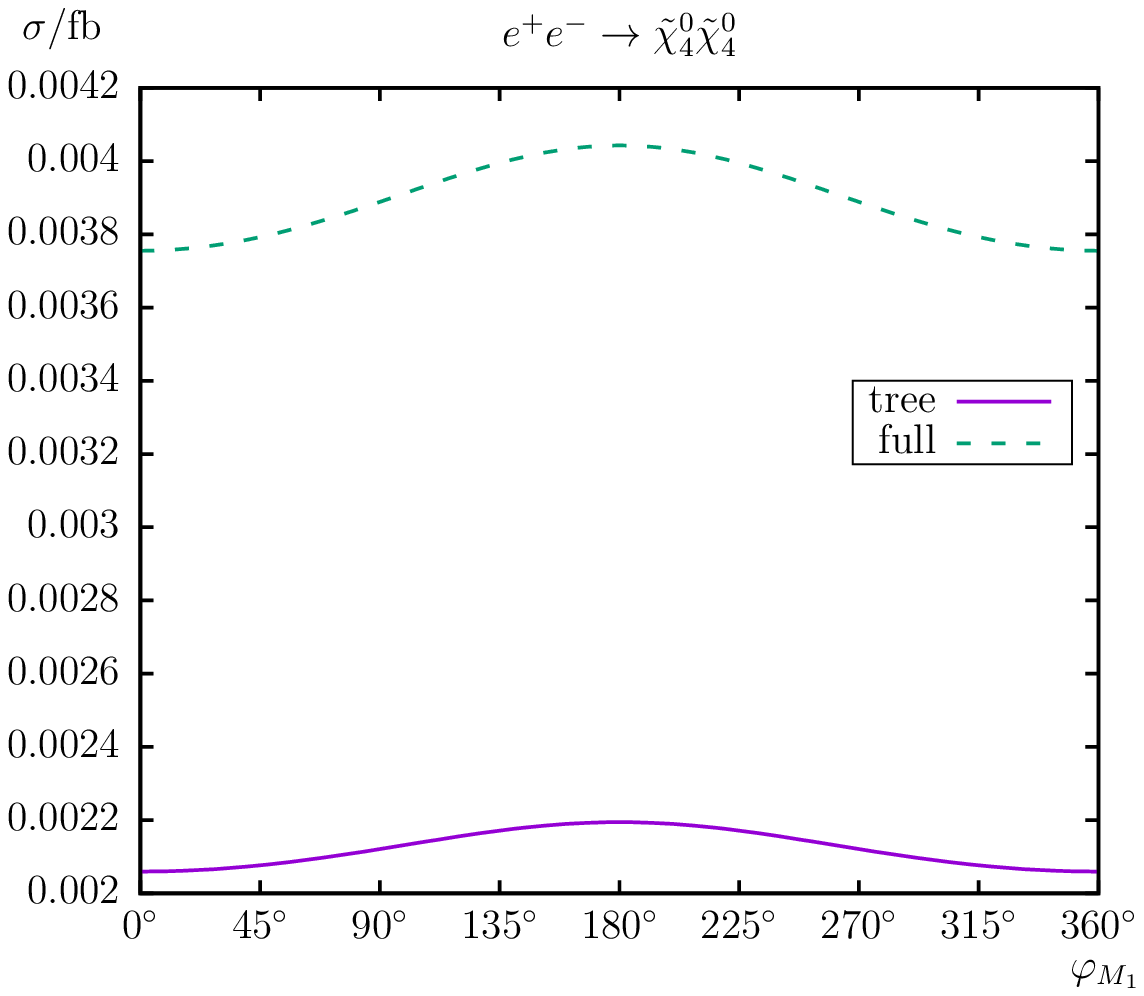}
\end{tabular}
\caption{\label{fig:een4n4}
  $\sig(\eenvnv)$.
  Tree-level and full one-loop corrected cross sections are shown with 
  parameters chosen according to \Scs; see \refta{tab:para}.
  The upper plots show the cross sections with $\sqrt{s}$ (left) and 
  $\mu$ (right) varied;  the lower plots show $\MSL = \MSE$ (left) and 
  $\phiMe$ (right) varied. 
  u denotes unpolarized, $+$ right-, and $-$ left-circular 
  polarized electrons and/or positrons (see text).
}
\end{center}
\end{figure}

In \reffis{fig:een1n1} -- \ref{fig:een4n4} we show the results for the 
processes \eenn\ ($n,\npri = 1,2,3,4$) as before as a function of 
$\sqrt{s}$, $\mu$, $\MSL = \MSE$ and $\phiMe$.  It should be noted that 
for $s \to \infty$ decreasing cross sections $\propto 1/s$ are expected;
see \citere{Bartl:1986hp}.
If not indicated otherwise, unpolarized electrons and positrons are assumed.

\medskip

We start with the process $\eenene$ shown in \reffi{fig:een1n1}.
Away from the production threshold, loop corrections of $\sim +13\%$ at 
$\sqrt{s} = 1000\gev$ are found in scenario \Scs\ (see \refta{tab:para}), 
with a maximum of nearly 7~fb at $\sqrt{s} \approx 2000\gev$. 
The relative size of the loop corrections increase with increasing 
$\sqrt{s}$ and reach $\sim +22\%$ at $\sqrt{s} = 3000\gev$.

With increasing $\mu$ in \Scs\ (upper right plot) we find a strong decrease 
of the production cross section, as can be expected from kinematics, 
discussed above.  The relative loop corrections reach $\sim +14\%$ at 
$\mu = 240\gev$ (at the border of the experimental exclusion bounds) and 
$\sim +13\%$ at $\mu = 450\gev$ (\ie \Scs).
The tree crossing takes place at $\mu \approx 1600\gev$.  For higher
$\mu$ values the loop corrections are negative, where the relative size
becomes large due to the (relative) smallness of the tree-level results,
which goes to zero for $\mu \approx 2000\gev$.

The cross sections are decreasing with increasing $\MSL$, \ie the (negative)
interference of the $t$-channel exchange decreases the cross sections, 
and the full one-loop result has its maximum of $\sim 100$~fb at 
$\MSL = 200\gev$. 
Analogously the relative corrections are decreasing from $\sim +27\%$ at 
$\MSL = 200\gev$ to $\sim +12\%$ at $\MSL = 2000\gev$.  For the other 
parameter variations one can conclude that a cross section larger by nearly 
one order of magnitude can be possible for very low $\MSL$ 
(which are not yet excluded experimentally).

Now we turn to the complex phase dependence. As for the chargino production, 
$\phiAt$ enters only via final state vertex corrections.  On the other hand, 
$\phiMe$ enters already at tree-level, and correspondingly larger effects 
are expected.
We find that the phase dependence $\phiMe$ of the cross section in \Scs\ 
is small (lower right plot), possibly not completely negligible, 
amounting up to $\sim 2.3\%$ for the full corrections.  
The loop corrections at the level of $\sim +13\%$ are found to be nearly 
independent of $\phiMe$, with a relative variation of $\sigloop/\sigtree$ 
at the level of $\sim +0.2\%$, 
(see the inlay in the lower right plot of \reffi{fig:een1n1}).
The loop effects of $\phiAt$ are found at the same level as the ones of 
$\phiMe$, \ie rather negligible.

\medskip

The relative corrections for the process $\eenenz$, as shown in 
\reffi{fig:een1n2}, are rather small for the parameter set chosen; 
see \refta{tab:para}.  In the upper left plot of \reffi{fig:een1n2} 
the peak (hardly visible in the dotted line) at 
$\sqrt{s} \approx 940\gev$ is again the production threshold 
$\mcha2 + \mcha2 = \sqrt{s}$.  The relative corrections are quasi 
constant below $\sim -2\%$ for $\sqrt{s} \gsim 1000\gev$.  

The dependence on $\mu$ with $M_2 = \mu/2$ is shown in the upper right 
plot (the case of $M_2 = 450\gev$ and the \Code{CNN[$c,n,n'$]} 
renormalization schemes are discussed below).  It is nearly linear, 
and decreasing from $\sim 2.4$~fb at small $\mu$ down to zero at 
$\mu \approx 1350\gev$ due to kinematics.  
The peak (hardly visible in the dotted line) at $\mu \approx 481\gev$ 
is the production threshold $\mcha2 + \mcha2 \approx \sqrt{s} = 1000\gev$.
The relative corrections are $\sim -2\%$ at $\mu = 230\gev$ and 
$\sim -1\%$ at $\mu = 450\gev$ (\ie \Scs).

The dependence on $\MSL$ is shown in the lower left plot of
\reffi{fig:een1n2} and follows the same pattern as for $\eenene$,
\ie a strong decrease with increasing $\MSL$.  Also in this case for 
the other parameter variations an order of magnitude increase could 
be possible for very low $\MSL$.

The phase dependence $\phiMe$ of the cross section in \Scs\ is shown in 
the lower right plot of \reffi{fig:een1n2}.
In this case it turns out to be substantial, changing the full cross
section by up to $26\%$. The tree crossings are at 
$\phiMe \approx 45^\circ, 315^\circ$.
The relative loop corrections ($\sigloop/\sigtree$) vary with $\phiMe$ 
between $\sim -0.7\%$ and $\sim +3\%$. 
The variation with $\phiAt$, on the other hand, is substantially smaller. 
The loop corrected cross section varies by less than $-1\%$, as can be 
seen in the inlay.

Finally, in addition we have calculated the process $\eenenz$ also 
within the \Code{CNN[1,1,3]}, \Code{CNN[1,1,4]}, \Code{CNN[2,1,2]}, 
and \Code{CNN[2,1,3]} renormalization schemes; see \citere{MSSMCT}.
The differences (compared to the \Code{CCN[1]} scheme including our
default choice of $M_2 = \mu/2$) for all parameters $\sqrt{s}$, $\mu$, 
$\MSL$, and $\phiMe$ varied with our input parameter set \Scs\ are very 
small (far below $1\%$).  The only exception here is the \Code{CNN[2,1,3]} 
renormalization scheme, where for $\mu > 1000\gev$ we found a slightly 
larger difference of $\sim 1\%$.  Because of these very small differences 
(within \Scs) we have omitted to show the results for the 
\Code{CNN[$c,n,n'$]} schemes in our \reffi{fig:een1n2}.

In order to analyze the differences between the various renormalization
schemes in more detail, we evaluated the process $\eenenz$ for a slightly 
different parameter set with fixed $M_2 = 450\gev$ and $\mu$ varied. 
In the upper right plot of \reffi{fig:een1n2} the corresponding results 
are shown for the \Code{CCN[1]}, \Code{CNN[1,1,3]}, and \Code{CNN[2,1,3]} 
schemes.%
\footnote{
  In the \Code{CNN[2,1,2]} scheme the mass splitting between the tree 
  level neutralino mass $\mneu3$ and the one-loop corrected mass 
  $\mneu3^\os$ (see \refeq{eq:minoOS}) is larger than 200\% for this 
  special parameter set (\ie $M_2 = 450\gev$) and therefore 
  unreliable.  The \Code{CNN[1,1,4]} scheme is even worse and 
  delivers a negative $\mneu3^\os$.
}
One can clearly see the expected breakdown of the \Code{CCN[1]} scheme 
for $\mu \approx M_2$, \ie in our case at 
$\mu \approx M_2 = 450\gev$ (see also \citeres{LHCxN,LHCxNprod})
and the smooth behavior of \Code{CNN[1,1,3]} and \Code{CNN[2,1,3]}
around $\mu \sim M_2 = 450\gev$. 
Outside the region of $\mu \sim M_2$ the scheme \Code{CCN[1]} is
expected to be reliable, since each of the three OS conditions is
strongly connected to one of the three input parameters, $M_1$, $M_2$ and
$\mu$.  Similarly, \Code{CNN[2,1,3]} (\Code{CNN[1,1,3]}) is expected to be
reliable for $\mu$ smaller (larger) than $M_2$, as in this case
again each of the three OS renormalization conditions is strongly
connected to the three input parameters.  Exactly this behavior can be
observed in the plot: for $\mu \le M_2 = 450\gev$ \Code{CNN[2,1,3]} is
nearly identical to \Code{CCN[1]}, whereas for $\mu > M_2 = 450\gev$ the 
other scheme, \Code{CNN[1,1,3]}, is very close to \Code{CCN[1]}.  
A rising deviation between the \Code{CNN[2,1,3]} and the other two 
schemes can be observed for $\mu > 1000\gev$.  Here the fact contributes 
that we have an increasing mass splitting of the one-loop corrected masses 
$\mneu{2}^\os$ between these schemes in the kinematics.%
\footnote{
  It should also be noted that for $\mu > 1092\gev$ within 
  \Code{CNN[2,1,3]} we find an (increasing) mass splitting 
  between the tree $\mneu{2}$ and corrected neutralino mass 
  $\mneu{2}^\os$ of $> 10\%$, pointing to a rather unreliable 
  scheme for this part of the parameter space.
}

\medskip

We now turn to the process $\eenend$ shown in \reffi{fig:een1n3},
which is found to be rather small of \order{1\, \fb}.
As a function of $\sqrt{s}$ (upper row, left plot) we find a small 
shift \wrt $\sqrt{s}$ directly at the production threshold, as well 
as a shift of $\sim +50\gev$ of the maximum cross section position.
The loop corrections range from $\sim +11\%$ at $\sqrt{s} = 1000\gev$ 
(\ie \Scs) to $\sim +28\%$ at $\sqrt{s} = 3000\gev$.

The dependence on $\mu$ (upper right plot) is rather small.  The relative 
corrections are $\sim +17\%$ at $\mu = 240\gev$, $\sim +11\%$ at 
$\mu = 450\gev$ (\ie \Scs), and have a tree crossing at
$\mu \approx 650\gev$. 
For larger $\mu$ the cross section goes to zero due to kinematics.

The cross section decreases with $\MSL$ (middle left plot), again
due to the negative interference of the $t$-channel
contribution. The full correction has a maximum of $\sim 2$~fb for 
$\MSL = 200\gev$, going down to $\sim 0.5$~fb at $\MSL = 2000\gev$.
Analogously the relative corrections are decreasing from $\sim +28\%$ 
at $\MSL = 200\gev$ to $\sim +8\%$ at $\MSL = 2000\gev$.

The phase dependence $\phiMe$ of the cross section in \Scs\ is shown in 
the middle right plot of \reffi{fig:een1n3}.  It is very pronounced
and can vary $\sigfull(\eenend)$ by 60\%.
The (relative) loop corrections are at the level of $\sim 10\%$ and 
vary with $\phiMe$ below $\pm 1\%$ \wrt the tree cross section. 

Here we also show the variation with $\TB$ in the lower plot of 
\reffi{fig:een1n3}. 
The loop corrected cross section decreases from $\sim 0.7$~fb at small
$\TB$ to $\sim 0.45$~fb at $\TB = 50$.  The relative corrections for the 
$\TB$ dependence are increasing from $\sim +9\%$ at $\TB = 3$ to 
$\sim +12\%$ at $\TB = 50$.

\medskip

The process $\eenenv$ is shown in \reffi{fig:een1n4}, which is found to be 
very small in \Scs\ at \order{0.1\, \fb}, but can be substantially 
larger by nearly one order of magnitude for small $\MSL$; see below.
Away from the production threshold, loop corrections of $\sim +17\%$ at 
$\sqrt{s} = 1000\gev$ (\ie \Scs) are found.  They reach their maximum of 
$\sim +19\%$ at $\sqrt{s} = 1250\gev$ and then decrease to $\sim +16\%$ at 
$\sqrt{s} = 3000\gev$.

With increasing $\mu$ in \Scs\ (upper right plot) we find again a decrease 
of the production cross section, as can be expected from kinematics.
The relative loop corrections reach $\sim +21\%$ at $\mu = 240\gev$ and go 
down to $\sim +17\%$ at $\mu = 450\gev$ (\ie \Scs).
The tree crossing is found at $\mu \approx 700\gev$, where the cross
section is already below the observable level.

The cross section depends strongly on $\MSL$.  It is decreasing with 
increasing $\MSL$ and the full corrections have their maximum of 
$\sim 2.8$~fb at $\MSL = 200\gev$, going down to $\sim 0.1$~fb at 
$\MSL = 2000\gev$.  The variation of the relative corrections are rather 
small, $\sim +14\%$ at $\MSL = 200\gev$, $\sim +17\%$ at $\MSL = 1500\gev$ 
and $\sim +14\%$ at $\MSL = 2000\gev$.

The phase dependence on $\phiMe$ of the cross section in \Scs\ is shown 
in the lower right plot.  The full cross section varies by more than 
40\%, and the (relative) loop corrections vary with $\phiMe$ between 
$\sim +17\%$ and $\sim +23\%$, \ie max.\ $\sim +6\%$.

\medskip

The process $\eenznz$ is shown in \reffi{fig:een2n2}.
Away from the production threshold we find large loop corrections of 
$\sim +43\%$ at $\sqrt{s} = 1000\gev$. The maximum cross section of
nearly 4~fb is shifted from $\sqrt{s} \approx 2100 \gev$ down to 
$\sqrt{s} \approx 1600 \gev$ due to the full one-loop corrections. 
They have a tree crossing at $\sqrt{s} \approx 1900 \gev$ and reach 
$\sim -20\%$ at $\sqrt{s} = 3000 \gev$.

With increasing $\mu$ in \Scs\ (upper right plot) we find a decrease 
of the production cross section, as can be expected from kinematics.
The relative loop corrections also decrease from $\sim +47\%$ at 
$\mu = 240\gev$ to $\sim +43\%$ at $\mu = 450\gev$ (\ie \Scs). 
The loop corrections go to zero for $\mu = 1000\gev$, where also the 
cross section goes to zero.

As for other neutralino production cross sections, $\sig(\eenznz)$ 
depends strongly on $\MSL$, where values one order of magnitude larger
than in \Scs\ with $\MSL = 1500\gev$ are possible for small $\MSL$.
One can see that the full corrections have their maximum of $\sim 50$~fb 
at $\MSL = 200\gev$.  The relative corrections are increasing from 
$\sim -3\%$ at $\MSL = 200\gev$ to $\sim +84\%$ at $\MSL = 2000\gev$ with 
a tree crossing at $\MSL \approx 550\gev$.

The phase dependence $\phiMe$ of the cross section in \Scs\ is shown in 
the lower right plot.  The full cross section varies by $\sim 7\%$, 
where loop corrections are found at the level of $\sim +43\%$ \wrt 
the tree cross section.
The relative corrections ($\sigloop/\sigtree$) vary up to $\sim +2.5\%$ 
as a function of $\phiMe$.

The dependence on $\TB$ (not shown) is qualitatively similar to $\eenend$. 
The relative corrections for the $\TB$ dependence are increasing from 
$\sim +41\%$ at $\TB = 3$ to $\sim +44\%$ at $\TB = 50$.

\medskip

Now we turn to the process $\eenznd$ shown in \reffi{fig:een2n3}.
The peak in the upper left plot of \reffi{fig:een2n3} (in the dotted line) 
at $\sqrt{s} \approx 940\gev$ is again the production threshold 
$\mcha2 + \mcha2 = \sqrt{s}$.
As a function of $\sqrt{s}$ we find relative corrections of 
$\sim +11\%$ at $\sqrt{s} = 1000\gev$ (\ie \Scs), and $\sim -2\%$ at
$\sqrt{s} = 3000\gev$ with a tree crossing at $\sqrt{s} \approx 2700 \gev$.

The dependence on $\mu$ is shown in the upper right plot.  
The peak (hardly visible in the dotted line) at $\mu \approx 481\gev$ is 
(again) the production threshold $\mcha2 + \mcha2 \approx \sqrt{s} = 1000\gev$.
The relative corrections are $\sim +16\%$ at $\mu = 240\gev$, $\sim +11\%$ 
at $\mu = 450\gev$ (\ie \Scs), and decreasing with a tree crossing
for $\mu \approx 550\gev$.  Due to kinematics the cross section goes to
zero for $\mu = 680\gev$.

In the analysis as a function of $\MSL$ (lower row, left plot) the cross 
section is decreasing with increasing $\MSL$, but varies (only) by a
factor of $\sim 2$ \wrt \Scs. The full correction has its maximum of 
$\sim 4.5$~fb at $\MSL = 200\gev$.  The relative corrections are increasing 
from $\sim -7\%$ at $\MSL = 200\gev$ to $\sim +12\%$ at $\MSL = 2000\gev$ 
with a tree crossing at $\MSL \approx 520\gev$.

The phase dependence $\phiMe$ of the cross section in \Scs\ is shown in 
the lower right plot of \reffi{fig:een2n3}.
The full cross section is found to be very at the per-cent level.
The loop corrections are $\sim 11\%$, but the (relative)
variation with $\phiMe$ stays below $0.4\%$.

\medskip

The process $\eenznv$ is shown in \reffi{fig:een2n4} and is found to be 
rather small, where as before an increase by an order of magnitude is 
possible for low $\MSL$; see below.
The peak in the upper left plot (not visible in the dotted line) at 
$\sqrt{s} \approx 940\gev$ is (again) the production threshold 
$\mcha2 + \mcha2 = \sqrt{s}$.
As a function of $\sqrt{s}$ we find loop corrections of $\sim +14\%$ 
at $\sqrt{s} = 1000\gev$ (\ie \Scs), a tree crossing at 
$\sqrt{s} \approx 1400\gev$ and $\sim -38\%$ at $\sqrt{s} = 3000\gev$.

The dependence on $\mu$ is shown in the upper right plot.  The peak 
(not visible in the dotted line) at $\mu \approx 481\gev$ is (again) 
the production threshold $\mcha2 + \mcha2 \approx \sqrt{s} = 1000\gev$.
The relative corrections are $\sim +20\%$ at $\mu = 240\gev$, $\sim +14\%$ 
at $\mu = 450\gev$ (\ie \Scs), and decrease further for larger $\mu$, 
crossing zero at $\mu \approx 600\gev$, where the cross section is below 
the observable level in \Scs.

In the analysis as a function of $\MSL$ (middle left plot) the cross 
sections are decreasing with increasing $\MSL$ and the full corrections 
have their maximum of $\sim 3$~fb at $\MSL = 200\gev$, about an order of 
magnitude larger than in \Scs.  The relative corrections are changing 
from $\sim -28\%$ at $\MSL = 200\gev$ to $\sim +28\%$ at 
$\MSL = 2000\gev$ with a tree crossing at $\MSL = 1100\gev$.

The phase dependence $\phiMe$ of the cross section in \Scs\ is shown 
in the middle right plot of \reffi{fig:een2n4}.  
The full correction is seen to vary up to $\sim +12\%$ with loop
corrections increasing the tree-level result by $\sim +14\%$.
The phase dependence of the relative loop correction is (again) rather 
small and found to be below $0.6\%$.

We show again in the lower row the dependence on $\TB$. 
Contrary to other neutralino production cross sections analyzed before, 
$\sigfull(\eenznv)$ increases with $\TB$ by up to $\sim 21\%$ going 
from the lowest to the highest $\TB$ values. 
The relative corrections for the $\TB$ dependence vary below $+0.8\%$, 
between $\sim +13.75\%$ at $\TB = 17$ and $\sim +14.5\%$ at $\TB = 50$.

\medskip

The process $\eendnd$ is shown in \reffi{fig:een3n3}.
The overall size of this cross section turns out to be very small,
including all analyzed parameter variations. Consequently, the loop
corrections have a sizable impact, as can be seen in all four panels
of \reffi{fig:een3n3}, but never lift the cross section above 0.08~fb.
For this reason we refrain from a more detailed discussion here.
However, we would like to remark that with polarized positrons 
($P(e^+) = +30\%$) and electrons ($P(e^-) = -80\%$) cross sections up to 
$\sim 0.1$~fb are possible in \Scs, as we show in the upper right plot.
This could result in an observable cross section for some parts of the 
allowed parameter range; see \citere{pol-report} for related discussions.

\medskip

We now turn to the process $\eendnv$ shown in \reffi{fig:een3n4}, 
which turns out to be sizable at the level of several 10~fb.
As a function of $\sqrt{s}$ (upper left plot) we find loop corrections 
of $\sim -20\%$ at $\sqrt{s} = 1000\gev$ (\ie \Scs), and $\sim +11\%$ at 
$\sqrt{s} = 3000\gev$, with a tree crossing at $\sqrt{s} \approx 1400\gev$.

The dependence on $\mu$ is shown in the upper right plot.  The relative 
corrections are $\sim +8\%$ at $\mu = 240\gev$, $\sim -22\%$ at 
$\mu = 450\gev$ (\ie \Scs), where the cross section goes to zero at 
$\mu = 490\gev$.  The tree crossing is found at $\mu \approx 320\gev$.

In the analysis as a function of $\MSL$ (lower left plot) the cross section
is nearly independent of $\MSL$, due to the strong higgsino admixture of 
the final state neutralinos.  This feature was already observable in 
\reffi{fig:een3n3}.  The loop corrected cross section is $\sim 18$~fb. 
The relative corrections are increasing from $\sim -18\%$ at 
$\MSL = 200\gev$ to $\sim -23\%$ at $\MSL = 2\tev$.

The phase dependence $\phiMe$ of the cross section in \Scs\ is shown in 
the lower right plot of \reffi{fig:een3n4} and is found to be negligible. 
This applies to the cross section as well as the absolute and the relative 
size of the loop corrections as a function of $\phiMe$.

\medskip

Finally we analyze the process $\eenvnv$, shown in \reffi{fig:een4n4}.
Again as for $\eendnd$ the overall size of this cross section is small, 
\order{10\, \ab}.  This holds again for all parameter variations.
For this reason we (again) skip a more detailed discussion here.
Using polarized electrons/positrons ($P(e^+) = +30\%$ and $P(e^-) = -80\%$),
as shown in the upper right plot could yield production cross sections up 
to $\sim 0.4$~fb, again possibly observable over some part of the relevant 
parameter space.

\medskip

Overall, for the neutralino pair production the leading order corrections 
can reach a level of \order{10\, \fb}, depending on the SUSY parameters,
but is very small for the production of two equal higgsino dominated
neutralinos at the \order{10\, \ab} level.  This renders these 
processes difficult to observe at an $e^+e^-$ collider.%
\footnote{
  The limit of $10$~ab corresponds to ten events at an integrated 
  luminosity of $\cL = 1\, \iab$, which constitutes a guideline 
  for the observability of a process at a linear collider.
}
Having both beams polarized could turn out to be crucial to yield a
detectable production cross section in this case; 
see \citere{pol-report} for related analyses.

The full one-loop corrections are very roughly 10-20\% of the tree-level 
results, but vary strongly on the size of $\mu$ and $\MSL$.
Depending on the size of in particular these two parameters the loop 
corrections can be either positive or negative.
This shows that the loop corrections, while being large, have to be
included point-by-point in any precision analysis.
The dependence on $\phiMe$ was found at the level of $\sim 15\%$, but
can go up to $\sim 40\%$ for the extreme cases.  
The relative loop corrections varied by up to $5\%$ with $\phiMe$.
Consequently, the complex phase dependence must be taken into account 
as well.


\section{Conclusions}
\label{sec:conclusions}

We have evaluated all chargino/neutralino production modes at $e^+e^-$
colliders with a two-particle final state, \ie \eecc\ and \eenn\ 
allowing for complex parameters. 
In the case of a discovery of charginos and neutralinos a subsequent
precision measurement of their properties will be crucial to determine
their nature and the underlying (SUSY) parameters. 
In order to yield a sufficient accuracy, one-loop corrections to the 
various chargino/neutralino production modes have to be considered. 
This is particularly the case for the high anticipated accuracy of the
chargino/neutralino property determination at $e^+e^-$
colliders~\cite{LCreport}.

The evaluation of the processes (\ref{eq:eecc}) and (\ref{eq:eenn})
is based on a full one-loop calculation, also including hard and soft 
QED radiation.  The renormalization is chosen to be identical as for 
the various chargino/neutralino decay calculations; see, \eg
\citeres{LHCxC,LHCxN,LHCxNprod} or chargino/higgsino production from 
heavy Higgs boson decay; see, \eg \citere{HiggsDecayIno}. 
Consequently, the predictions for the production and decay can be used 
together in a consistent manner (\eg in a global phenomenological 
analysis of the chargino/neutralino sector at the one-loop level).

We first very briefly reviewed the relevant sectors including some
details of the one-loop renormalization procedure of the cMSSM, which are 
relevant for our calculation.  In most cases we follow \citere{MSSMCT}. 
We have discussed the calculation of the one-loop diagrams, the
treatment of UV, IR, and collinear divergences that are canceled by the 
inclusion of (hard, soft, and collinear) QED radiation. 
As far as possible we have checked our result against the literature, 
and in most cases we found good agreement, 
where parts of the differences can be attributed to problems with input 
parameters and/or different renormalization schemes (conversions).

For the analysis we have chosen a standard parameter set (see
\refta{tab:para}), that allows the production of all combinations of
charginos/neutralinos at an $e^+e^-$ collider with a center-of-mass energy up
to $\sqrt{s} = 1000\gev$.
In the analysis we investigated the variation of the various production
cross sections with the center-of-mass energy $\sqrt{s}$, the Higgs mixing
parameter $\mu$, the slepton soft SUSY-breaking parameter $\MSL$ and the
complex phases $\phiMe$ and $\phiAt$ of the gaugino mass parameter $M_1$ and
the trilinear Higgs-stop coupling, $\At$, respectively. Where relevant we also
showed the variation with $\TB$. 

In our numerical scenarios we compared the tree-level 
production cross sections with the full one-loop corrected cross sections. 
The numerical results we have shown are, of course, dependent on the choice 
of the SUSY parameters. Nevertheless, they give an idea of the relevance
of the full one-loop corrections.
For the chargino pair production, \eecc, we observed an decreasing 
cross section $\propto 1/s$ for $s \to \infty$. 
The full one-loop corrections are very roughly 10-20\% of the
tree-level results, but depend strongly on the size of $\mu$, where
larger values result even in negative loop corrections.
The cross sections are largest for $\eecece$ and $\eeczcz$ and roughly
smaller by one order of magnitude for $\eececz$ due to the absence of the  
$\gamma\, \chapm{1} \champ{2}$ coupling at tree level in the MSSM.
The variation of the cross sections and of the $\CP$~asymmetry $A_{12}$
with $\phiMe$ or $\phiAt$ is found extremely small and the 
dependence on other phases were found to be roughly at the same level and 
have not been shown explicitely.

For the neutralino pair production, \eenn, the cross section can 
reach a level of \order{10\, \fb}, depending on the SUSY parameters,
but is very small for the production of two equal higgsino dominated
neutralinos at the \order{10\, \ab}.  This renders these 
processes difficult to observe at an $e^+e^-$ collider.%
\footnote{
  The limit of $10$~ab corresponds to ten events at an integrated 
  luminosity of $\cL = 1\, \iab$, which constitutes a guideline 
  for the observability of a process at a linear collider.
}
Having both beams polarized could turn out to be crucial to yield a
detectable production cross section in this case. 
The full one-loop corrections are very roughly 10-20\% of the tree-level 
results, but vary strongly on the size of $\mu$ and $\MSL$.
Depending on the size of in particular these two parameters the loop 
corrections can be either positive or negative.
The dependence on $\phiMe$ was found to reach up to $\sim 15\%$,
but can go up to $\sim 40\%$ for the extreme cases.  The (relative)
loop corrections varied by up to $5\%$ with $\phiMe$. 
This shows that the loop corrections, including the complex phase dependence,
have to be included point-by-point in any precision analysis, or any 
precise determination of (SUSY) parameters from the production of
cMSSM charginos/neutralinos at $e^+e^-$ linear colliders.
We emphasize again that our full one-loop calculation can readily be
used together with corresponding full one-loop corrections to
chargino/neutralino decays~\cite{LHCxC,LHCxN,LHCxNprod} or other
chargino/neutralino production modes~\cite{HiggsDecayIno}.


\subsection*{Acknowledgements}

We thank A.~Bharucha, T.~Blank, T.~Hahn and F.~von~der~Pahlen for
helpful discussions.  
The work of S.H.\ is supported in part by CICYT (Grant FPA 2013-40715-P),
in part by the MEINCOP Spain under contract FPA2016-78022-P, 
in part by the ``Spanish Agencia Estatal de Investigación'' (AEI) and the EU
``Fondo Europeo de Desarrollo Regional'' (FEDER) through the project
FPA2016-78022-P, 
and by the Spanish MICINN's Consolider-Ingenio 2010 Program under Grant 
MultiDark CSD2009-00064.



\newcommand\jnl[1]{\textit{\frenchspacing #1}}
\newcommand\vol[1]{\textbf{#1}}


\begin{thebibliography}{99} 

\begingroup \raggedright

%
%

\bibitem{mssm}
H.~Nilles, 
\jnl{Phys. Rept.} \vol{110} (1984) 1; \\
R.~Barbieri, 
\jnl{Riv. Nuovo Cim.} \vol{11} (1988) 1. 

\bibitem{HaK85}
H.~Haber, G.~Kane,
\jnl{Phys. Rept.} \vol{117} (1985) 75.

\bibitem{GuH86}
J.~Gunion, H.~Haber,
\jnl{Nucl. Phys.} \vol{B 272} (1986) 1.

\bibitem{ATLASdiscovery} 
G.~Aad et al.\ [ATLAS Collaboration],
\jnl{Phys. Lett.} \vol{B 716} (2012) 1
[arXiv:1207.7214 [hep-ex]].

\bibitem{CMSdiscovery} 
S.~Chatrchyan et al.\ [CMS Collaboration],
\jnl{Phys. Lett.} \vol{B 716} (2012) 30 
[arXiv:1207.7235 [hep-ex]].

\bibitem{EHNOS} 
H.~Goldberg,
\jnl{Phys. Rev. Lett.} \vol{50} (1983) 1419;\\
J.~Ellis, J.~Hagelin, D.~Nanopoulos, K.~Olive and M.~Srednicki,
\jnl{Nucl. Phys.} \vol{B 238} (1984) 453.

\bibitem{mhiggsCPXgen} A.~Pilaftsis,
\jnl{Phys. Rev.} \vol{D 58} (1998) 096010
[arXiv:hep-ph/9803297];\\
A.~Pilaftsis,
\jnl{Phys. Lett.} \vol{B 435} (1998) 88
[arXiv:hep-ph/9805373].

\bibitem{Demir}
D.~Demir,
\jnl{Phys. Rev.} \vol{D 60} (1999) 055006
[arXiv:hep-ph/9901389].

\bibitem{mhiggsCPXRG1} 
A.~Pilaftsis and C.~Wagner, 
\jnl{Nucl. Phys.} \vol{B 553} (1999) 3
[arXiv:hep-ph/9902371].

\bibitem{mhiggsCPXFD1} 
S. Heinemeyer,
\jnl{Eur. Phys. J.} \vol{C 22} (2001) 521
[arXiv:hep-ph/0108059].

\bibitem{ILC-TDR}
H.~Baer et al.,
\textit{The International Linear Collider Technical Design Report - Volume 2:
  Physics},
arXiv:1306.6352 [hep-ph].

\bibitem{teslatdr} 
TESLA Technical Design Report [TESLA Collaboration] Part~3, 
\textit{Physics at an $e^+e^-$ Linear Collider},
arXiv:hep-ph/0106315, \\
see: \url{http://tesla.desy.de/new_pages/TDR_CD/start.html}\,;\\
K.~Ackermann et al.,
DESY-PROC-2004-01.

\bibitem{ilc}
J.~Brau et al.\  [ILC Collaboration],
\textit{ILC Reference Design Report Volume 1 - Executive Summary},
arXiv:0712.1950 [physics.acc-ph];\\
G.~Aarons et al.\  [ILC Collaboration],
\textit{International Linear Collider Reference Design Report Volume 2:
  Physics at the ILC},
arXiv:0709.1893 [hep-ph].

\bibitem{LCreport}
G.~Moortgat-Pick et al.,
\jnl{Eur. Phys. J.} \vol{C 75} (2015) 8, 371
[arXiv:1504.01726 [hep-ph]].

\bibitem{CLIC} 
L.~Linssen, A.~Miyamoto, M.~Stanitzki and H.~Weerts,
arXiv:1202.5940 [physics.ins-det];\\
H.~Abramowicz et al.\ [CLIC Detector and Physics Study Collaboration],
\textit{Physics at the CLIC $e^+e^-$ Linear Collider -- 
Input to the Snowmass process 2013}, 
arXiv:1307.5288 [hep-ex].

\bibitem{lhcilc} 
G.~Weiglein et al.\ [LHC/ILC Study Group],
\jnl{Phys. Rept.} \vol{426} (2006) 47
[arXiv:hep-ph/0410364];\\
A.~De Roeck et al.,
\jnl{Eur. Phys. J.} \vol{C 66} (2010) 525
[arXiv:0909.3240 [hep-ph]];\\
A.~De Roeck, J.~Ellis and S.~Heinemeyer,
\jnl{CERN Cour.} \vol{49N10} (2009) 27.

\bibitem{LHCxC} 
S.~Heinemeyer, F.~von~der~Pahlen and C.~Schappacher,
\jnl{Eur. Phys. J.} \vol{C 72} (2012) 1892
[arXiv:1112.0760 [hep-ph]];\\
S.~Heinemeyer, F.~von~der~Pahlen and C.~Schappacher,
arXiv:1202.0488 [hep-ph].

\bibitem{LHCxN} 
A.~Bharucha, S.~Heinemeyer, F.~von~der~Pahlen and 
C.~Schappacher,
\jnl{Phys. Rev.} \vol{D 86} (2012) 075023
[arXiv:1208.4106 [hep-ph]].

\bibitem{LHCxNprod}
A.~Bharucha, S.~Heinemeyer and F.~von~der~Pahlen,
\jnl{Eur. Phys. J.} \vol{C 73} (2013) 2629
[arXiv:1307.4237 [hep-ph]].

\bibitem{HiggsDecayIno} 
S.~Heinemeyer and C.~Schappacher,
\jnl{Eur. Phys. J.} \vol{C 75} (2015) 5, 230
[arXiv:1503.02996 [hep-ph]].

\bibitem{Bartl:1985fk}
A.~Bartl, H.~Fraas and W.~Majerotto,
\jnl{Z. Phys.} \vol{C 30} (1986) 441.

\bibitem{Bartl:1986hp}
A.~Bartl, H.~Fraas and W.~Majerotto,
\jnl{Nucl. Phys.} \vol{B 278} (1986) 1.

\bibitem{Gounaris:2002pj}
G.~J.~Gounaris and C.~Le Mouel,
\jnl{Phys. Rev.} \vol{D 66} (2002) 055007
[arXiv:hep-ph/0204152].

\bibitem{Osland:2007xw}
P.~Osland and A.~Vereshagin,
\jnl{Phys. Rev.} \vol{D 76} (2007) 036001
[arXiv:0704.2165 [hep-ph]].

\bibitem{RoKa2007}
K.~Rolbiecki and J.~Kalinowski,
\jnl{Phys. Rev.} \vol{D 76} (2007) 115006
[arXiv:0709.2994 [hep-ph]].

\bibitem{OsKaRoVe2007}
P.~Osland, J.~Kalinowski, K.~Rolbiecki and A.~Vereshagin,
arXiv:0709.3358 [hep-ph].

\bibitem{Diaz:1997kv}
M.~A.~Diaz, S.~F.~King and D.~A.~Ross,
\jnl{Nucl. Phys.} \vol{B 529} (1998) 23
[arXiv:hep-ph/9711307].

\bibitem{Kiyoura:1998yt}
S.~Kiyoura, M.~M.~Nojiri, D.~M.~Pierce and Y.~Yamada,
\jnl{Phys. Rev.} \vol{D 58} (1998) 075002
[arXiv:hep-ph/9803210].

\bibitem{Blank:2000uc}
T.~Blank and W.~Hollik,
arXiv:hep-ph/0011092.

\bibitem{Diaz:2002rr}
M.~A.~Diaz and D.~A.~Ross,
arXiv:hep-ph/0205257.

\bibitem{Kilian:2006cj}
W.~Kilian, J.~Reuter and T.~Robens,
\jnl{Eur. Phys. J.} \vol{C 48} (2006) 389
[arXiv:hep-ph/0607127].

\bibitem{Robens:2006np}
T.~Robens,
arXiv:hep-ph/0610401.

\bibitem{Blank:2000fqa}
T.~Blank,
``Strahlungskorrekturen zur Chargino- und Neutralinoproduktion in $e^+e^-$ 
und Hadronkollisionen,'' PhD thesis, Karlsruhe, Germany, 2000.

\bibitem{Diaz:2001vm}
M.~A.~Diaz and D.~A.~Ross,
\jnl{JHEP} \vol{0106} (2001) 001
[arXiv:hep-ph/0103309].

\bibitem{Oller:2004br}
W.~\"Oller, H.~Eberl and W.~Majerotto,
\jnl{Phys. Lett.} \vol{B 590} (2004) 273
[arXiv:hep-ph/0402134].

\bibitem{Oller:2005xg}
W.~\"Oller, H.~Eberl and W.~Majerotto,
\jnl{Phys. Rev.} \vol{D 71} (2005) 115002
[arXiv:hep-ph/0504109].

\bibitem{FrHo2004}
T.~Fritzsche and W.~Hollik,
\jnl{Nucl. Phys. Proc. Suppl.} \vol{135} (2004) 102
[arXiv:hep-ph/0407095].

\bibitem{Fritzsche:2004ek}
T.~Fritzsche,
arXiv:hep-ph/0408307.

\bibitem{dissTF}
T.~Fritzsche,
PhD thesis, Cuvillier-Verlag, G\"ottingen 2005, ISBN 3-86537-577-4.

\bibitem{Diaz:2009um}
M.~A.~Diaz, M.~A.~Rivera and D.~A.~Ross,
\jnl{JHEP} \vol{1004} (2010) 098
[arXiv:0911.4403 [hep-ph]].

\bibitem{Bharucha:2012nx}
A.~Bharucha, A.~Fowler, G.~Moortgat-Pick and G.~Weiglein,
\jnl{JHEP} \vol{1305} (2013) 053
[arXiv:1211.3134 [hep-ph]].

\bibitem{Bharucha:2012ya}
A.~Bharucha, J.~Kalinowski, G.~Moortgat-Pick, K.~Rolbiecki and G.~Weiglein,
\jnl{Eur. Phys. J.} \vol{C 73} (2013) 6,  2446
[arXiv:1211.3745 [hep-ph]].

\bibitem{feynarts}
J.~K\"ublbeck, M.~B\"ohm and A.~Denner, 
\jnl{Comput. Phys. Commun.} \vol{60} (1990) 165;\\
T.~Hahn, 
\jnl{Comput. Phys. Commun.} \vol{140} (2001) 418
[arXiv:hep-ph/0012260];\\
T.~Hahn and C.~Schappacher, 
\jnl{Comput. Phys. Commun.} \vol{143} (2002) 54
[arXiv:hep-ph/0105349].\\
Program, user's guide and model files
are available via: \url{http://www.feynarts.de}\,.

\bibitem{MSSMCT} 
T.~Fritzsche, T.~Hahn, S.~Heinemeyer, F.~von~der~Pahlen, H.~Rzehak 
and C.~Schappacher,   
\jnl{Comput. Phys. Commun.} \vol{185} (2014) 1529
[arXiv:1309.1692 [hep-ph]].

\bibitem{formcalc}
T.~Hahn and M.~P\'erez-Victoria,
\jnl{Comput. Phys. Commun.} \vol{118} (1999) 153
[arXiv:hep-ph/9807565].

\bibitem{HiggsDecaySferm} 
S.~Heinemeyer and C.~Schappacher,
\jnl{Eur. Phys. J.} \vol{C 75} (2015) 5, 198
[arXiv:1410.2787 [hep-ph]].

\bibitem{SbotRen} 
S.~Heinemeyer, H.~Rzehak and C.~Schappacher,
\jnl{Phys. Rev.} \vol{D 82} (2010) 075010
[arXiv:1007.0689 [hep-ph]];\\
S.~Heinemeyer, H.~Rzehak and C.~Schappacher,
\jnl{PoSCHARGED} \vol{2010} (2010) 039
[arXiv:1012.4572 [hep-ph]].

\bibitem{Stop2decay}
T.~Fritzsche, S.~Heinemeyer, H.~Rzehak and C.~Schappacher, 
\jnl{Phys. Rev.} \vol{D 86} (2012) 035014
[arXiv:1111.7289 [hep-ph]].

\bibitem{Gluinodecay} 
S.~Heinemeyer and C.~Schappacher,
\jnl{Eur. Phys. J.} \vol{C 72} (2012) 1905
[arXiv:1112.2830 [hep-ph]].

\bibitem{Stau2decay} 
S.~Heinemeyer and C.~Schappacher,
\jnl{Eur. Phys. J.} \vol{C 72} (2012) 2136
[arXiv:1204.4001 [hep-ph]].

\bibitem{HiggsProd}
S.~Heinemeyer and C.~Schappacher,
\jnl{Eur. Phys. J.} \vol{C 76} (2016) 4, 220
[arXiv:1511.06002 [hep-ph]].

\bibitem{HpProd}
S.~Heinemeyer and C.~Schappacher,
\jnl{Eur. Phys. J.} \vol{C 76} (2016) 10, 535
[arXiv:1606.06981 [hep-ph]].

\bibitem{mhcMSSMlong}
M.~Frank, T.~Hahn, S.~Heinemeyer, W.~Hollik, H.~Rzehak and G.~Weiglein,
\jnl{JHEP} \vol{0702} (2007) 047
[arXiv:hep-ph/0611326].

\bibitem{dissAF}
A.~Fowler, PhD thesis:
``Higher-order and CP-violating effects in the neutralino and 
Higgs-boson sectors of the MSSM'',
Durham University, UK, September 2010;\\
A.~Fowler and G.~Weiglein,
\jnl{JHEP} \vol{1001} (2010) 108
[arXiv:0909.5165 [hep-ph]].

\bibitem{onshellCNmasses}
A.~Chatterjee, M.~Drees, S.~Kulkarni and Q.~Xu,
\jnl{Phys. Rev.} \vol{D 85} (2012) 075013
[arXiv:1107.5218 [hep-ph]].

\bibitem{complexmassscheme}
A.~Denner, S.~Dittmaier, M.~Roth and D.~Wackeroth,
\jnl{Nucl. Phys. B} \vol{560} (1999) 33
[arXiv:hep-ph/9904472].

\bibitem{cdr}
F.~del Aguila, A.~Culatti, R.~Mu\~noz Tapia and 
M.~P\'erez-Victoria,
\jnl{Nucl. Phys.} \vol{B 537} (1999) 561
[arXiv:hep-ph/9806451].

\bibitem{dred}
W.~Siegel, 
\jnl{Phys. Lett.} \vol{B 84} (1979) 193; \\
D.~Capper, D.~Jones, and P.~van Nieuwenhuizen,
\jnl{Nucl. Phys.} \vol{B 167} (1980) 479. 

\bibitem{dredDS}
D.~St\"ockinger,
\jnl{JHEP} \vol{0503} (2005) 076
[arXiv:hep-ph/0503129].

\bibitem{dredDS2}
W.~Hollik and D.~St\"ockinger,
\jnl{Phys. Lett.} \vol{B 634} (2006) 63
[arXiv:hep-ph/0509298].

\bibitem{denner}
A.~Denner,
\jnl{Fortsch. Phys.} \vol{41} (1993) 307
[arXiv:0709.1075 [hep-ph]].

\bibitem{slicing}
K.~Fabricius, I.~Schmitt, G.~Kramer and G.~Schierholz, 
\jnl{Zeit. Phys.} \vol{C 11} (1981) 315;\\
G.~Kramer and B.~Lampe, 
\jnl{Fortschr. Phys.} \vol{37} (1989) 161;\\
H.~Baer, J.~Ohnemus and J.~Owens, 
\jnl{Phys. Rev.} \vol{D 40} (1989) 2844;\\
B.~Harris and J.~Owens,
\jnl{Phys. Rev.} \vol{D 65} (2002) 094032
[arXiv:hep-ph/0102128].

\bibitem{cuba}
T.~Hahn, 
\jnl{Comput. Phys. Commun.} \vol{168} (2005) 78
[arXiv:hep-ph/0404043];\\
T.~Hahn,
arXiv:1408.6373 [physics.comp-ph].\\
The program is available via: \url{http://www.feynarts.de/cuba/}\,.

\bibitem{SPS1a}
J.~Aguilar-Saavedra et al.,
\jnl{Eur. Phys. J.} \vol{C 46} (2066) 43
[arXiv:hep-ph/0511344].

\bibitem{eennHWiener}
H.~Eberl, W.~Majerotto and V.~Spanos,
\jnl{Phys. Lett.} \vol{B 538} (2002) 353
[arXiv:hep-ph/0204280];\\
H.~Eberl, W.~Majerotto and V.~Spanos,
\jnl{Nucl. Phys.} \vol{B 657} (2003) 378
[arXiv:hep-ph/0210038].

\bibitem{eennH}
T.~Hahn, S.~Heinemeyer and G.~Weiglein,
\jnl{Nucl. Phys.} \vol{B 652} (2003) 229
[arXiv:hep-ph/0211204];\\
T.~Hahn, S.~Heinemeyer and G.~Weiglein,
\jnl{Nucl. Phys. Proc. Suppl.} \vol{116} (2003) 336
[arXiv:hep-ph/0211384].

\bibitem{aoife}
A.~Bharucha,
\textit{private communication}, 05.03.2015 and 11.04.2017.

\bibitem{pdg}
C.~Patrignani et al.\ (Particle Data Group), 
\jnl{Chin. Phys.} \vol{C 40} (2016) 100001.   

\bibitem{ccb}
J.~Fr\`ere, D.~Jones and S.~Raby,
\jnl{Nucl. Phys.} \vol{B 222} (1983) 11;\\
M.~Claudson, L.~Hall and I.~Hinchliffe,
\jnl{Nucl. Phys.} \vol{B 228} (1983) 501;\\
C.~Kounnas, A.~Lahanas, D.~Nanopoulos and M.~Quiros,
\jnl{Nucl. Phys.} \vol{B 236} (1984) 438;\\
J.~Gunion, H.~Haber and M.~Sher,
\jnl{Nucl. Phys.} \vol{B 306} (1988) 1;\\
J.~Casas, A.~Lleyda and C.~Mu\~noz,
\jnl{Nucl. Phys.} \vol{B 471} (1996) 3
[arXiv:hep-ph/9507294];\\
P.~Langacker and N.~Polonsky,
\jnl{Phys. Rev.} \vol{D 50} (1994) 2199
[arXiv:hep-ph/9403306];\\
A.~Strumia,
\jnl{Nucl. Phys.} \vol{B 482} (1996) 24
[arXiv:hep-ph/9604417].

\bibitem{ATLAS-CN}
D.~Costanzo,
talk given at ``SUSY16'', July 2016, Melbourne, Australia, see:\\
\url{https://indico.cern.ch/event/443176/contributions/2193015/attachments/}\\
\url{1302695/1945477/DC_SUSY2016_ATLAS.pdf}\,.

\bibitem{CMS-CN} 
A.~Askew,
talk given at ``SUSY16'', July 2016, Melbourne, Australia, see:\\
\url{https://indico.cern.ch/event/443176/contributions/2192974/attachments/}\\
\url{1302694/1945476/Askew_SUSY16.pdf}\,.

\bibitem{chargedmhiggs2L}
M.~Frank et al., 
\jnl{Phys.\ Rev.} \vol{D 88} (2013) 5, 055013
[arXiv:1306.1156 [hep-ph]].

\bibitem{MSSMcomplphasen}
S.~Dimopoulos and S.~Thomas,
\jnl{Nucl. Phys.} \vol{B 465} (1996) 23
[arXiv:hep-ph/9510220].

\bibitem{SUSYphases}
M.~Dugan, B.~Grinstein and L.~Hall,
\jnl{Nucl. Phys.} \vol{B 255} (1985) 413.

\bibitem{EDMrev2}
D.~Demir, O.~Lebedev, K.~Olive, M.~Pospelov and A.~Ritz,
\jnl{Nucl. Phys.} \vol{B 680} (2004) 339
[arXiv:hep-ph/0311314].

\bibitem{EDMPilaftsis}
D.~Chang, W.~Keung and A.~Pilaftsis,
\jnl{Phys. Rev. Lett.} \vol{82} (1999) 900
[Erratum-ibid.\ \vol{83} (1999) 3972]
[arXiv:hep-ph/9811202];\\
A.~Pilaftsis,
\jnl{Phys. Lett.} \vol{B 471} (1999) 174
[arXiv:hep-ph/9909485].

\bibitem{EDMRitz}
O.~Lebedev, K.~Olive, M.~Pospelov and A.~Ritz,
\jnl{Phys. Rev.} \vol{D 70} (2004) 016003
[arXiv:hep-ph/0402023].

\bibitem{EDMDoink}
W.~Hollik, J.~Illana, S.~Rigolin and D.~St\"ockinger,
\jnl{Phys. Lett.} \vol{B 416} (1998) 345
[arXiv:hep-ph/9707437];\\
W.~Hollik, J.~Illana, S.~Rigolin and D.~St\"ockinger,
\jnl{Phys. Lett.} \vol{B 425} (1998) 322
[arXiv:hep-ph/9711322].

\bibitem{EDMheavy}
P.~Nath,
\jnl{Phys. Rev. Lett.} \vol{66} (1991) 2565;\\
Y.~Kizukuri and N.~Oshimo,
\jnl{Phys. Rev.} \vol{D 46} (1992) 3025.

\bibitem{EDMmiracle}
T.~Ibrahim and P.~Nath,
\jnl{Phys. Lett.} \vol{B 418} (1998) 98
[arXiv:hep-ph/9707409];\\
T.~Ibrahim and P.~Nath,
\jnl{Phys. Rev.} \vol{D 57} (1998) 478 
[Erratum-ibid.\ \vol{D 58} (1998) 019901] 
[Erratum-ibid.\ \vol{D 60} (1998) 079903] 
[Erratum-ibid.\ \vol{D 60} (1999) 119901]
[arXiv:hep-ph/9708456];\\
M.~Brhlik, G.~Good and G.~Kane,
\jnl{Phys. Rev.} \vol{D 59} (1999) 115004
[arXiv:hep-ph/9810457].

\bibitem{EDMrev1}
S.~Abel, S.~Khalil and O.~Lebedev,
\jnl{Nucl. Phys.} \vol{B 606} (2001) 151
[arXiv:hep-ph/0103320].

\bibitem{EDMrev3}
Y.~Li, S.~Profumo and M.~Ramsey-Musolf,
\jnl{JHEP} \vol{1008} (2010) 062
[arXiv:1006.1440 [hep-ph]].

\bibitem{Ya2013}
N.~Yamanaka,
\jnl{Phys. Rev.} \vol{D 87} (2013) 011701
[arXiv:1211.1808 [hep-ph]].

\bibitem{plehnix}
V.~Barger, T.~Falk, T.~Han, J.~Jiang, T.~Li and T.~Plehn,
\jnl{Phys. Rev.} \vol{D 64} (2001) 056007
[arXiv:hep-ph/0101106].

\bibitem{pol-report}
G.~Moortgat-Pick et al.,
\jnl{Phys. Rept.} \vol{460} (2008) 131
[arXiv:hep-ph/0507011].

\endgroup

\end{thebibliography}
\end{document}